\documentclass[structabstract]{aa}  
\usepackage{graphicx} 
\usepackage{txfonts}
\usepackage{natbib}
\usepackage{breqn}
\usepackage{lscape}
\usepackage{hyperref}
\usepackage{siunitx}  
\newcommand{\HII}{\mbox{H\,{\sc ii}}}
\newcommand{\HeI}[1]{\mbox{He\,{\sc i}~$\lambda${#1}}}
\newcommand{\HeII}[1]{\mbox{He\,{\sc ii}~$\lambda${#1}}}

\newcommand{\CIII}[1]{\mbox{C\,{\sc iii}~$\lambda${#1}}}

\newcommand{\NVd}[1]{\mbox{N\,{\sc v}~$\lambda\lambda${#1}}}

\newcommand{\GG}{\mbox{$G$}}
\newcommand{\GBP}{\mbox{$G_{\rm BP}$}}
\newcommand{\GRP}{\mbox{$G_{\rm RP}$}}
\newcommand{\GGc}{\mbox{$G^\prime$}}

\newcommand{\mci}[1]{\multicolumn{1}{c}{#1}}
\newcommand{\mcii}[1]{\multicolumn{2}{c}{#1}}
\newcommand{\dCC}{\mbox{$d_{\rm CC}$}}
\newcommand{\pmra}{\mbox{$\mu_{\alpha *}$}}
\newcommand{\pmdec}{\mbox{$\mu_{\delta}$}}
\newcommand{\pmrag}{\mbox{$\mu_{\alpha *,{\rm g}}$ }}
\newcommand{\pmdecg}{\mbox{$\mu_{\delta,{\rm g}}$}}
\newcommand{\pigz}{\mbox{$\varpi_{\rm g,0}$}}

\newcommand{\spic}{\mbox{$\sigma_{\varpi_{\rm c}}$}}
\newcommand{\pig}{\mbox{$\varpi_{\rm g}$}}
\newcommand{\spig}{\mbox{$\sigma_{\varpi_{\rm g}}$}}
\newcommand{\pigc}{\mbox{$\varpi_{\rm g,c}$}}
\newcommand{\spigc}{\mbox{$\sigma_{\varpi_{\rm g,c}}$}}
\newcommand{\Nf}{\mbox{$N_{\rm f}$}}

\newcommand{\EBV}{\mbox{$E(4405-5495)$}}
\newcommand{\RV}{\mbox{$R_{5495}$}}
\newcommand{\AV}{\mbox{$A_V$}}
\newcommand{\VO}[1]{Villafranca~O-{#1}}
\newcommand{\dr}{\mbox{$d_{\rm r}$}}
\newcommand{\dn}{\mbox{$d_{\rm n}$}}
\newcommand{\st}{\mbox{$\sigma_{\rm t}$}}

\begin{document}

   \title{The Villafranca catalog of Galactic OB groups: \linebreak
          I. Systems with O2-O3.5 stars}
   \titlerunning{Villafranca OB groups: I. Systems with O2-O3.5 stars.}

   \author{J. Ma\'{\i}z Apell\'aniz \inst{1}
           \and
           P. Crespo Bellido\inst{1,2}
           \and
           R. H. Barb\'a\inst{3}
           \and
           R. Fern\'andez Aranda\inst{1,2}
           \and
           A. Sota\inst{4}
           }
   \authorrunning{J. Ma\'{\i}z Apell\'aniz et al.}

   \institute{Centro de Astrobiolog\'{\i}a, CSIC-INTA. Campus ESAC. 
              C. bajo del castillo s/n. 
              E-\num{28692} Villanueva de la Ca\~nada, Madrid, Spain\linebreak
              \email{jmaiz@cab.inta-csic.es} \\
              \and
              Departamento de Astrof{\'\i}sica y F{\'\i}sica de la Atm\'osfera, Universidad Complutense de Madrid. 
              E-\num{28040} Madrid, Spain. \\
              \and
              Departamento de F{\'\i}sica y Astronom{\'\i}a, Universidad de La Serena.
              Av. Cisternas 1200 Norte.
              La Serena, Chile \\
              \and
              Instituto de Astrof\'{\i}sica de Andaluc\'{\i}a-CSIC. 
              Glorieta de la Astronom\'{\i}a s/n. 
              E-\num{18008} Granada, Spain.
             }

   \date{Received XX XXX 2020 / Accepted XX XXX 2020}

% \abstract{}{}{}{}{} 
% 5 {} token are mandatory
 
  \abstract
   {The spectral classifications of the Galactic O-Star Spectroscopic Survey (GOSSS) and the astrometric and photometric data from \textit{Gaia} 
    have significantly improved our ability to measure distances and determine memberships of stellar groups (clusters, associations, or parts 
    thereof) with OB stars. In the near future the situation will further improve with more \textit{Gaia} data releases and new photometric and 
    spectroscopic surveys.}
   {We have started a program to identify, measure distances, and determine the membership of Galactic stellar groups with OB stars. Given the data
    currently available, we start with the identification and distance determinations of groups with O stars. In this paper we concentrate on 
    groups
    that contain stars with the earliest spectral subtypes.}
   {We use GOSSS to select Galactic stellar groups with O2-O3.5 stars and the method described 
%   by Ma\'{\i}z~Apell{\'a}niz~(2019) 
    in paper 0 of this series % REFEREE
    that combines \textit{Gaia} DR2 G + \GBP\ + \GRP\ photometry, positions, proper motions, and parallaxes to assign robust
    memberships and measure distances.
    We also include Collinder~419 and NGC~2264, the clusters in that paper, to 
    generate our first list of 16 O-type Galactic stellar groups.}
   {We derive distances, determine the membership, and analyze the structure of sixteen Galactic stellar groups with O stars, \VO{001} to \VO{016},
    including the fourteen groups with the earliest-O-type optically-accessible stars known in the Milky Way.
    We compare our distance with previous literature results and establish that the best consistency is with (the small number of) VLBI parallaxes
    and the worst is with kinematic distances. Our results indicate that 
    very
    massive stars can form in relatively low-mass clusters or even in near-isolation,
    as is the case for the Bajamar star in the North America nebula. 
    This lends support to the hierarchical scenario of star formation, where some stars are born in well-defined bound clusters but others are born in associations that 
    are unbound from the beginning: groups of newborn stars come in many shapes and sizes.
    We propose that HD~\num{64568} and HD~\num{64315}~AB could have been ejected simultaneously from 
    Haffner~18 (\VO{012}~S). % REFEREE
    Our results are consistent with a difference of $\approx$20~$\mu$as in the {\it Gaia}~DR2 parallax zero point between bright and faint stars.
    }
   {}
   \maketitle
%
%________________________________________________________________

\section{Introduction}

\subsection{Clusters, associations, and groups}

$\,\!$\indent Traditionally, stellar clusters are defined as stellar groups with at least several tens of members and a stellar density large enough (with a threshold
around 1~M$_\odot$\,pc$^{-3}$) to keep the system in a bound state i.e. to have a total (kinetic + potential) negative energy. According to the often-cited work 
of \citet{LadaLada03}, most or all stars are formed in bound clusters but as they evolve the vast majority of them become unbound as the total kinetic energy
of the stars does not change much but their potential energy becomes less negative (closer to zero) when the associated molecular gas is dispersed. In the 
\citet{LadaLada03} view, most clusters evolve into low-density OB associations \citep{Amba58} with sizes of tens of parsecs that eventually dissolve into the Galactic 
disc. In an alternative view  
to the ``most stars form in clusters'' scenario,
star formation is a hierarchical process that can take place in both bound and unbound clouds with a wide range of scales \citep{Elme10,Wardetal20}. 
In that scenario, OB associations may be born that way (the effect of nature) or they may be the consequence or gas dispersal (the effect of nurture). 
Note that current models have not yet provided the final answer to how clusters form \citep{KrumMcKe20}.

In order to decide which of the two views is correct one needs to precisely measure the total energy of stellar groups (which can be either bound stellar clusters or 
unbound stellar associations) but, as it often happens in astronomy, the data are not always clear for several reasons:

\begin{itemize}
 \item Velocities in the plane of the sky require accurate proper motions, which until the advent of {\it Gaia} were difficult to precisely measure unless the stellar group
       was close. Here the limitation is the need to observe the system with a large enough time baseline.
 \item Velocities in the radial direction require (usually multi-object) spectroscopy but OB stars have a large binary and higher-order multiplicity fraction
       \citep{Maizetal19b} that can artificially inflate the velocity dispersion \citep{HenBetal12a}. This requires the spectroscopy to be not only multi-object but also
       multi-epoch.
 \item What counts as a group member and what does not? In sightlines with long paths across the Galactic Plane this is far 
       from obvious and even with good data one cannot be completely sure for some stars.
 \item Even if one solves the previous issues, stellar groups cannot be always easily divided into clusters and associations. Some stellar groups show two compact cores
       instead of one (double clusters, \citealt{dLFMdLFM09}), some associations have bound cores at their centers (e.g. \citealt{Maiz01b}), and the limits between 
       two neighbor associations may not be well defined.
\end{itemize}

Given those limitations and considering that most stellar groups that contain OB stars (i.e. the young and massive end of the distribution) are located several kpc away and are
obscured by the dust close to the Galactic Plane, a priori it is not possible to produce a clear cut between clusters, associations, or even subassociations. In other words,
lacking excellent multi-epoch multi-type 
data, the division into strict categories is not possible at this time for most of a reasonably large sample. For that reason, in this
work we will refer to stellar groups in general acknowledging that to some point the limits between one group and the next one will be arbitrary but, at the same time, we will
present our reasons why we think such a division is a reasonable one for that particular case. The reasons may be different for one stellar group than for the next one but such
a case-by-case analysis is the only alternative we have to do a thorough job. For stellar groups, diversity reigns and no single size fits all.

\subsection{The Villafranca catalog of Galactic OB groups}

$\,\!$\indent For over a decade we have been conducting the Galactic O-Star Spectroscopic Survey (GOSSS, \citealt{Maizetal11}), which has collected 
high-quality $R\sim2500$ blue-violet spectroscopy of several thousand stars of spectral types O and B. The availability of the data has
allowed us to produce uniform spectral classifications for a large sample of objects and correct many misclassifications in the literature. The survey
has produced three major papers \citep{Sotaetal11a,Sotaetal14,Maizetal16} and several others to date, of which the most recent are 
\citet{Maizetal18a,Maizetal18b,Maizetal19b}. 
In the process, the GOSSS papers have become some of the most cited articles on O stars of the last decade.
The data and spectral classifications from GOSSS are available from the Galactic O-Star Catalog 
(GOSC, \url{https://gosc.cab.inta-csic.es}, \citealt{Maizetal04b}).
There are several 
existing multi-fiber
large-scale Galactic spectroscopic surveys such as Gaia-ESO \citep{Gilmetal12} or LAMOST \citep{Luoetal15} but they do not 
have spectra for many O stars published for different reasons
(Blomme et al. in preparation will soon become an exception).
The most significant one is that they have complex scheduling strategies designed to satisfy many simultaneous 
goals, with O stars not being among their highest priorities. 
Therefore, even when a project like Gaia-ESO observed clusters, most of them were of low mass or too old to contain O stars and in those that were massive and young enough
to do so only in one case (Trumpler~14) were fibers allocated to O stars \citep{Jacketal20}. % REFEREE
GOSSS, on the other hand, spends over 50\% of its time on O stars.
\hyphenation{results}
Several recent single-fiber \'echelle surveys have laso spent a significant fraction of their time observing O stars
(e.g. \citealt{SimDetal11c,Maizetal12,Barbetal17,Neguetal15a}) and they have led to important 
results on the physical properties, multiplicity, and intervening ISM for such systems.

The {\it Gaia} mission \citep{Prusetal16} produced its second data release (DR2) in April 2018 \citep{Browetal18}.  {\it Gaia} DR2 includes five-parameter
astrometry (positions, parallaxes, and proper motions) and optical photometry in the three bands \GG, \GBP, and \GRP\ for over $1.3 \cdot 10^9$~sources. 
Those data constitute the largest ever collection of high-precision photometry and astrometry, allowing for a huge improvement in the identification and 
distance measurement of stellar 
groups. 
Soon after DR2 a large study on its application to stellar clusters was published \citep{CanGetal18} and since then
several other papers have followed suit (e.g. \citealt{Soubetal18,CasGetal19,CasGetal20,CanGetal19a}). With future data releases {\it Gaia} will undoubtedly keep making a huge impact
on the field. 

Why are OB stars (and their descendants) important for the study of stellar groups that contain or contained them? There are several reasons: they irradiate with UV photons 
the nearby ISM, creating a feedback (which can be positive or negative) for star formation within the group; they inject energy into their surroundings in the form of stellar winds 
and supernova explosions that can make the primordial gas disperse faster; they are the fastest polluters of the nearby ISM with the product of nuclear reactions in their interiors, 
thus creating the possibility of differences in chemical composition within the stellar group; and they can interact with other stars and among themselves to dynamically perturb 
trajectories through close encounters. In summary, no other stellar type has such a large influence in the evolution of a stellar group and the presence of a single OB star may alter
its whole evolution. Therefore, to understand stellar groups (including those that had OB stars in the past but have already lost them) in general, the study of how
OB stars associate with lower-mass objects is crucial. Is the massive-star initial mass function (IMF) constant or does it depend on the environment? Is it always populated according
to the same mechanisms? Can massive stars form in isolation or near-isolation? We specify that in this last question by ``isolation'' we mean that a massive star may be formed just 
by itself (possibly with one or several bound companions in a multiple system) but without belonging to a cluster or a small-scale ($\sim$3~pc) association (for the time-scales 
involved in star formation, material located at distances longer than $\sim$3~pc cannot have a significant influence in the process, see \citealt{Bresetal12}). By ``near-isolation'' 
we mean that (usually with some effort) a group can be located around the massive star but the combined mass of the rest of the stars is so low as to have an atypical IMF with a 
single object having a significant fraction of the total mass. % REFEREE

With this paper we start a project that combines GOSSS and {\it Gaia} data to catalog the Galactic stellar groups that contain OB 
stars. The availability of these new high-quality data makes this time an opportune moment to revisit this issue in a more complete manner than previous studies.
In the future we will add information from the subsequent {\it Gaia} data releases and from other surveys, either photometric 
(e.g. GALANTE, \citealt{LorGetal19}) or spectroscopic (e.g. WEAVE, \citealt{Dalt16}). We name the catalog resulting from the project ``Villafranca'' 
after the location of the European Space Astronomy Centre (ESAC), formerly the Villafranca del Castillo Satellite Tracking Station (Vilspa), where the 
institution of three of the authors is located. In this way we follow the tradition of the stellar cluster community of naming catalogs after 
geographical locations (Berkeley, Alicante, Bochum, Escorial\ldots) and we honor two relationships between Villafranca and the study of OB stellar groups:
on the one hand, the International Ultraviolet Explorer (IUE) was part-time controlled from Vilspa between 1978 and 1996 and played an important historical 
role for these objects, which are among the brightest Galactic targets in the UV; on the other hand, the {\it Gaia} Science Data Center and Archive are located at 
ESAC and without {\it Gaia} this work could not have been possible.
The zeroth paper of the series was \citet{Maiz19}, from now on MA19, where we describe the method 
used to establish the membership and determine the distance to stellar groups (see next section). 
In this first paper we search for stellar groups around the optically-accessible earliest Galactic O stars but we also add the two clusters in MA19 for completeness. % REFEREE
The future papers in the series will add the rest of stellar groups with O stars but, given their large number, this 
should be considered a long-term project that will take a decade or so to complete, especially when one takes into account that new data will likely lead to new 
distance measurements and membership analysis of the already studied sample. Such is the curse of the devoted cataloger. If funding and human resources are available, 
we may also consider adding groups with B stars (but no O types) to the mix, but in that case the sample would have to be significantly incomplete, given the much larger
number of such older and/or lower mass objects.

The structure of this paper is as follows. In section 2 of this paper we present the methods and data: how we have selected the sample, determined the membership and
distances, processed complementary information for some objects, and searched the literature for previous distance measurements. In section 3 we analyze each of the
objects in our sample, combining new and literature information. In the last section we compare our distances with previous ones, discuss the validity of different
criteria used to evaluate the membership for OB stellar groups, analyze the internal motions and the nature of the IMF for our objects, study the dependence of
the {\it Gaia}~DR2 parallax zero point on magnitude, and present our future plans.

\begin{table*}[ht!]
 \caption{Sample of Galactic O-type stellar groups in this paper. The other IDs may refer to the cluster or the associated H\,{\sc ii} region.}
\centerline{
\begin{tabular}{lllll}
ID                & Other ID(s)           & O- and WR-type stars             & Spectral type                & Ref.                                  \\
\hline
O-001             & NGC 3603              & NGC 3603 HST-A1                  & WN6ha                        & C98                                   \\
                  & RCW 57                & NGC 3603 HST-B                   & WN6ha                        & C98                                   \\
                  &                       & NGC 3603 HST-C                   & WN6ha                        & C98                                   \\
                  &                       & NGC 3603 MTT 58                  & O2 If*/WN6 + O3 If           & R13b                                  \\
                  &                       & NGC 3603 MTT 31                  & O2 V                         & R16                                   \\
                  &                       & NGC 3603 HST-48                  & O3.5 If*                     & M16                                   \\
                  &                       & many others                      &                              & D95,W02,M08,R12,R13a,R16 \vspace{1mm} \\
O-002             & Trumpler 14           & HD \num{93129} Aa,Ab             & O2 If* + O2 If* + OB?        & M17                                   \\
                  &                       & HD \num{93129} B                 & O3.5 V((f))z                 & M16                                   \\
                  &                       & HD \num{93128}                   & O3.5 V((fc))z                & S14                                   \\
                  &                       & many others                      &                              & S14,M16                  \vspace{1mm} \\
O-003             & Trumpler 16~W         & HD \num{93162}                   & O2.5 If*/WN6 + OB            & S14                                   \\
                  &                       & ALS \num{15210}                  & O3.5 If* Nwk                 & S14                                   \\
                  &                       & HD \num{93205}                   & O3.5 V((f)) + O8 V           & S14,TW                                \\
                  &                       & HD \num{93204}                   & O5.5 V((f))                  & S14                      \vspace{1mm} \\
O-004             & Westerlund 2          & V712 Car                         & O3 If*/WN6                   & TW                                    \\
                  & RCW 49                & WR 20b                           & WN6ha                        & R11                                   \\
                  &                       & Westerlund 2-199                 & O3 V                         & N08                                   \\
                  &                       & many others                      &                              & N08,R11,V13              \vspace{1mm} \\
O-005             & Pismis 24             & Pismis 24-1 A,B                  & O3.5 If*                     & S14                                   \\
                  & NGC 6357 W            & Pismis 24-17                     & O3.5 III(f*)                 & S14                                   \\
                  & RCW 131 W             & WR 93                            & WC7 + O7/9                   & V01                                   \\
                  &                       & several others                   &                              & S14,M16                  \vspace{1mm} \\
O-006             & Gum 35                & THA 35-II-153                    & O3.5 If*/WN7                 & M16                                   \\
                  & Majaess 133           & ALS 2063                         & O5 Ifp                       & S14                                   \\
                  &                       & ALS \num{18551}                  & O4.5 V(n)z + O4.5 V(n)z      & M16                                   \\
                  &                       & several others                   &                              & S14,M16                  \vspace{1mm} \\
O-007             & Cyg OB2-22 cluster    & Cyg OB2-22 A                     & O3 If*                       & S11                                   \\
                  & Bica 1                & Cyg OB2-9                        & O4 If + O5.5 III(f)          & M19                                   \\
                  &                       & Cyg OB2-22 B                     & O6 V((f))                    & S14                                   \\
                  &                       & several others                   &                              & S11,S14                  \vspace{1mm} \\
O-008             & Cyg OB2-8 cluster     & Cyg OB2-7                        & O3 If*                       & S11                                   \\
                  & Bica 2                & Cyg OB2-8 C                      & O4.5 (fc)p var               & S14                                   \\
                  &                       & Cyg OB2-8 A                      & O6 Ib(fc) + O4.5 III(fc)     & M19                                   \\
                  &                       & several others                   &                              & S11,S14,M91              \vspace{1mm} \\
O-009             & M17                   & ALS \num{19613} A                & O2/4 V                       & TW                                    \\
                  & Omega nebula          & ALS \num{19618} A                & O4 V(n)((fc))                & M16                                   \\
                  & NGC 6618              & ALS \num{19617}                  & O5 V                         & H08                                   \\
                  & Sh 2-45               & several others                   &                              & C78,H08                               \\
                  & RCW 160               &                                  &                              &                          \vspace{1mm} \\
O-010             & NGC 6193              & HD \num{150136} Aa,Ab            & O3.5-4 III(f*) + O6 IV       & S14                                   \\
                  & RCW 108               & HD \num{150135} Aa,Ab            & O6.5 V((f))z                 & S14                      \vspace{1mm} \\
O-011             & Berkeley 90           & LS III +46 11                    & O3.5 If* + O3.5 If*          & M16                                   \\
                  & Sh 2-115              & LS III +46 12                    & O4.5 IV(f)                   & M16                      \vspace{1mm} \\
O-012             & NGC 2467              & CPD $-$26 2704                   & O7 V(n)                      & M16                                   \\
                  & Sh 2-311              & $\dagger$ HD \num{64568}         & O3 V((f*))z                  & S14                                   \\
                  & RCW 16                & $\dagger$ HD \num{64315} A,B     & O5.5 V + O7 V                & M16                                   \\
O-012~N           & Haffner 19            & ---                              &                              &                                       \\
O-012~S           & Haffner 18            & CPD $-$26 2704                   & O7 V(n)                      & M16                      \vspace{1mm} \\
O-013             & Sh 2-158              & Sh 2-158 1                       & O3.5 V((f*)) + O9.5: V       & M16                                   \\
                  & NGC 7358              & Sh 2-158 2                       & O9.5: V + B0.5: V            & M16                      \vspace{1mm} \\
O-014             & North America nebula  & Bajamar star                     & O3.5 III(f*) + O8:           & M16                                   \\
                  & NGC 7000              & $\dagger\dagger$ HD \num{199579} & O6.5 V((f))z                 & S11                                   \\
                  & Sh 2-117              &                                  &                              &                          \vspace{1mm} \\
O-015             & Collinder 419         & HD \num{193322} Aa,Ab            & O9~Vnn + O8.5~III + B2.5: V: & T11                      \vspace{1mm} \\
O-016             & NGC 2264              & 15 Mon Aa,Ab                     & O7~V((f))z + B1: Vn          & M20                                   \\
                  & Sh 2-273              &                                  &                              &                                       \\
O-016~N           & NGC 2264 N            & 15 Mon Aa,Ab                     & O7~V((f))z + B1: Vn          & M20                                   \\
O-016~S           & NGC 2264 S            & ---                              &                              &                                       \\
\hline
\multicolumn{5}{l}{$\dagger$: Outside region, possible runaway. $\;\;\;$ $\dagger\dagger$: In nebula but chance alignment.}                       \\
\multicolumn{5}{l}{Ref.:           C78:  \citet{Crametal78},  C98:  \citet{CrowDess98},  D95:  \citet{Drisetal95},  H08:  \citet{Hoffetal08},  }  \\
\multicolumn{5}{l}{\phantom{Ref.:} M91:  \citet{MassThom91},  M08:  \citet{Meleetal08},  M16:  \citet{Maizetal16},  M17:  \citet{Maizetal17a}, }  \\
\multicolumn{5}{l}{\phantom{Ref.:} M19:  \citet{Maizetal19b}, M20:  \citet{MaizBarb20},  N08:  \citet{Nazeetal08a}, R11:  \citet{Rauwetal11},  }  \\
\multicolumn{5}{l}{\phantom{Ref.:} R12:  \citet{RomL12},      R13a: \citet{RomL13a},     R13b: \citet{RomL13b},     R16:  \citet{RomLetal16},  }  \\
\multicolumn{5}{l}{\phantom{Ref.:} S11:  \citet{Sotaetal11a}, S14:  \citet{Sotaetal14},  T11:  \citet{tenBetal11},  TW:   This work,           }  \\
\multicolumn{5}{l}{\phantom{Ref.:} V01:  \citet{vadH01},      V13:  \citet{VarAetal13},  W02:  \citet{Walbetal02b}.}
\end{tabular}
}
\label{sample}                   
\end{table*}

\section{Methods and data}

\subsection{Sample selection}

$\,\!$\indent The latest version of GOSC includes 594 O~stars with GOSSS spectral types which, depending on how one defines them, can be 
collected into 200-300 stellar groups. 
An exact number cannot be given for the reasons given in the introduction: does a double cluster count as a group or as two? Is a cluster within an association
a separate group? Do we divide a 
large association such as Scorpius-Centaurus into its subassociations (Upper Centaurus Lupus, Lower Centaurus Crux, and 
Upper Scorpius) or not?  If all stellar groups were well defined clusters a precise number could be given but, alas, they are not.

For this first installment of our Villafranca catalog we start by finding all the stars in GOSC with the earliest spectral types, from O2 to O3.5, and we select the groups, 
previously well defined or not, to which they belong. Given the diversity among stellar groups, one has to go case by case, to see whether all such stars are in 
clusters or not. As we will see in the last section, the answer to that question is relevant to competing theories about the initial mass function (IMF).
Such a selection based on spectral type favors the youngest and most massive groups, as in older ones the earliest-type stars would have already evolved and 
low-mass groups are less likely to form stars with a mass large enough to be of those spectral types at birth (but see below).
We note that the most massive stars
of all are not born as O-type but rather as WNh~stars \citep{Crowetal10}, a point that will be discussed for some individual groups below. 

The selection criterion above yields 14 stellar groups with at least one O-type star, which we name \VO{001} to \VO{014} sorting them (roughly) 
by richness, from well-populated, concentrated, and well-defined clusters to poorly-defined clusters or stellar groups with just one or two O type-systems. 
As we have already analyzed two clusters with this technique in MA19, we include in our list \VO{015} and \VO{016}, Collinder~419 and NGC~2264, respectively. 
They will be considered separately regarding their lack of early-type O stars but as an integral part of the Villafranca catalog, as indeed most groups with O stars
do not have such subtypes and will constitute the rest of the sample in subsequent articles of this series.
The sample is shown in Table~\ref{sample}. 
To alleviate the degree of arbitrariness implicit in the definition of stellar groups, we define the groups in the catalog in a ``cluster scale'' (sizes of a 
few~pc), independently of their true nature as clusters, rather than in an ``association scale'' (sizes of 10~pc or more). In other words, we divide associations into 
subassociations and clusters whenever possible. However, in each case we indicate the membership of the group to an association, where relevant.
More specifically, each of the two OB associations richer in O stars in the solar
neighborhood (the Carina nebula and Cygnus~OB2 associations) has two stellar groups with O2-O3.5 stars and to leave this clear they are joined in the same
subsections in section~3.

\subsection{Distances to and membership of stellar groups}

$\,\!$\indent The supervised method used to determine the distances to and membership of stellar groups is the one described in MA19, which we summarize 
here. We start selecting a star or stars representative of the 
groups
from GOSSS and calculate an extinguished isochrone for a \GBP$-$\GRP\ vs. \GGc\ 
CMD (see \citealt{MaizWeil18} for a definition of \GGc), using the family of extinction laws of \citet{Maizetal14a} and the extinction parameters from 
\citet{MaizBarb18}. We download the {\it Gaia}~DR2 data around the reference star from the archive and filter them first by RUWE (Renormalized Unit Weight Error), 
\dCC\ (see \citealt{MaizWeil18}), and \spic\ (the external or corrected parallax uncertainty). We then filter the remaining stars by (a) distance $r$ in the plane of
the sky to the group center $\alpha$+ $\delta$ in equatorial J2000 coordinates, (b) distance $r_\mu$ to a central proper motion \pmra\ + \pmdec, and 
(c) position in the \GBP$-$\GRP\ vs. \GGc\ CMD using as reference the previously calculated extinguished isochrone. This results in a preliminary sample with 
$N_{*,0}$ objects which is further culled eliminating outliers in the normalized parallax space, leaving $N_*$ final objects. The whole process is iterated to 
assess the effect of changes in the selection parameters and establish the robustness of the group average parallax \pig. The final selection parameters are 
chosen to maximize both the cleanliness and number of the sample.

Once we have the final sample we calculate the group average parallax uncertainty \spig\ using Eqn.~5 in \citet{Campetal19}, noting that the covariance
term is usually the dominant one in the error budget. To correct for the parallax zero point and determine the final group average we use:

\begin{equation}
\pigc = \pig + 0.040, \;\; \spigc^2 = \spig^2 + 0.010^2,
\end{equation}

\noindent where the values above are given in milliarcseconds (see below for a discussion on the value of the zero point). The final group 
distance is then calculated using the Bayesian prior described 
by \citet{Maiz01a,Maiz05c} with the updated Galactic (young) disk parameters from \citet{Maizetal08a}. 

The above supervised method is applied to calculate the distances and membership of all groups but one in this paper simply by varying the region of the 
sky and the selection parameters. In one case, 
\VO{014} (a stellar group defined here to be associated with the North America nebula), % REFEREE
we followed a different strategy due to the peculiarities of that stellar group, as 
described below. The field sizes and filters used for each group are given in Table~\ref{filters}, where \Nf\ is the number of {\it Gaia}~DR2 sources in the
field.

\begin{table*}
\caption{Field sizes and filters applied to the O-type stellar groups and subgroups in this paper. RUWE is filtered as $<$1.4 and \spic\ as $<$0.1~mas in all 
         cases. For \VO{014} the coordinates correspond to those of the Bajamar star.}
\centerline{
\begin{tabular}{lrccccrcccc}
\hline
ID      & \mci{\Nf}    & field                        & \dCC   & $\alpha$ & $\delta$ & \mci{$r$}       & \pmra   & \pmdec  & $r_\mu$ & $\Delta(\GBP-\GRP)$ \\
        &              &                              &        & (deg)    & (deg)    & \mci{(\arcsec)} & (mas/a) & (mas/a) & (mas/a) &                     \\
\hline
O-001   &  \num{27873} & $ 20\arcmin\times 20\arcmin$ & $<$0.4 &   168.79 & $-$61.26 &             206 & $-$5.61 & $+$1.97 &    0.44 &            $>-$0.34 \\
O-002   &  \num{14340} & $ 20\arcmin\times 20\arcmin$ & $<$0.4 &   160.95 & $-$59.56 &             166 & $-$6.45 & $+$2.25 &    0.70 &            $>-$0.69 \\
O-003   &  \num{18441} & $ 20\arcmin\times 20\arcmin$ & $<$0.4 &   161.09 & $-$59.73 &             103 & $-$6.93 & $+$2.61 &    0.73 &            $>-$1.30 \\
O-004   &  \num{60144} & $ 40\arcmin\times 40\arcmin$ & $<$0.4 &   155.99 & $-$57.76 &             400 & $-$5.10 & $+$2.82 &    0.70 &            $>-$0.65 \\
O-005   &  \num{10374} & $ 30\arcmin\times 30\arcmin$ & $<$0.4 &   261.18 & $-$34.21 &             605 & $-$1.10 & $-$2.20 &    1.40 &            $>-$1.00 \\
O-006   &  \num{48211} & $ 20\arcmin\times 20\arcmin$ & $<$0.4 &   164.68 & $-$61.18 &             200 & $-$5.50 & $+$2.30 &    0.30 &            $>-$0.90 \\
O-007   &         5241 & $ 20\arcmin\times 20\arcmin$ & $<$0.4 &   308.30 & $+$41.22 &             186 & $-$2.65 & $-$4.49 &    0.72 &            $>-$0.60 \\
O-008   &         6136 & $ 20\arcmin\times 20\arcmin$ & $<$0.4 &   308.32 & $+$41.31 &             126 & $-$2.67 & $-$4.15 &    0.50 &            $>-$0.80 \\
O-009   &         5671 & $ 20\arcmin\times 20\arcmin$ & $<$0.4 &   275.12 & $-$16.18 &             236 & $+$0.10 & $-$1.52 &    1.10 &            $>-$0.70 \\
O-010   &  \num{33424} & $ 20\arcmin\times 20\arcmin$ & $<$0.4 &   250.30 & $-$48.76 &             285 & $+$1.57 & $-$3.92 &    1.04 &            $>-$0.70 \\
O-011   &         6513 & $ 20\arcmin\times 20\arcmin$ & $<$0.4 &   308.83 & $+$46.84 &             341 & $-$2.78 & $-$4.31 &    0.50 &                 --- \\
O-012   &  \num{25632} & $ 30\arcmin\times 30\arcmin$ & $<$0.4 &   118.18 & $-$26.33 &             390 & $-$2.50 & $+$2.60 &    0.40 &            $>-$0.15 \\
O-012~N &  \num{25632} & $ 30\arcmin\times 30\arcmin$ & $<$0.4 &   118.19 & $-$26.28 &             200 & $-$2.50 & $+$2.60 &    0.40 &            $>-$0.15 \\
O-012~S &  \num{25632} & $ 30\arcmin\times 30\arcmin$ & $<$0.4 &   118.18 & $-$26.38 &             200 & $-$2.50 & $+$2.60 &    0.40 &            $>-$0.15 \\
O-013   &         3235 & $ 20\arcmin\times 20\arcmin$ & $<$0.4 &   348.43 & $+$61.50 &             109 & $-$3.70 & $-$2.14 &    0.73 &                 --- \\
O-014   & \num{396261} & $120\arcmin\times120\arcmin$ & $<$0.4 &   313.96 & $+$43.87 &             --- & $+$0.01 & $-$4.52 &    0.90 &            $>+$0.00 \\
O-015   &  \num{19049} & $ 30\arcmin\times 30\arcmin$ & $<$0.2 &   304.60 & $+$40.78 &             800 & $-$2.60 & $-$6.40 &    0.75 &            $>-$0.30 \\
O-016   &  \num{25177} & $ 60\arcmin\times 60\arcmin$ & $<$0.2 &   100.25 & $+$09.75 &            1500 & $-$1.80 & $-$3.70 &    1.50 &            $>-$0.20 \\
O-016~N &  \num{25177} & $ 60\arcmin\times 60\arcmin$ & $<$0.2 &   100.20 & $+$09.88 &             540 & $-$1.80 & $-$3.70 &    1.50 &            $>-$0.20 \\
O-016~S &  \num{25177} & $ 60\arcmin\times 60\arcmin$ & $<$0.2 &   100.28 & $+$09.53 &             540 & $-$1.80 & $-$3.70 &    1.50 &            $>-$0.20 \\
\hline
\end{tabular}
}
\label{filters}                  
\end{table*}

In addition to the strategy described in MA19, we have also searched in the vicinity for possible runaways from each group. We have done that by looking for 
objects with parallaxes compatible with the distance to the group and proper motions that point away from its center. The search has been carried out only for
the brightest targets in each field i.e. the possible massive runaways.

\subsection{Spectral classifications and AstraLux data}

$\,\!$\indent We use GOSSS spectral classifications whenever possible and literature ones otherwise. In some cases we present new GOSSS spectra and spectral 
classifications.  The reader is referred to the GOSSS papers cited in the introduction for details on how they are obtained and processed. We also use 
high-resolution spectra from LiLiMaRlin \citep{Maizetal19a}, a {\it Li}brary of {\it Li}braries of {\it Ma}ssive-Star High-{\it R}eso{\it l}ut{\it i}o{\it n} 
Spectra built by collecting data from four different surveys (CAF\'E-BEANS, \citealt{Neguetal15a}; IACOB, \citealt{SimDetal15b}; NoMaDS, \citealt{Maizetal12}; and 
OWN \citealt{Barbetal10,Barbetal17}) and with additional spectra from other programs led by us and from public archives. In order to do spectral classification 
with such data we degrade its spectral resolution to the GOSSS value of 2500 in order to compare them with the GOSSS standards. For two dim objects we also
use OSIRIS/GTC spectra obtained with $R\sim 2000$, lower than the standard $R\sim 2500$ value. The new spectra in this paper are shown in Fig.~\ref{spectra}.

In one case we also present the binary data
(separation, position angle, and $\Delta m$) for one star based on lucky imaging obtained with AstraLux at Calar Alto. See \citet{Maiz10a,Maiz19,Maizetal19b} for
details on AstraLux data.

\subsection{Literature distances}

$\,\!$\indent We have searched the literature for previous distance measurements to the \VO{001} to \VO{016} stellar groups and show them in 
Table~\ref{literature}. For each measurement we provide the distance in pc, its uncertainty (when available), the target used (group, subgroup, individual star,
or ISM object), the method used, and the reference. For groups with an associated H\,{\sc ii} region (e.g. \VO{009} and M17) we designate both as the same target in
Table~\ref{literature}. For simplicity, we denote methods that use a combination of spectroscopy and photometry as 
``spectro-photometry''. Many (but not all) of our targets have {\it Gaia}~DR2 parallax-based distances from \citet{CanGetal18} with
very small uncertainties. As already pointed out in MA19, those uncertainties do not include the effect of the spatial covariance of the {\it Gaia}~DR2
parallaxes \citep{Lindetal18a}, which is the dominant error source.

\section{Results}

$\,\!$\indent The membership and distance results which constitute the main output of this paper are given in Table~\ref{results}, see \citet{Maiz19} for the definition
of $t_\varpi$, $t_{\mu_{\alpha *}}$, $t_{\mu_{\delta}}$, \pmrag, and \pmdecg. The plots used to select the parameters of each group are shown in 
Figs.~\ref{NGC_3603_Gaia}~to~\ref{North_America_Gaia}. Possible runaways from each group are given in Table~\ref{runaways}.
Each stellar group is analyzed in detail below, where we compare the {\it Gaia}~DR2 distances with the 
previous measurements. We already point out here that all of our values are within one sigma of the \citet{CanGetal18} ones but see the note at the end of the
previous section on the effect of the spatial covariance.

\subsection{\VO{001} = NGC~3603 = RCW~57}

$\,\!$\indent This is the densest and richest cluster in the sample and is similar to R136, the core of 30 Doradus, in several aspects
\citep{Walb73b,Moff83,Moffetal94}. It has three WNh stars with masses above 100~M$_\odot$ \citep{Drisetal95,Crowetal10} and a very rich population of O stars
(see the references in Table~\ref{sample}). One of them, NGC 3603 HST-A1, is a very massive eclipsing binary \citep{Schnetal08a}. At one point it was thought that 
there were few massive stars around the cluster but recent discoveries have changed that picture \citep{RomL13a,Kalaetal19,Drewetal19}.

There are many distance measurements to 
NGC~3603 (\VO{001}) % \VO{001} REFEREE
going back over fifty years ago, see Table~\ref{literature}. Most measurements are in the 6.0-8.5~kpc range with some outliers
(such as the first two measurements) and a significant scatter, making this the most distant group in our sample. The \citet{CanGetal18} value is significantly larger and with 
small uncertainties, a result of not including the spatial covariance term\footnote{
Other things being equal, the effect of the covariance term on the distance uncertainties is larger for 
a simultaneously more distant and rich cluster such as NGC~3603. % \VO{001} REFEREE
}. 
Our measurement of 8.0$^{+2.6}_{-1.7}$~kpc is more compatible with the literature results but the 
error bars are large, a situation that will likely improve with DR3 data. The cluster is not seen as a peak in the parallax distribution of Fig.~\ref{NGC_3603_Gaia}, where 
we see a nearly flat distribution of {\it Gaia}~DR2 sources between 2~kpc and the cluster distance. If we are able to see this deep into the Galaxy in this direction is 
because this is a mostly interarm sightline with an exceptionally low amount of dust. Indeed, 
NGC~3603 % \VO{001} REFEREE
is an object in the fourth quadrant at a similar distance as the 
Galactic center and with $A_V\sim 5$~mag, as opposed to $A_V\sim 30$~mag for the GC. Still, most of the foreground population has an even lower extinction and that is the 
main criterion used to differentiate the cluster stars in Fig.~~\ref{NGC_3603_Gaia}, as the differences in proper motion are small. 

The most outstanding issue regarding the distance to 
NGC~3603 % \VO{001} REFEREE
in the literature is the extinction correction. Most papers assume a standard extinction law 
(\RV\ = 3.0-3.2) and some detect a variable extinction across the face of the cluster. \citet{Pandetal00}, on the other hand, use an extinction model towards 
NGC~3603 % \VO{001} REFEREE
with two components, a Galactic one with an assumed \RV\ = 3.1 and a cluster one for which they measure \RV\ = 4.3. More recently, \citet{MaizBarb18} have 
measured the extinction towards four O stars in the cluster and obtained a partially consistent result with \citet{Pandetal00}. The value of \RV\ is indeed intermediate 
between 3.1 and 4.3 ($\sim 3.9$) but the variable reddening (\EBV = 1.2-1.4) does not show a clear correlation with \RV. We think this issue requires further study in order 
to be fully resolved.

Recently, \citet{Drewetal19} have searched for potential runaway stars from 
NGC~3603 % \VO{001} REFEREE
using {\it Gaia}~DR2 data and found nine candidates that have been likely 
ejected in the last million years. The field we have used for our search is smaller than theirs and we do not go as deep in magnitude but we also pick up the three 
candidates in the sample in common (Table~\ref{runaways}). In addition, we find another three bright potential candidates that may have been excluded from their list due
to their slightly larger impact parameter.

\subsection{The Carina nebula association: $\;\;\;\;\;\;\;\;\;\;\;\;\;\;\;\;\;\;\;\;\;\;\;\;$ \VO{002} = Trumpler~14 
            and $\;\;\;\;\;\;\;\;\;\;\;\;\;\;\;\;\;\;\;\;\;\;\;\;$ \VO{003} = Trumpler~16~W}

$\,\!$\indent The Carina nebula (NGC~3372 = RCW~53) is the site of the most intense burst of recent star formation within 3~kpc of the Sun \citep{Walb73c,Walb95,Smit06b}. 
Indeed, the stars that were used to define the spectral type O3 were in the Carina nebula \citep{Walb71b}. The standard division into groups has Trumpler~14, a
compact very young cluster with a halo around it that makes it look like a less massive version of 
NGC~3603 % \VO{001} REFEREE
\citep{Asceetal07b};
Trumpler 15, an older and less massive cluster to the
NE of Trumpler 14; and four additional regions, Trumpler~16, Collinder~228, Collinder~232, and Bochum~11, without a clearly defined cluster-like structure, likely making 
them  just different parts of a larger group, the Carina nebula association (or Car~OB1). In this paper we deal with Trumpler~14 (\VO{002}) and the western part of 
Trumpler~16 (\VO{003}), as those
are the two regions with very early O-type systems. Note that $\eta$~Car, the most famous star in the nebula, is in the eastern part of Trumpler~16. The whole
region is a bright \HII\ region powered mostly by its early-O type and WR stars \citep{Smit06b} but it must have experienced previous generations of massive stars, 
as evidenced by the existence of multiple expanding shells in the region with velocities of hundreds of km/s \citep{Walb82d,Walbetal02c}. It is important to note
that the extinction to the Carina nebula is surprisingly low for an object inside the Solar circle farther away than 2~kpc, as most of the space between us seems to lie 
in the interarm region, which was also the case for 
NGC~3603. % \VO{001} REFEREE

Trumpler~14 % \VO{002} REFEREE
is dominated by the HD~\num{93129} system, a hierarchical system composed of two O2 supergiants in a centuries-long eccentric orbit (one of them with a likely 
short-period companion) and a farther away component (B) of spectral type O3.5~V((f))z \citep{Maizetal16,Maizetal17a}. Inside the core there is another very early type
star, HD~\num{93128}, and many more O and early-B stars 
\citep{Morretal88,Smit06b,Sotaetal14}, 
for a total of 91 stars detected by our method (including all the just mentioned).
The halo around the cluster is extended towards the SW, where the closest remnants
of the molecular cloud are located, with a pillar pointing towards the cluster core (Fig.~\ref{Trumpler_14_Gaia} 
and \citealt{Tapietal03}).
Trumpler~14 % \VO{002} REFEREE
is easily detected in both the source density 
and proper motion plots in Fig.~\ref{Trumpler_14_Gaia}. A well formed isochrone is also seen in the CMD, indicating a relatively uniform extinction (\AV$\sim$2), with a
secondary sequence to the right that could be either objects with a higher extinction or, more likely, PMS stars 
\citep{Rochetal11,Damietal17a}. In the parallax histogram the peak of 
the overall {\it Gaia}~DR2 density coincides with that of the cluster, likely due not only to the presence of 
Trumpler~14 % \VO{002} REFEREE
itself but of other stars in the Carina nebula association.

Trumpler~16~W (\VO{003}), % \VO{003} REFEREE
on the other hand, is a looser group defined by the ``hot slash'' star HD~\num{93162} and the early-O supergiant ALS~\num{15210} on its western side and 
HD~\num{93205} and HD~\num{93204} on its eastern side. No stars earlier than O4 are found in the rest of Trumpler~16 which, considering that $\eta$~Car is an evolved
object, points towards a small age difference between 
Trumpler~16~W % \VO{003} REFEREE
and Trumpler~16~E. HD~\num{93205} \citep{Morretal01} is an SB2 classified in \citet{Sotaetal14} as 
O3.5~V((f))~+~O8~V using a GOSSS spectrum. We have found a LiLiMaRlin epoch where the two components are well separated in velocity and have found the same spectral types 
(Fig.~\ref{spectra}). Note that the ``defiant finger'' is ionized by HD~\num{93162} and ALS~\num{15210} despite its proximity in the sky to $\eta$~Car (Fig.~1 in 
\citealt{Smitetal04}). 
Trumpler~16~W % \VO{003} REFEREE
does not appear as a significant overdensity in the source density plot of Fig.~\ref{Trumpler_16_W_Gaia} but its stars are clearly
concentrated at the peak of the proper motion distribution, with our method detecting 20 members (including the four O stars mentioned here). 
It should be noted that 
Trumpler~16~W % \VO{003} REFEREE
is crossed by the edge of the V-shaped dust lane that is one of the defining features of the \HII\ region, with HD~\num{93162} and 
ALS~\num{15210} in the dust lane region (higher extinction) and HD~\num{93205} and HD~\num{93204} outside of it (lower extinction). This likely explains 
not only the lack of a significant spatial concentration of sources but also the large spread of extinctions in the CMD. Compare this situation with the near uniform 
extinction of 
Trumpler~14. % \VO{002} REFEREE
As it happened for 
Trumpler~14, % \VO{002} REFEREE
in the parallax histogram the peak of the overall {\it Gaia}~DR2 density coincides with that of 
Trumpler~16~W. % \VO{003} REFEREE
In this case the role of the rest of the Carina nebula association
is larger in achieving that, given the small number of objects in the analyzed group.

%\paragraph{HD~93\,205 = V560~Car = CPD~$-$59~2587 = ALS~1849.}          
%\object[HD 93205]{}         % GOS 287.57-00.71_02
%GOSSS-II:     O3.5   V((f))        + O8     V                                 Ma    
%Villa-I:      O3.5   V((f))        + O8     V                                 Ma      LiLiMaRlin fits/FEROS/HD_93_205_090503_F_V48000.fits

Trumpler~14 and Trumpler~16~W % \VO{002} and \VO{003} REFEREE
are kinematically distinct: their differences in both declination proper motion and right ascension proper motion are large and
significant at more than the 4$\sigma$ level. The two systems are moving in a close-to-radial direction approaching one another, with the closest approach taking
place 1~Ma from now or a bit less than that. The values of $t_{\mu_{\alpha *}}$ and $t_{\mu_{\delta}}$ are significantly larger for 
\VO{002}, 
the likely result of a detection of the internal motions of the cluster in the {\it Gaia}~DR2 proper motions, as expected for a true compact cluster such as Trumpler~14. 

The distance to the Carina nebula and its parts has a long literature that extends over at least fifty years (see Table~\ref{literature})
Before {\it Gaia}~DR2 the most reliable result was the
geometric value of \citet{Smit06a} of 2350$\pm$50~pc based on the 3-D expansion of the Homunculus nebula around $\eta$~Car. The {\it Gaia}~DR2 results from 
different authors \citep{CanGetal18,BindPovi18,ShulDanf19,Kuhnetal19,Limetal19,Zucketal20} and in this paper are consistent with that value, 
indicating that 
Trumpler~14, Trumpler~16~W, % \VO{002}, \VO{003}, REFEREE
and $\eta$~Car are at the same or similar distance. The only discrepant {\it Gaia}~DR2 paper is \citet{Davietal18a}, which places
Trumpler~14 % \VO{002} REFEREE
$450\pm 200$~pc beyond Trumpler~16, but that work does not properly take into account the spatial correlations and systematic errors in the data. Other
results in Table~\ref{literature} tend to find distances that are longer and with discrepancies between Trumpler~14 and Trumpler~16. For example, \citet{Carretal04}
finds the second one 1.5~kpc farther away than the first one and \citet{MassJohn93} finds that 
Trumpler~14 % \VO{002} REFEREE
is 3.61~kpc away. We believe that most of the discrepancies are due to 
unaccounted variations in the extinction law and to the presence of hidden binaries in the sample. 
As shown by \citet{MaizBarb18} not only is the amount of extinction variable in the Carina nebula but so is
the extinction law itself, with \RV\ being lower than 4 in some regions and higher than 6 in others (see the top left panel of Fig.~7 in that paper). As pointed out by 
several authors \citep{Feinetal73,Theetal80,Walb95} errors in the value of \RV\ lead to errors in the distance to this object. A different issue is that the values derived by 
\citet{ShulDanf19} using {\it Gaia}~DR2 data are also slightly higher (but within 1$\sigma$), something we discuss later on. The discrepant value found by 
\citet{Megietal09} deserves a special mention. Those authors use the ISM Ca\,{\sc ii} absorption lines and note that for Trumpler~16 they have complex kinematic profiles
which they erroneously associate with circumstellar envelopes. They are actually produced by a number of expanding shells produced by SN explosions 
\citep{Walb82d,Walbetal02c}, which are more prominent in Ca\,{\sc ii} than in Na\,{\sc i} \citep{RoutSpit51,RoutSpit52}. Given this effect, it is not a good idea to use
Ca\,{\sc ii} to measure distances to stars in regions where SN explosions have already taken place unless that effect is taken into consideration in the analysis.

We find three candidate runaways ejected from 
Trumpler~14 % \VO{002} REFEREE
(Table~\ref{runaways}). One of them, HDE~\num{303313}, is a B2~V~+~B2~V SB2 \citep{Alexetal16}. 
We also find a possible runaway ejected from 
Trumpler~16~W. % \VO{003} REFEREE

\subsection{\VO{004} = Westerlund~2 = RCW~49}

$\,\!$\indent Westerlund~2 (\VO{004}) % \VO{004} REFEREE
is a massive young cluster that was not identified as such until the 1960s due to its heavy extinction \citep{West61}
despite being associated with the bright \HII\ region RCW~49. It has a rich population of massive stars, including the spectroscopic binary V712~Car, which
is composed of two of the most massive known stars in the Galaxy 
\citep{Rauwetal04}. \citet{CrowWalb11} classified the pair as O3~If*/WN6~+~O3~If*/WN6. In Fig.~\ref{spectra} we show a GOSSS spectrum with the system caught close 
to conjunction (the \NVd{4604,4620} lines are single) where H$\beta$ has a P-Cygni-like profile, thus confirming the integrated spectrum is of an early Of/WN or 
``hot slash'' nature \citep{Sotaetal14}. 
Westerlund~2 % \VO{004} REFEREE
also includes a massive WNh star, WR~20b \citep{Rauwetal11}.

%\paragraph{V712~Car.}
%\object[V712~Car]{}         % GOS 284.27-00.34_02
%Villa-I:      O3     If*/WN6ha                                                Ma 

As shown in Table~\ref{literature}, the distance to 
Westerlund~2 % \VO{004} REFEREE
has highly discrepant values in the literature, with a range that spans almost a factor of three between the
minimum and the maximum. Our value lies in the middle or the range and is in reasonable agreement with the more recently published results, as most of the shorter and
longer distances correspond to older papers. The extinction is higher (\AV$\sim$6) than that of 
NGC~3603 % \VO{001} REFEREE
despite being close to half the distance and both located in Carina.  The difference is mainly caused by the sightline to 
NGC~3603 % \VO{001} REFEREE
being mostly through interarm space, as opposed to the sightline to 
Westerlund~2 % \VO{004} REFEREE
having a significant fraction
inside the Carina-Sagittarius arm. The high extinction likely contributes to the differences in distance measured by different authors.

In the top left panel of Fig~\ref{Westerlund_2_Gaia} 
Westerlund~2 % \VO{004} REFEREE
shows its previously known core+halo structure. We identify 174 objects as cluster members, including many with
early-type spectral classifications, the highest number so far in our sample. The peak in the distribution of {\it Gaia}~DR2 is in
the foreground at about $\sim$1~kpc shorter distances. This is consistent with the high extinction to the cluster, whose stars can be easily observed only because of 
their high luminosity. The proper motion diagram puts 
Westerlund~2 % \VO{004} REFEREE
at the lower left extreme of an elongated distribution that likely reflects the velocity differences 
along the spiral arm structure from the foreground population at the point where the sightline reaches the arm around $\sim$2~kpc until where we get to the distance 
of the cluster. 

We detect two possible runaways from 
Westerlund~2. % \VO{004} REFEREE
The first one is THA~35-II-42 (also known as WR~21a), an O2~If*/WN5 star \citep{Maizetal16} with an 
early-O companion \citep{Niemetal08,Trametal16} whose runaway nature was already suggested by \citet{RomLetal11}. The second one is SS~215 (also known as WR~20aa),
another O2~If*/WN5 star \citep{Maizetal16} with a runaway nature also already suggested by \citet{RomLetal11}. Outside of our field of view we find WR~20c, also suggested by 
\citet{RomLetal11} as a star possibly ejected from the cluster, and whose {\it Gaia}~DR2 proper motion is indeed compatible with that hypothesis. See \citet{Drewetal18} for a 
more complete study of the possible runaways from 
Westerlund~2. % \VO{004} REFEREE

\subsection{\VO{005} = Pismis~24 = NGC~6357~W = RCW~131~W}

\begin{figure*}
\centerline{\includegraphics*[width=\linewidth]{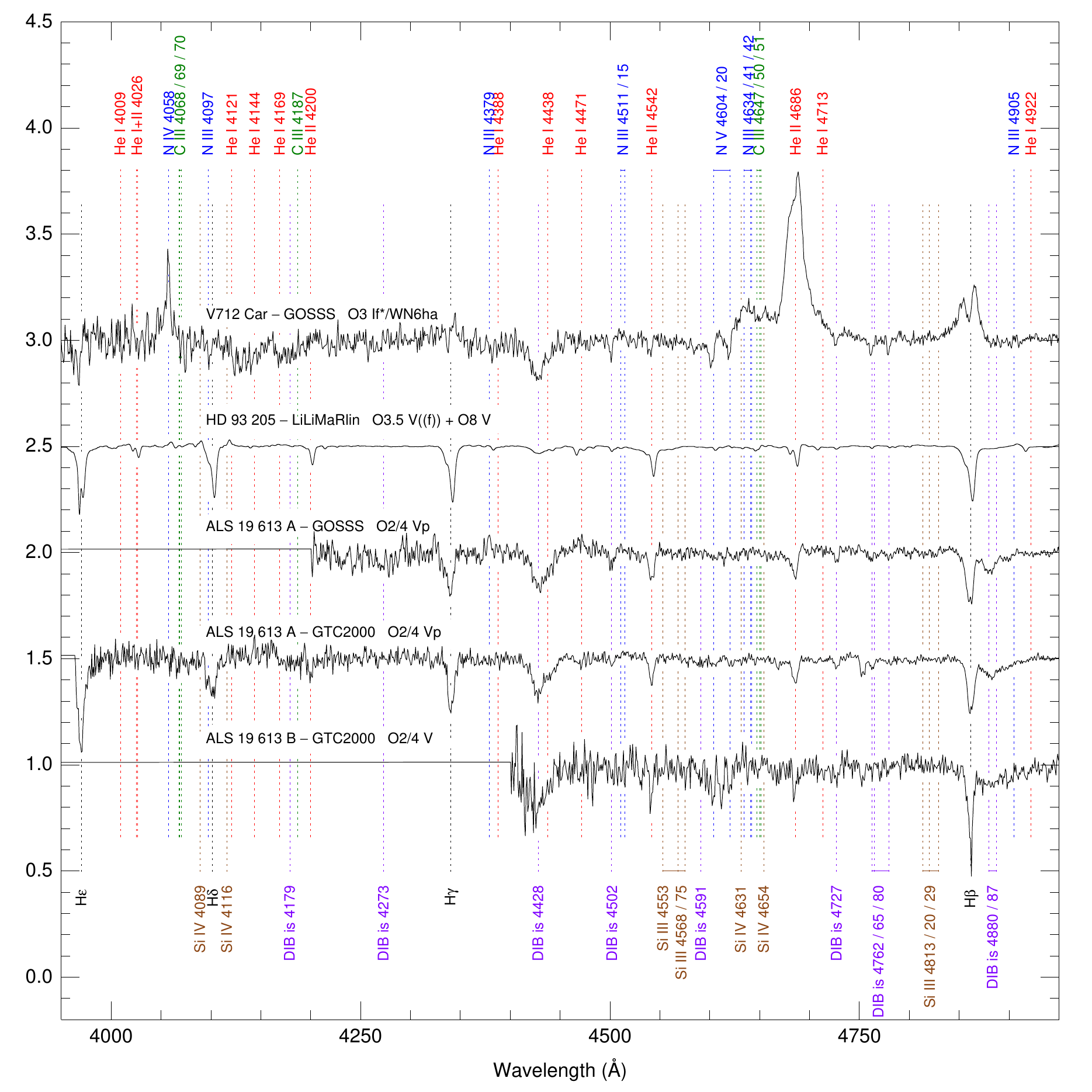}}
\caption{New spectra in this paper.}
% \caption{TBD. Spectra of HD~93\,205 (LiLiMaRlin), V712~Car (GOSSS), ALS~19\,613~A (GOSSS+GTC-2000), ALS~19\,613~B (GTC-2000),  
%         HD~150\,135~Aa,Ab (GOSSS+LiLiMaRlin), HD~150\,136~Aa,Ab (LiLiMaRlin), HD~64\,315 (LiLiMaRlin), and Sh~2-158~1 (GOSSS+LILiMaRlin).}
\label{spectra}
\end{figure*}   

\addtocounter{figure}{-1}

\begin{figure*}
\centerline{\includegraphics*[width=\linewidth]{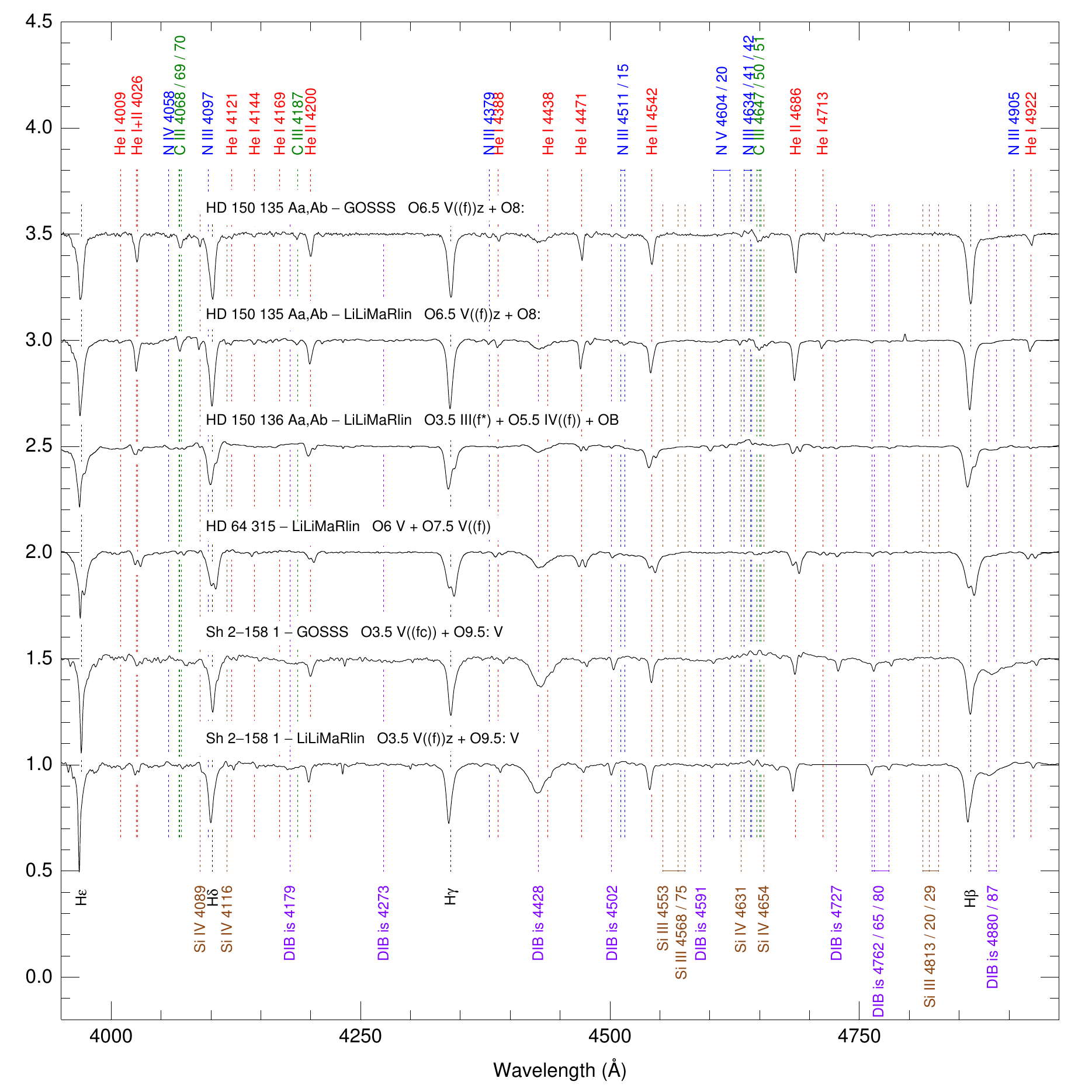}}
\caption{(Continued).}
\end{figure*}   

$\,\!$\indent Pismis~24 (\VO{005}) % \VO{005} REFEREE
is a cluster with a compact core in the western part of the \HII\ region NGC~6357 \citep{Pism59}, which itself is likely associated with the
nearby \HII~region NGC~6334 \citep{Fukuetal18}. It has two O3.5-type systems, Pismis~24-1 
and Pismis~24-17, the first one with a supergiant luminosity classification and the second one with a giant luminosity classification 
\citep{Walbetal02b,Sotaetal14}. Pismis~24-1 is both a visual and a spectroscopic binary \citep{Maizetal07}, with the two visual components having a small 
$\Delta m$, so it is likely that the system contains at least two very early type stars. However, at the present time there are no published spectral
classifications in which the components are either spatially or kinematically resolved.

This is the closest cluster so far on the list, the first one under 2~kpc. There is a general good agreement between literature distances (Table~\ref{literature})
except for two of the papers which use spectro-photometry (overestimates) and the four papers which give kinematic values (underestimates). This sightline is close to the 
direction of the Galactic center, so kinematic distances are expected to be of poor quality. 
Pismis~24 % \VO{005} REFEREE
has a very high extinction (\AV$\sim$6) for its short distance, a likely consequence of its location in the Carina-Sagittarius arm close to the direction of the Galactic center.

The upper left panel of Fig.~\ref{Pismis_24_Gaia} shows a well-defined core with another concentration towards the south and an extended halo (this region has a complex ISM,
see \citealt{Cappetal11}). Pismis~24-1 is not 
detected as a cluster member because it has no parallax or proper motions in {\it Gaia}~DR2, but other bright objects in the cluster core such as Pismis~24-17 are.
The brightest detected star is WR~93, a WC7~+~O7/9 spectroscopic binary, that is the central object of a small subcluster 5\arcmin\ to the East of the core. We detect 197 
cluster members, even more than for 
Westerlund~2, % \VO{004} REFEREE
due to the combination of proximity and richness. Nevertheless, the proper motion statistical tests ($t_{\mu_{\alpha *}}$ and 
$t_{\mu_{\delta}}$) are relatively high, indicating the detection of significant internal motions by {\it Gaia}~DR2.

The proper motion panel of Fig.~\ref{Pismis_24_Gaia} provides little discrimination, which is expected in a direction close to the Galactic center due to the nearly flat 
Galactic rotation curve. Most of the discrimination of cluster members is done using the CMD thanks to the higher extinction of the cluster compared to the field population. 
An important difference with previous objects in our sample is the proximity of that field population, as most of it seems to be in the foreground between a distance of 1~kpc
and the cluster itself. The proportion of stars beyond 2~kpc is significantly lower than for the other groups we have analyzed so far, indicating the existence of an
extinction wall at the cluster distance or slightly beyond.

We detect eight possible runaway stars from 
Pismis~24 % \VO{005} REFEREE
(Table~\ref{runaways}). Of those, Pismis 24-18 is likely to be an early-type B star. See \citet{Gvaretal11b} for a previous study on potential runaways from 
Pismis~24. % \VO{005} REFEREE

\subsection{\VO{006} = Gum~35 = Majaess~133}

$\,\!$\indent This is the most overlooked object in our sample, with only a few significant references in the literature \citep{Dutretal03a,Maja13,Mohretal17}.
Its high extinction caused its discovery to be produced in the IR but it is the likely source of the visible \HII\ region Gum~35 \citep{Gum55} and its brightest stars 
have been studied in the visible. ALS~2067 was already classified as a ``hot slash'' star by \citet{GomeNiem87}, a classification that was changed to O supergiant by 
\citet{WalbFitz00} and later revised in \citet{Sotaetal14} to O5~Ifp. THA~35-II-153 is the earliest star in the cluster and this one is actually a ``hot slash'' 
star but its nature was only recently noticed by \citet{Maizetal16}, where its proximity to ALS~2067 was noted but it was erroneously assigned to 
Collinder~228. ALS~\num{18551} is an early O-type SB2 system also identified for the first time by \citet{Maizetal16} but also erroneously assigned to Collinder~228
there. It was not until \citet{Mohretal17} that the relevance of this cluster was recognized.

The only previous distance measurements to \VO{006} are kinematic distances between 6~and~9~kpc.
A cluster at that position is clearly detected in the {\it Gaia}~DR2 data 6.4~kpc away with relatively large uncertainties.
That value makes it the second most distant target in the sample. It is close to 
NGC~3603 % \VO{001} REFEREE
in position in the sky ($2^{\rm o}$ away) and extinction (\AV$\sim$5) and the two 
distance error bars overlap, so it is possible the two objects are physically 
associated, as suggested by \citet{Mohretal17}. The cluster has a core/halo structure but the core itself has a complex, filamentary structure. Most of the field population is
in the foreground one to several kpc closer. In the proper motion diagram of Fig.~\ref{Gum_35_Gaia} \VO{006} lies at the end of an elongated structure that likely traces the
foreground objects along the Carina-Sagittarius arm. The cluster is more easily distinguished from the field population by its higher extinction. A relatively large number of 
cluster members, 98, is identified, with THA~35-II-153 and ALS~\num{18551} among them but not ALS~2067 due to its large RUWE.

There are two possible runaways in the {\it Gaia}~DR2 data (Table~\ref{runaways}).
However, for objects so far away one is in dire straits determining distances for individual stars, so in this case there is a higher chance (compared to our other candidates) that
they may be rejected by a subsequent analysis using better data and techniques (e.g. \citealt{Tetzetal11}).
The first candidate, 2MASS J10584671$-$6105512, is interesting because it was classified as O8~Iabf by \citet{Maizetal16}.
However, its proper motion is just outside our selection circle, so it could be just an unrelated object in this complex line of sight.

\begin{table*}
\caption{Membership and distance results.}
\centerline{
\renewcommand{\arraystretch}{1.2}
\begin{tabular}{lrrcccr@{$\pm$}lr@{$\pm$}lr@{$\pm$}lr@{$\pm$}lr@{}l}
\hline
ID      & $N_{*,0}$ & $N_*$ & $t_\varpi$ & $t_{\mu_{\alpha *}}$ & $t_{\mu_{\delta}}$ & \mcii{\pig}  & \mcii{\pmrag}  & \mcii{\pmdecg} & \mcii{\pigc} & \mcii{$d$}              \\
        &           &       &            &                      &                    & \mcii{(mas)} & \mcii{(mas/a)} & \mcii{(mas/a)} & \mcii{(mas)} & \mcii{(pc)}             \\
\hline
O-001   &       143 &   137 &       1.13 &                 1.97 &               1.65 & 0.074&0.043  & $-$5.551&0.067 & $+$1.974&0.067 & 0.114&0.044  & 8000&$^{+2600}_{-1700}$ \\
O-002   &        93 &    91 &       0.88 &                 3.98 &               3.29 & 0.387&0.042  & $-$6.540&0.067 & $+$2.042&0.067 & 0.427&0.043  & 2430&$^{+290}_{-230}$   \\
O-003   &        20 &    20 &       0.86 &                 2.42 &               1.88 & 0.401&0.042  & $-$7.073&0.067 & $+$2.641&0.066 & 0.441&0.043  & 2380&$^{+270}_{-220}$   \\
O-004   &       178 &   174 &       1.05 &                 1.89 &               1.76 & 0.188&0.042  & $-$5.190&0.064 & $+$2.927&0.064 & 0.228&0.043  & 4730&$^{+1130}_{-780}$  \\
O-005   &       208 &   197 &       1.20 &                 2.45 &               3.71 & 0.562&0.040  & $-$0.918&0.059 & $-$2.216&0.059 & 0.602&0.041  & 1690&$^{+130}_{-110}$   \\
O-006   &        99 &    98 &       1.06 &                 1.38 &               1.52 & 0.112&0.043  & $-$5.513&0.067 & $+$2.270&0.067 & 0.152&0.044  & 6400&$^{+1800}_{-1200}$ \\
O-007   &        82 &    81 &       0.98 &                 2.10 &               1.67 & 0.554&0.043  & $-$2.697&0.067 & $-$4.482&0.068 & 0.594&0.044  & 1720&$^{+140}_{-120}$   \\
O-008   &        45 &    45 &       1.20 &                 1.62 &               1.92 & 0.581&0.043  & $-$2.667&0.067 & $-$4.371&0.067 & 0.621&0.044  & 1640&$^{+130}_{-110}$   \\
O-009   &        30 &    30 &       1.15 &                 3.26 &               3.12 & 0.587&0.043  & $+$0.166&0.069 & $-$1.618&0.069 & 0.627&0.044  & 1630&$^{+130}_{-110}$   \\
O-010   &       115 &   111 &       1.12 &                 2.39 &               3.00 & 0.813&0.043  & $+$1.307&0.066 & $-$3.972&0.066 & 0.853&0.044  & 1185&$^{+65}_{-59}$     \\
O-011   &        82 &    79 &       1.22 &                 1.67 &               1.38 & 0.287&0.042  & $-$2.840&0.066 & $-$4.316&0.066 & 0.327&0.043  & 2990&$^{+390}_{-340}$   \\
O-012   &       206 &   194 &       1.15 &                 1.54 &               1.63 & 0.178&0.041  & $-$2.523&0.061 & $+$2.586&0.061 & 0.218&0.042  & 4850&$^{+1160}_{-800}$  \\
O-012~N &        88 &    87 &       1.14 &                 1.41 &               1.18 & 0.164&0.043  & $-$2.544&0.067 & $+$2.487&0.067 & 0.204&0.044  & 5190&$^{+1380}_{-930}$  \\
O-012~S &        67 &    62 &       1.10 &                 1.75 &               1.57 & 0.180&0.043  & $-$2.503&0.067 & $+$2.663&0.067 & 0.220&0.044  & 4830&$^{+1210}_{-830}$  \\
O-013   &        11 &    11 &       0.95 &                 1.96 &               3.95 & 0.316&0.043  & $-$3.708&0.068 & $-$2.214&0.068 & 0.356&0.044  & 2930&$^{+440}_{-340}$   \\
O-014   &        12 &    12 &       0.94 &                 2.98 &               2.63 & 1.364&0.035  & $-$0.092&0.055 & $-$4.176&0.055 & 1.404&0.036  &  714&$^{+19}_{-18}$     \\
O-015   &        93 &    75 &       0.98 &                 3.20 &               2.82 & 0.957&0.034  & $-$2.605&0.048 & $-$6.390&0.048 & 0.997&0.035  & 1006&$^{+37}_{-34}$     \\
O-016   &       340 &   286 &       1.12 &                 5.15 &               3.56 & 1.354&0.029  & $-$1.885&0.041 & $-$3.716&0.041 & 1.394&0.031  &  719&$^{+16}_{-16}$     \\
O-016~N &       102 &    99 &       1.04 &                 3.99 &               3.25 & 1.357&0.040  & $-$1.716&0.059 & $-$3.705&0.059 & 1.397&0.041  &  719&$^{+22}_{-21}$     \\
O-016~S &        94 &    90 &       1.19 &                 5.15 &               3.43 & 1.350&0.040  & $-$2.077&0.057 & $-$3.788&0.057 & 1.390&0.041  &  722&$^{+22}_{-21}$     \\
\hline
\end{tabular}
\renewcommand{\arraystretch}{1.0}
}
\label{results}                  
\end{table*}

\subsection{The Cygnus~OB2 association: $\;\;\;\;\;\;\;\;\;\;\;\;\;\;\;\;\;\;\;\;\;\;\;\;$ \VO{007} = Cyg~OB2-22 cluster = Bica~1 
            and \VO{008} = Cyg~OB2-8 cluster = Bica~2}

%\begin{figure*}
%\centerline{\includegraphics*[width=\linewidth]{Cyg_OB2_GALANTE.pdf}}
%\caption{RGB mosaic of three GALANTE fields \citep{LorGetal19} in the Cyg~OB2 region. The red channel is F861M (calcium triplet band), the green channel is F660N (H$\alpha$), 
%         and the blue channel is F515N (Str\"omgren~$y$-like but narrower). Each channel is a composite of several 50~s, 10~s, 1~s, and 0.1~s exposures to yield a high-dynamic 
%         range image that includes faint stars around 19th magnitude without saturating any bright star. The two cyan circles mark \VO{007} (bottom) and \VO{008} (top).}
%\label{GALANTE}
%\end{figure*}   
%
%\begin{figure*}
%\centerline{\includegraphics*[width=\linewidth]{Cyg_OB2_GALANTE_zoom.pdf}}
%\caption{Zoom of Fig~\ref{GALANTE} around the \VO{007} and \VO{008} groups.}
%\label{GALANTE_zoom}
%\end{figure*}   

$\,\!$\indent The first six Villafranca groups are all located in the southern hemisphere and this is not a coincidence. Given our position in the Galaxy,
it is easier to find massive young clusters towards the inner two Galactic quadrants than towards the two outer ones. The first two Villafranca 
groups in the northern hemisphere are located in Cygnus~OB2, in the first quadrant but close to the second one. Cygnus~OB2 is the northern OB association
with the highest number of O stars and is sometimes presented as the northern equivalent of the Carina nebula association, given their relatively similar
sizes, stellar contents, and distances and their symmetric positions with respect to the Galactic center \citep{Knod00,Comeetal02,ComePasq12,Berletal18a}. 

One difference that is sometimes pointed out between the Carina nebula and Cyg~OB2 is the lack of significant clusters in the latter but that is not actually
true as there are two stellar groups in Cyg~OB2 that would stand out as significant clusters if they were isolated objects: those around the multiple systems
Cyg~OB2-22 and Cyg~OB2-8 \citep{Bicaetal03,dLFMdLFM09,Maiz10a}, hereafter 
Bica~1 (\VO{007}) and Bica~2 (\VO{008}). % \VO{007} and \VO{008} REFEREE
Bica~2 % \VO{008} REFEREE
includes Cyg~OB2-7, the second O3 supergiant anywhere and the first object earlier than O4 in the northern hemisphere to be identified \citep{Walb73d}. 
Bica~1 % \VO{007} REFEREE
includes Cyg~OB2-22~A, the second object earlier than O4 in the northern hemisphere to be identified \citep{Walbetal02b}. Other interesting objects are Cyg~OB2-9 in 
Bica~1, % \VO{007} REFEREE
a massive highly eccentric SB2 system \citep{Nazeetal12c,Maizetal19b}; Cyg~OB2-8~C in 
Bica~2, % \VO{008} REFEREE
an early-type Ofc star \citep{Walbetal10a}; and Cyg~OB2-8~A in 
Bica~2, % \VO{008} REFEREE
another massive SB2 system composed of a supergiant and a giant stars. Cyg~OB2-22~B is not selected by our algorithm as a member of 
Bica~1 % \VO{007} REFEREE
due to its large RUWE but 
it is highly likely that it is a cluster member because of the combination of its early spectral type, small magnitude difference, and proximity to Cyg~OB2-22~A
\citep{Maiz10a}. In any case, as it is not selected by the algorithm its {\it Gaia}~DR2 measurements are not taken into account for the distance measurement.
On the other hand, Cyg~OB2-22~I (=~ALS~\num{15161}) has RUWE~=~1.1 and $\varpi = 1.040\pm 0.054$~mas, so it appears to be a foreground object.

The distance measurements to Cyg~OB2 go back to the association discovery by \citet{JohnMorg54} and cluster around a short distance of $\sim$1.4~kpc
and a long one of $\sim$1.7~kpc. The recent Gaia~DR2 analysis by \citet{Berletal19} suggests that one possible explanation for this would be the existence of 
two subassociations: the main one, located at the long distance, and another one with only $\sim$10\% of the stars, located at the short distance. The latter 
would be younger and include the four masers of \citet{Rygletal12} and the four eclipsing binaries of \citet{Kimietal15a}. However, one should not discard that
some results are still affected by extinction, as the column density of dust towards Cyg~OB2 is at the same time very high for a group so close 
and highly variable, as the dust is mostly associated with the region rather than being spread uniformly along the line
of sight. Our measurements for both 
Bica~1 and Bica~2 % \VO{007} and \VO{008} REFEREE
are compatible with the long $\sim$1.7~kpc distance, with a difference of less than one sigma between them. Therefore, the hypothesis of \citet{Bicaetal03} that 
they are the core of the association may remain valid if most of Cyg~OB2 is at the long distance.

Both 
Bica~1 and Bica~2 % \VO{007} and \VO{008} REFEREE
distinguish themselves in the {\it Gaia} CMD as reddened sequences to the right of the low-extinction Galactic population sequence.
They are also seen as density concentrations in the plane of the sky but are hard to distinguish in proper motion. The proper motions in right ascension of the
two clusters are very similar while those in declination differ by a little over one sigma. The source density over the whole field is significantly lower than
for the previous groups, a consequence of the larger angular distance from the Galactic center and the existence of strong extinction at short distances: 
Cyg~OB2 is closer than the Carina nebula but the dust column is several times higher. Overall, the available information points towards 
Bica~1 and Bica~2 % \VO{007} and \VO{008} REFEREE
being a double cluster separated by $\sim$2.7~pc in the plane of the sky. We also point out that the field parallax histogram (black line in the bottom middle 
panels of Figs.~\ref{Cyg_OB2-22_Gaia}~and~\ref{Cyg_OB2-8_Gaia}) peaks at the same value as the clusters, a sign of the richness of the Cyg~OB2 association (which 
increases the numbers at its parallax) and of the large amount of dust associated with it (which decreases the numbers at smaller parallaxes). 

There are two possible runaways in the {\it Gaia} data. Cyg~OB2-24 is a fast-rotating O8~Vn star \citep{Sotaetal11a} that appears to have been ejected from 
Bica~1 % \VO{007} REFEREE
in the general direction of 
Bica~2. % \VO{008} REFEREE
[MT91]~453 is a B5:~V star \citep{Kimietal07} that may have been ejected from 
Bica~2 (or even Bica~1) % \VO{008} (or even \VO{007})  REFEREE
towards the north.

\subsection{\VO{009} = M17 = Omega~nebula = NGC~6618 = Sh~2-45 = RCW~160}

$\,\!$\indent \VO{009} is immersed in one of the most famous nearby \HII\ regions, M17, which is characterized by the highly variable and strong extinction associated 
with it. There are several early-O stars in the cluster.
ALS~\num{19613} was classified as O5~V in the optical and O3-4 in the $K$ band by \citet{Hansetal97}. \citet{Hoffetal08} separates the A and B components and gives O4~V
spectral classifications for both but indicate that the two of them are spectroscopic binaries. Both A and B (but especially the second one) are very weak targets for
which we have only been able to obtain relatively noisy spectra with GOSSS (for A) and with a reduced spectral resolution of 2000 with GTC (for both A and B) and even
then with only the bluest wavelengths of the classification range (Fig.~\ref{spectra}). Both are indeed early-type O stars, as indicated by the weak or absent 
\HeI{4471} absorption, but the low S/N precludes an accurate classification. Other early-O stars are ALS~\num{19618}~A and ALS~\num{19617}.

%\paragraph{ALS~19\,613~A.}          
%\object[ALS 19613 A]{}      % GOS 015.06-00.69_02
%Villa-I:      O2/4   Vp                                                       Ma    
%Villa-I:      O2/4   Vp                                                       Ma      GTC 2000
%\paragraph{ALS~19\,613~B.}          
%\object[ALS 19613 B]{}      % GOS 015.06-00.69_03
%Villa-I:      O2/4                                                            Ma      GTC 2000

We have obtained lucky images of the core of M17 with AstraLux on two different dates, 2012-10-02 and 2019-06-15, and measured the separation, position angle, and 
$\Delta m$ in two filters ($z$ and $i$) for ALS~\num{19613}~A,B. We obtain a separation of $1.651\pm0.006$\arcsec, a position angle of $227.17\pm0.10$\degr, a $\Delta z$
of $0.585\pm0.037$~mag, and a $\Delta i$ of $0.894\pm0.064$~mag, with no appreciable motion in the span of almost seven years. 

The distance measurements in the literature are concentrated between our value and ones $\sim50\%$ higher, with an extreme outlier that places M17 beyond the Galactic 
center \citep{Quiretal06}
and three measurements around 1.3~kpc.
Some of the differences may be caused by uncorrected extinction effects but others may be caused by a different location along the line of 
sight, as the two water maser measurements \citep{Xuetal11,Chibetal16} give a distance around 2~kpc. Kinematic distances are overestimated, to some point expected as 
this sightline is close to the direction of the Galactic center. The proper motion statistical tests ($t_{\mu_{\alpha *}}$ and $t_{\mu_{\delta}}$) are relatively high, 
indicating the detection of significant internal motions by {\it Gaia}~DR2.

The proper motion diagram in Fig.~\ref{M17_Gaia} does not allow the discrimination of cluster sources easily, as expected from the small angle difference with 
the Galactic center. The concentration of sources and nebular emission seen towards the center
(with an extension towards the East) in the top panels of Fig.~\ref{M17_Gaia} is caused mostly by the lower extinction in those regions. The horizontal spread among
cluster members in the CMD (bottom left panel of Fig.~\ref{M17_Gaia}) is another sign of the strong differential extinction. It is quite likely that the cluster is
much richer in luminous OB stars than the 30 members indicated in Table~\ref{results} \citep{Hoffetal08}. {\it Gaia}~DR2 resolves ALS~\num{19613} into A and B but their
astrometric results are discrepant (and one of them with a high RUWE), so it is possible that there is cross-contamination in their data.

We detect seven possible runaways from M17. One of them is BD~$-$16~4826, an SB2 classified as O7~III((f))~+~O9/B0~V by \citet{Maizetal19a}. Another one,
NGC~6618~B-373 was classified as O8~V by \citet{Povietal09}, who noted the star may be a binary based on its luminosity. A third one, 2MASS~J18200299$-$1602068,
was also classified by \citet{Povietal09} as O9~V.

\subsection{\VO{010} = NGC~6193 = RCW~108}

$\,\!$\indent NGC~6193 (\VO{010}) % \VO{010} REFEREE
is a cluster in the Ara~OB1a association that ionizes the adjacent RCW~108 \HII\ region \citep{Arnaetal87,Baumetal11}. Its core is dominated by
the triple system HD~\num{150136}~Aa,Ab \citep{NiemGame05,SanBetal13,Sanaetal13a}. \citet{Sotaetal14} classified two of the components as O3.5-4~III(f*)~+~O6~IV. The other
O-type system in the cluster is HD~\num{150135}~Aa,Ab, classified in GOSSS-II as O6.5~V((f))z but note that the OWN project had previously identified it as an SB2.

We have obtained a LiLiMaRlin spectrum of HD~\num{150136}~Aa,Ab and we have caught the system with clear double lines and a weak third component seen in He\,{\sc i}
(Fig.~\ref{spectra}). The corresponding spectral classification is O3.5~III(f*)~+~O5.5~IV((f))~+~OB, where the slightly earlier spectral type of the secondary can be 
explained by contamination from the tertiary in the GOSSS-II spectrum. We have also obtained GOSSS and LiLiMaRlin spectra of HD~\num{150135}~Aa,Ab and in both cases we see 
it as an SB2 with spectral types O6.5~V((f))z~+~O8:.

%\paragraph{HD~150\,135~Aa,Ab = CPD~$-$48~8703~Aa,Ab = ALS~3700~Aa,Ab.}          
%object[HD 150135]{}        % GOS 336.71-01.57_01
%GOSSS-II:     O6.5   V((f))z                                                  Ma    
%Villa-I:      O6.5   V((f))z       + O8:                                      Ma      GOSSS
%Villa-I:      O6.5   V((f))z       + O8:                                      Ma      LiLiMaRlin fits/FEROS/HD_150_135_AaAb_160730_F_V48000.fits
%\paragraph{HD~150\,136~Aa,Ab = CPD~$-$48~8703~Ca,Cb = ALS~3701~Aa,Ab.}          
%object[HD 150136]{}        % GOS 336.71-01.57_02
%GOSSS-II:     O3.5-4 III(f*)       + O6     IV                                Ma    
%Villa-I:      O3.5   III(f*)       + O5.5   IV((f))    + OB                   Ma      LiLiMaRlin fits/FEROS/HD_150_136_AaAb_040505_F_V48000.fits

This is the cluster in the sample that is more easily distinguished from the field population in proper motion, which can be explained by its proximity compared to most 
of the stars in the field. The cluster has a well defined core centered on the two O stars and the background population has a non-uniform spatial distribution centered
several arcminutes to the west of the cluster core. The explanation of this effect lies in the geometry created by the destruction of the parent molecular cloud of the
cluster. The ionizing radiation and stellar winds of 
NGC~6193 % \VO{010} REFEREE
have created a cavity around it that has only partially ruptured the molecular cloud around it. The optically thicker remains are located just at the right edge and beyond 
the frame of the top left panel in Fig.~\ref{NGC_6193_Gaia} and contain an infrared cluster \citep{Straetal87}. The only place where the cloud
appears to have been completely ruptured along the line of sight is the region between the cluster core and those optically thicker remains, and indeed there we are able
to see the Galactic disc through a hole up to several times the distance to the cluster itself. The extinction to the 
NGC~6193 % \VO{010} REFEREE
itself is relatively low. \citet{MaizBarb18} give a mean \EBV\ of 0.445 for the two O stars with a high \RV\ close to 4.0, which is typical of H\,{\sc ii} regions and 
suggests that a significant part of the dust affecting the line of sight is associated with the cluster itself.
 
The pre-{\it Gaia}~DR2 distance measurements to 
NGC~6193 % \VO{010} REFEREE
cluster around 1.4~kpc but we find a value about 200~pc smaller. One possible origin for the discrepancy is the
anomalous value of \RV. We have also tested whether the dark cloud to the west of the cluster that blocks most of the stars behind it is indeed at the same distance as 
NGC~6193. % \VO{010} REFEREE
To do that we have downloaded the {\it Gaia}~DR2 data in a $6\arcmin\times6\arcmin$ region in the cloud and checked the parallax distribution of the sources
there. Indeed, we find a steep drop in the number of sources when we get to the value of \pig\ (compare this with the center left panel of Fig.~\ref{NGC_6193_Gaia})
and the few sources with smaller parallax values become significantly redder than the ones with larger values (compare this with the central panel in 
Fig.~\ref{NGC_6193_Gaia}). Therefore, we confirm that the dark cloud is associated with the cluster, as expected. This technique is a less sophisticated version of the one
used by \citet{Zucketal20} to measure the distance to the molecular cloud to the south of the cluster, for which they find a slightly lower distance by 100-150~pc, 
indicating that other cloud may be also associated with the cluster but located slightly closer to us.

We have found one possible runaway from 
NGC~6193. % \VO{010} REFEREE
2MASS~J16403254$-$4846296 is moving towards the west from the cluster.

\subsection{\VO{011} = Berkeley~90 = Sh~2-115}

$\,\!$\indent This cluster is dominated by two O-type systems, LS~III~+46~11 and LS~III~+46~12 \citep{Maizetal15a,Maizetal15b,MarcNegu17} and is associated with
the faint H\,{\sc ii} region Sh~2-115 \citep{HartFell80}.
The early type nature of its two central sources was first recognized by \citet{Motcetal97}. LS~III~+46~11 is a massive near-twin binary \citep{Maizetal15a} 
and the cluster experiences a significant differential extinction \citep{Maizetal15b}. 
Berkeley~90 (\VO{011}) % \VO{011} REFEREE
is the group located at a higher Galactic latitude and at a larger physical distance from the Galactic plane in our sample.

Berkeley~90 % \VO{011} REFEREE
presents a well-defined core surrounded by a halo. It does not distinguish itself well in proper motion and distance from the surrounding population, 
which concentrates at similar values. The most efficient separation filter is the CMD, as the local extinction affects the cluster more than the surrounding area.
Therefore, it is likely that we are discarding some objects associated with the cluster but with lower extinction. 

The Gaia~DR2 distance to 
Berkeley~90 % \VO{011} REFEREE
of $\sim$3~kpc is somewhat 
longer than the two early estimates by \citet{MayeMaca73} and \citet{Motcetal97} but consistent with the more recent values of \citet{Maizetal15a} and \citet{MarcNegu17}, 
especially the former. 
As with most groups in this paper, our distance agrees with the \citet{CanGetal18} value but with a much larger uncertainty caused by the covariance term.
The reason for the differences 
with the two early estimates
lays in the incorrect spectral classifications of 
the two central O-type systems, the undetected binary nature of LS~III~+46~11, and the complex extinction \citep{Maizetal15b}. The Gaia~DR2 parallax of 
LS~III~+46~12 is compatible with that of the cluster, which according to the discussion in \citet{Maizetal15a} indicates that the star is overluminous
for its spectral type or that it is a yet undetected binary. We confirm that the B8~III star 2MASS~J20352201$+$4651518, the B9~III star 2MASS~J20352097$+$4648368, 
the F6~V star 2MASS~J20350745+4651367, the F6~III star 2MASS~J20351813+4650525, the F8~IV star 2MASS~J20350955$+$4652199, and the G2~III star 2MASS~J20351026+4651364 
are foreground objects according to their parallaxes and have proper motions significantly different to that of the cluster \citep{MarcNegu17}.

From its parallax, proper motion, and CMD position the B0.5~V star 2MASS~J20351422$+$4650118 \citep{MarcNegu17} appears to be either a background object or to have 
been ejected towards the south after an interaction with LS~III~+46~11. We also detect four additional possible objects ejected from the cluster, all with $G$ 
magnitudes that would correspond to B-type stars. The first two are to the north and are bluer (likely experience less extinction) than the stars identified in the 
group while the last two are to the south and are redder (likely experience more extinction) than the other group stars. This is possibly another consequence of the 
strong differential extinction around 
Berkeley~90. % \VO{011} REFEREE

\subsection{\VO{012} = NGC~2467 = Sh~2-311  = RCW 16}

$\,\!$\indent This group is a double cluster composed of a northern component (Haffner~19, \VO{012}~N) and a southern one (Haffner~18, \VO{012}~S). Its inclusion on this paper 
needs an explanation. There is only one confirmed O star that meets the requirements of our method, CPD~$-$26~2704 in 
Haffner~18, % \VO{012}~S REFEREE
which is an O7~V(n) \citep{Maizetal16} i.e. 
significantly later than O4. However, there is indeed a star earlier than O4 associated with the cluster: HD~\num{64568}, an O3~V((f*))z \citep{Sotaetal14} that appears 
in the {\it Gaia}~DR2 data as a likely runaway from \VO{012}, probably 
Haffner~18, % \VO{012}~S REFEREE
with a flight time around 400~ka. In addition, there is another O star in the
vicinity, HD~\num{64315}~AB, classified by \citet{Maizetal16} as O5.5~V~+~O7~V. HD~\num{64315}~AB has a RUWE of 1.8 and a negative parallax, possibly due to the presence of two
unresolved visible components (each one of them a binary itself, see \citealt{Loreetal17}) in the Gaia data, 
but its position in the CMD indicates that it is likely to be at a similar distance as HD~\num{64568} and \VO{012}.
Interestingly, HD~\num{64315}~AB is moving away from 
Haffner~18 % \VO{012}~S REFEREE
in a direction nearly opposite to that of HD~\num{64568} and with a similar flight time, leading to the
possibility that the two O-type systems were simultaneously ejected from the cluster. 
For those reasons, we consider that most likely 
Haffner~18, % \VO{012}~S REFEREE
is the origin of the O3 star HD~\num{64568} and 
include it in the sample of this paper.
The Sh~2-311 H\,{\sc ii} region has a bright core close to HD~\num{64315}~AB, its likely
main ionizing source, and an extended halo. There is little H$\alpha$ emission close to HD~\num{64568} despite its higher ionizing flux, another indication that the star is
already far from its primordial cloud.

The literature distances for \VO{012} (Table~\ref{literature}) show a high dispersion, making it one of the objects in our sample where {\it Gaia} data is more useful. Indeed,
there are discrepancies as to whether 
Haffner~18 and Haffner~19 % \VO{012}~N and \VO{012}~S REFEREE
(and other nearby regions) are at the same distance or not (e.g. \citealt{Yadaetal15} claims that the first  
one is twice as far as the second one). Our results yield a difference of less than one sigma between the two distances, indicating that it is possible they are physically
associated, but the error bars are relatively large so we cannot discard that they are not. In this case future {\it Gaia} data releases may provide a more conclusive answer. 
For the time being, we assign a single catalog number to 
Haffner~18 and Haffner~19. % \VO{012}~N and \VO{012}~S REFEREE
The proper motions of the two clusters are similar, with a hint of the two approaching each other.

The double core structure is clearly seen in the top left panel of Fig.~\ref{NGC_2467_Gaia}. In the proper motion diagram \VO{012} is located at the end of an elongated 
structure that likely traces the proper motion changes as a function of distance. We are able to see a long way through this sightline, placed in a low extinction hole 
towards the outer galaxy (this is the target located at a larger Galactocentric radius in our sample). \citet{MaizBarb18} measured $\AV = 1.388\pm0.018$ for HD~\num{64568}
and higher values for the other two stars, indicating there is little dust in the sightline and that the internal contribution is significant. The {\it Gaia} CMD gives a 
higher average extinction (as well as a larger dispersion in its values) for the southern cluster. 

We have obtained a LiLiMaRlin spectrum of HD~\num{64315}~AB. The spectral type we derive from it is O6~V~+~O7.5~V((f)), that is, the spectral types for both the primary and
secondary are later by half a spectral type than in the GOSSS result.

%\paragraph{HD~64\,315~A,B = V402~Pup~A,B = CPD~$-$26~2698~A,B = ALS~827~A,B.}         
%\object[HD 64315]{}         % GOS 243.16+00.36_01
%GOSSS-III:    O5.5   V             + O7     V                                 Ma    
%Villa-I:      O6     V             + O7.5   V((f))                            Ma      LiLiMaRlin fits/FEROS/HD_64_315_AB_061205_F_V48000.fits

In addition to the two cases already mentioned (HD~\num{64568} and HD~\num{64315}~AB) there are another five possible runaways from \VO{012}. As for the other groups, they are
listed in Table~\ref{runaways}.

\subsection{\VO{013} = Sh~2-158 = NGC~7538}

$\,\!$\indent This group is a poorly defined cluster embedded in the bright \HII\ region Sh~2-158. The majority of the ionizing photons originate in the primary star of the 
central binary system Sh~2-158~1 with a small contribution from the late-type O star Sh~2-158~2 \citep{Maizetal16}. A third bright object near the two, Tyc 4279-01349-1 is 
actually a foreground K star \citep{Wynnetal74}. Differential extinction is very strong and is likely to be hiding additional cluster members, especially to the south.

In GOSSS-III we classified Sh~2-158~1 as O3.5~V((f))~+~O9.5:~V. In Fig.~\ref{spectra} we present two new spectra of this object, one from GOSSS and the other one from
LiLiMaRlin, selected from several tens we have obtained with those projects. The primary is classified as O3.5~V and the secondary as O9.5:~V in all three cases. The 
suffix of the primary changes between spectra as a consequence of the varying strength of \CIII{4650} in emission and \HeII{4686} in absorption. We also observe rapid 
velocity changes from night to night in our spectra, a sign that this spectroscopic binary has a short period. One possibility is that the two stars are close enough 
for the object to be an eclipsing or ellipsoidal variable and that the suffix changes are being caused by different cross sections of the stars being exposed.
We checked the photometry in the ASAS-SN project \citep{Kochetal17} and we found no signs of eclipses.

%\paragraph{Sh~2-158~1 = Tyc~4279-01463-1.}      
%\object[Tyc 4279-01463-1]{} % GOS 111.53+00.82_02
%GOSSS-III:    O3.5   V((f))        + O9.5:  V                                 Ma    
%Villa-I:      O3.5   V((fc))       + O9.5:  V                                 Ma      GOSSS
%Villa-I:      O3.5   V((f))z       + O9.5:  V                                 Ma      LiLiMaRlin fits/NoMaDS/Sh_2-158_1_121203_H_B30000.fits

It is difficult to distinguish \VO{013} from the surrounding population, which is mostly located at distances similar to that of the cluster. Its proper
motion is somewhat more negative in RA but its most differentiating characteristic is its higher extinction, which is associated with the cluster and \HII\ 
region. This is clearly the poorest stellar cluster so far in our sample, with only eleven confirmed members. One reason why we do not detect more
cluster members in the {\it Gaia}~DR2 data is that the presence of strong nebulosity that contaminates \GBP\ and \GRP\ pushes some \dCC\ values above our selection
threshold.

The literature values in Table~\ref{literature} can be divided in two types: high kinematic distances and low values similar to the one found using our method.
In this case it is clear that the kinematic values are wrong, pointing to a possible peculiar velocity of the cluster, as it is not close to the Galactic center 
or anticenter. Both O-type systems are among the eleven confirmed members.

There are five possible runaways from \VO{013} in the {\it Gaia}~DR2 data from the positions and proper motions. The brightest one is red and bright, so it is
either a luminous, highly extinguished OB star or a fast-moving red giant coincidentally at the distance of the cluster. The other four candidates are bluer and
with positions in the CMD consistent with being B stars. One of them, [MO2001]~77, is listed as an object with H$\alpha$ emission by \citet{MikaOgur01}.

\subsection{\VO{014} = North~America~nebula = NGC~7000 = Sh 2-117}

$\,\!$\indent The North America nebula is one of the most famous ionized nebulae in the sky and also one of the closest and largest (in angular size) ones. It is 
located to the east of the Pelican nebula and the two of them are thought to be a single \HII\ region obscured at its center by a molecular cloud with the shape of the 
Atlantic Ocean and Gulf of Mexico (the latter part is called L935) that gives the North America nebula its distinct shape and name \citep{ReipSchn08,Zhanetal14}.

The ionizing star of the North America nebula was unknown until \citet{ComePasq05} 
used 2MASS photometry and visible+infrared spectroscopy to identify % REFEREE
a highly reddened early-O object and give it a preliminary spectral type of O5~V,
noting that they could not exclude an earlier spectral subtype. \citet{Maizetal16} observed it as part of GOSSS, discovered it is an SB2 with spectral types 
O3.5~III(f*)~+~O8:, and named it Bajamar star based on its location with respect to the nebula\footnote{As \citet{ComePasq05} put it, ``just East of the Florida 
Peninsula'', where they meant East in the geographical sense, not in the astronomical one. Note that Bajamar, low tide in Spanish, was the original name given to the 
Bahamas.}.

The literature on the distance to the North America nebula is long and goes back to the 1950s (Table~\ref{literature}). Previous results yielded wildly varying distances 
from 150~pc to 2~kpc, in good part due to the different assumptions about the ionizing source which, as mentioned above, was not identified until 2005. For example, 
\citet{Necketal80b} gave the shortest distance above based on the identification of a star (2MASS~J20535282$+$4424015) as the ionizing source but that object is actually a 
late-type object that {\it Gaia}~DR2 places at a distance of about 2~kpc.
The more recent, pre-{\it Gaia} results yielded distances that cluster around 500-600~pc and were mostly based on detecting the blocking
effect of the molecular clouds on the background stars. A similar technique using {\it Gaia}~DR2 \citep{Zucketal20} yields significantly longer distances around 
800~pc (with slightly different values for different parts of the molecular cloud). An analysis of the stars in the nearby Pelican nebula with {\it Gaia}~DR2 
\citep{Bharetal19} gives also a longer distance of $858\pm 56$~pc.

The Bajamar star has a {\it Gaia}~DR2 $\varpi = 1.473\pm 0.097$~mas, which after applying the same zero point and prior as for the rest of the targets in this
paper leads to a distance of $675^{+47}_{-42}$~pc. That distance places it between the pre-{\it Gaia} results and the {\it Gaia} ones for the molecular cloud (which have
a significant spread for different parts) and the Pelican nebula. This confirms that it is the main ionizing source of the region and makes it 
the only massive star earlier than O4 within 1~kpc, significantly closer than HD~\num{150136}~Aa,Ab, which is the second such star in terms of distance to the Sun.

The question now is what about the stellar group that the Bajamar star belongs to? \citet{ComePasq05} already noted the isolation of the star. They used a search
radius of half a degree and determined there were no companions earlier than B2~V. \citet{Damietal17c} searched for X-ray sources and found no concentration of young
low mass stars, specifically stating that ``Unlike most star-forming regions, this most massive star appears isolated even in X-ray images''. Given that, we used a
conservative approach of downloading from the {\it Gaia} archive a very large area of 2\degr$\times$2\degr\ around $\alpha=314.55\degr$, $\delta=+44.14\degr$. The 
top left panel of Fig.~\ref{North_America_Gaia} shows that the {\it Gaia} source density traces well the shape of North America, which indicates that the molecular cloud
stands in front of the vast majority of the stars in the field and that the ionized gas is optically thin, allowing us to see well beyond its location. Indeed, this is 
what can be seen in the panels where parallax is plotted, where the source density increases up to $\approx$3~kpc. Those circumstances preclude the functioning of our
standard method, as the alleged group would be a minor contaminant and it would be strongly affected by extinction, with the possible exception of some stars that could be
present on the near side of the molecular cloud. Furthermore, using a circular aperture around the Bajamar star soon starts including the region of the southeastern part
of the United States, where the source density is much higher.

To estimate the distance to the North America nebula we change the method in two ways: we substitute the circular aperture by an ad-hoc narrow polygonal aperture that traces
the core of the molecular cloud (effectively shielding most of the background population) and we establish a \GGc\ magnitude cut at 18.5. Also, we use HD~\num{199579} as 
our reference star for the isochrone, as the Bajamar star is not included in \citet{MaizBarb18}. The resulting sample are the 
12 stars in Table~\ref{North_America_sample}. The sample is clearly separated in proper motion from the field population, as expected from the difference in distance.
The Bajamar star is the most extinguished object among the 12, which could be partly a selection effect (low-mass objects with that extinction would be dimmer than 
\GGc\ = 18.5) but cannot be the whole story, as BA stars with its extinction should still be detected. The distance to \VO{014} from the 12 stars is 
$714^{+19}_{-16}$~pc, which is within one sigma of the distance to the Bajamar star, and indicates that pre-{\it Gaia} measurements from the last two decades in general
underestimated the distance to the nebula by 10-30\%. With that distance, the second and third stars in Table~\ref{North_America_sample}
have $\it Gaia$~DR2~+~2MASS photometry roughly consistent with being extinguished MS B-type stars embedded in the Gulf of Mexico + Atlantic Ocean molecular cloud (but
we cannot discard them being of late spectral type due to the intrinsic color-reddening degeneracy for {\it Gaia} colors). Note, 
however, that both stars are considerably far away in the plane of the sky from the Bajamar star (9~pc and 6~pc, respectively) so they cannot be part of the same bound
cluster. Instead, they are just stars being born from the same extended cloud. 

\begin{table}
\caption{{\it Gaia} DR2-selected stars in \VO{014}.}
\centerline{
\begin{tabular}{lr@{$\pm$}ll}
\hline
\mci{Gaia DR2 ID}         & \mcii{$\varpi$} & \mci{other ID}            \\
                          & \mcii{(mas)}    &                           \\
\hline
\num{2162889493831375488} & 1.473&0.097     & Bajamar star              \\
\num{2162977317322486144} & 1.334&0.054     & ---                       \\ % 2MASS J20532229$+$4427482 \\
\num{2162091866866267008} & 1.211&0.084     & ---                       \\ % 2MASS J20563353$+$4323385 \\
\num{2162128013311282304} & 1.458&0.065     & ---                       \\ % 2MASS J20581082$+$4353082 \\
\num{2162967421717817984} & 1.430&0.080     & ---                       \\ % 2MASS J20535299$+$4431234 \\
\num{2162965531932197504} & 1.372&0.112     & ---                       \\ % 2MASS J20534213$+$4426536 \\
\num{2162121244442829568} & 1.379&0.104     & ---                       \\ % 2MASS J20563412$+$4339036 \\
\num{2162902413092613248} & 1.317&0.134     & ---                       \\ % 2MASS J20555220$+$4409047 \\
\num{2162963019373410432} & 1.206&0.143     & ---                       \\ % 2MASS J20541776$+$4425275 \\
\num{2162123198648385024} & 1.433&0.152     & ---                       \\ % 2MASS J20570431$+$4344293 \\
\num{2162910930018064512} & 1.298&0.138     & ---                       \\ % 2MASS J20543040$+$4407118 \\
\num{2162909108944432256} & 1.262&0.165     & ---                       \\ % 2MASS J20551151$+$4409517 \\
\hline
\end{tabular}
}
\label{North_America_sample}                 
\end{table}

Previous works (e.g. \citealt{Laugetal06,StraLaug08a,Armoetal11,Damietal17c}) have suggested some stars are associated with the North America and Pelican nebulae.
Here we discuss them based on their {\it Gaia}~DR2 data:

\begin{itemize}
 \item HD~\num{199579} is an O6.5~V((f))z star \citep{Sotaetal11a} with a faint B-type companion \citep{Willetal01} that was proposed in the past as the ionizing source
       of the region \citep{SharOste52}. Its parallax ($1.0633\pm 0.0589$~mas) puts it beyond the nebula and its proper motion also differs from those in \VO{014}. 
 \item V1057~Cyg is an FU~Ori star whose parallax ($1.0864\pm 0.0388$~mas) places it a distance similar to that of HD~\num{199579} but with a quite different proper motion.
 \item V2493~Cyg, on the other hand, has a parallax ($\varpi = 1.2973\pm 0.0313$~mas) consistent with that of the 12 stars identified in \VO{014} and the only reason why 
       it was previously discarded was its proper motion being slightly outside the proper motion circle. Therefore, it is another likely member. 
 \item V1539~Cyg, LkHA~186, and LkHA~188 are three stars in a situation similar to V2493~Cyg, so they are additional likely members of \VO{014}. Indeed, those four
       objects are in the same region in the Gulf of Mexico and they are the brightest members of the closest structure to a cluster in the region, which also includes
       Gaia~DR2~\num{2162128013311282304}, one of the objects in Table~\ref{North_America_sample}.
 \item On the other hand, LkHA~187 and LkHA~189, despite being in the same region of the Gulf of Mexico, have parallaxes that put them beyond the nebula. Their proper 
       motions are also closer to those of the (more distant) field population.
 \item Objects 1, 4, 7, and 9 of \citet{StraLaug08a}, who they suspected could be extinguished O stars, have {\it Gaia}~DR2 parallaxes with relatively large
       uncertainties but they all appear to be significantly beyond \VO{014}.
\end{itemize}

For \VO{014} it is complicated to search for runaways in the same manner as for the rest of the sample, as there is no defined center to run away from. 
For the sake of completeness we have searched in the low-extinction regions of the field for bright stars with parallaxes compatible with that of the North America nebula.
Here are the most significant cases:

\begin{itemize}
 \item HD~\num{200102} is a sixth-magnitude star with a Simbad classification of G1~Ib in the upper-midwest United States part of the nebula. It has a proper motion significantly different 
       from those of the stars in Table~\ref{North_America_sample} moving towards the NW, which combined with its spectral type (which suggests a non-coeval age with an early-O star) 
       points towards being a field star. 
 \item V354~Cyg is listed as a long period variable in Simbad and is located in the southwestern United States part of the nebula. Its proper motion points away from the Gulf of Mexico 
       so it could be an ejected PMS star.
 \item Simbad gives a spectral classification of K4~III for Tyc~3179-00416-1, which is located in the Central America region of the nebula and moving southward away from the Gulf of
       Mexico. Therefore, this is another potentially ejected PMS star.
 \item Another interesting case is Tyc~3179-00023-1, located close to the previous one, but moving away from the Caribbean Sea or Atlantic Ocean regions, possibly even the location
       of the Bajamar star. Simbad gives a spectral type of B9~IV, so if it is a young runaway it would be already close to the main sequence.
 \item Tyc~3179-01439-1 is located near the Yucatan peninsula but moving towards the Gulf of Mexico instead of away from it, so it is likely a field star. Simbad gives a classification
       of A4~V.
 \item ALS~\num{11602} is near the edge of the field, close to the Pelican nebula. Its parallax and its proper motion are similar to those of the stars in Table~\ref{North_America_sample},
       and it is a B2~Vn star according to \citet{Straetal99}, so it is a likely member of the association, though a distant one from the Bajamar star.
 \item Finally, [SKV93]~2-72 is close to the previous object but has a very different proper motion. It is a fast moving object and its proper motion traces back to a region close to
       the Bajamar star, making it another possible runaway star. \citet{Straetal93} classify it as K0~III-IV.
\end{itemize}

In summary, the Bajamar star seems to be a genuine case of a massive binary star born in near-isolation 
i.e. with only a small number of intermediate- and low-mass stars around it. 
There are other stars being born from the same molecular cloud but they are considerably less massive and not bound to the Bajamar star in a cluster-like fashion.
What we call \VO{014} in this paper is a poor stellar association instead and the North America and Pelican nebulae have no clearly defined, concentrated cluster despite 
being significantly larger (in size and number of ionizing photons) than other nearby H~{\sc ii} regions such as the Orion nebula or NGC~2264, whose central clusters are well 
established. % REFEREE

After this paper was submitted to the journal, an independent analysis of the North America and Pelican nebulae using {\it Gaia}~DR2 data was published and the referee dutifully 
pointed it to us \citep{Kuhnetal20}. Most of the results in that paper agree with what we present here but there are some small differences that we discuss now.
First, the authors put into doubt that the Bajamar star is a double-lined spectroscopic binary based on a single-epoch spectroscopic observation.
We have obtained multi-epoch spectroscopy of the system and we can indeed confirm that the absorption lines move as expected for such a binary. We are currently working on a paper with an 
orbit for the system. Regarding the distance, \citet{Kuhnetal20} find a mean parallax for their group E (objects in the Gulf of Mexico) of 1.27$\pm$0.02~mas, which can be compared to our
non-zero-point-corrected value of 1.354$\pm$0.029~mas. Their value for the dispersion apparently does not include the spatial covariance term (their section 7.4) but including it would leave
the difference between our two results at a two-sigma level. Finally, the paper suggests that the Bajamar star may have originated from the region of the Pelican nebula (their group D) and 
they calculate that it is moving away from there with a relative velocity of 6.6$\pm$0.5~km/s. That scenario is plausible but we note that it is far below the standard threshold for runaway
stars and that some refer to such objects as ``walkaway stars'' \citep{Renzetal19b}. Note that \citet{Kuhnetal20} place their group D beyond their group E and that the parallax for the 
Bajamar star places it even closer albeit with a relatively large uncertainty. Hopefully, future {\it Gaia} releases will shed some light into these (relatively small) distance discrepancies.
In any case, if the walkaway scenario from group D were true it would not change our basic conclusion, as the Bajamar star would still be the only O-type system in the region and we would 
still have a case of a very massive star forming in near-isolation.

%Stars in \citet{StraLaug08a}:
% 1 (2MASS J20495723+4422538) has $\varpi =  0.8067\pm 0.7300$ suspected O, far from Bajamar star.
% 2 (2MASS J20503505+4426296) has $\varpi =  1.2822\pm 0.2586$ suspected C rich object.
% 3 (2MASS J20524679+4402308) has $\varpi =  0.6813\pm 0.8029$.
% 4 (2MASS J20531582+4432570) has $\varpi =  0.4749\pm 0.4282$ suspected O, far from Bajamar star.
% 5 (2MASS J20541342+4402589) has $\varpi =  0.0778\pm 0.1727$.
% 6 (2MASS J20541626+4343091) has $\varpi =  0.2310\pm 0.1658$.
% 7 (2MASS J20552516+4418144) has $\varpi =  0.2191\pm 0.1664$ suspected O.
% 9 (2MASS J20555270+4353242) has $\varpi =  0.4819\pm 0.3048$ suspected O.
%10 (2MASS J20573647+4404559) has $\varpi = -0.1038\pm 0.2352$.
%11 (2MASS J20580673+4355141) is not in {\it Gaia} DR2.
%12 (2MASS J20582424+4356386) is not in {\it Gaia} DR2.
%13 (2MASS J20582622+4342385) has $\varpi =  0.5011\pm 0.1076$.

\subsection{\VO{015} = Collinder~419}

$\,\!$\indent This cluster was studied in MA19, here we just summarize what was said there. 
As previously mentioned, we include it here as one of the (many future) members of the Villafranca catalog of groups with O stars even though it has no objects 
of the O2-3.5 subtypes.
Collinder~419 (\VO{015}) % \VO{015} REFEREE
is a relatively unstudied cluster in Cygnus
dominated by the O-type system HD~\num{193322}. It is quite poor, as MA19 only confirmed 75 members, and can be described as a small concentration around the
O-type system surrounded by an asymmetric halo. It is much better defined in proper motion than in position, as it sits in front of a rich background
Galactic population at a distance of 2-5~kpc. The cluster experiences low extinction and MA19 derives a distance of $1.006^{+0.037}_{-0.034}$~kpc.
HD~\num{193322} is a complex system with at least two O stars, but of late subtype compared to the much earlier types present in \VO{001} to \VO{014}.

MA19 noted the existence of a late-type giant, 2MASS~J20175763$+$4044373, with a parallax compatible with that of 
Collinder~419 % \VO{015} REFEREE
and a peculiar NE-SW motion that indicated a possible ejection from a high-extinction region to the NE. Here we list in Table~\ref{runaways} another five stars 
with similar parallaxes and an anomalous motion in the opposite direction and away from 
Collinder~419. % \VO{015} REFEREE
They could be runaways from the cluster or constitute an independent moving group.

\subsection{\VO{016} = NGC~2264 = Sh~2-273}

$\,\!$\indent This object was also studied in MA19 and here we provide a summary of those results. 
As previously mentioned, we include it here as one of the (many future) members of the Villafranca catalog of groups with O stars even though it has no objects 
of the O2-3.5 subtypes.
NGC~2264 (\VO{016}) % \VO{016} REFEREE
is a well-known cluster and a favorite target
of amateur astronomers due to its associated \HII\ region. It has a double-cluster structure, with the northern core centered around the O-type multiple
system 15~Mon and the southern core around the Cone nebula, with possibly more embedded cores. It is a richer cluster than 
Collinder~419, % \VO{015} REFEREE
with 286 members confirmed by MA19, and also clearly defined in proper motion. The molecular cloud associated with the \HII\ region acts as a screen blocking the
background population, located at much longer distances that the value of 719$\pm$16~pc determined by MA19, with no significant differences in distance between the 
two cores. The interesting new result by \citet{Zucketal20} finds distances to the associated molecular clouds consistent with our distance result but with the 
clouds around the edges somewhat closer to us than the one located at the same position as the cluster core. This would be consistent with the cluster having 
carved a hole on the near side of the molecular cloud that lets us see the cluster with little extinction with the cloud still blocking the view of the background. 
NGC~2264 % \VO{016} REFEREE
appears to be very young, as indicated by the \HII\ region and its associated structures, the embedded cores, and the z
suffix in the O7~V((f))z spectral classification of 15~Mon~Aa. Note that since the publication of MA19, 15~Mon has been spectroscopically separated into its 
Aa and Ab components by \citet{MaizBarb20}.

For this object we have also searched for possible runaways and detected two (Table~\ref{runaways}). In both cases their proper motions indicate a more likely
ejection from the northern core than from the southern one.

\section{Analysis and future work}

\subsection{Comparison with previous distances}

\begin{figure*}
\centerline{\includegraphics*[width=\linewidth]{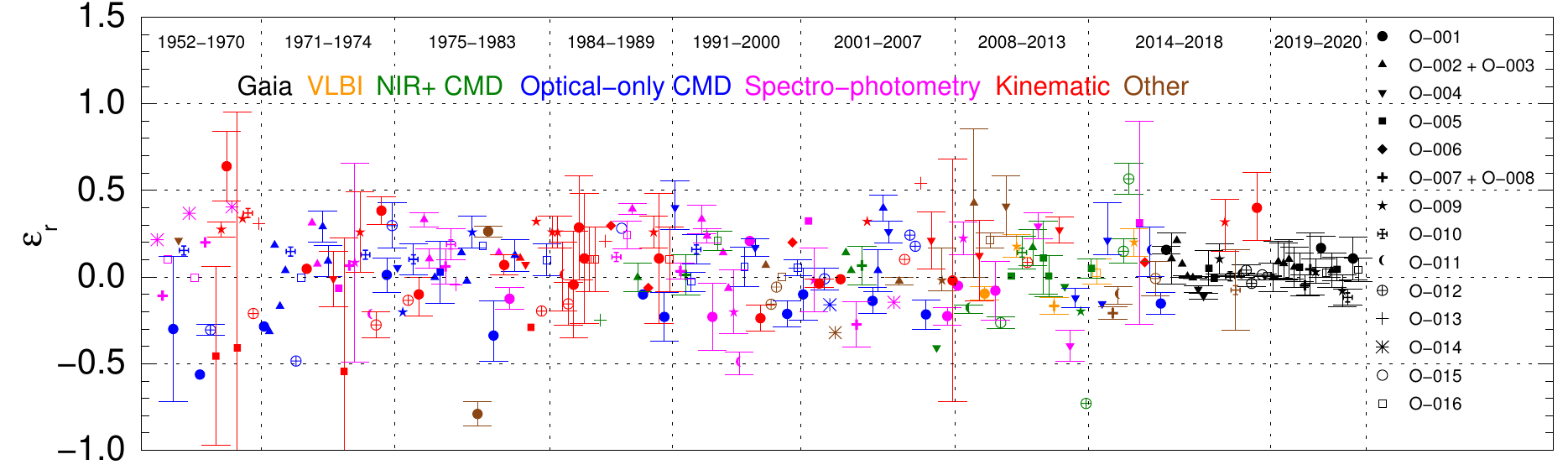}}
\caption{Fractional distance difference of the literature distance measurements $d$ with respect to the values reported in this paper \dr. Colors are used 
         to encode the method used and symbols to encode the group. Symbols without error bars correspond to measurements without uncertainties and those with them
         reflect only the uncertainty in $d$, not in \dr.}
\label{reldis}
\end{figure*}   

\begin{table}[h!]
\caption{Distance statistics as a function of year range, stellar group, method, and first author. For the latter, only those with four or more measurements are listed.}
\centerline{
\renewcommand{\arraystretch}{1.2}
\begin{tabular}{lrrrrrr}
\hline
Type               & \mci{$\overline{\varepsilon}$} & \mci{$\sigma_\varepsilon$} & \mci{$N_\varepsilon$} & \mci{$\overline{d_{\rm n}}$} & \mci{$\sigma_{d_{\rm n}}$} & \mci{$N_{d_{\rm n}}$} \\
\hline
Year range         &                                &                            &                       &                              &                            &                       \\
\hline
1952-1970          &                           0.06 &                       0.34 &                    20 &                         0.63 &                       2.32 &                     6 \\
1971-1974          &                           0.00 &                       0.25 &                    25 &                         0.74 &                       1.43 &                    10 \\
1975-1983          &                           0.01 &                       0.23 &                    29 &                         0.40 &                       3.70 &                    15 \\
1984-1989          &                           0.10 &                       0.18 &                    23 &                         1.04 &                       1.74 &                    16 \\
1991-2000          &                           0.03 &                       0.21 &                    24 &                         0.44 &                       1.84 &                    16 \\
2001-2007          &                           0.03 &                       0.22 &                    29 &                         0.00 &                       1.48 &                    16 \\
2008-2013          &                           0.01 &                       0.26 &                    25 &                         0.28 &                       2.08 &                    19 \\
2014-2018          &                           0.07 &                       0.16 &                    34 &                         0.37 &                       1.26 &                    30 \\
2019-2020          &                           0.04 &                       0.07 &                    17 &                         0.42 &                       0.68 &                    17 \\
\hline
Group              &                                &                            &                       &                              &                            &                       \\
\hline
O-001              &                        $-$0.10 &                       0.18 &                    18 &                      $-$0.31 &                       0.51 &                    11 \\
O-002 + O-003      &                           0.11 &                       0.15 &                    41 &                         1.30 &                       1.58 &                    23 \\
O-004              &                           0.07 &                       0.23 &                    20 &                         0.66 &                       1.18 &                    14 \\
O-005              &                        $-$0.05 &                       0.25 &                    16 &                         0.10 &                       0.50 &                    12 \\
O-006              &                           0.13 &                       0.15 &                     4 &                          --- &                        --- &                     0 \\
O-007 + O-008      &                        $-$0.03 &                       0.14 &                    12 &                      $-$0.57 &                       1.23 &                     9 \\
O-009              &                           0.13 &                       0.18 &                    22 &                         1.31 &                       1.37 &                    17 \\
O-010              &                           0.11 &                       0.13 &                    12 &                         0.14 &                       0.25 &                     7 \\
O-011              &                        $-$0.12 &                       0.21 &                     7 &                      $-$0.83 &                       1.57 &                     6 \\
O-012              &                        $-$0.05 &                       0.28 &                    21 &                         0.34 &                       1.73 &                    10 \\
O-013              &                           0.10 &                       0.28 &                    11 &                         0.41 &                       1.83 &                     6 \\
O-014              &                           0.01 &                       0.33 &                    20 &                      $-$0.25 &                       3.91 &                    14 \\
O-015              &                           0.01 &                       0.27 &                     3 &                      $-$2.46 &                       3.99 &                     2 \\
O-016              &                           0.08 &                       0.08 &                    19 &                         1.20 &                       1.44 &                    14 \\
\hline
Method             &                                &                            &                       &                              &                            &                       \\
\hline
{\it Gaia}         &                           0.03 &                       0.07 &                    34 &                         0.30 &                       0.57 &                    32 \\
VLBI               &                           0.03 &                       0.16 &                     5 &                         0.34 &                       1.71 &                     5 \\
NIR+ CMD           &                        $-$0.02 &                       0.26 &                    21 &                         0.42 &                       2.32 &                    13 \\
Optical-only CMD   &                           0.01 &                       0.22 &                    54 &                         0.33 &                       1.43 &                    31 \\
Spectro-photometry &                           0.06 &                       0.22 &                    43 &                         0.61 &                       2.17 &                    22 \\
Kinematic          &                           0.09 &                       0.25 &                    52 &                         0.77 &                       1.36 &                    31 \\
Other              &                        $-$0.01 &                       0.29 &                    17 &                         0.02 &                       4.46 &                    11 \\
\hline
First author       &                                &                            &                       &                              &                            &                       \\
\hline
Binder             &                           0.05 &                       0.08 &                     4 &                         0.41 &                       0.55 &                     4 \\
Becker             &                        $-$0.05 &                       0.29 &                     8 &                          --- &                        --- &                     0 \\
Carraro            &                           0.08 &                       0.35 &                     4 &                         1.04 &                       2.47 &                     4 \\
Cantat-Gaudin      &                           0.01 &                       0.06 &                    11 &                         0.07 &                       0.32 &                    11 \\
Fich               &                           0.05 &                       0.14 &                     6 &                         0.42 &                       1.07 &                     6 \\
Georgelin          &                           0.08 &                       0.32 &                    10 &                         0.48 &                       1.83 &                     5 \\
Humphreys          &                           0.17 &                       0.12 &                     4 &                         1.50 &                       1.75 &                     4 \\
Kuhn               &                           0.06 &                       0.03 &                     5 &                         0.54 &                       0.15 &                     5 \\
Moffat             &                           0.09 &                       0.19 &                     5 &                         0.59 &                       1.40 &                     3 \\
Massey             &                           0.23 &                       0.14 &                     4 &                         2.05 &                       1.43 &                     3 \\
Russeil            &                           0.10 &                       0.23 &                     5 &                      $-$0.18 &                       1.21 &                     3 \\
Shaver             &                           0.01 &                       0.22 &                     6 &                         0.27 &                        --- &                     1 \\
Stark              &                           0.19 &                       0.09 &                     6 &                         1.20 &                       0.99 &                     6 \\
Th\'e              &                           0.00 &                       0.13 &                     4 &                          --- &                        --- &                     0 \\
Walborn            &                           0.11 &                       0.12 &                     6 &                          --- &                        --- &                     0 \\
Zucker             &                           0.00 &                       0.09 &                     5 &                         0.16 &                       0.62 &                     5 \\
\hline
\end{tabular}
\renewcommand{\arraystretch}{1.0}
}
\label{diststat}
\end{table}

$\,\!$\indent In this paper we have presented {\it Gaia}~DR2 distances to 16 stellar groups with O stars, \VO{001} to \VO{016}. Two of those (\VO{012} and \VO{016}) are
double clusters located at the same distance (within our measurement errors) and another two pairs (\VO{002} + \VO{003} and \VO{007} + \VO{008}) have been analyzed
separately but we have also determined they are also likely physically related, so a common single distance can be adopted for each pair 
(besides the long list of references in the corresponding subsections and in Table~\ref{literature}, the interested reader can find additional information on such 
physical associations in \citealt{TurnMoff80,Piatetal10,Huretal12,ReitPark19}).
This leaves us with a total of 14
distances to stellar groups \dr\ (or pairs of them) for which we have collected 226 literature distances $d$ (i.e. an average of 16.1 measurements per stellar 
group)\footnote{We exclude from the sample the kinematic distance to 
M17 % REFEREE \VO{009} 
by \citealt{Quiretal06} because it is an extreme outlier that distorts the analysis.}. In
this subsection we analyze the accuracy of the literature distances according to different parameters. In Fig.~\ref{reldis} we plot the fractional distance difference 
$\varepsilon = (d-\dr)/{\rm max}(d,\dr)$ in chronological order using colors and symbols to encode methods and groups, respectively\footnote{We define $\varepsilon$ with
the maximum of $d$ and \dr\ in the denominator to avoid outliers in the resulting distribution caused by one value being many times greater than the other.}. Of the 226 
literature distances, 145 have uncertainty measurements and for those we have computed the uncertainty for the difference $d-\dr$, 
$\st = \sqrt{\sigma_d^2 + \sigma_{d_{\rm r}}^2}$, and calculated 
the normalized deviation of the literature distance from our value $\dn = (d-\dr)/\st$, which encodes not only the accuracy of the literature values but also of their 
uncertainty estimates. In Table~\ref{diststat} we display the basic statistics of $\varepsilon$ and \dn\ (average, standard deviation, and number of measurements) as a function 
of publication year range, stellar group, method, and first author. Ideally, both averages should be zero for unbiased measurements, $\sigma_\varepsilon$ should be as low as 
possible for a better precision of the average distance, and $\sigma_{d_{\rm n}}$ should be close to 1 for independent measurements with correctly estimated uncertainties.

An overall feature of Table~\ref{diststat} is that positive values of $\overline{\varepsilon}$ and $\overline{\dn}$ dominate over negative ones. Therefore, our distances
tend to be shorter than literature values. The difference is small when compared to other {\it Gaia}~DR2 results, for which $\varepsilon$ is just $3\pm 7\%$ overall and 
$1\pm 6\%$ for the \citet{CanGetal18} values, showing the extent of the small effect of sample selection and different parallax zero points. Also note that 
$\sigma_{d_{\rm n}}$ is lower than 1 for {\it Gaia}~DR2 results, as the measurements are not independent. 

The evolution with time shows no significant change of $\overline{\varepsilon}$ i.e. overall distances have not been getting shorter or longer. 
However, there is an
evolution of the dispersion. The highest value is for results older than fifty years. Then comes a long period with little change in the dispersion, which is only
significantly reduced when {\it Gaia}~DR2 results appear. We could ascribe that to the better quality of {\it Gaia}~DR2 results but one should keep in mind that the comparison
is not between independent data, so the final verdict should come from a future confirmation.  Comparing different methods results in few differences. 
VLBI parallaxes have a lower dispersion than other non-{\it Gaia} methods but run into the problem of the sources not being necessarily at the same distance as the 
group.   
Using the NIR (with or without the optical) to build CMDs does not provide a significant advantage over optical-only equivalents. Spectro-photometry and, especially, 
kinematic distances overestimate (on average) distances more than other methods but in all cases the dispersion is significantly larger than $\overline{\varepsilon}$.

Using as a distance to a group the average of the literature values produces a result that is at worst 13\% off from our value. The best predictor of how large the
dispersion in the literature measurements is comes from the richness of the group: those with few stars or with low contrast with the field population (i.e. \VO{012} to 
\VO{014}) have a larger scatter than concentrated clusters with many stars 
(e.g. NGC~3603 or Trumpler~14) % REFEREE (i.e. \VO{001} or \VO{002}) 
or those that are easily differentiated from the field population
(e.g. NGC~6193 or NGC~2244). % REFEREE (i.e. \VO{010} or \VO{016}). 
As for individual first authors, those that use primarily {\it Gaia}~DR2 distances (Binder, Cantat-Gaudin, Kuhn, and Zucker) have values of 
$\overline{\varepsilon}$ close to zero and low values of $\sigma_\varepsilon$, as expected. There are significant differences among the rest of first authors. Fich, 
Th\'e, and Walborn produce the best-quality results. In the other extreme, the distances by Becker, Carraro, and Georgelin yield the largest dispersions and those by 
Humphreys, Stark, and especially Massey produce the largest overestimates.

\subsection{Identifying group members}

$\,\!$\indent Many 
group-finding 
algorithms (e.g. \citealt{CanGetal18}) give a large weight to proper motions in the identification of members of a stellar 
group. 
The results in this paper indicate that proper motions are indeed quite useful for that task when the distance is less than 1.5~kpc but for higher values they become less so, as the
differences in proper motion decrease as distance 
increases. For massive young clusters, such as the ones analyzed here, giving more weight to the CMD is recommended, as the
extinction usually associated with the cluster allows for a better discrimination. This is not necessarily a general result, as astrometry with better precision may
increase the distance range at which proper motions are good discriminants and as older 
groups 
should not have a significant associated extinction, but it appears to be true
under the circumstances described here. We plan to test this hypothesis in future papers. 

Even though not used here, radial velocities are another potential discriminant for 
group 
membership, especially as the only dependence of their quality with distance is that 
given by S/N and their values do not tend to zero at infinity as proper motions do. Ongoing multifiber ground-based surveys and future {\it Gaia}~DR2 data releases should be 
useful in this respect. Note, however, two important caveats. O and B stars (especially the former) have few useful lines for radial velocities in the Calcium triplet region used
by the {\it Gaia} RVS instrument, so for those stars one needs to resort to ground-based surveys. Also, a large fraction of OB stars are spectroscopic binaries so single-epoch
spectroscopy will not do the job correctly. 

Finally, we point out a problem with {\it Gaia} photometric data. \GG\ is obtained from PSF fitting of image-like data (actually, one coordinate is spatial and the other one is temporal
as the spacecraft sweeps the sky) but \GBP\ and \GRP\ are obtained through aperture photometry of slitless spectrophotometric data. This makes the latter two quite sensitive to stellar 
crowding and nebular 
contamination. 
In some stellar groups (disperse and with no nebulosity) the effect is small but in others, such as \VO{013}, it is not and causes our algorithm to reject 
many stars based on the \dCC\ criterion. This is one case where ground-based photometric surveys such as GALANTE can complement {\it Gaia} data and improve on the sample selection. In 
addition, {\it Gaia}~DR2 (and the future EDR3) provides little information on the spectral energy distribution (SED) to the left of the Balmer jump, given the low sensitivity of \GBP\ at 
those wavelengths \citep{MaizWeil18}, and that part of the SED is crucial to correctly measure the effective temperature of OB stars without resorting to expensive spectroscopy 
\citep{MaizSota08,Maizetal14a}. That is another aspect where ground-based surveys can complement {\it Gaia} with a $u$-like filter. In principle, {\it Gaia}~DR3 should help with this issue
when it releases the \GBP-associated spectrophotometry but we note that there may be some calibration issues for a significant number of OB stars given the combination of the \GBP\
sensitivity profile and the reddened character of most of the Galactic OB SEDs \citep{Weiletal20}.

\subsection{Internal motions}

\begin{figure}
\centerline{\includegraphics*[width=\linewidth]{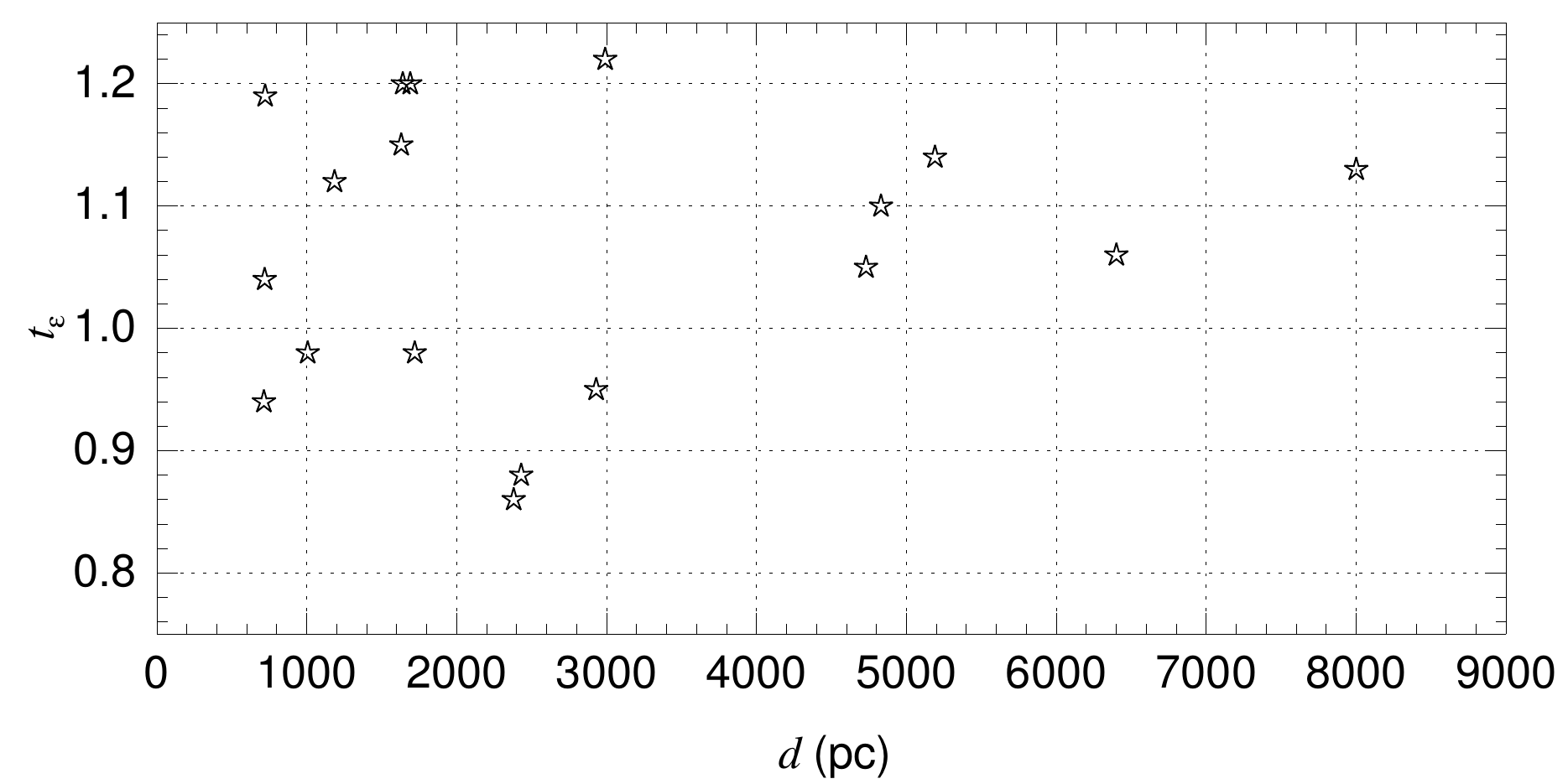}}
\centerline{\includegraphics*[width=\linewidth]{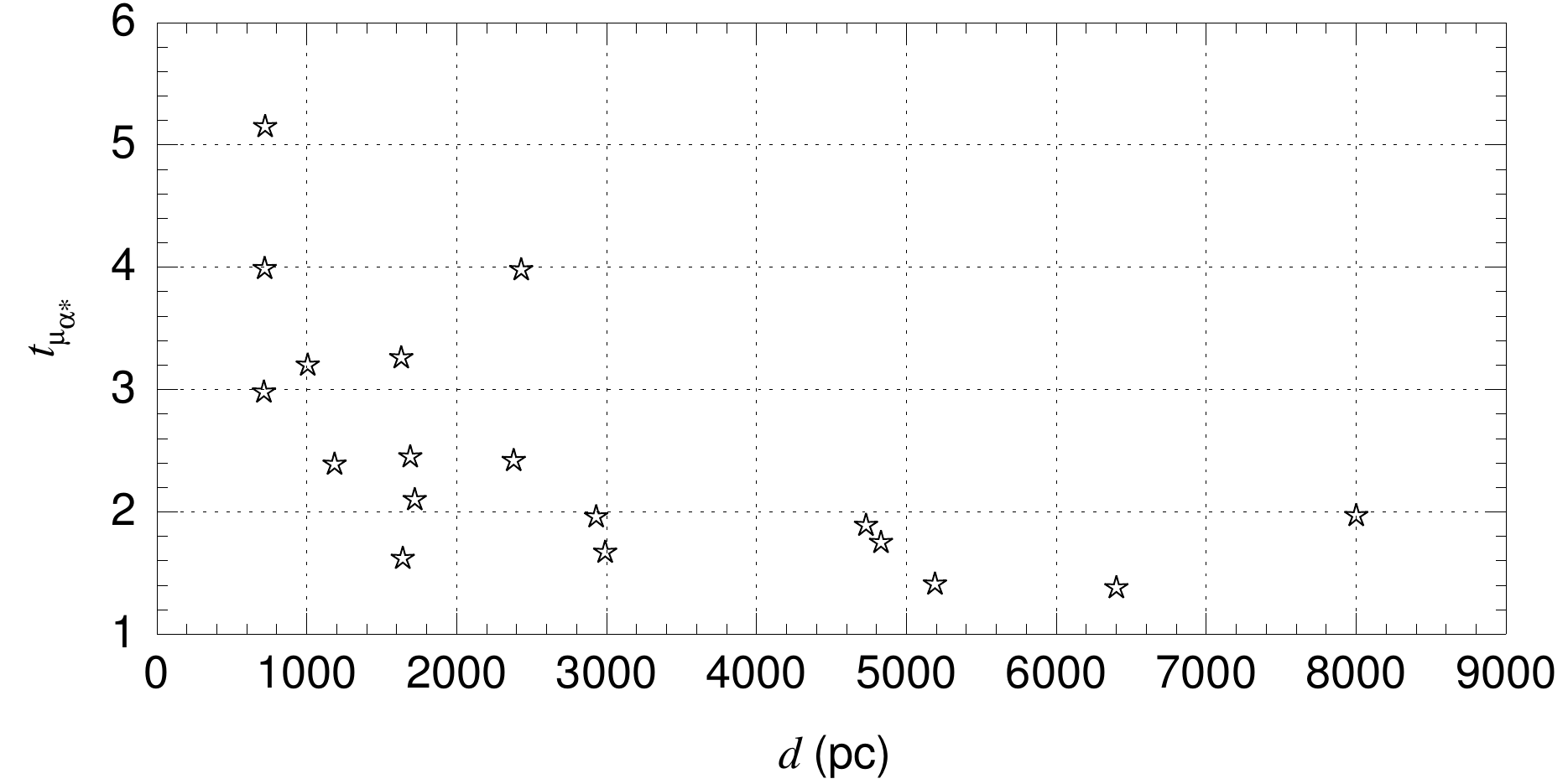}}
\centerline{\includegraphics*[width=\linewidth]{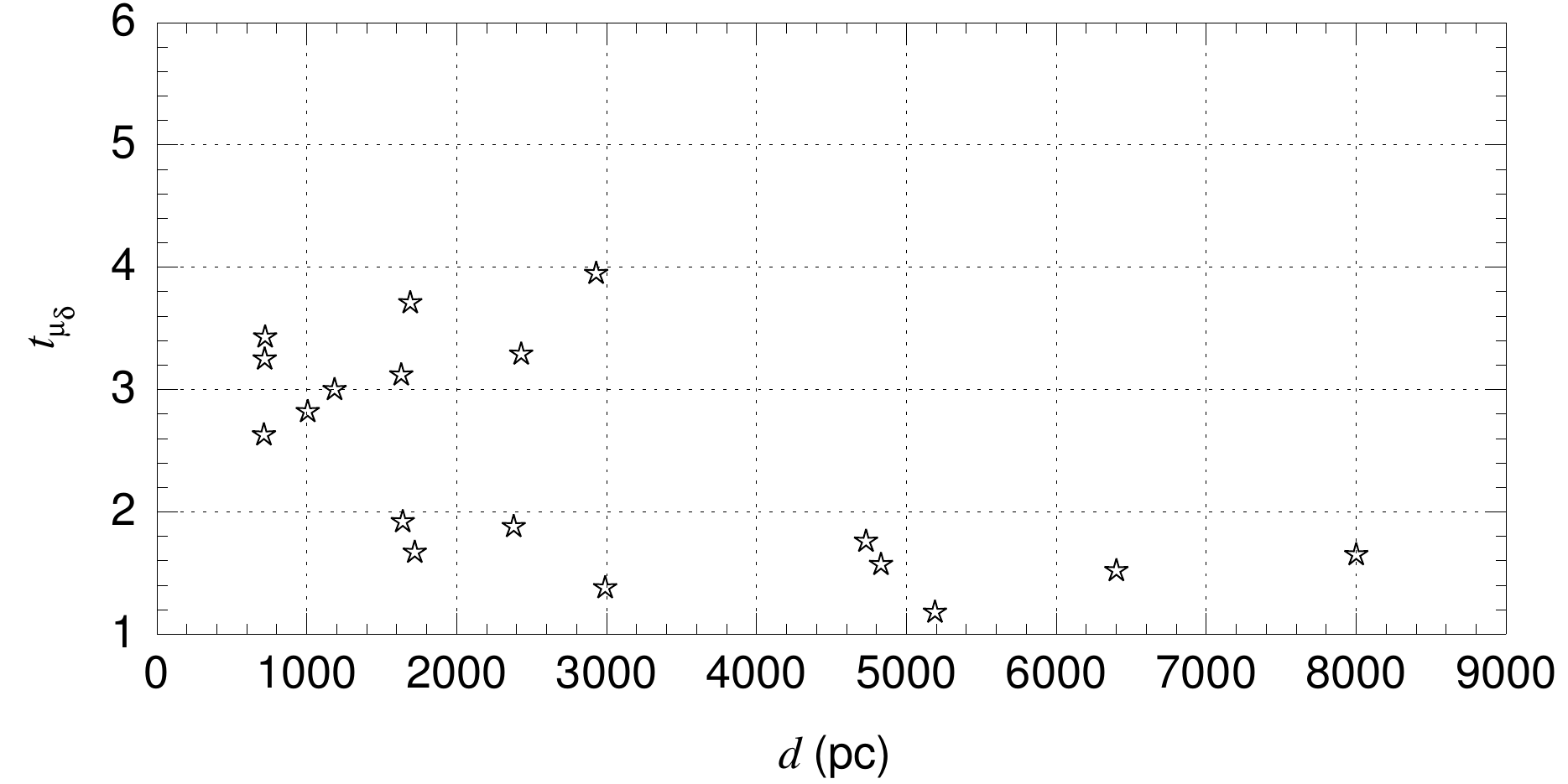}}
\caption{Statistical tests $t_\varpi$, $t_{\mu_{\alpha *}}$, and $t_{\mu_{\delta}}$ as a function of distance for the groups in this paper including \VO{012}~N, 
         \VO{012}~S, \VO{016}~N, and \VO{016}~S separately.}
\label{tstat}
\end{figure}   

$\,\!$\indent In Fig.~\ref{tstat} we plot the statistical tests $t_\varpi$, $t_{\mu_{\alpha *}}$, and $t_{\mu_{\delta}}$ as a function of distance for the groups studied here. 
$t_\varpi$ shows no trend with distance and the values cluster around 1, indicating the sample selection process works well and the final result should have few contaminants 
(non-group members).  The plots for the proper motion statistical tests are different. All values are above 1, indicating that there are likely internal group 
motions affecting the stellar proper motions. A trend with distance is also clear: for distances longer than 4~kpc the two proper motion tests stay below 2 while
for shorter distances the two tests (especially $t_{\mu_{\alpha *}}$) increase as we move towards zero. This is an effect of the proper motions being inversely proportional
to the distance and shows that {\it Gaia}~DR2 is limited in the detection of internal proper motions. The two exceptions to the trend are 
Pismis~24 % \VO{005} REFEREE
and, especially,
Trumpler~14. % \VO{002} REFEREE
The likely explanation is that those two are the most massive compact clusters in our sample within 4~kpc.

\subsection{Richness and the IMF}

$\,\!$\indent What is the relationship between the richness of a cluster, defined as the concentration of many stars in a small volume, and its initial mass function (IMF)?
In the view of \citet{Krou04} the IMF is universal but the mass of a cluster correlates with the maximum stellar mass, as clusters of small mass will
not be able to form very massive stars. Therefore, as \citet{WeidKrou06} put it, ``$10^4$ clusters of mass $10^2$~M$_\odot$ will not produce the same IMF as one cluster 
with a mass of $10^6$~M$_\odot$'', implying that the second option should have a larger proportion of massive stars. A corollary of this hypothesis is that massive stars 
should not be able to form in isolation and that any such object found without nearby massive companions should be a runaway \citep{Gvaretal12a}.

There are several observational facts that seem to contradict that hypothesis. The first one is that massive stars are seen in associations \citep{Amba58}, which agrees
with star formation being a hierarchical process that happens in both bound and unbound clouds with a wide range of scales \citep{Elme10}. Associations can be rich in
massive stars even if they do not have well-defined clusters in them and they can be scaled up to large masses, with two objects with structures as different as 30~Dor 
and NGC~604 having formed similar numbers of massive stars \citep{Maiz01b}. One past critique of this has been that associations form as clusters that lose their gas 
rapidly and disperse until we see them as unbound structures. However, this critique has been disproven with modern data for the northern association with the largest 
number of O stars (\citealt{Wrigetal14}, the title of the paper says it clearly: ``Cygnus~OB2 was always an association'') and the results in this paper confirm that:
Cyg~OB2 retains a double core, each with a normal velocity dispersion as determined from the proper motions (see previous subsection) and no abnormal relative velocity 
between them, in line with the \citet{Wrigetal16} analysis that reveals no global expansion pattern. 

Another contradictory observational fact is the existence of massive stars that appear to be truly isolated while not being runaways. \citet{Bresetal12} found 15 such
objects in 30~Dor, for which they noted they could not be line-of-sight runaways based on their radial velocities. A second study \citep{Platetal18} discarded the 
possibility that most of them could be plane-of-the-sky runaway stars, indicating they are true cases of isolated massive-star formation. Of course, 30~Dor is 50~kpc away
and even though \citet{Bresetal12} used HST images to try to discard the existence of a multiple system or mini-cluster around their targets, the spatial resolution at that
distance could not rule that out. In this paper, however, we have presented the case of the Bajamar star, which is located almost two orders of magnitude closer than
30~Doradus and, furthermore, with a primary of earliest type than any of the \citet{Bresetal12} objects. It is located near the center of the molecular cloud from where it
appears to have been born, 
shows only a small relative velocity with respect to its natal cloud \citep{Kuhnetal20},
and there are no other ejected O stars in the vicinity seen in the {\it Gaia}~DR2 data.
It is a short-period spectroscopic binary with approximate masses (as estimated from their spectral types) of $\gtrsim$50~M$_\odot$ and $\sim$25~M$_\odot$, there is no
cluster around it, and the few relatively nearby association members are at most of intermediate mass. Therefore, it is a true counterexample to the hypothesis that massive
stars (or 
massive
binary systems) cannot form in relative isolation.

Another attempt at salvaging the \citet{WeidKrou06} hypothesis is the idea that very massive stars can exist in not so massive clusters because they form by collisions 
resulting from three-body interactions after the formation of the cluster \citep{OhKrou18}. Such three-body encounters indeed take place in clusters and this is one of 
the two classic mechanisms that produce O-type runaways \citep{Poveetal67,Hoogetal01,Maizetal18b}. However, for a runaway to be produced in such a way one needs a very 
compact cluster to start with, as otherwise there is little chance that the three-body interaction will take place. No such cluster exists in the case of the Bajamar star in 
the North America nebula, so that is not the way that system could have formed. Note that in that case we would also have to explain that what we currently have is a 
short-period spectroscopic binary, implying that at least four stars should have been involved in the interaction (the two progenitors of the primary, the secondary, and 
the ejected star). We already made some of these points regarding LS~III~$+$46~11 and LS~III~$+$46~12 in 
Berkeley~90 % REFEREE \VO{011} 
in \citet{Maizetal15a}. With the results in this paper and the non-detection of ejected O stars from the cluster we confirm that 
Berkeley~90 % REFEREE \VO{011} 
is an example of a low-mass cluster with two very massive systems and no other O stars that does not have the high stellar density at its core required for mergers from 
three-body interactions to be likely. Finally, we point to the case of \VO{013}, which is rather similar to 
Berkeley~90 % REFEREE \VO{011} 
but with an even poorer stellar density at its core: one very massive spectroscopic system, another O star, no other good candidates for being
massive stars above 15~M$_\odot$, and no massive runaways. In summary, the results in this paper clearly point in the direction that {\it massive stars can form in 
near-isolation or in relatively low-mass clusters}
and that star formation is a hierarchical process.

A corollary to this conclusion is that a fixed criterion to search for a stellar group around an O star cannot be established, as we had already anticipated in the sample 
selection subsection above. If the star(s) is/are immersed in a rich cluster, the task is easy. If not and a poorer cluster exists, one may still be detect the cluster if 
the contrast with the field population is high enough. As a third option, if a cluster is not seen, the star may be located in an association but identifying its members may be hard if it 
extends over a large region of the sky. Finally, there is the chance that the O star is isolated or nearly so, which could be caused by [a] being a runaway, in which case one 
needs to analyze its kinematics to search for an origin in a known stellar group, or by [b] having truly formed in near-isolation, in which case one can develop an ad-hoc method 
to see if any lower-mass stars can be detected in the vicinity with comparable distances and kinematics.

\subsection{The Gaia DR2 parallax zero point}

$\,\!$\indent \citet{Lindetal18a} presented the astrometric results from {\it Gaia}~DR2 and detected the existence of a parallax zero point
of $\approx$30~$\mu$as in the sense that {\it Gaia}~DR2 parallaxes are too small and that is the value that should be added to correct for
the zero point. To establish that value, \citet{Lindetal18a} used a sample of quasars, which in general are faint sources compared to typical
{\it Gaia}~DR2 sources. Other authors \citep{Riesetal18,Zinnetal19,Khanetal19,ChanBovy20} have found that for brighter (stellar) sources the zero 
point is $\approx$50~$\mu$as. Given the variability, in this work we used a zero point of 40$\pm$10~$\mu$as, as already mentioned above.

\begin{figure}
\centerline{\includegraphics*[width=\linewidth]{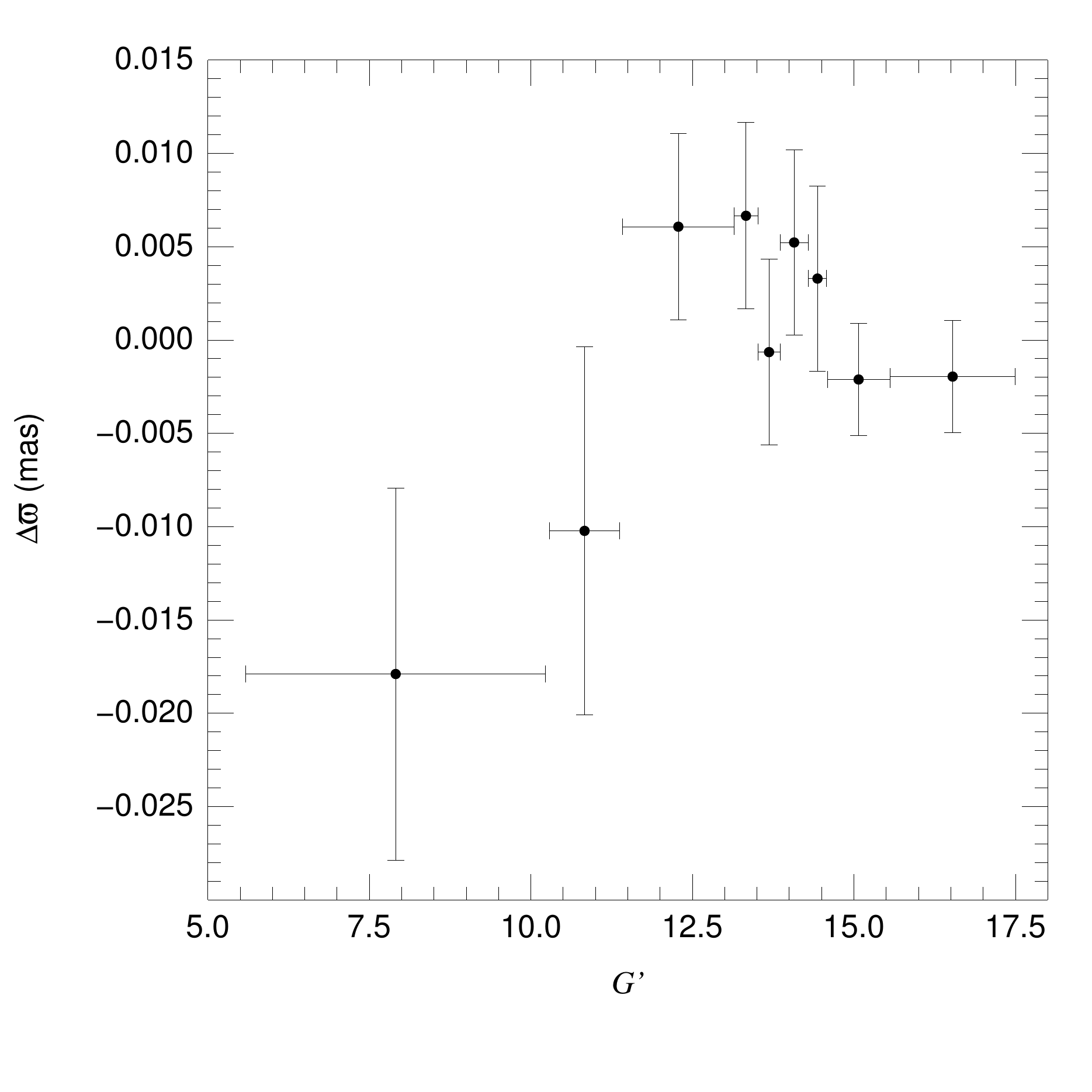}}
\caption{Difference between individual stellar parallaxes and the group parallax as a function of \GGc\ for the stars in \VO{001} to 
         \VO{013} plus \VO{015} and \VO{016}. The data have been binned to see the effect as a function of magnitude. The horizontal error bars
         show the extent of \GGc\ magnitudes binned and the vertical error bars show the weighted standard deviation of the mean using as inputs
         the values with the external uncertainties.}
\label{bias}
\end{figure}   

Our data give us the opportunity of studying the magnitude dependence of the parallax zero point in a relative sense. We do that by plotting in
Fig.~\ref{bias} the difference between the stellar parallax of the members of all the groups in our sample (excluding the special case of
\VO{014}) and the group parallax itself as a function of \GGc. The data are binned in magnitude to reduce the uncertainties but note that the
binning covers larger ranges for bright stars due to their relative scarcity. Even though the error bars are relatively large, the results are
consistent with a difference of $\approx$20~$\mu$as between bright and dim stars found by other authors. The transition takes place around
\GGc\ between 11 and 12.

We think this effect is the reason for the small difference in the distance to Trumpler~14 and Trumpler~16 between our results and those of \citet{ShulDanf19}.
Most of our stars in those groups are in the dim range as defined above while those authors use objects mostly in the bright region. They also apply a parallax 
zero point of 30~$\mu$as while we use one of 40~$\mu$as. Both differences move their distances towards higher values.

\subsection{Future work}

$\,\!$\indent In the future we will keep using {\it Gaia} to analyze more Galactic stellar groups with O stars and, if resources allow it, extend the catalog to groups 
without them but with B stars. We will also revise the results once the early third data release becomes available in 
late 2020, as currently expected. 
Later on, we will also incorporate the
new types of output from {\it Gaia}~DR3 such as the spectrophotometry. The larger sample will allow us to study the spatial distribution of the groups containing O stars.

In addition to {\it Gaia} data we will also incorporate results from GALANTE \citep{Maizetal19c,LorGetal19}, a photometric survey that is imaging the northern Galactic
Plane in seven narrow- and intermediate-band filters using the JAST/T80 telescope at Javalambre, Teruel, Spain. Each single-CCD-chip field has a size of 2 square degrees 
with a pixel size of 0\farcs55. The filter set has been especially tailored to measure the effective temperatures and extinction of OB stars and the survey includes 
different exposure times to achieve a large dynamic range. GALANTE will be used as a complement to Gaia data to study the Villafranca groups. 
%An example mosaic that includes two of the groups in this paper is shown in Figs.~\ref{GALANTE}~and~\ref{GALANTE_zoom}. 
The survey will be extended to the southern Galactic Plane in the future 
using the T-80S telescope at Cerro Tololo, Chile. GALANTE will allow us to overcome one of the limitations of {\it Gaia}, the photometric study of crowded 
and nebular
regions, where the results for $G$ are reliable but those for \GBP\ and \GRP\ are not. In turn, that will allow us to study extinction variations and derive the IMF more accurately.

Finally, we will keep obtaining new spectroscopy using GOSSS and adding new data to our optical+NIR high-resolution spectroscopic database LiLiMaRlin
\citep{Maizetal19a} to characterize the stars in the Villafranca stellar groups. To those surveys we will add WEAVE \citep{Dalt16}, a multi-fiber instrument
that will be mounted at the William Herschel Telescope at La Palma, Spain in 2020. One of the WEAVE projects, SCIP (Stellar, Circumstellar, and Interstellar
Physics), will obtain intermediate-resolution spectroscopy of a large number of OB stars in stellar groups in the northern Galactic Plane and their results will be 
used to improve our knowledge of their membership.

\begin{acknowledgements}
We dedicate this paper to the memory of Nolan R. Walborn, of whom the first and third authors were postdoctoral researchers and who sparkled their interest in several 
of these stellar groups.
We thank Danny Lennon for useful discussions on this topic.
J.M.A. and A.S. acknowledge support from the Spanish Government Ministerio de Ciencia through grants AYA2016-\num{75931}-C2-2-P and PGC2018-\num{095049}-B-C22. 
R.H.B. acknowledges support from DIDULS Project \num{18143} and the ESAC visitors program.
This work has made use of data from the European Space Agency (ESA) mission {\it Gaia} ({\tt https://www.cosmos.esa.int/gaia}), 
processed by the {\it Gaia} Data Processing and Analysis Consortium (DPAC, {\tt https://www.cosmos.esa.int/web/gaia/dpac/consortium}).
Funding for the DPAC has been provided by national institutions, in particular the institutions participating in the {\it Gaia} 
Multilateral Agreement. Additionally, this paper includes data obtained with the MPG/ESO 2.2~m Telescope at the Observatorio~de~La~Silla, Chile; 
the 2.5~m du Pont Telescope at the Observatorio de Las Campanas, Chile; the 10~m Hobby-Eberly Telescope at McDonald Observatory, Texas, U.S.A.; 
the 4.2~m William Herschel Telescope and the 10.4~m Gran Telescopio Canarias at the Observatorio del Roque de los Muchachos, La Palma, Spain; and 
the 2.2~m Telescope at the Centro Astron\'omico Hispano Andaluz, Almer{\'\i}a, Spain.
%and the JAST/T80 telescope at the Observatorio Astrof{\'\i}sico de Javalambre, Teruel, Spain (owned, managed, and operated by the Centro de 
%Estudios de F{\'\i}sica del Cosmos de Arag\'on). 
We thank the staff at those observatories for their support.
This research has made extensive use of the SIMBAD and VizieR databases, operated at CDS, Strasbourg, France. 
\end{acknowledgements}

%------------------------------------------------------------------

%\vfill
%
%\eject

\bibliographystyle{aa} % style aa.bst
\bibliography{general} % your references references.bib

\begin{appendix}

\section{Additional tables and figures}

$\,\!$\indent In this appendix we present additional material: a list of the existing literature distances for the stellar groups in this paper,
the plots used to analyze the {\it Gaia}~DR2 results for \VO{001} to \VO{014} (see MA19 for the equivalent plots for \VO{015} and \VO{016}), and
a list of the candidate runaways detected.

\begin{table*}
\caption{Literature distances. The specific target and a brief description of the method is given. Note that ``spectro-photometry'' indicates a combination 
         of spectroscopic and photometric methods. For simplicity, random and systematic uncertainties have been combined.}
\centerline{
\renewcommand{\arraystretch}{1.2}
\begin{tabular}{lr@{}llll}
\hline
ID            & \mcii{distance}            & target                    & description                                                          & Ref.             \\
              & \mcii{(pc)}                &                           &                                                                      &                  \\
\hline
O-001         &    9490&$^{+190}_{-210}$   & NGC~3603                  & {\it Gaia} DR2 parallax, no covariance term                          & C18              \\
              &    7600&                   & NGC~3603                  & Visible spectro-photometry                                           & M08              \\
              &    6900&$^{+600}_{-600}$   & NGC~3603                  & Visible CMD                                                          & S04              \\
              &    7900&                   & H\,{\sc ii} gas           & Kinematic distance                                                   & R03              \\
              &    7700&$^{+200}_{-200}$   & H\,{\sc ii} gas           & Kinematic distance                                                   & N02              \\
              &    7200&$^{+1200}_{-1200}$ & NGC~3603                  & Visible CMD                                                          & S01              \\
              &    6300&$^{+600}_{-600}$   & NGC~3603                  & Visible CMD                                                          & P00a             \\
              &    6100&$^{+600}_{-600}$   & H\,{\sc ii} gas           & Kinematic distance                                                   & d99              \\
              & \num{10100}&               & NGC~3603                  & Visible spectro-photometry                                           & C98              \\
              &    7200&                   & NGC~3603                  & Visible CMD                                                          & M89              \\
              &    8600&$^{+500}_{-500}$   & H\,{\sc ii} gas           & Kinematic distance                                                   & S83a             \\
              &    7000&$^{+500}_{-500}$   & NGC~3603                  & Visible spectro-photometry                                           & M83              \\
              &    5300&$^{+1600}_{-1200}$ & NGC~3603                  & Visible CMD                                                          & M82              \\
              &    7200&$^{+800}_{-1000}$  & H\,{\sc ii} gas           & Kinematic distance                                                   & v78              \\
              &    8100&$^{+800}_{-800}$   & NGC~3603                  & Visible CMD                                                          & M74              \\
              &    8400&                   & H\,{\sc ii} gas           & Kinematic distance                                                   & G72              \\
              &    5720&                   & NGC~3603                  & Visible CMD                                                          & B71              \\
              &    3500&                   & NGC~3603                  & Visible CMD                                                          & S65  \vspace{2mm}\\
O-002 + O-003 &    2500&$^{+250}_{-250}$   & molecular clouds          & {\it Gaia} DR2 parallax                                              & Z20              \\
              &    2400&$^{+200}_{-200}$   & Car OB1                   & {\it Gaia} DR2 parallax                                              & L19              \\
              &    2640&$^{+310}_{-250}$   & Trumpler~14               & {\it Gaia} DR2 parallax                                              & K19              \\
              &    2610&$^{+310}_{-250}$   & Trumpler 16               & {\it Gaia} DR2 parallax                                              & K19              \\
              &    2680&$^{+310}_{-310}$   & Trumpler~14               & {\it Gaia} DR2 parallax                                              & S19              \\
              &    2550&$^{+300}_{-300}$   & Trumpler 16               & {\it Gaia} DR2 parallax                                              & S19              \\
              &    2690&$^{+400}_{-400}$   & Car OB1                   & {\it Gaia} DR2 parallax                                              & B18              \\
              &    3050&                   & Trumpler~14               & {\it Gaia} DR2 parallax                                              & D18              \\
              &    2600&                   & Trumpler 16               & {\it Gaia} DR2 parallax                                              & D18              \\
              &    2414&$^{+9}_{-9}$       & Trumpler~14               & {\it Gaia} DR2 parallax, no covariance term                          & C18              \\
              &    2395&$^{+12}_{-11}$     & Trumpler 16               & {\it Gaia} DR2 parallax, no covariance term                          & C18              \\
              &    2900&$^{+300}_{-300}$   & Trumpler~14 + Trumpler 16 & Visible + NIR CMD                                                    & H12              \\
              &    4200&$^{+1800}_{-1800}$ & Trumpler 16               & Ca\,{\sc ii} ISM lines                                               & M09a             \\
              &    2350&$^{+50}_{-50}$     & $\eta$ Car                & Geometric distance from Homunculus expansion                         & S06              \\
              &    2500&$^{+300}_{-300}$   & Trumpler~14               & Visible CMD                                                          & C04a             \\
              &    4000&$^{+300}_{-300}$   & Trumpler 16               & Visible CMD                                                          & C04a             \\
              &    2800&                   & Trumpler~14               & Visible + NIR CMD                                                    & T03              \\
              &    2500&                   & Trumpler 16               & Visible + NIR CMD                                                    & T03              \\
              &    2580&                   & Car OB1                   & Statistical parallaxes                                               & R99              \\
              &    2800&                   & Car OB1                   & Visible spectro-photometry with \RV\ = 3.0                           & W95              \\
              &    2250&                   & Car OB1                   & Visible spectro-photometry with \RV\ = 4.0                           & W95              \\
              &    3610&$^{+300}_{-300}$   & Trumpler~14               & Visible spectro-photometry                                           & M93              \\
              &    3150&$^{+130}_{-130}$   & Trumpler 16               & Visible spectro-photometry                                           & M93              \\
              &    3960&$^{+130}_{-130}$   & Trumpler~14               & Visible spectro-photometry                                           & M88              \\
              &    2400&$^{+200}_{-200}$   & Car OB1                   & Visible + NIR CMD                                                    & T88              \\
              &    2750&$^{+250}_{-250}$   & Trumpler~14               & Visible CMD                                                          & F83              \\
              &    2700&                   & H\,{\sc ii} gas           & Kinematic distance                                                   & S83a             \\
              &    2800&                   & Trumpler 16               & Visible spectro-photometry                                           & W82              \\
              &    2800&                   & Car OB1                   & Visible CMD with \RV\ = 3.3                                          & T80              \\
              &    2350&                   & Car OB1                   & Visible CMD with \RV\ = 4.0                                          & T80              \\
              &    3600&$^{+150}_{-150}$   & Trumpler~14               & Visible spectro-photometry                                           & H78              \\
              &    2690&$^{+150}_{-150}$   & Trumpler 16               & Visible spectro-photometry                                           & H78              \\
              &    2400&                   & Car OB1                   & Visible CMD                                                          & F78              \\
              &    3500&                   & Trumpler~14               & Visible spectro-photometry                                           & W73a             \\
              &    2600&                   & Trumpler 16               & Visible spectro-photometry                                           & W73a             \\
\hline
\end{tabular}
\renewcommand{\arraystretch}{1.0}
}
\label{literature}               
\end{table*}

\addtocounter{table}{-1}

\begin{table*}
\caption{(Continued).}
$\,\!$
\centerline{
\renewcommand{\arraystretch}{1.2}
\begin{tabular}{lr@{}llll}
\hline
ID            & \mcii{distance}            & target                    & description                                                          & Ref.             \\
              & \mcii{(pc)}                &                           &                                                                      &                  \\
\hline
O-002 + O-003 &    3390&$^{+300}_{-300}$   & Trumpler~14 + Trumpler 16 & Visible CMD with \RV\ = 3.0                                          & F73              \\
(continued)   &    2650&$^{+240}_{-240}$   & Trumpler~14 + Trumpler 16 & Visible CMD with \RV\ = 4.0                                          & F73              \\
              &    1650&                   & Trumpler~14               & Visible CMD                                                          & B71              \\
              &    2950&                   & Trumpler 16               & Visible CMD                                                          & B71              \\
              &    2000&                   & Trumpler~14               & Visible CMD                                                          & T71              \\
              &    2500&                   & Trumpler 16               & Visible CMD                                                          & T71  \vspace{2mm}\\
O-004         &    4400&$^{+100}_{-100}$   & Westerlund~2              & {\it Gaia} DR2 parallax                                              & B18              \\
              &    4208&$^{+74}_{-80}$     & Westerlund~2              & {\it Gaia} DR2 parallax, no covariance term                          & C18              \\
              &    4000&                   & Westerlund~2              & Visible CMD                                                          & Z15              \\
              &    6000&$^{+1300}_{-500}$  & Westerlund~2              & Visible CMD                                                          & H15              \\
              &    6500&$^{+500}_{-500}$   & molecular clouds          & Kinematic distance                                                   & B13              \\
              &    4469&                   & Westerlund~2              & Visible + NIR CMD                                                    & K13              \\
              &    2850&$^{+430}_{-430}$   & Westerlund~2              & Visible spectro-photometry                                           & C13              \\
              &    4160&$^{+270}_{-270}$   & Westerlund~2              & Visible CMD                                                          & V13              \\
              &    6698&$^{+512}_{-475}$   & Westerlund~2              & Visible spectro-photometry                                           & K12              \\
              &    8000&$^{+1400}_{-1400}$ & three EBs                 & Spectroscopic eclipsing binaries                                     & R11              \\
              &    5400&$^{+1100}_{-1400}$ & molecular clouds          & Kinematic distance                                                   & F09              \\
              &    6000&$^{+1000}_{-1000}$ & molecular clouds          & Kinematic distance                                                   & D07              \\
              &    2800&                   & Westerlund~2              & NIR CMD                                                              & A07              \\
              &    6400&$^{+400}_{-400}$   & Westerlund~2              & Visible CMD                                                          & C04b             \\
              &    5700&$^{+300}_{-300}$   & Westerlund~2              & Visible CMD                                                          & P98              \\
              &    7900&$^{+1200}_{-1000}$ & Westerlund~2              & Visible CMD                                                          & M91a             \\
              &    5100&                   & H\,{\sc ii} gas           & Kinematic distance                                                   & S83a             \\
              &    5000&                   & Westerlund~2              & Visible CMD                                                          & M75              \\
              &    4680&$^{+750}_{-750}$   & H\,{\sc ii} gas           & Kinematic distance                                                   & G73              \\
              &    6000&                   & Westerlund~2              & Association with spiral arm                                          & W61  \vspace{2mm}\\
O-005         &    1770&$^{+120}_{-120}$   & NGC 6357                  & {\it Gaia} DR2 parallax                                              & R20              \\
              &    1790&$^{+150}_{-150}$   & Pismis~24                 & {\it Gaia} DR2 parallax                                              & K19              \\
              &    1780&$^{+180}_{-180}$   & NGC 6357                  & {\it Gaia} DR2 parallax                                              & B18              \\
              &    1678&$^{+18}_{-16}$     & NGC 6357                  & {\it Gaia} DR2 parallax, no covariance term                          & C18              \\
              &    2460&$^{+1440}_{-1440}$ & Pismis~24                 & Visible spectro-photometry                                           & R17              \\
              &    1780&$^{+100}_{-100}$   & NGC 6357                  & NIR CMD                                                              & L14              \\
              &    1900&$^{+400}_{-400}$   & NGC 6357                  & Visible + NIR CMD                                                    & R12a             \\
              &    1700&$^{+200}_{-200}$   & Pismis~24                 & Visible + NIR CMD                                                    & F12              \\
              &    1700&                   & Pismis~24                 & Visible + NIR CMD                                                    & G11              \\
              &    2500&                   & Pismis~24                 & Visible spectro-photometry                                           & M01              \\
              &    1200&                   & H\,{\sc ii} gas           & Kinematic distance                                                   & S83a             \\
              &    1740&$^{+310}_{-310}$   & NGC 6357 + NGC 6334       & Visible CMD                                                          & N78              \\
              &    1580&                   & NGC 6357                  & Visible spectro-photometry                                           & J73              \\
              &     770&$^{+1300}_{-1300}$ & H\,{\sc ii} gas           & Kinematic distance                                                   & G73              \\
              &    1000&$^{+2300}_{-2300}$ & H\,{\sc ii} gas           & Kinematic distance                                                   & W70              \\
              &     920&$^{+870}_{-870}$   & H\,{\sc ii} gas           & Kinematic distance                                                   & D67  \vspace{2mm}\\
O-006         &    7000&                   & H\,{\sc ii} gas           & Kinematic distance                                                   & M17a             \\
              &    8000&                   & H\,{\sc ii} gas           & Kinematic distance                                                   & G00              \\
              &    6000&                   & H\,{\sc ii} gas           & Kinematic distance                                                   & A89              \\
              &    9100&                   & H\,{\sc ii} gas           & Kinematic distance                                                   & C87  \vspace{2mm}\\
O-007 + O-008 &    1600&$^{+100}_{-100}$   & Cyg OB2                   & {\it Gaia} DR2 parallax                                              & L19              \\
              &    1760&$^{+370}_{-261}$   & Cyg OB2                   & {\it Gaia} DR2 parallax, foreground group at 1350$^{+210}_{-160}$ pc & B19a             \\
              &    1330&$^{+60}_{-60}$     & four EBs                  & Spectroscopic eclipsing binaries                                     & K15              \\
              &    1400&$^{+80}_{-80}$     & four masers               & VLBI parallaxes                                                      & R12b             \\
              &    1220&$^{+220}_{-220}$   & Cyg OB2                   & Visible spectro-photometry                                           & H03              \\
              &    1800&$^{+200}_{-200}$   & Bica 1 + 2                & NIR CMD                                                              & B03              \\
              &    1740&$^{+80}_{-80}$     & Cyg OB2                   & Visible spectro-photometry                                           & M91b             \\
              &    1700&$^{+200}_{-200}$   & Cyg OB2                   & Visible + NIR CMD                                                    & T91              \\
              &    1790&$^{+180}_{-180}$   & Cyg OB2                   & Visible spectro-photometry                                           & H78              \\
\hline
\end{tabular}
\renewcommand{\arraystretch}{1.0}
}
\end{table*}

\addtocounter{table}{-1}

\begin{table*}
\caption{(Continued).}
$\,\!$
\centerline{
\renewcommand{\arraystretch}{1.2}
\begin{tabular}{lr@{}llll}
\hline
ID            & \mcii{distance}            & target                    & description                                                          & Ref.             \\
              & \mcii{(pc)}                &                           &                                                                      &                  \\
\hline
O-007 + O-008 &    1800&                   & Cyg OB2                   & Visible spectro-photometry                                           & W73b             \\
(continued)   &    2100&                   & Cyg OB2                   & Visible spectro-photometry                                           & R66              \\
              &    1500&                   & Cyg OB2                   & Visible spectro-photometry                                           & J54  \vspace{2mm}\\
O-009         &    1500&$^{+150}_{-150}$   & molecular clouds          & {\it Gaia} DR2 parallax                                              & Z20              \\
              &    1680&$^{+130}_{-110}$   & M17                       & {\it Gaia} DR2 parallax                                              & K19              \\
              &    1820&$^{+160}_{-160}$   & M17                       & {\it Gaia} DR2 parallax                                              & B18              \\
              &    2390&$^{+310}_{-410}$   & H\,{\sc ii} gas           & Kinematic distance                                                   & W18              \\
              &    2040&$^{+160}_{-170}$   & one maser                 & VLBI parallax                                                        & C16              \\
              &    1308&                   & M17                       & Visible + NIR CMD                                                    & K13              \\
              &    1980&$^{+140}_{-120}$   & one maser                 & VLBI parallax                                                        & X11              \\
              &    2100&$^{+200}_{-200}$   & M17                       & Visible + NIR spectro-photometry                                     & H08              \\
              &    1600&$^{+300}_{-100}$   & M17                       & Luminosity budget                                                    & P07              \\
              & \num{14550}&               & H\,{\sc ii} gas           & Kinematic distance                                                   & Q06              \\
              &    2400&                   & H\,{\sc ii} gas           & Kinematic distance                                                   & R03              \\
              &    1600&$^{+300}_{-300}$   & M17                       & Visible + NIR spectro-photometry                                     & N01              \\
              &    1300&$^{+400}_{-200}$   & M17                       & Visible + NIR spectro-photometry                                     & H97              \\
              &    2200&$^{+200}_{-200}$   & molecular clouds          & Kinematic distance                                                   & S89a             \\
              &    2200&$^{+200}_{-200}$   & molecular clouds          & Kinematic distance                                                   & F84              \\
              &    2200&$^{+200}_{-200}$   & molecular clouds          & Kinematic distance                                                   & S84              \\
              &    2400&                   & H\,{\sc ii} gas           & Kinematic distance                                                   & S83a             \\
              &    2200&$^{+200}_{-200}$   & M17                       & Visible CMD                                                          & C80              \\
              &    1300&                   & M17                       & Visible CMD                                                          & O76              \\
              &    1780&$^{+1020}_{-1020}$ & M17                       & Visible spectro-photometry                                           & J73              \\
              &    2200&$^{+510}_{-510}$   & H\,{\sc ii} gas           & Kinematic distance                                                   & G73              \\
              &    2460&                   & H\,{\sc ii} gas           & Kinematic distance                                                   & G70              \\
              &    2250&$^{+100}_{-100}$   & H\,{\sc ii} gas           & Kinematic distance                                                   & D67  \vspace{2mm}\\
O-010         &    1046&$^{+54}_{-54}$     & southern molecular cloud  & {\it Gaia} DR2 parallax                                              & Z20              \\
              &    1190&$^{+4}_{-4}$       & NGC~6193                  & {\it Gaia} DR2 parallax, no covariance term                          & C18              \\
              &    1096&$^{+274}_{-274}$   & HD~\num{150136}           & Spectroastrometric orbit                                             & M18              \\
              &    1380&$^{+130}_{-130}$   & NGC~6193                  & Visible + NIR CMD                                                    & B11a             \\
              &    1410&$^{+120}_{-120}$   & NGC~6193                  & Visible CMD                                                          & V92              \\
              &    1340&                   & Ara OB1                   & Visible spectro-photometry                                           & F87              \\
              &    1460&$^{+130}_{-130}$   & Ara OB1a                  & Visible spectro-photometry                                           & H78              \\
              &    1320&$^{+120}_{-120}$   & Ara OB1                   & Visible CMD                                                          & H77              \\
              &    1360&                   & NGC~6193                  & Visible CMD                                                          & M73b             \\
              &    1390&                   & NGC~6193                  & Visible CMD                                                          & B71              \\
              &    1880&                   & H\,{\sc ii} gas           & Kinematic distance                                                   & G70              \\
              &    1400&                   & Ara OB1                   & Visible CMD                                                          & W63  \vspace{2mm}\\
O-011         &    3016&$^{+41}_{-36}$     & Berkeley~90               & {\it Gaia} DR2 parallax, no covariance term                          & C18              \\
              &    3500&$^{+500}_{-500}$   & Berkeley~90               & Visible CMD                                                          & M17b             \\
              &    2670&$^{+150}_{-150}$   & Berkeley~90               & CHORIZOS analysis of LS~III~+46~11 and LS~III~+46~12                 & M15              \\
              &    2430&$^{+70}_{-70}$     & Berkeley~90               & NIR CMD                                                              & T08              \\
              &    1500&$^{+200}_{-200}$   & LS III $+$46 11           & Visible spectro-photometry                                           & M97              \\
              &    3000&$^{+600}_{-600}$   & molecular clouds          & Kinematic distance                                                   & F84              \\
              &    2320&                   & LS III $+$46 12           & Visible spectro-photometry                                           & M73a \vspace{2mm}\\
O-012         &    5050&$^{+170}_{-160}$   & Haffner~19                & {\it Gaia} DR2 parallax, no covariance term                          & C18              \\
              &    4673&$^{+92}_{-92}$     & Haffner~18                & {\it Gaia} DR2 parallax, no covariance term                          & C18              \\
              &    4810&$^{+480}_{-480}$   & two EBs                   & Spectroscopic eclipsing binaries                                     & L17              \\
              &    5700&$^{+400}_{-400}$   & Haffner~19                & Visible + NIR CMD                                                    & Y15              \\
              & \num{11200}&$^{+1000}_{-1000}$ & Haffner~18            & Visible + NIR CMD                                                    & Y15              \\
              &    1313&                   & \VO{012}                  & Visible + NIR CMD                                                    & K13              \\
              &    5300&                   & H\,{\sc ii} gas           & Kinematic distance                                                   & B11b             \\
              &    5400&                   & H\,{\sc ii} gas           & Kinematic distance                                                   & Q06              \\
\hline
\end{tabular}
\renewcommand{\arraystretch}{1.0}
}
\end{table*}

\addtocounter{table}{-1}

\begin{table*}
\caption{(Continued).}
$\,\!$
\centerline{
\renewcommand{\arraystretch}{1.2}
\begin{tabular}{lr@{}llll}
\hline
ID            & \mcii{distance}            & target                    & description                                                          & Ref.             \\
              & \mcii{(pc)}                &                           &                                                                      &                  \\
\hline
O-012         &    6400&                   & Haffner~19                & Visible CMD                                                          & G06              \\
(continued)   &    5900&                   & Haffner~18                & Visible CMD                                                          & G06              \\
              &    4800&$^{+300}_{-300}$   & Haffner~19                & Visible CMD                                                          & M02              \\
              &    4090&                   & Haffner~19                & Statistical parallaxes                                               & R99              \\
              &    4570&                   & Haffner~18                & Statistical parallaxes                                               & R99              \\
              &    4100&$^{+600}_{-600}$   & molecular clouds          & Kinematic distance                                                   & F84              \\
              &    3900&                   & H\,{\sc ii} gas           & Kinematic distance                                                   & S83a             \\
              &    4200&                   & H\,{\sc ii} gas           & Kinematic distance                                                   & P76              \\
              &    6900&$^{+900}_{-900}$   & \VO{012}                  & Visible CMD                                                          & F74              \\
              &    3510&$^{+370}_{-370}$   & H\,{\sc ii} gas           & Kinematic distance                                                   & G73              \\
              &    2500&                   & \VO{012}                  & Visible CMD                                                          & B71              \\
              &    3830&                   & H\,{\sc ii} gas           & Kinematic distance                                                   & G70              \\
              &    3370&$^{+160}_{-160}$   & \VO{012}                  & Visible CMD                                                          & L66  \vspace{2mm}\\
O-013         &    4880&$^{+1000}_{-920}$  & H\,{\sc ii} gas           & Kinematic distance                                                   & W18              \\
              &    2700&$^{+500}_{-500}$   & \VO{013}                  & NIR spectro-photometry                                               & P10              \\
              &    2650&$^{+120}_{-110}$   & two masers                & VLBI parallaxes                                                      & M09b             \\
              &    2270&$^{+150}_{-150}$   & \VO{013}                  & Visible spectro-photometry                                           & R07              \\
              &    6400&                   & H\,{\sc ii} gas           & Kinematic distance                                                   & Q06              \\
              &    2200&                   & \VO{013}                  & NIR CMD                                                              & M86              \\
              &    3700&                   & H\,{\sc ii} gas           & Kinematic distance                                                   & M86              \\
              &    2800&$^{+900}_{-900}$   & molecular clouds          & Kinematic distance                                                   & F84              \\
              &    2800&                   & two stars                 & Visible spectro-photometry                                           & C78              \\
              &    4750&$^{+380}_{-380}$   & H\,{\sc ii} gas           & Kinematic distance                                                   & G73              \\
              &    4230&                   & H\,{\sc ii} gas           & Kinematic distance                                                   & G70  \vspace{2mm}\\
O-014         &     800&$^{+100}_{-100}$   & molecular clouds          & {\it Gaia} DR2 parallax                                              & Z20              \\
              &     858&$^{+56}_{-56}$     & Pelican nebula stars      & {\it Gaia} DR2 parallax                                              & B19b             \\
              &     605&$^{+45}_{-45}$     & molecular clouds          & Visible CMD                                                          & D17              \\
              &     700&$^{+500}_{-500}$   & North America nebula      & Kinematic distance                                                   & C07              \\
              &     560&$^{+60}_{-60}$     & molecular clouds          & Visible CMD                                                          & L06              \\
              &     610&                   & Bajamar star              & Visible + NIR spectro-photometry                                     & C05              \\
              &     600&                   & molecular clouds          & Visible CMD                                                          & L02              \\
              &     485&                   & molecular clouds          & Star counts                                                          & C02              \\
              &     550&$^{+140}_{-140}$   & molecular clouds          & Visible spectro-photometry                                           & S93              \\
              &     800&$^{+300}_{-300}$   & molecular clouds          & Kinematic distance                                                   & S89a             \\
              &     550&$^{+100}_{-100}$   & molecular clouds          & Visible CMD                                                          & S89b             \\
              &    1000&$^{+300}_{-300}$   & molecular clouds          & Kinematic distance                                                   & S84              \\
              &     800&$^{+300}_{-300}$   & molecular clouds          & Kinematic distance                                                   & F84              \\
              &     970&$^{+30}_{-30}$     & molecular clouds          & Star counts                                                          & A81              \\
              &     150&$^{+50}_{-50}$     & 2MASS J20535282$+$4424015 & Stellar parameters estimation                                        & N80              \\
              &    1200&                   & HD~\num{199579}           & Visible spectro-photometry                                           & M68              \\
              &    1980&$^{+400}_{-400}$   & H\,{\sc ii} gas           & Kinematic distance                                                   & D67              \\
              &    1130&                   & HD~\num{199579}           & Visible spectro-photometry                                           & B63              \\
              &     500&$^{+300}_{-300}$   & molecular clouds          & Visible CMD                                                          & H58              \\
              &     910&                   & HD~\num{199579}           & Visible spectro-photometry                                           & S52  \vspace{2mm}\\
O-015         &    1019&$^{+7}_{-6}$       & Collinder~419             & {\it Gaia} DR2 parallax, no covariance term                          & C18              \\
              &     741&$^{+36}_{-36}$     & Collinder~419             & Visible + NIR CMD                                                    & R10              \\
              &    1400&                   & Collinder~419             & Visible CMD                                                          & H87              \\
\hline
\end{tabular}
\renewcommand{\arraystretch}{1.0}
}
\end{table*}

\addtocounter{table}{-1}

\begin{table*}
\caption{(Continued).}
$\,\!$
\centerline{
\renewcommand{\arraystretch}{1.2}
\begin{tabular}{lr@{}llll}
\hline
ID            & \mcii{distance}            & target                    & description                                                          & Ref.             \\
              & \mcii{(pc)}                &                           &                                                                      &                  \\
\hline
O-016         &     750&$^{+50}_{-50}$     & molecular clouds          & {\it Gaia} DR2 parallax                                              & Z20              \\
              &     738&$^{+23}_{-21}$     & NGC~2244                  & {\it Gaia} DR2 parallax, no covariance term                          & K19              \\
              &     723&$^{+2}_{-2}$       & NGC~2244                  & {\it Gaia} DR2 parallax                                              & C18              \\
              &     738&$^{+57}_{-50}$     & two masers                & VLBI parallaxes                                                      & K14              \\
              &     913&$^{+40}_{-40}$     & NGC~2244                  & $v\sin i$ and rotation periods                                       & B09              \\
              &     759&$^{+35}_{-35}$     & NGC~2244                  & Visible CMD                                                          & P00b             \\
              &     720&                   & NGC~2244                  & Statistical parallaxes                                               & R99              \\
              &     764&$^{+87}_{-87}$     & NGC~2244                  & Visible CMD                                                          & S97              \\
              &     910&$^{+50}_{-50}$     & NGC~2244                  & Visible + NIR CMD                                                    & N93              \\
              &     700&$^{+40}_{-40}$     & NGC~2244                  & Visible CMD                                                          & F91              \\
              &     800&$^{+150}_{-150}$   & molecular clouds          & Kinematic distance                                                   & S89a             \\
              &     950&$^{+75}_{-75}$     & NGC~2244                  & Visible spectro-photometry                                           & P87              \\
              &     800&$^{+150}_{-150}$   & molecular clouds          & Kinematic distance                                                   & S84              \\
              &     800&$^{+150}_{-150}$   & molecular clouds          & Kinematic distance                                                   & F84              \\
              &     798&$^{+73}_{-73}$     & NGC~2244                  & Visible CMD                                                          & S83b             \\
              &     875&                   & NGC~2244                  & Visible CMD                                                          & M80              \\
              &     715&                   & NGC~2244                  & Visible CMD                                                          & B71              \\
              &     715&                   & NGC~2244                  & Visible spectro-photometry                                           & B63              \\
              &     800&                   & NGC~2244                  & Visible spectro-photometry                                           & W56              \\
\hline
Ref.:         & \multicolumn{5}{l}{A81:  \citet{ArmaHerb81},  A89:  \citet{AvedPalo89},  A07:  \citet{Asceetal07a}, B63:  \citet{BeckFenk63}, } \\
              & \multicolumn{5}{l}{B71:  \citet{BeckFenk71},  B03:  \citet{Bicaetal03},  B09:  \citet{Baxtetal09},  B11a: \citet{Baumetal11}, } \\
              & \multicolumn{5}{l}{B11b: \citet{Balsetal11},  B13:  \citet{Benaetal13},  B18:  \citet{BindPovi18},  B19a: \citet{Berletal19}, } \\
              & \multicolumn{5}{l}{B19b: \citet{Bharetal19},  C05:  \citet{ComePasq05},  C78:  \citet{Crametal78},  C80:  \citet{Chinetal80}, } \\
              & \multicolumn{5}{l}{C87:  \citet{CaswHayn87},  C98:  \citet{CrowDess98},  C02:  \citet{Cambetal02},  C04a: \citet{Carretal04}, } \\
              & \multicolumn{5}{l}{C04b: \citet{CarrMuna04},  C07:  \citet{Cersetal07},  C13:  \citet{Carretal13},  C16:  \citet{Chibetal16}, } \\
              & \multicolumn{5}{l}{C18:  \citet{CanGetal18},  D67:  \citet{Diet67},      d99:  \citet{dePretal99},  D07:  \citet{Dame07},      D17:  \citet{Damietal17c},} \\
              & \multicolumn{5}{l}{D18:  \citet{Davietal18a}, F73:  \citet{Feinetal73},  F74:  \citet{FitzMoff74},  F78:  \citet{Fort78},      F83:  \citet{Fein83},     } \\
              & \multicolumn{5}{l}{F84:  \citet{FichBlit84},  F87:  \citet{Fitz87},      F91:  \citet{FeldvanG91},  F09:  \citet{Furuetal09}, } \\
              & \multicolumn{5}{l}{F12:  \citet{Fangetal12},  G70:  \citet{GeorGeor70},  G72:  \citet{Gossetal72},  G73:  \citet{Georetal73}, } \\
              & \multicolumn{5}{l}{G00:  \citet{Georetal00},  G06:  \citet{Gameetal06b}, G11:  \citet{Gvaretal11b}, H58:  \citet{Herb58},      H77:  \citet{HerbHavl77}, } \\
              & \multicolumn{5}{l}{H78:  \citet{Hump78},      H87:  \citet{Hron87},      H97:  \citet{Hansetal97},  H03:  \citet{Hans03},      H08:  \citet{Hoffetal08}, } \\
              & \multicolumn{5}{l}{H12:  \citet{Huretal12},   H15:  \citet{Huretal15},   J54:  \citet{JohnMorg54},  J73:  \citet{John73},      K12:  \citet{KaltGole12}, } \\
              & \multicolumn{5}{l}{K13:  \citet{Kharetal13},  K14:  \citet{Kameetal14},  K15:  \citet{Kimietal15a}, K19:  \citet{Kuhnetal19},  L66:  \citet{Lode66},     } \\
              & \multicolumn{5}{l}{L02:  \citet{LaugStra02},  L06:  \citet{Laugetal06},  L14:  \citet{Limaetal14},  L17:  \citet{Loreetal17}, } \\
              & \multicolumn{5}{l}{L19:  \citet{Limetal19},   M68:  \citet{Mill68},      M73a: \citet{MayeMaca73},  M73b: \citet{MoffVogt73},  M74:  \citet{Moff74b},    } \\
              & \multicolumn{5}{l}{M75:  \citet{MoffVogt75b}, M80:  \citet{MendGome80},  M82:  \citet{MelnGros82},  M83:  \citet{Moff83},     } \\
              & \multicolumn{5}{l}{M86:  \citet{MoreChav86},  M88:  \citet{Morretal88},  M89:  \citet{Melnetal89},  M91a: \citet{Moffetal91}, } \\
              & \multicolumn{5}{l}{M91b: \citet{MassThom91},  M93:  \citet{MassJohn93},  M97:  \citet{Motcetal97},  M01:  \citet{Massetal01}, } \\
              & \multicolumn{5}{l}{M02:  \citet{Moreetal02},  M08:  \citet{Meleetal08},  M09a: \citet{Megietal09},  M09b: \citet{Moscetal09}, } \\
              & \multicolumn{5}{l}{M15:  \citet{Maizetal15a}, M17a: \citet{Mohretal17},  M17b: \citet{MarcNegu17},  M18:  \citet{Mahyetal18}, } \\
              & \multicolumn{5}{l}{N78:  \citet{Neck78},      N80:  \citet{Necketal80b}, N01:  \citet{Nieletal01},  N02:  \citet{Nurnetal02},  N93:  \citet{Nerietal93}, } \\
              & \multicolumn{5}{l}{O76:  \citet{OgurIshi76},  P76:  \citet{PismMore76},  P87:  \citet{Pereetal87},  P98:  \citet{Piatetal98}, } \\
              & \multicolumn{5}{l}{P00a: \citet{Pandetal00},  P00b: \citet{Parketal00},  P07:  \citet{Povietal07},  P10:  \citet{Pugaetal10}, } \\
              & \multicolumn{5}{l}{Q06:  \citet{Quiretal06},  R66:  \citet{Reddetal66},  R99:  \citet{Rastetal99},  R03:  \citet{Russ03},     } \\
              & \multicolumn{5}{l}{R07:  \citet{Russetal07},  R10:  \citet{Robeetal10},  R11:  \citet{Rauwetal11},  R12a: \citet{Russetal12}, } \\
              & \multicolumn{5}{l}{R12b: \citet{Rygletal12},  R17:  \citet{Russetal17b}, R20:  \citet{RamTetal20},  S52:  \citet{SharOste52}, } \\
              & \multicolumn{5}{l}{S65:  \citet{Sher65},      S83a: \citet{Shavetal83},  S83b: \citet{SagaJosh83},  S84:  \citet{Star84},      S89a: \citet{StarBran89}, } \\
              & \multicolumn{5}{l}{S89b: \citet{Straetal89},  S93:  \citet{Straetal93},  S97:  \citet{Sungetal97},  S01:  \citet{Sagaetal01}, } \\
              & \multicolumn{5}{l}{S04:  \citet{SungBess04},  S06:  \citet{Smit06a},     S19:  \citet{ShulDanf19},  T71:  \citet{TheVlee71},  } \\
              & \multicolumn{5}{l}{T80:  \citet{Theetal80},   T88:  \citet{Tapietal88},  T91:  \citet{TorDetal91},  T03:  \citet{Tapietal03}, } \\
              & \multicolumn{5}{l}{T08:  \citet{Tadr08},      v78:  \citet{vadB78},      V92:  \citet{VazqFein92},  V13:  \citet{VarAetal13}, } \\
              & \multicolumn{5}{l}{W56:  \citet{Walk56},      W61:  \citet{West61},      W63:  \citet{Whit63},      W70:  \citet{Wilsetal70},  W73a: \citet{Walb73c},    } \\
              & \multicolumn{5}{l}{W73b: \citet{Walb73d},     W82:  \citet{Walb82a},     W95:  \citet{Walb95},      W18:  \citet{Wengetal18},  X11:  \citet{Xuetal11},   } \\
              & \multicolumn{5}{l}{Y15:  \citet{Yadaetal15},  Z15:  \citet{Zeidetal15},  Z20:  \citet{Zucketal20}. } \\
\end{tabular}
\renewcommand{\arraystretch}{1.0}
}
\end{table*}

\begin{figure*}
\centerline{\includegraphics*[width=0.34\linewidth, bb=0 0 538 522]{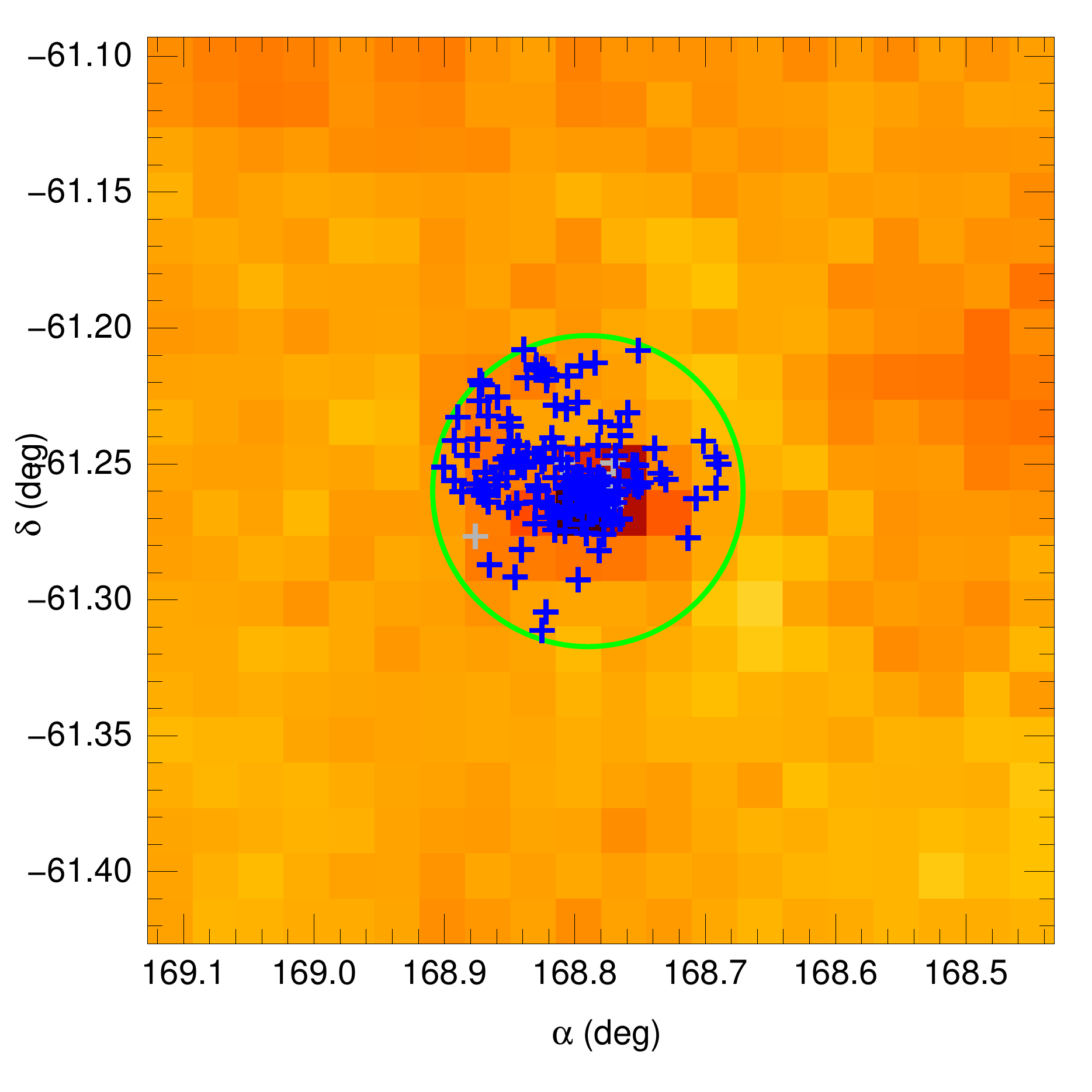} \
            \includegraphics*[width=0.34\linewidth, bb=0 0 538 522]{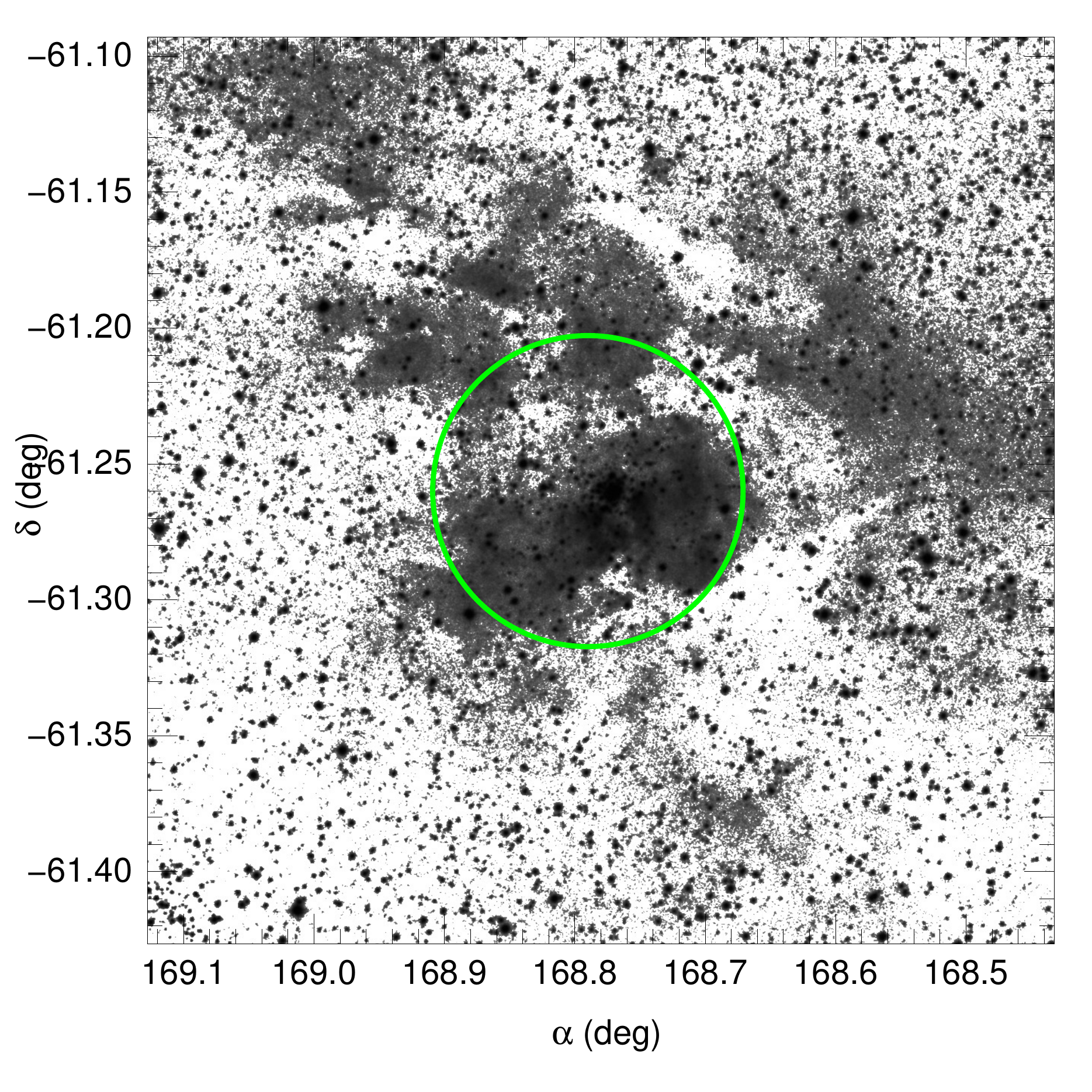} \
            \includegraphics*[width=0.34\linewidth, bb=0 0 538 522]{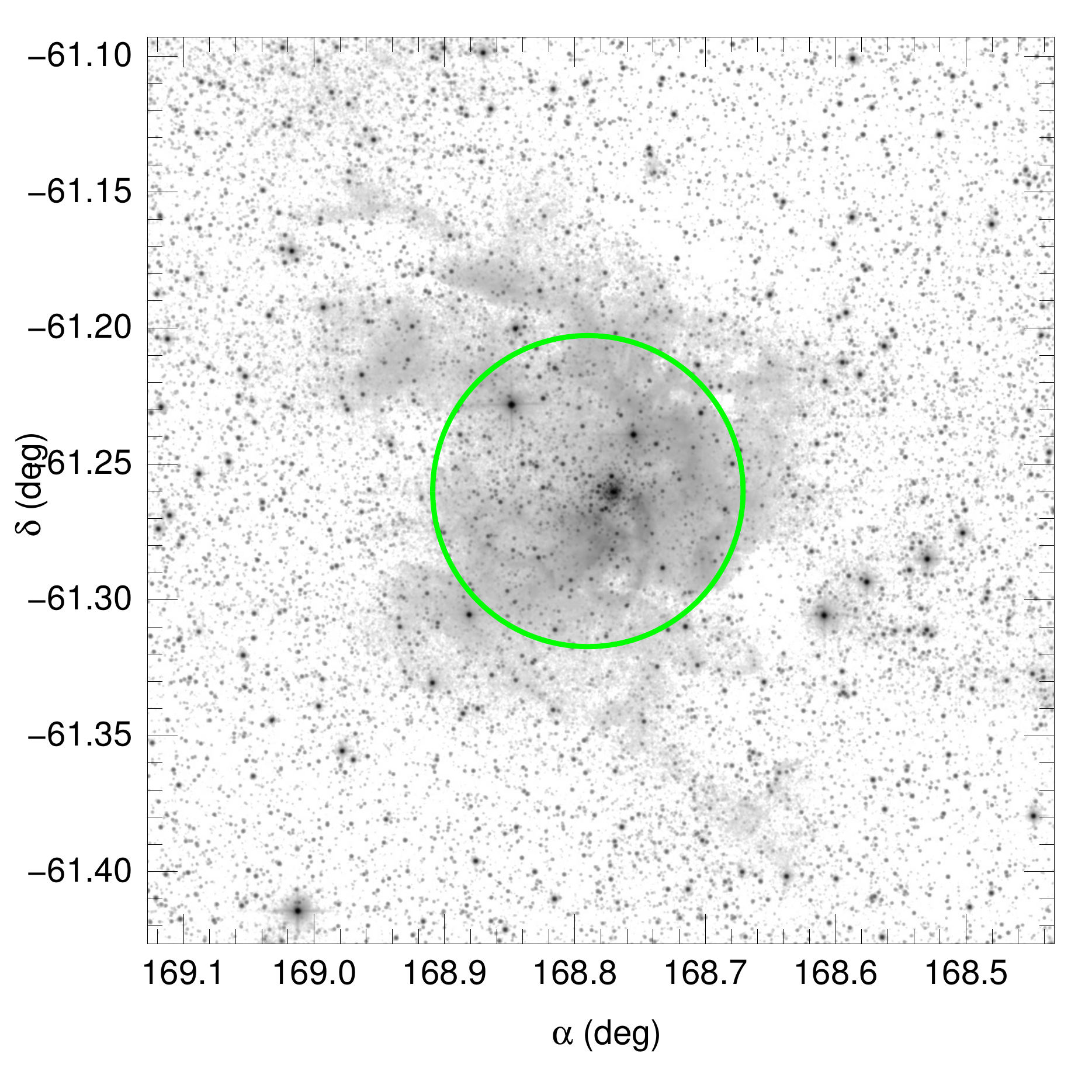}}
\centerline{\includegraphics*[width=0.34\linewidth, bb=0 0 538 522]{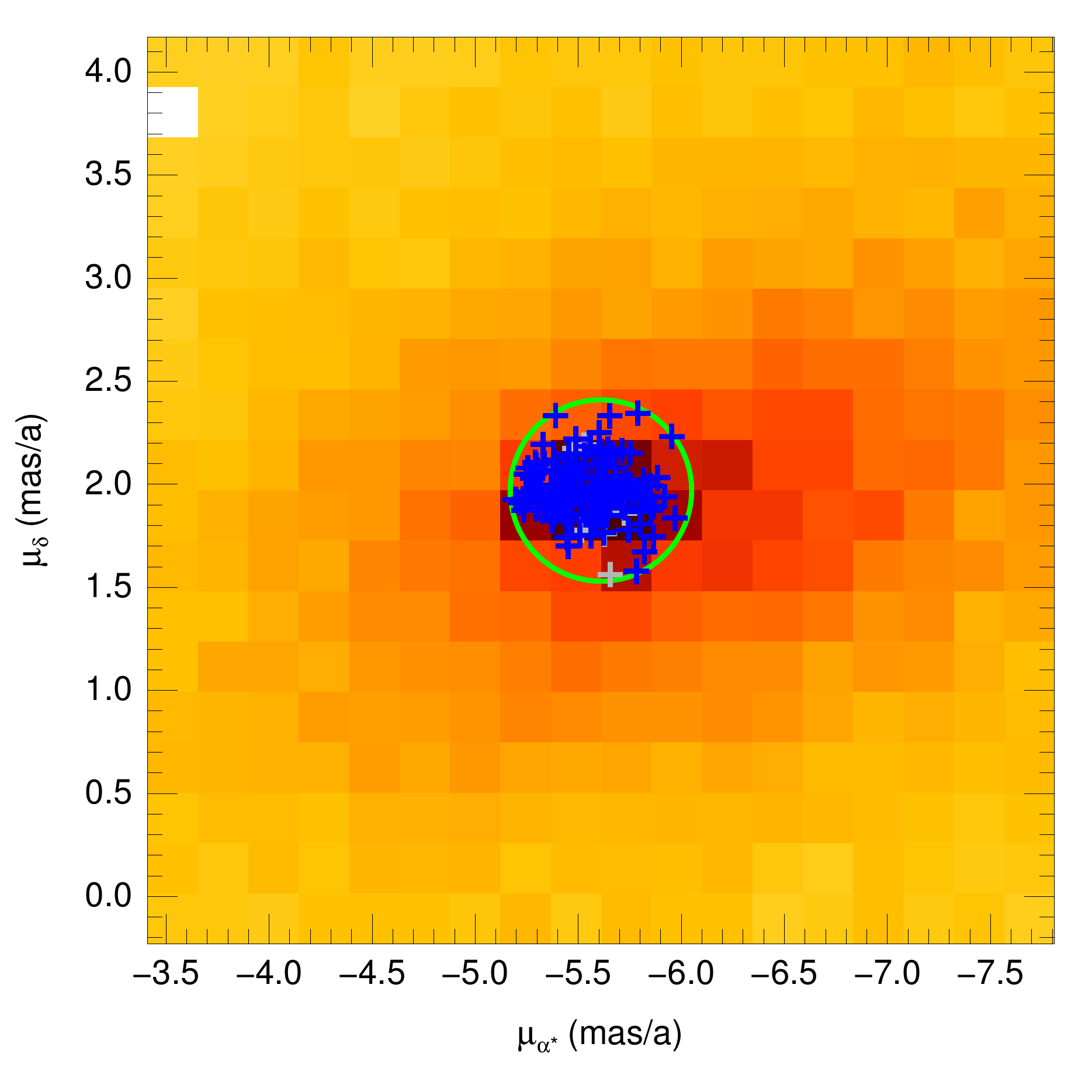} \
            \includegraphics*[width=0.34\linewidth, bb=0 0 538 522]{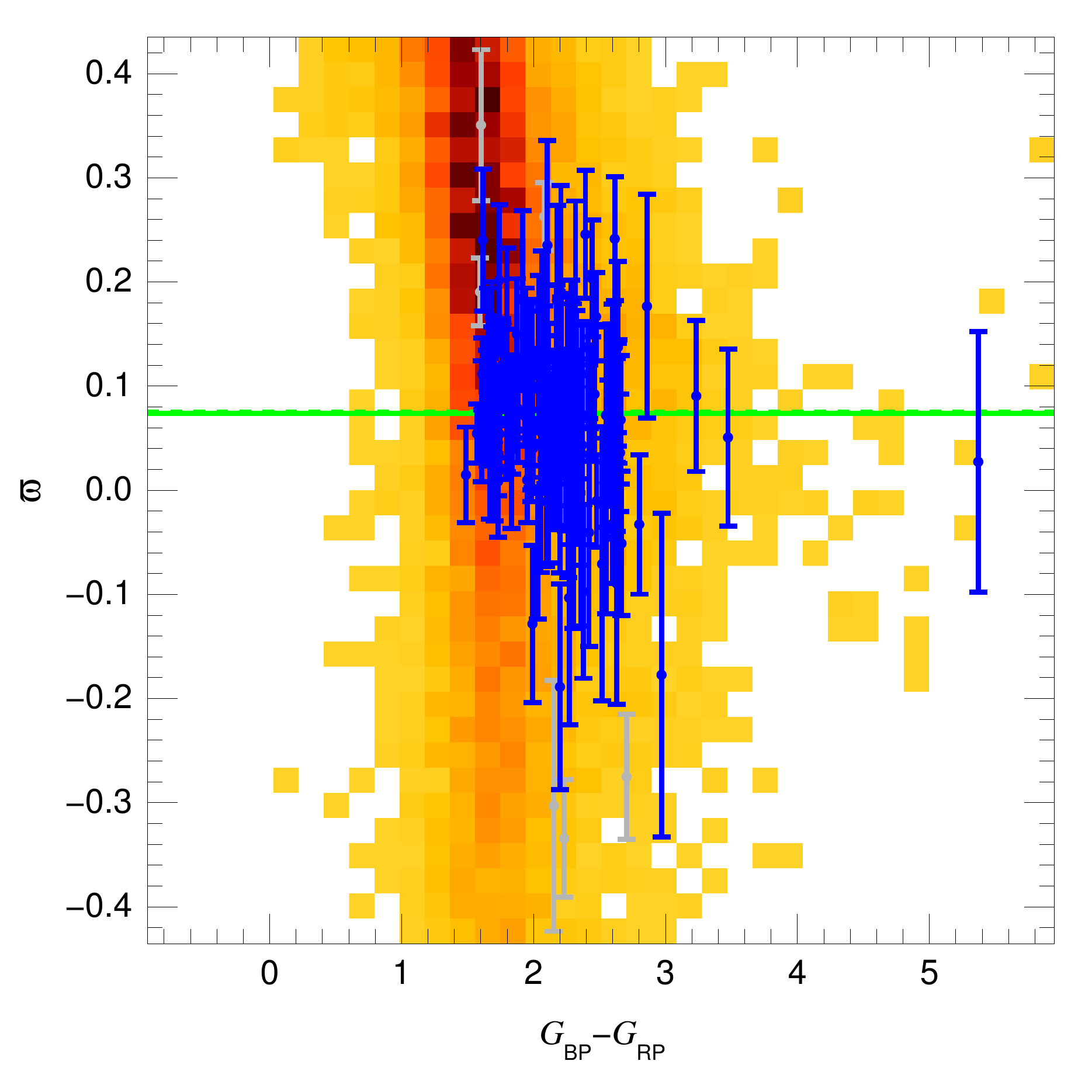} \
            \includegraphics*[width=0.34\linewidth, bb=0 0 538 522]{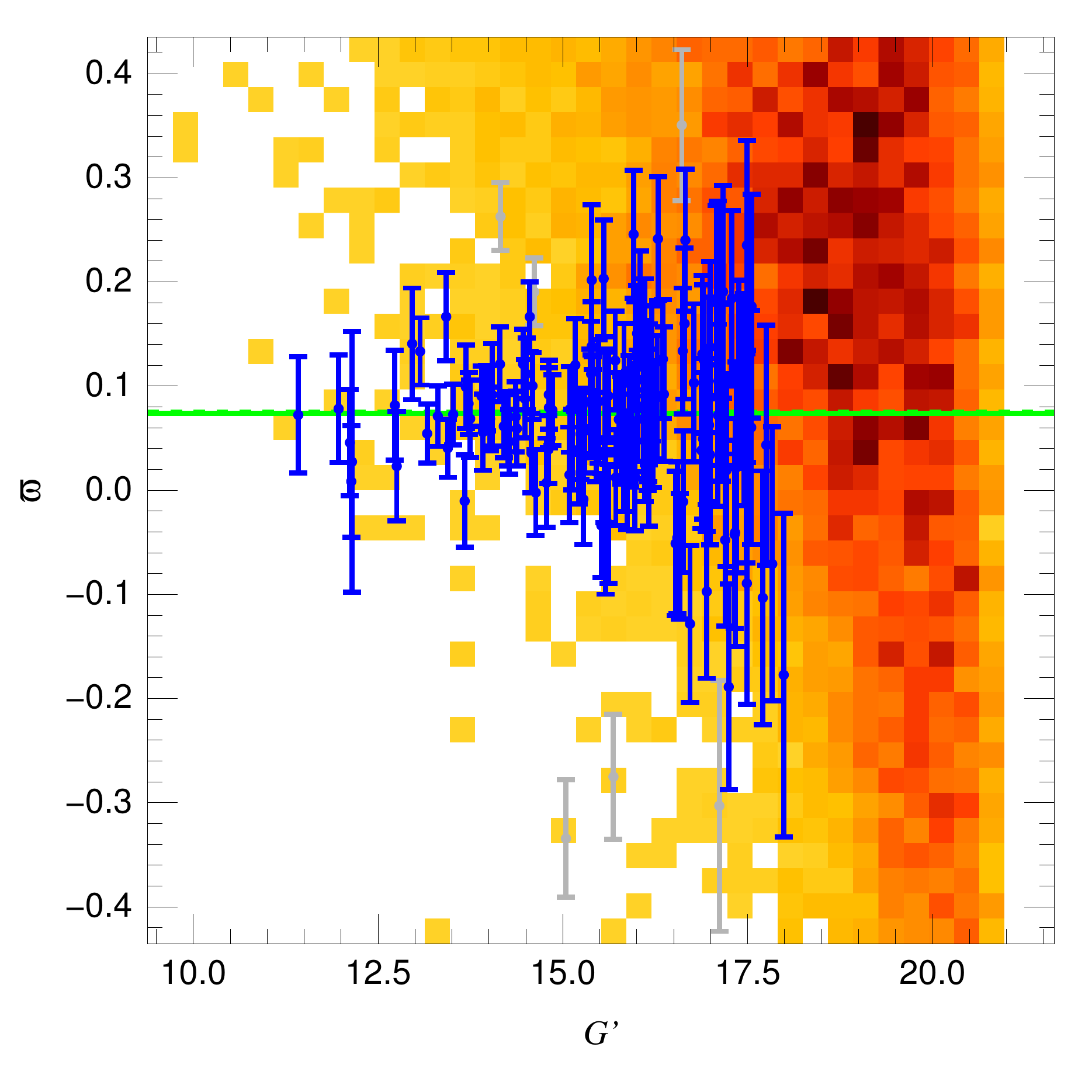}}
\centerline{\includegraphics*[width=0.34\linewidth, bb=0 0 538 522]{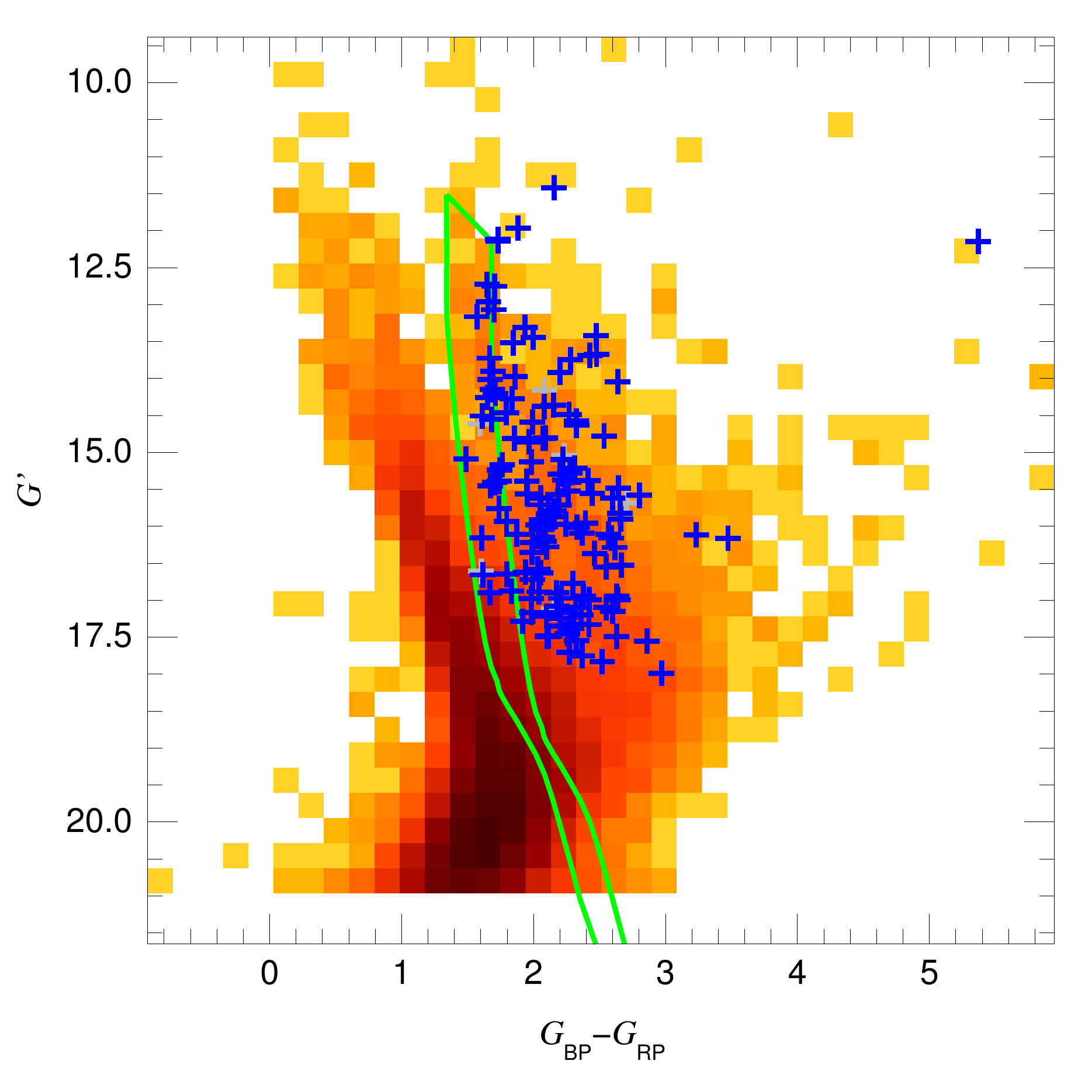} \
            \includegraphics*[width=0.34\linewidth, bb=0 0 538 522]{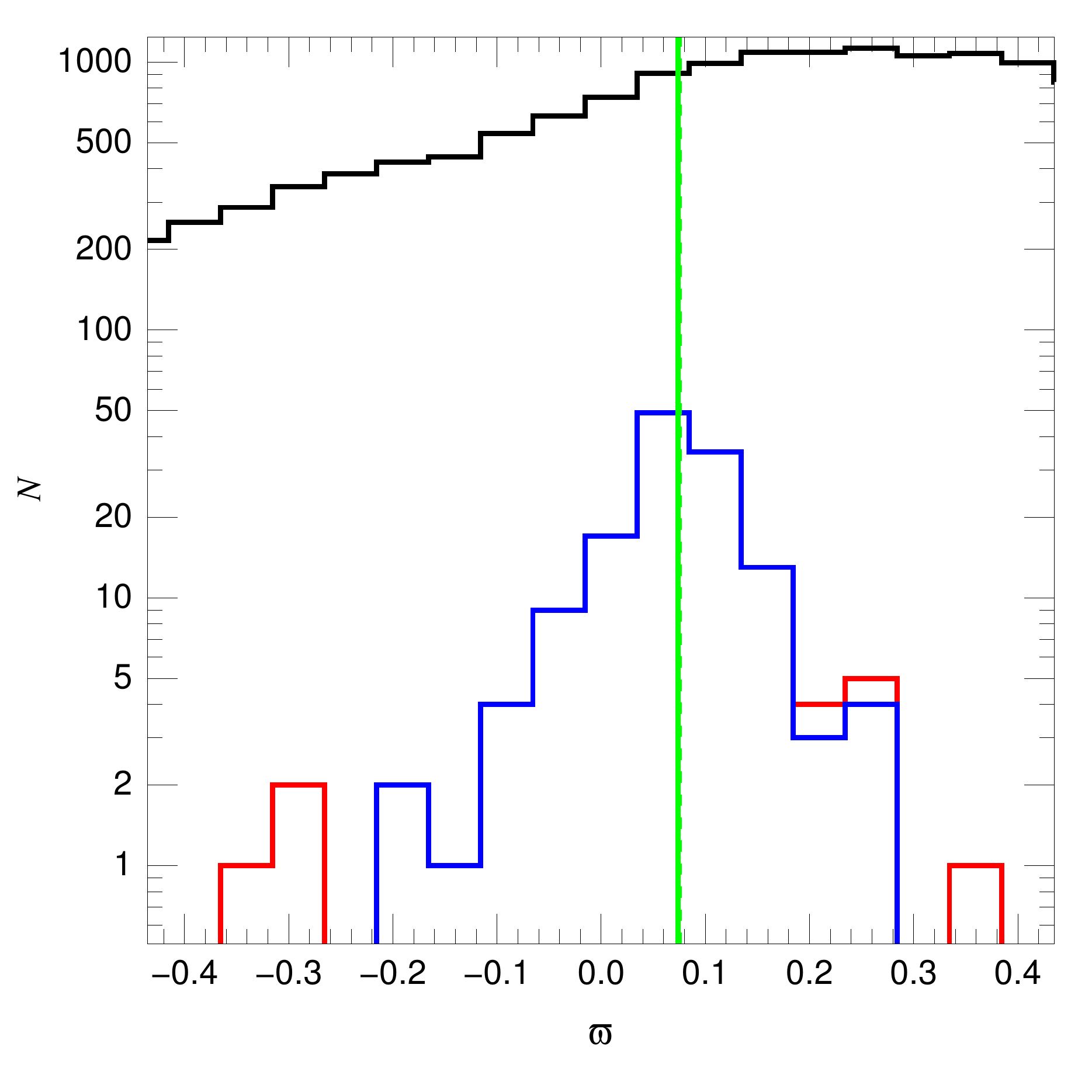} \
            \includegraphics*[width=0.34\linewidth, bb=0 0 538 522]{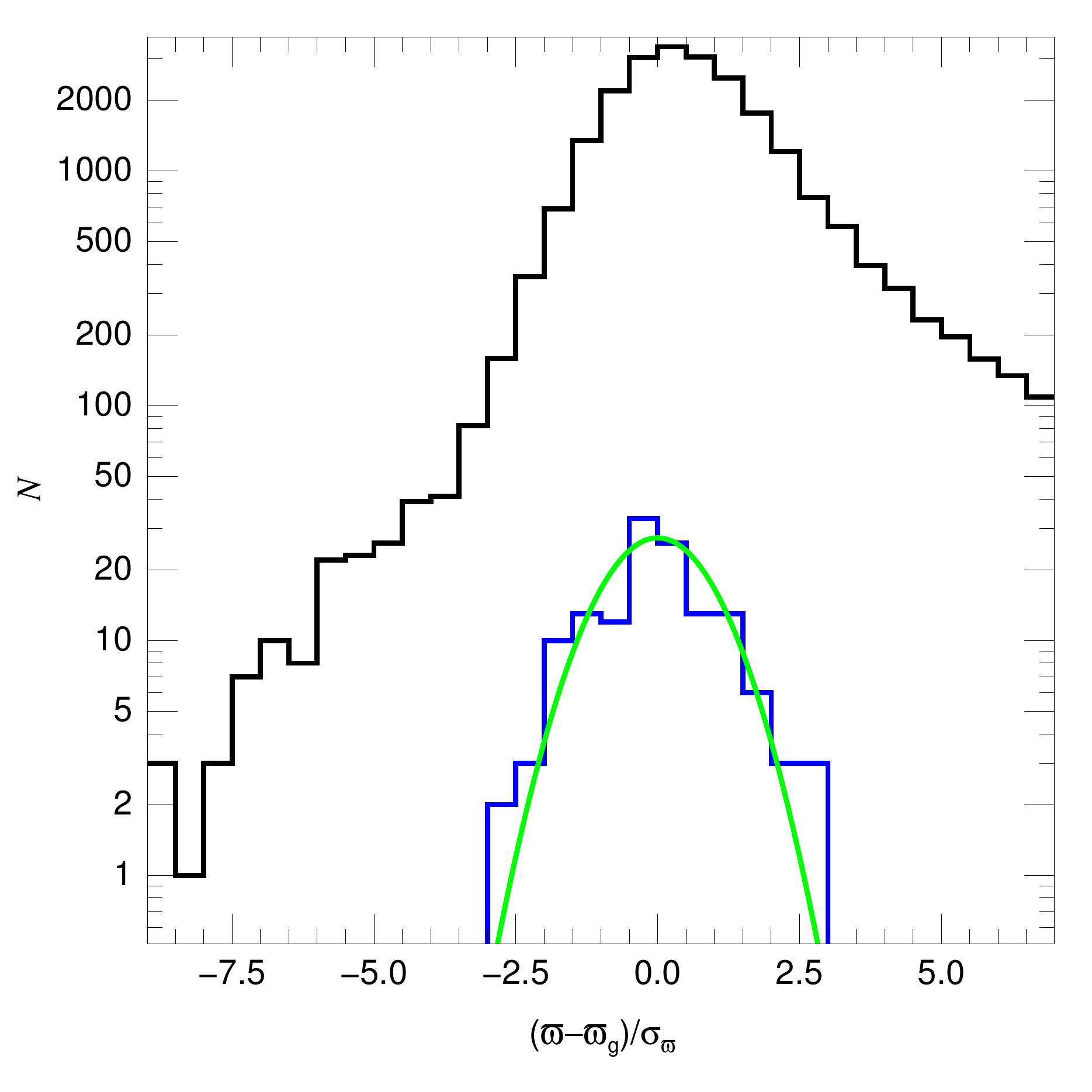}}
 \caption{NGC~3603 (\VO{001}) % REFEREE \VO{001}
         {\it Gaia}~DR2 distances and membership results. {\it Top row (left to right):} source density diagram, DSS2~red image, and 
         2MASS~$J$ image. {\it Middle row (left to right):} proper motions, color-parallax, and magnitude-parallax diagrams. {\it Bottom row
         (left to right):} color-magnitude diagram, parallax histogram, and normalized-parallax histogram. In all diagrams a heat-type scale
         (increasing as white-yellow-orange-red-black) is used to indicate the total {\it Gaia}~DR2 density in a linear scale (except in the
         CMD, where a log scale is used). In the first four panels the green circle indicates the coordinates or proper motion constraints. In
         the CMD the green lines show the reference extinguished isochrone (right) and the displaced isochrone used as constraint (left), joined at
         the top by the extinction trajectory. In all diagrams the blue symbols indicate the objects used in the final sample and the gray symbols 
         those rejected by the normalized parallax criterion. The plotted parallax uncertainties are the external ones. In the parallax 
         histogram black indicate the total {\it Gaia~DR2} density, red the sample prior to the application of the normalized parallax 
         criterion, and blue the final sample, while the two green vertical lines delineate the weighted-mean parallax: dotted for \pigz\ 
         and solid for \pig. Black and blue have the same meaning in the normalized parallax histogram, where the green line shows the 
         expected normal distribution.}
\label{NGC_3603_Gaia}
\end{figure*}   

\begin{figure*}
\centerline{\includegraphics*[width=0.34\linewidth, bb=0 0 538 522]{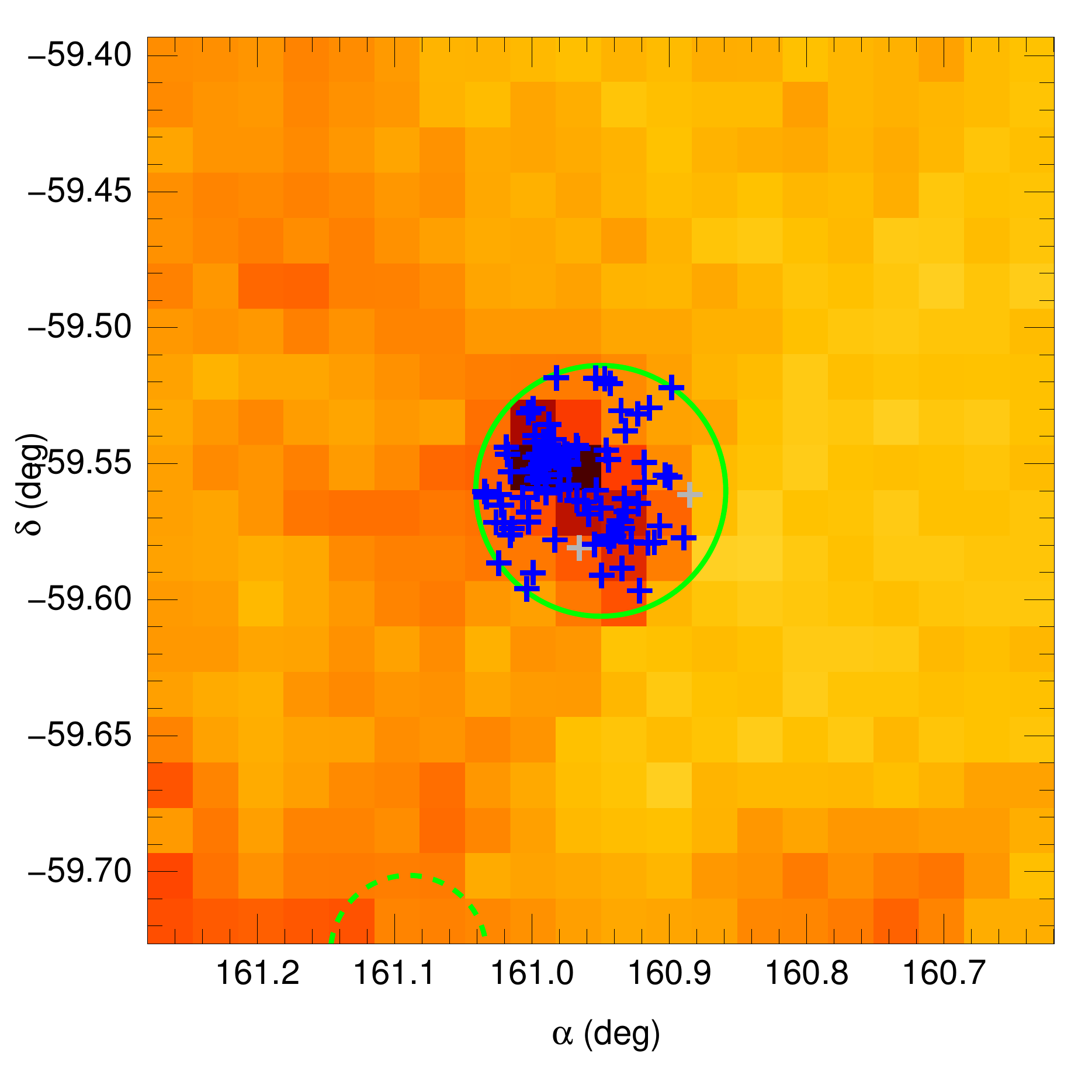} \
            \includegraphics*[width=0.34\linewidth, bb=0 0 538 522]{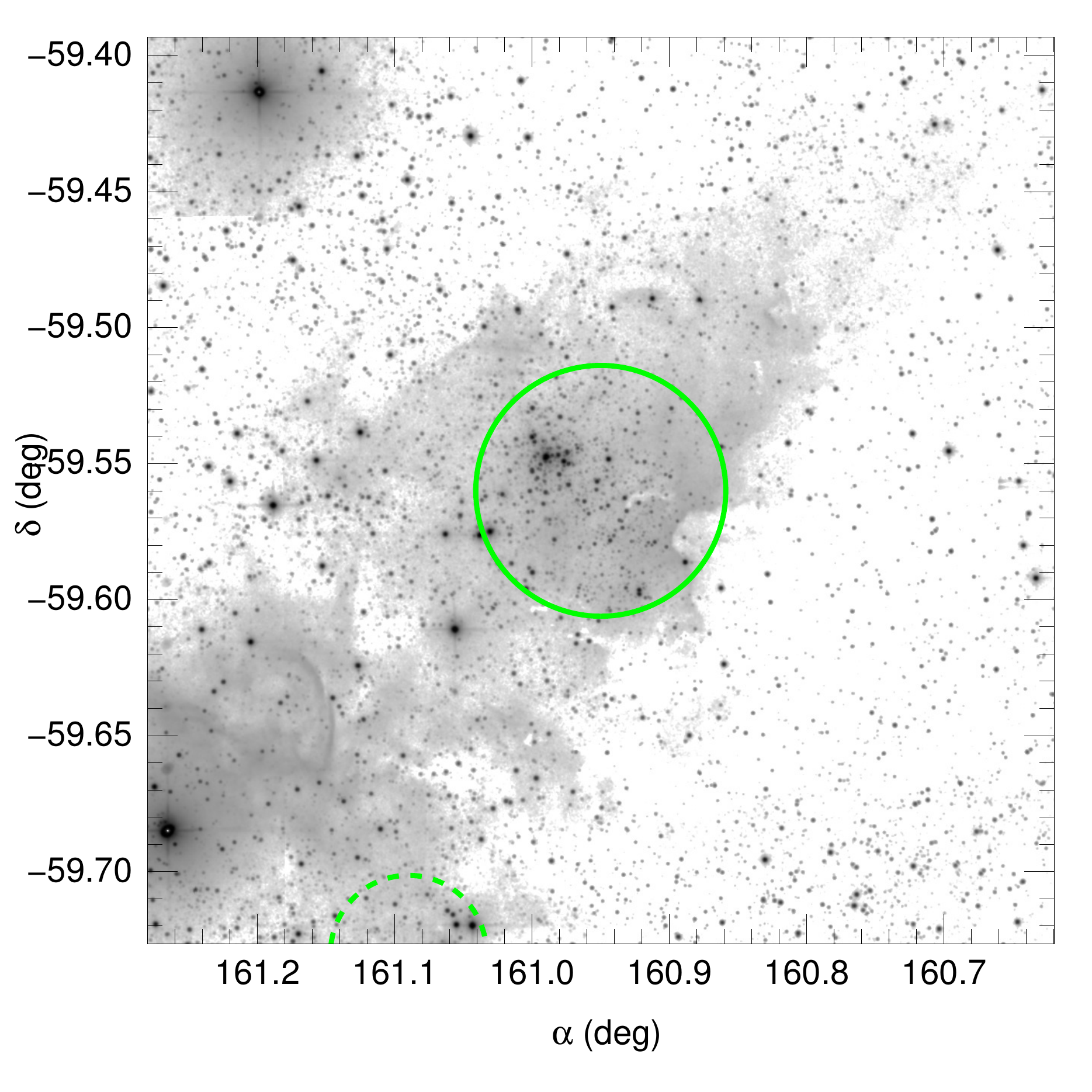} \
            \includegraphics*[width=0.34\linewidth, bb=0 0 538 522]{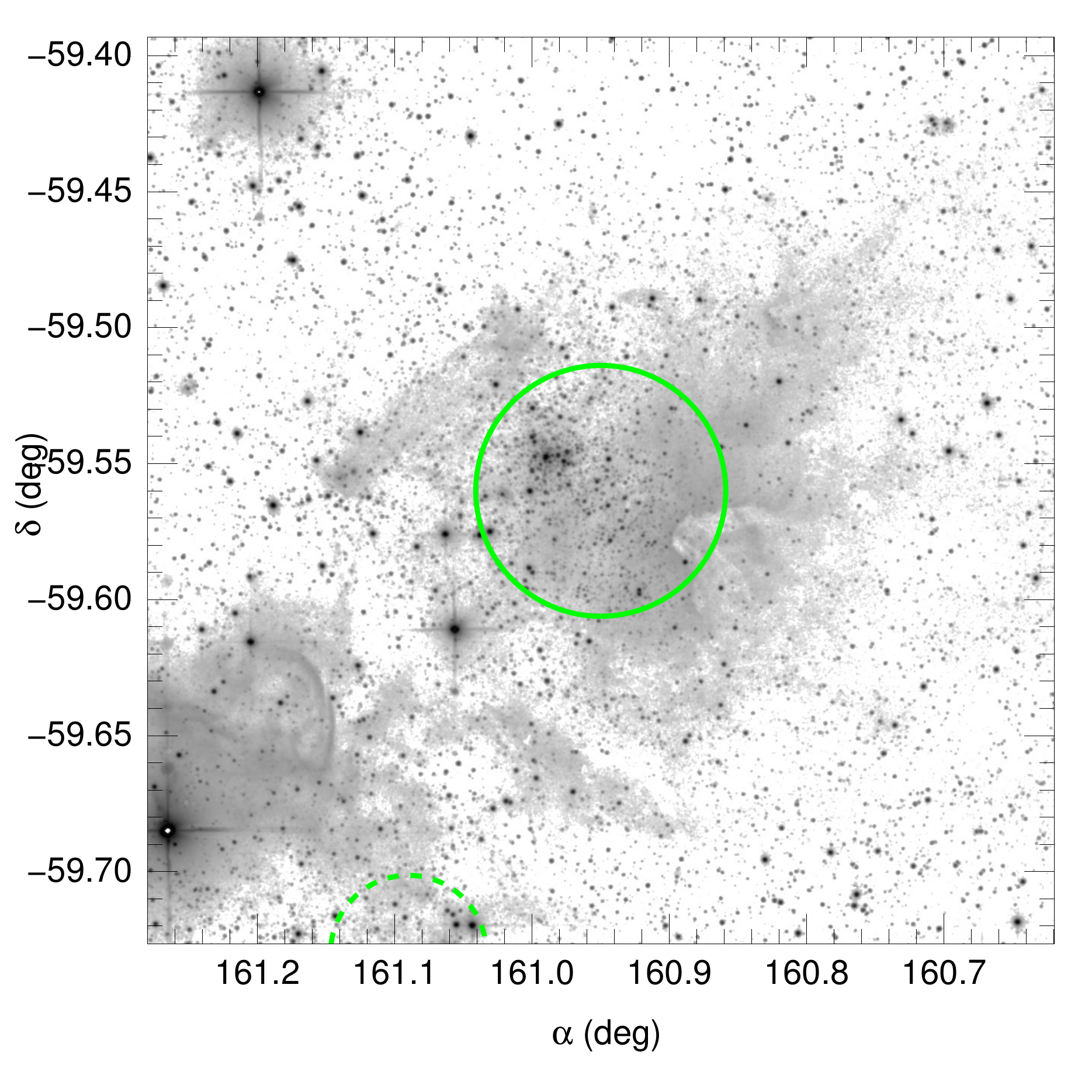}}
\centerline{\includegraphics*[width=0.34\linewidth, bb=0 0 538 522]{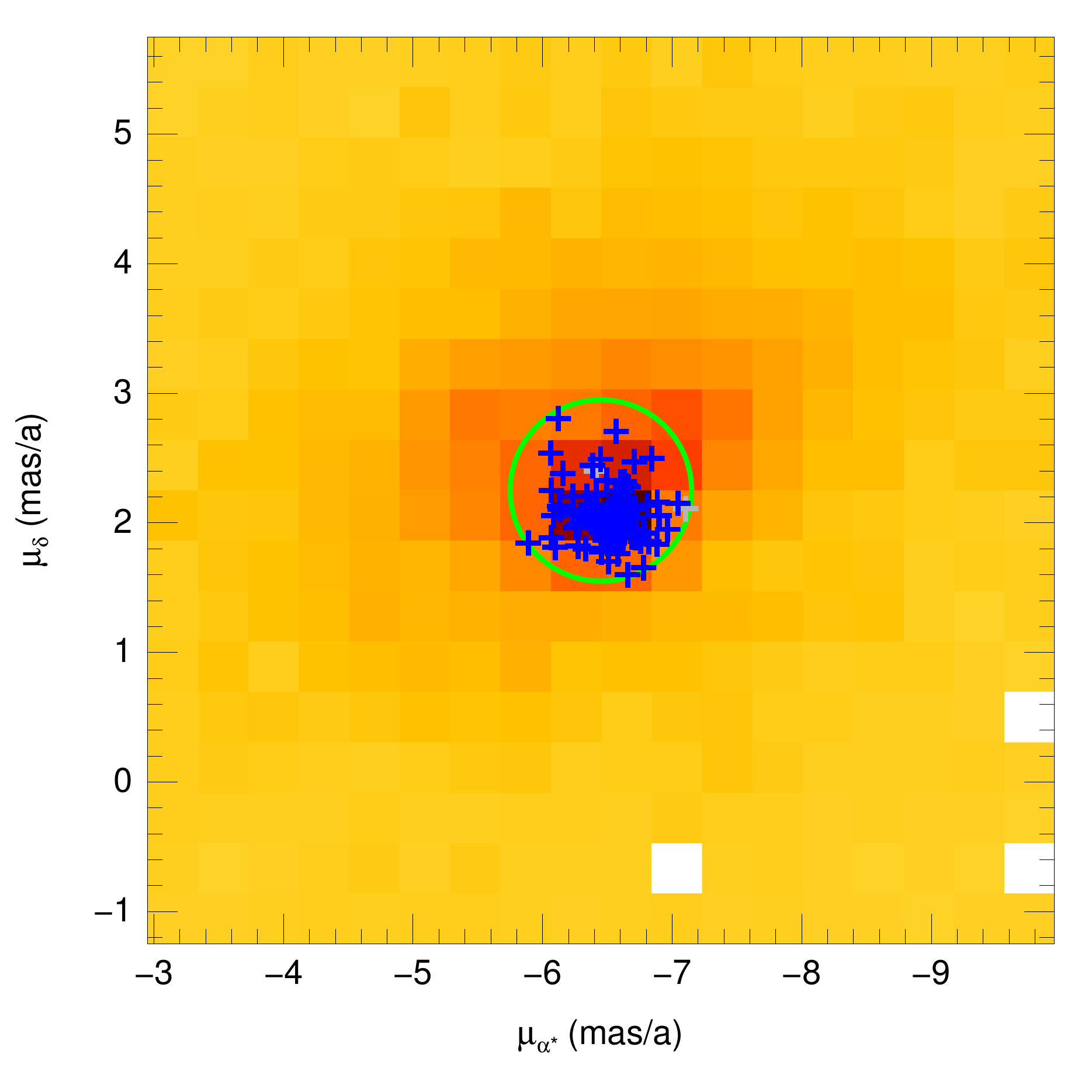} \
            \includegraphics*[width=0.34\linewidth, bb=0 0 538 522]{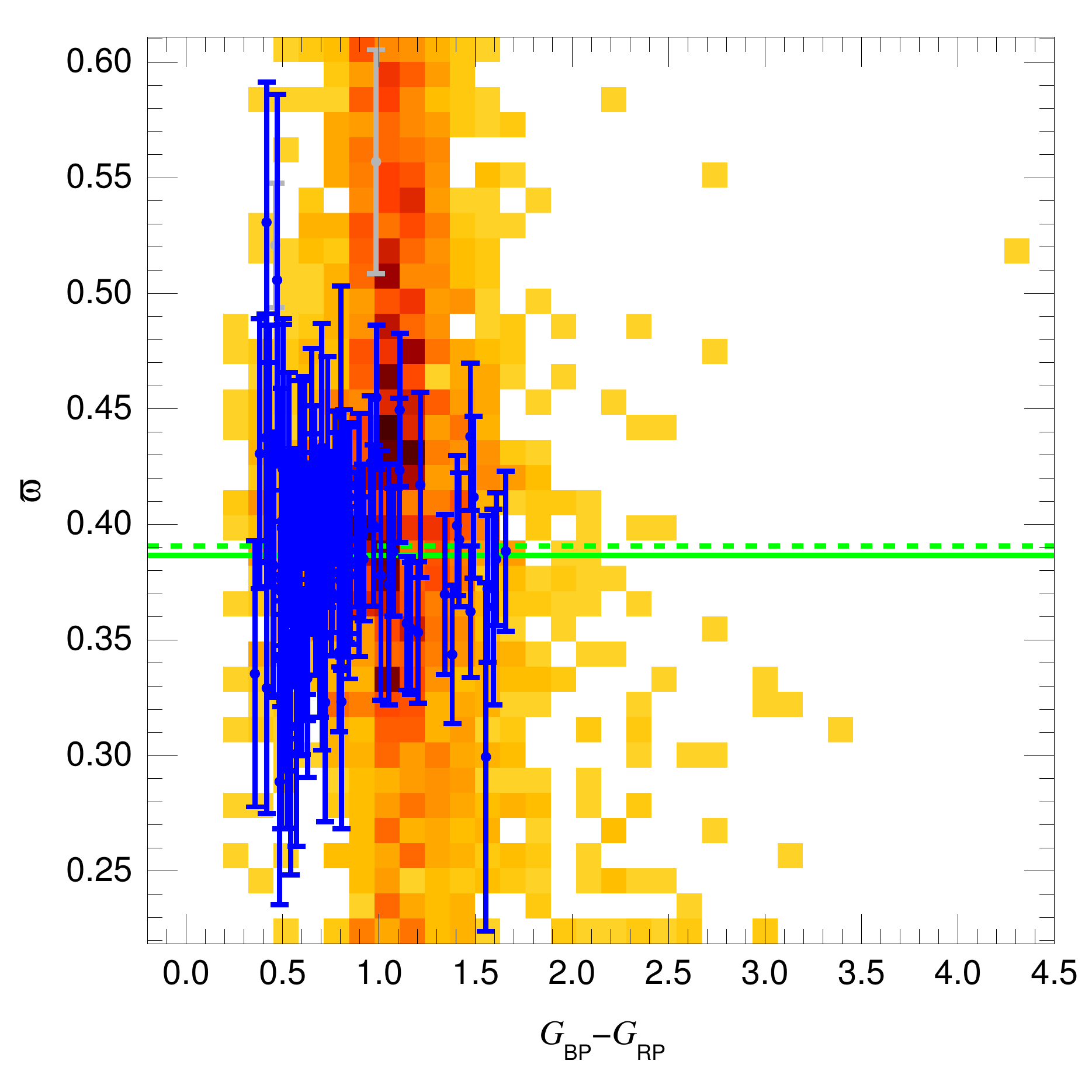} \
            \includegraphics*[width=0.34\linewidth, bb=0 0 538 522]{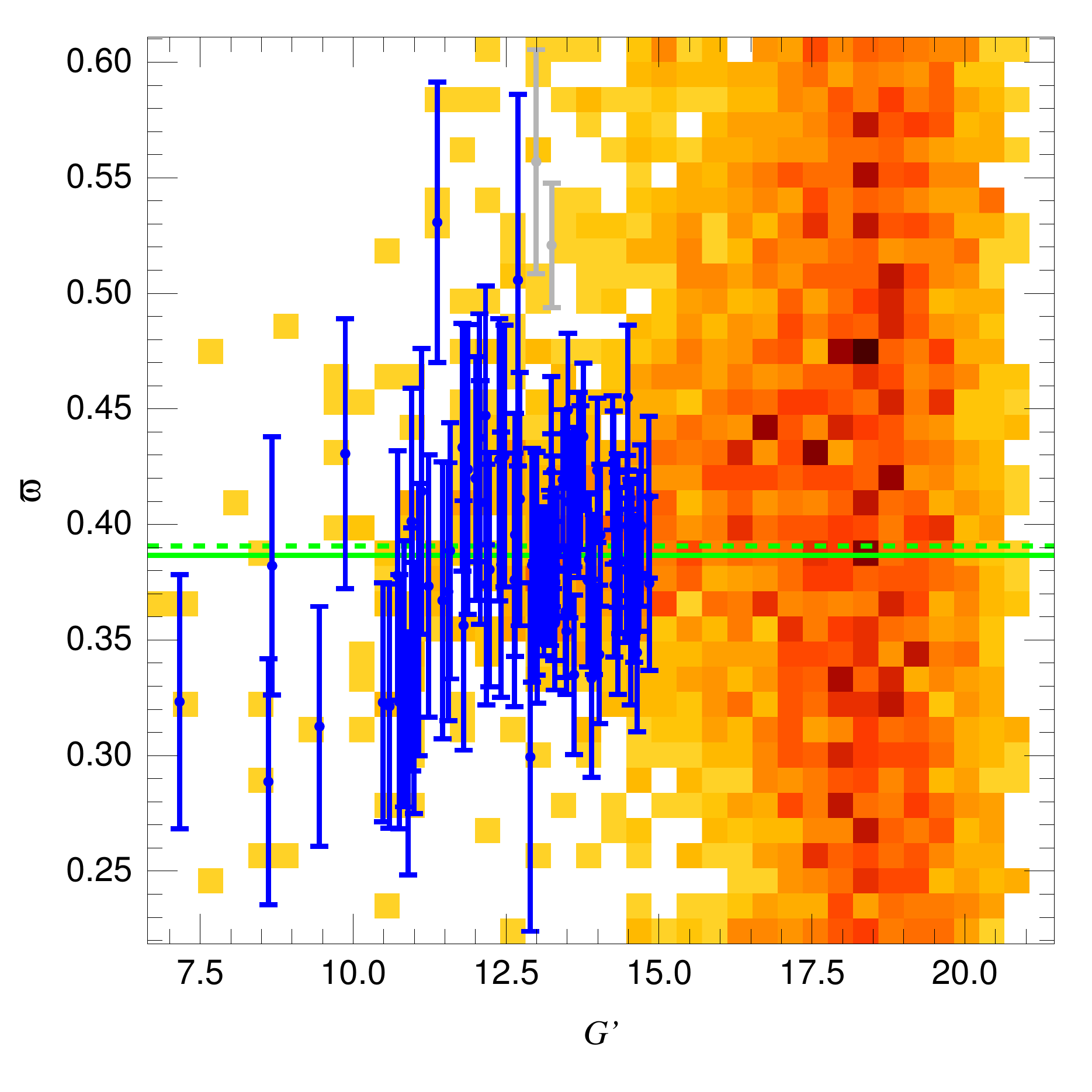}}
\centerline{\includegraphics*[width=0.34\linewidth, bb=0 0 538 522]{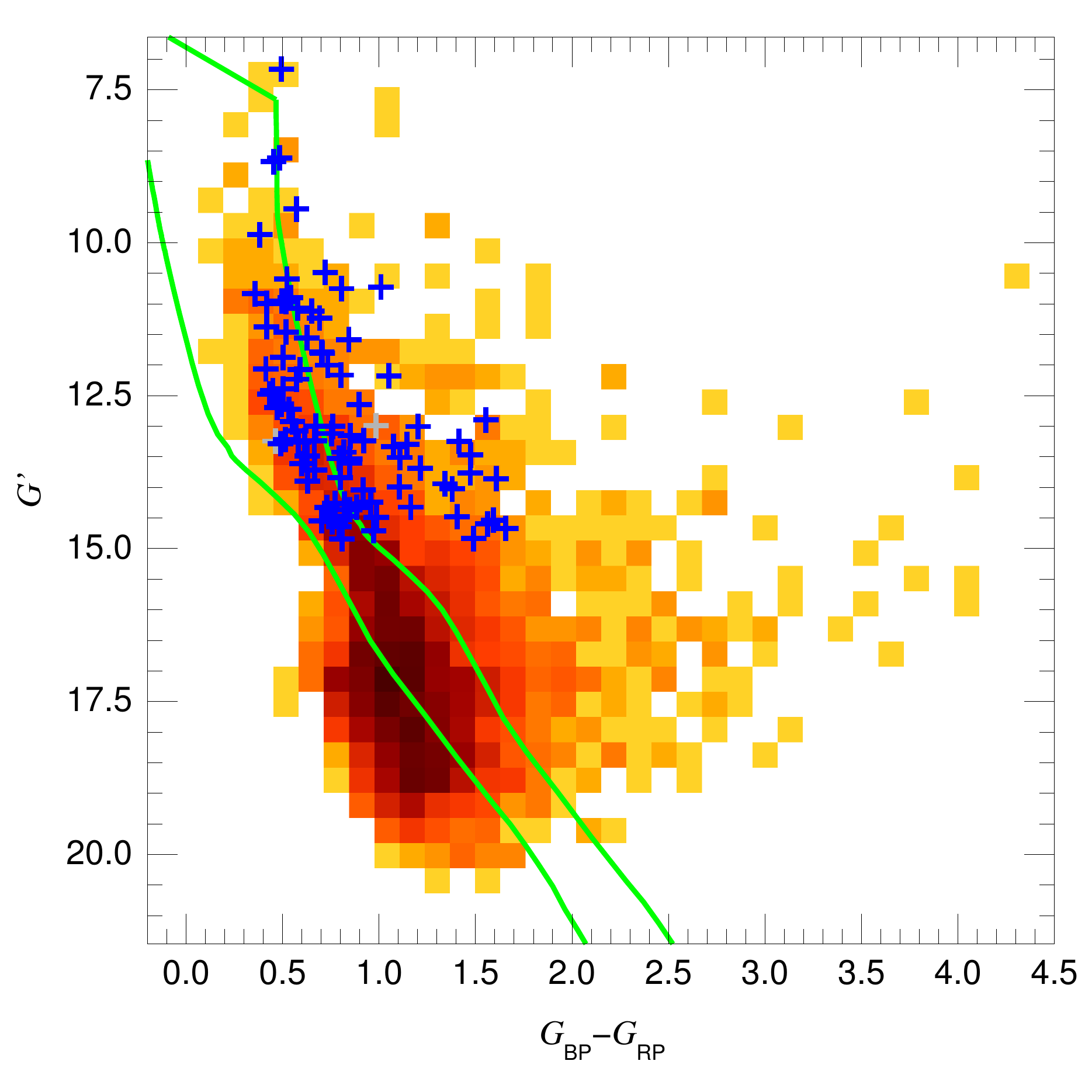} \
            \includegraphics*[width=0.34\linewidth, bb=0 0 538 522]{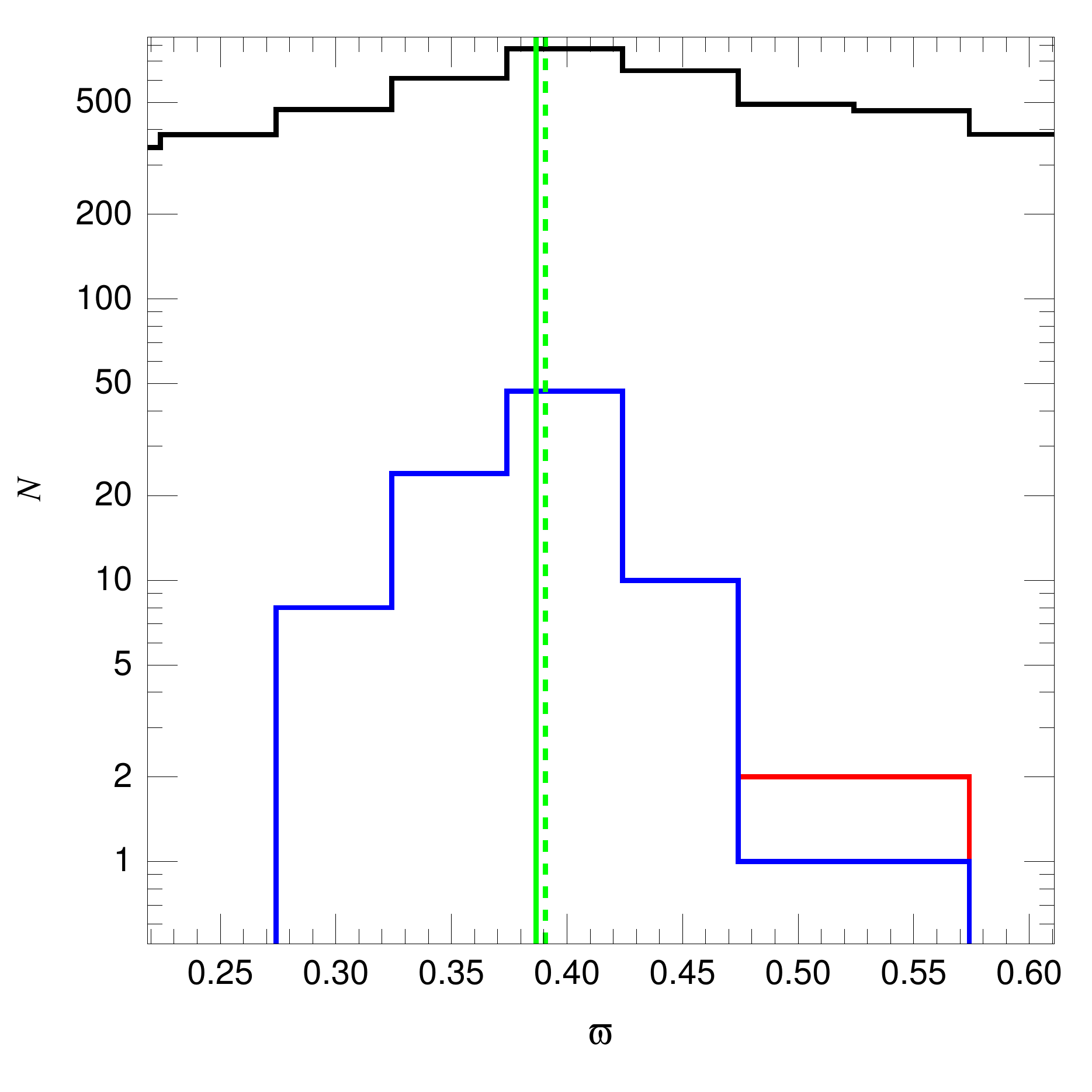} \
            \includegraphics*[width=0.34\linewidth, bb=0 0 538 522]{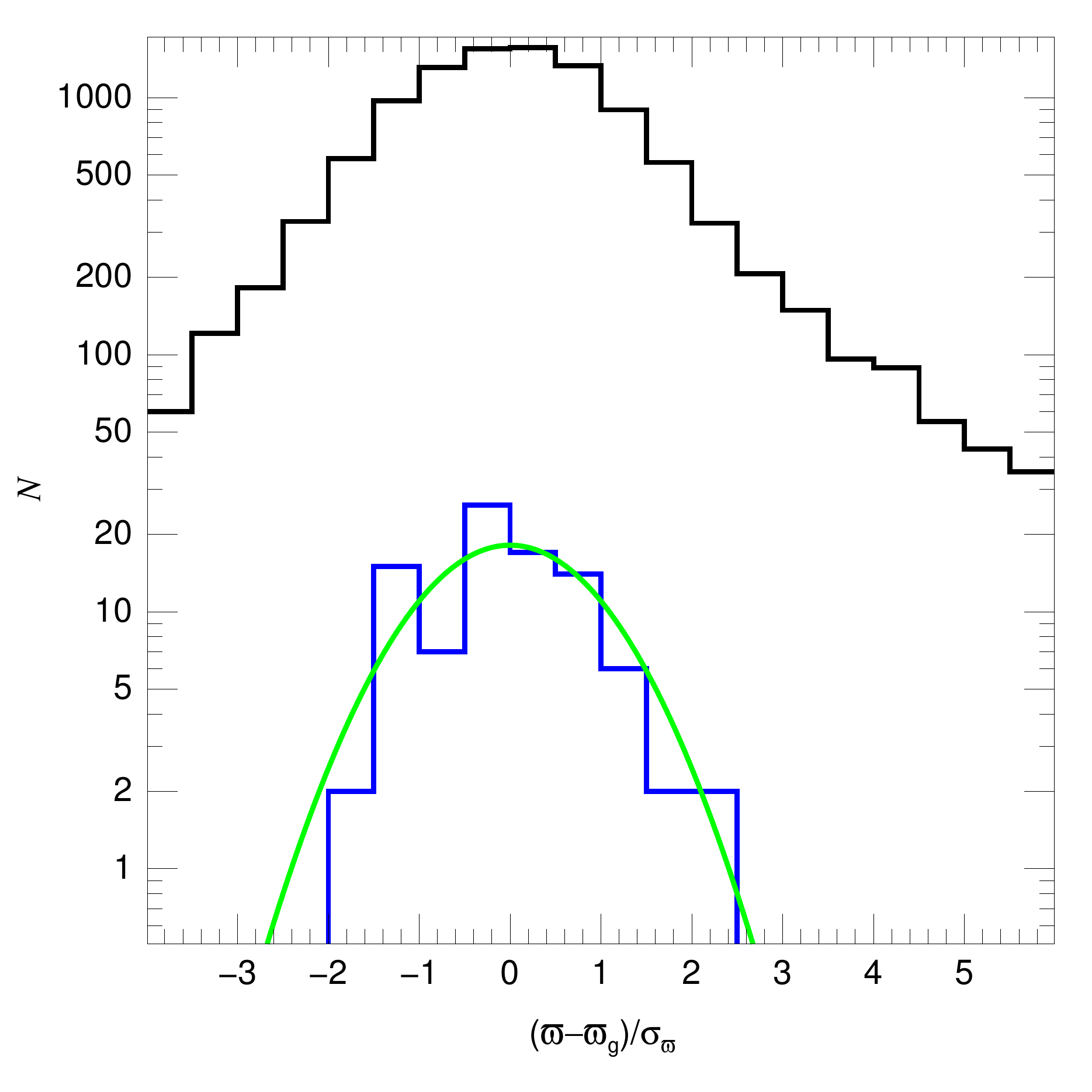}}
\caption{Same as Fig.~\ref{NGC_3603_Gaia} for 
         Trumpler~14 (\VO{002}) % REFEREE \VO{002}
         but with the 2MASS~$J$ image in the top center panel and the 2MASS~$K$ image in the top right one.
         The partial dashed green circle in the top three panels shows the position of the neighbor 
         Trumpler~16~W (\VO{003}). % REFEREE \VO{003}
         }
\label{Trumpler_14_Gaia}
\end{figure*}   

\begin{figure*}
\centerline{\includegraphics*[width=0.34\linewidth, bb=0 0 538 522]{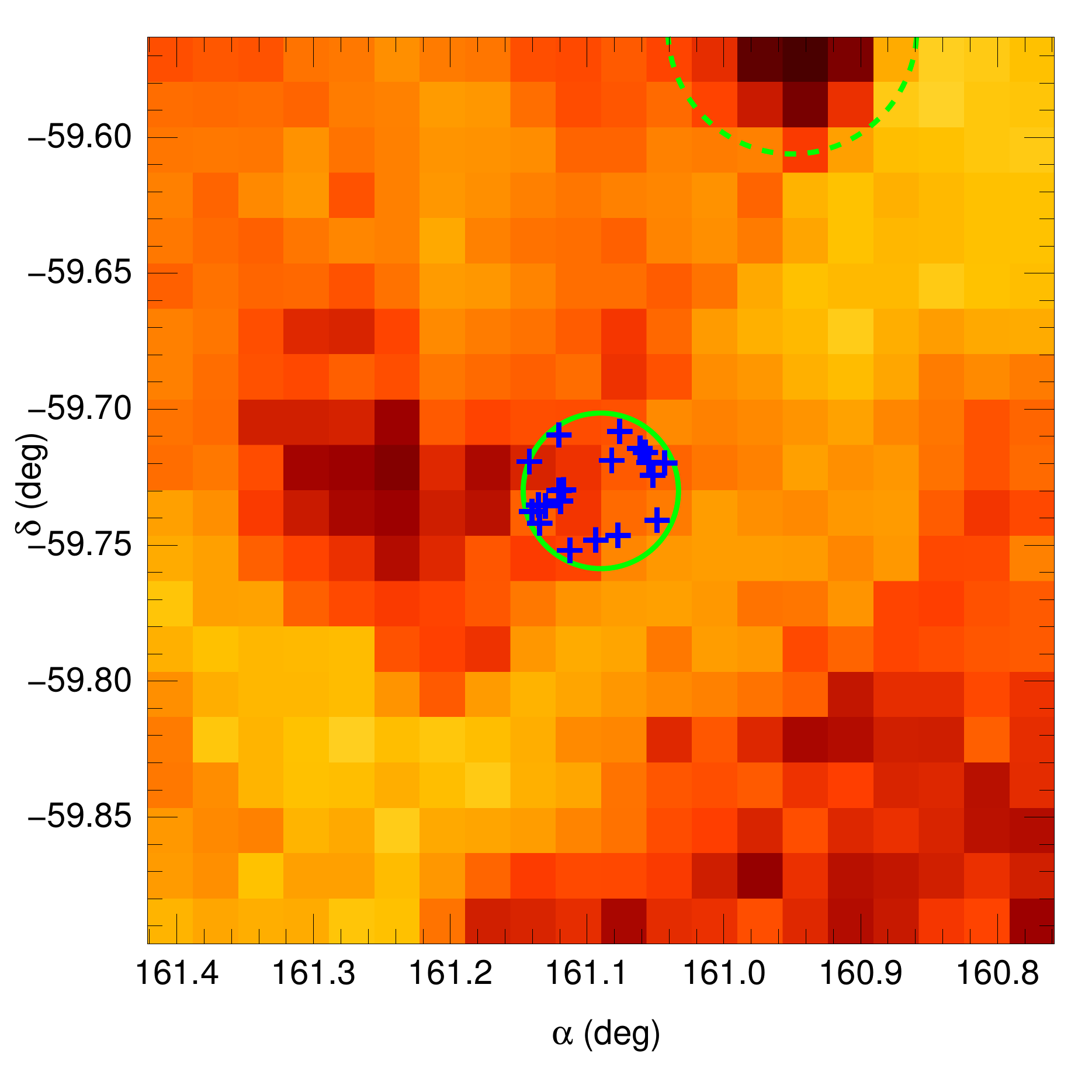} \
            \includegraphics*[width=0.34\linewidth, bb=0 0 538 522]{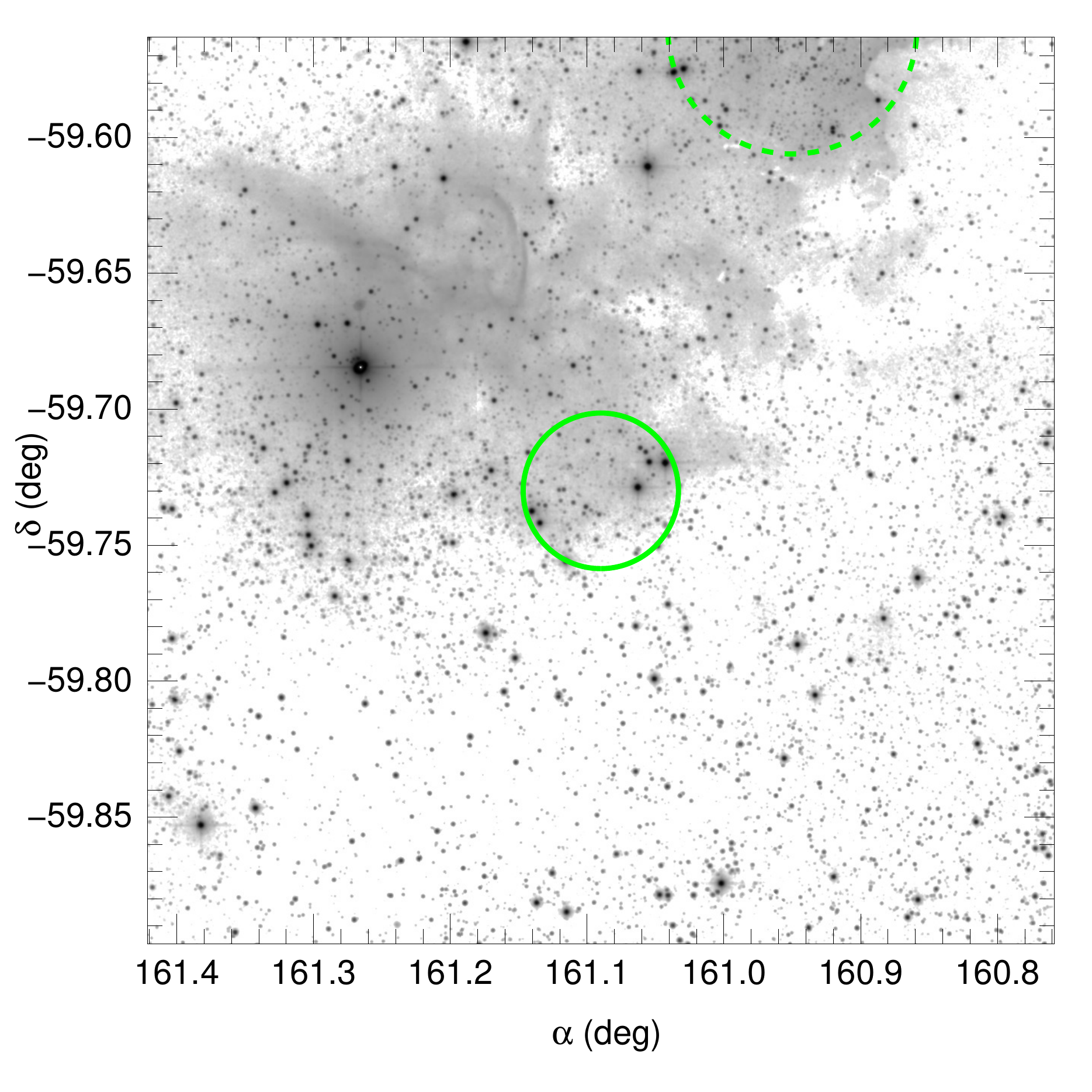} \
            \includegraphics*[width=0.34\linewidth, bb=0 0 538 522]{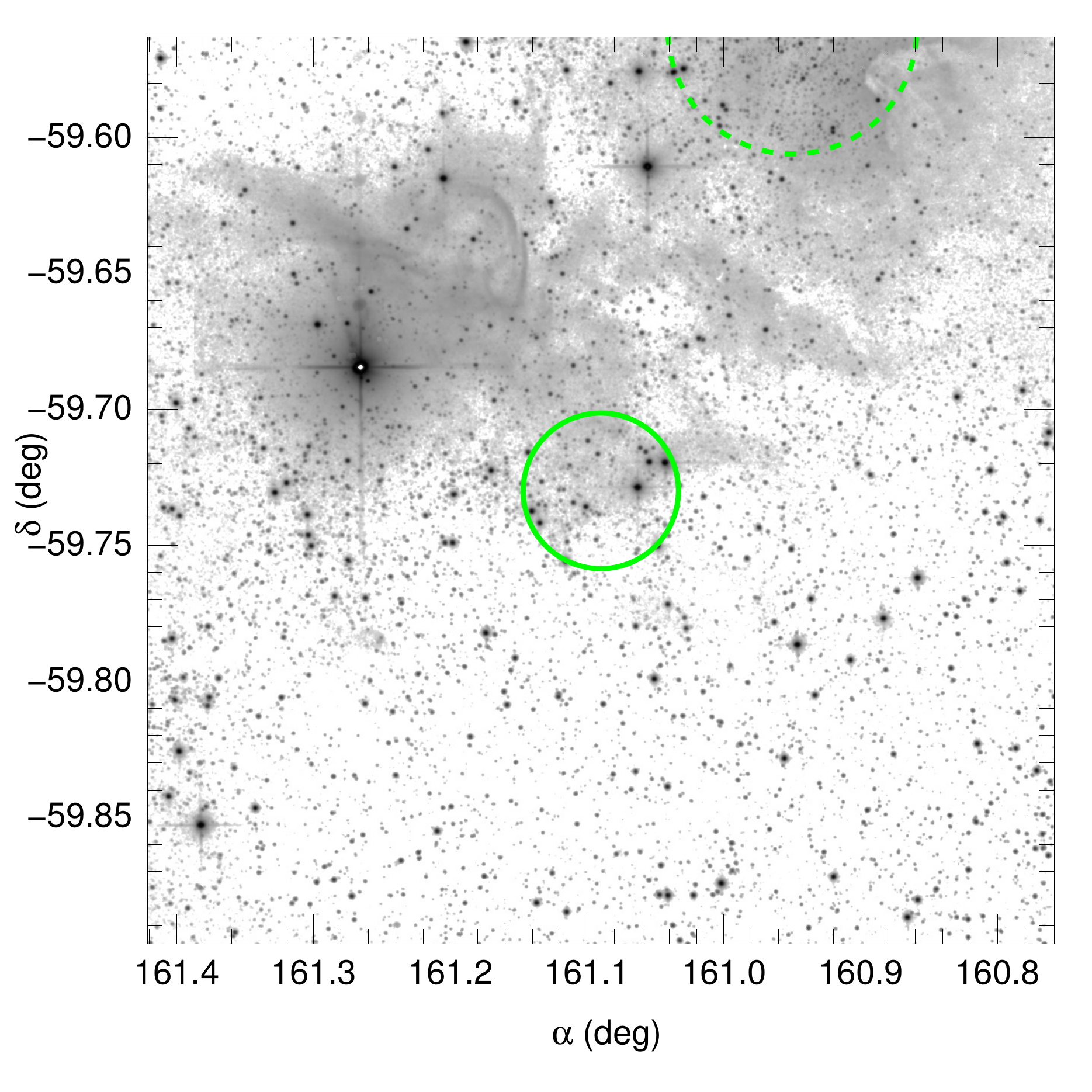}}
\centerline{\includegraphics*[width=0.34\linewidth, bb=0 0 538 522]{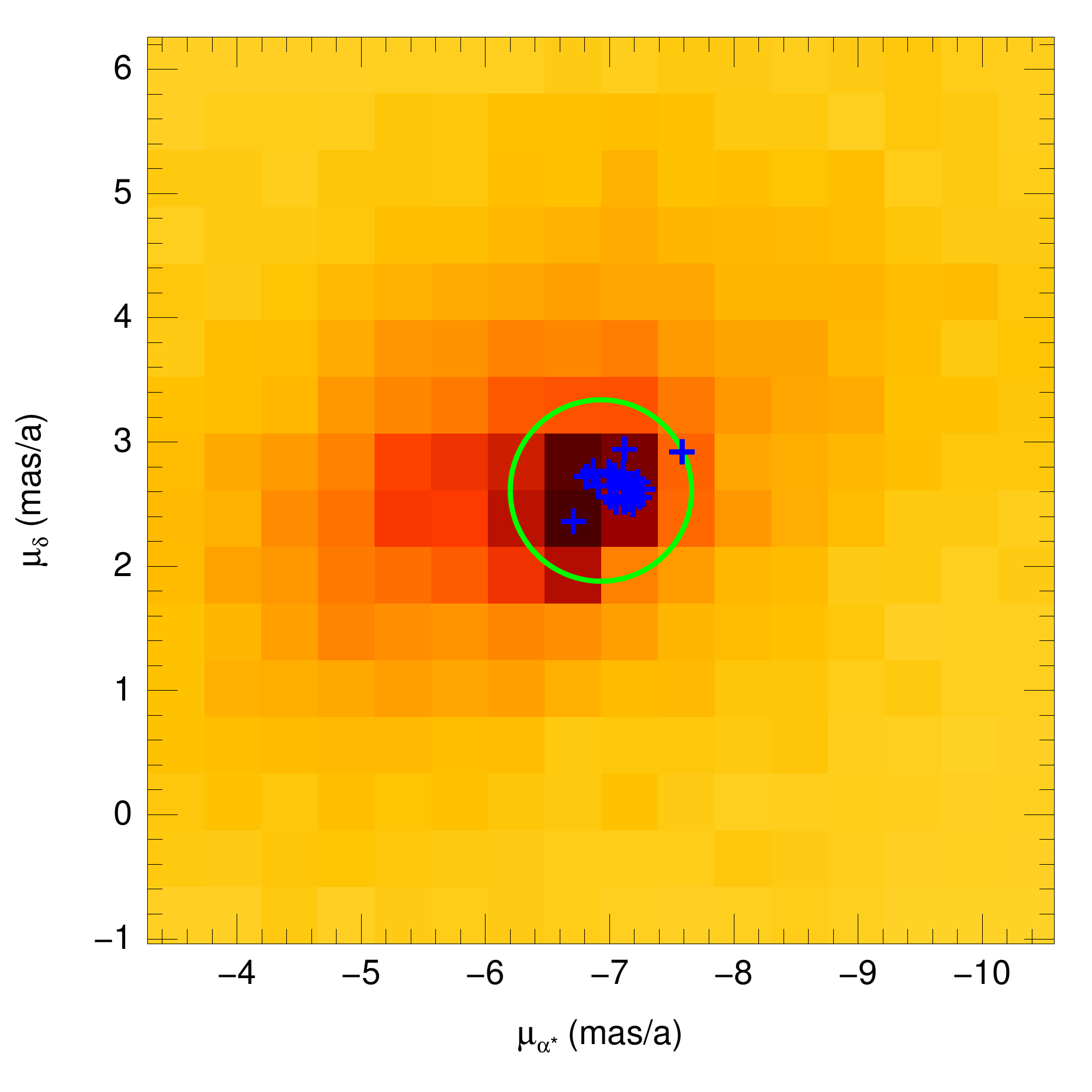} \
            \includegraphics*[width=0.34\linewidth, bb=0 0 538 522]{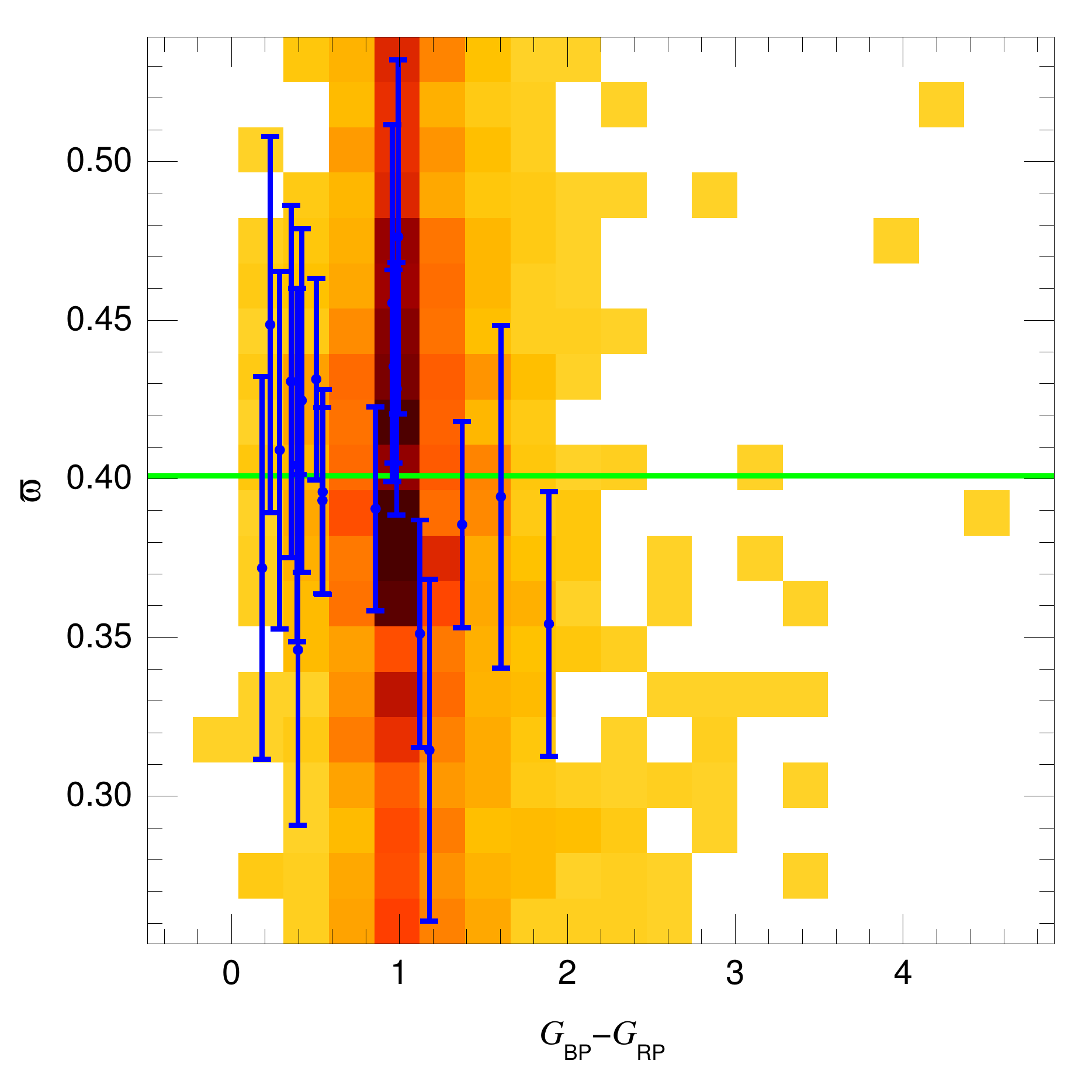} \
            \includegraphics*[width=0.34\linewidth, bb=0 0 538 522]{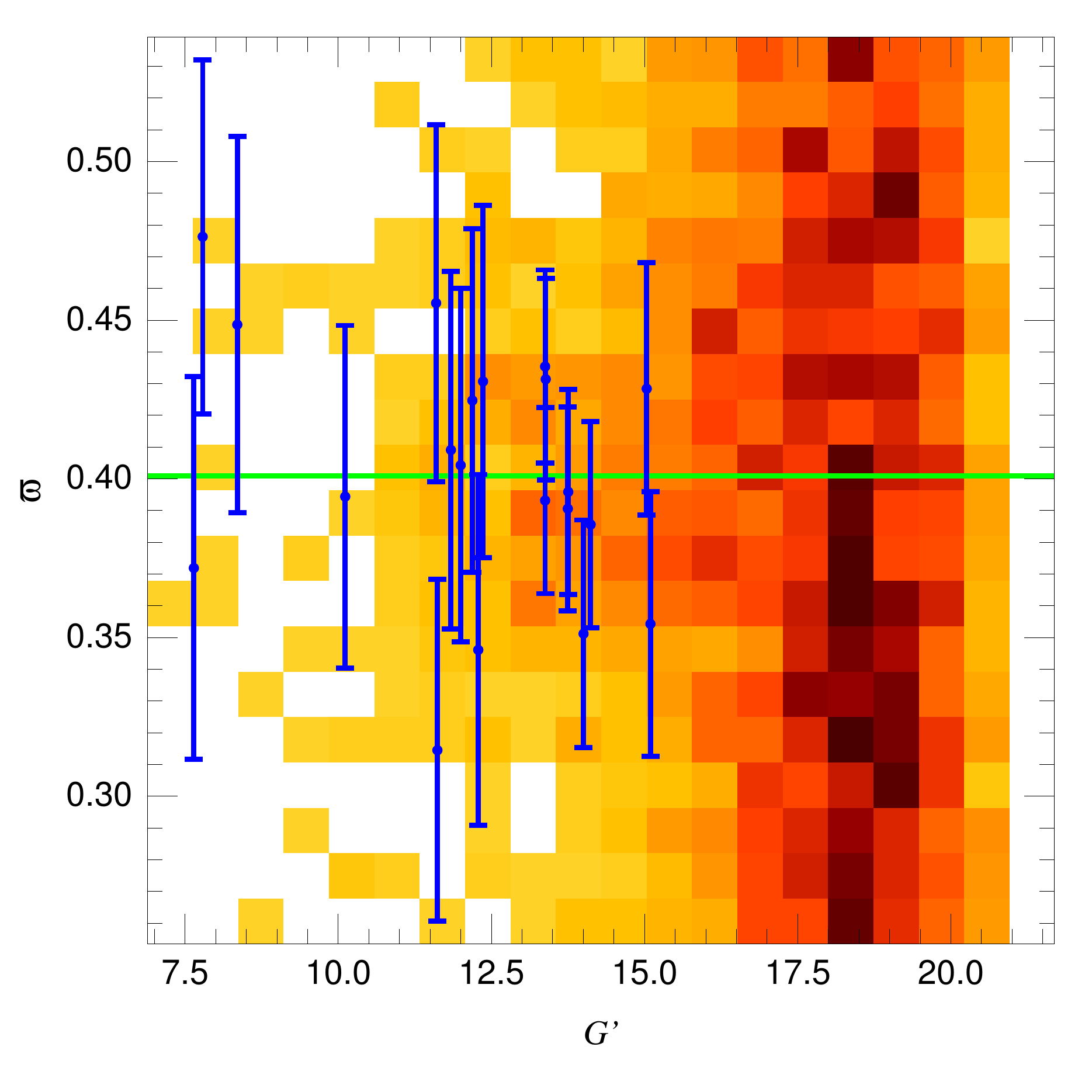}}
\centerline{\includegraphics*[width=0.34\linewidth, bb=0 0 538 522]{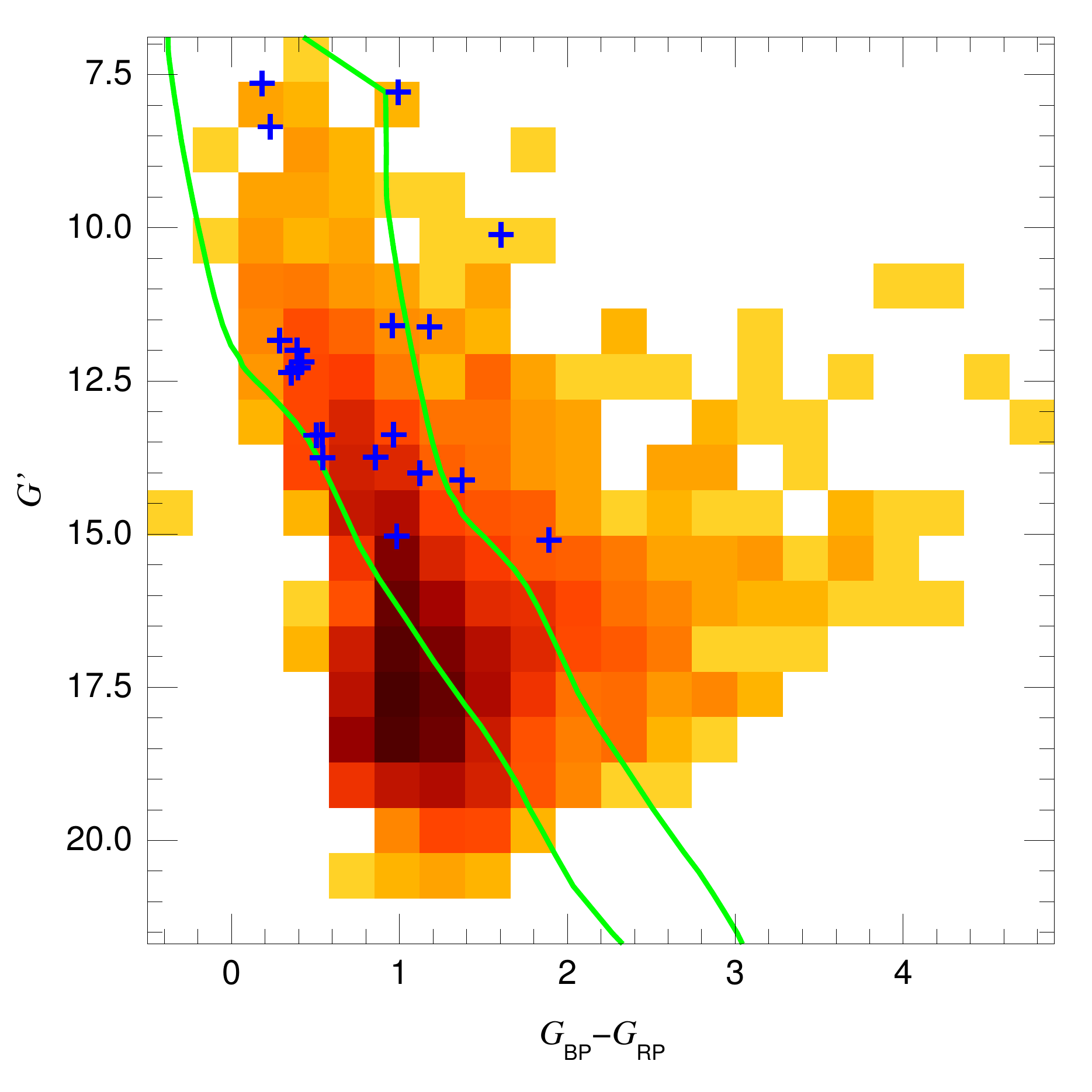} \
            \includegraphics*[width=0.34\linewidth, bb=0 0 538 522]{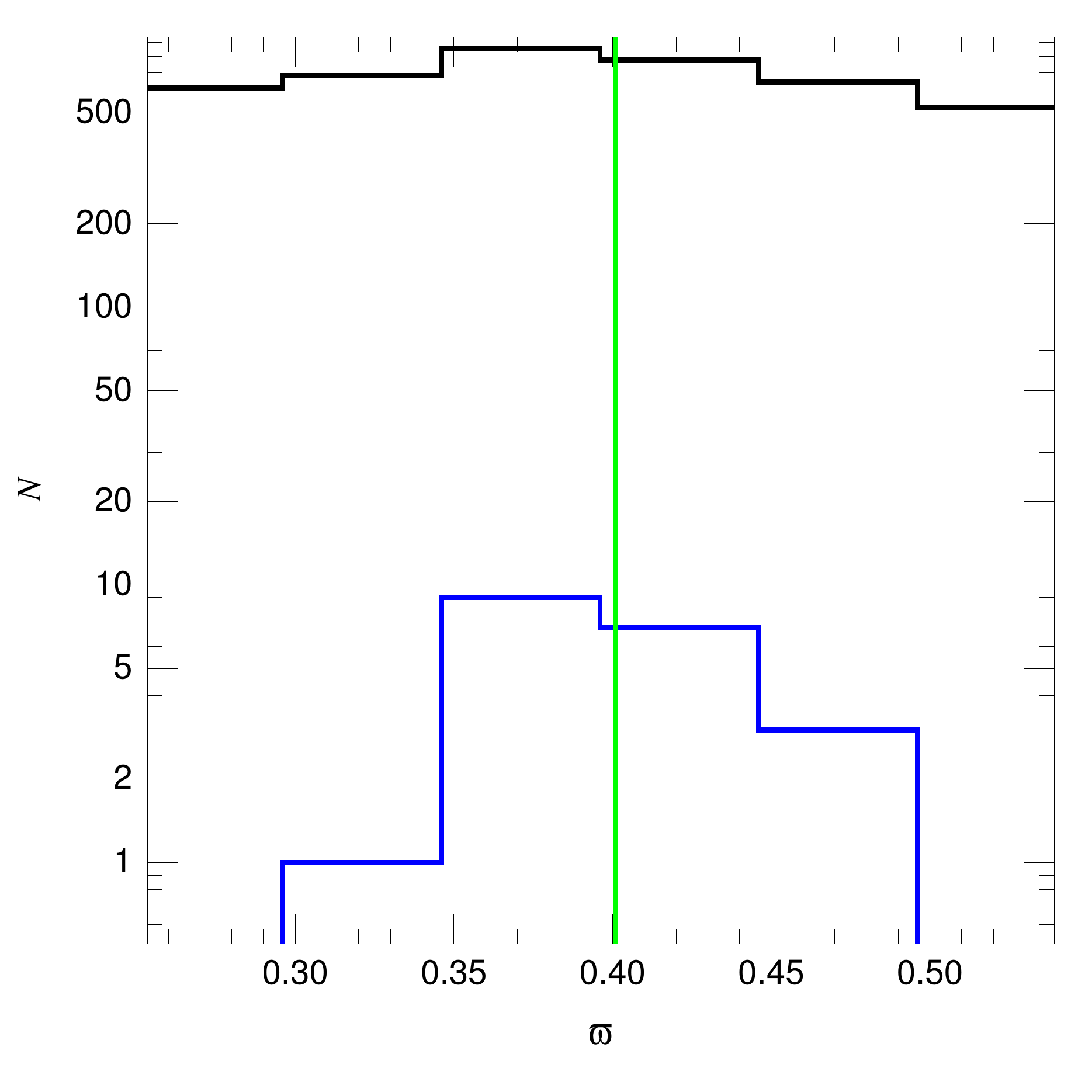} \
            \includegraphics*[width=0.34\linewidth, bb=0 0 538 522]{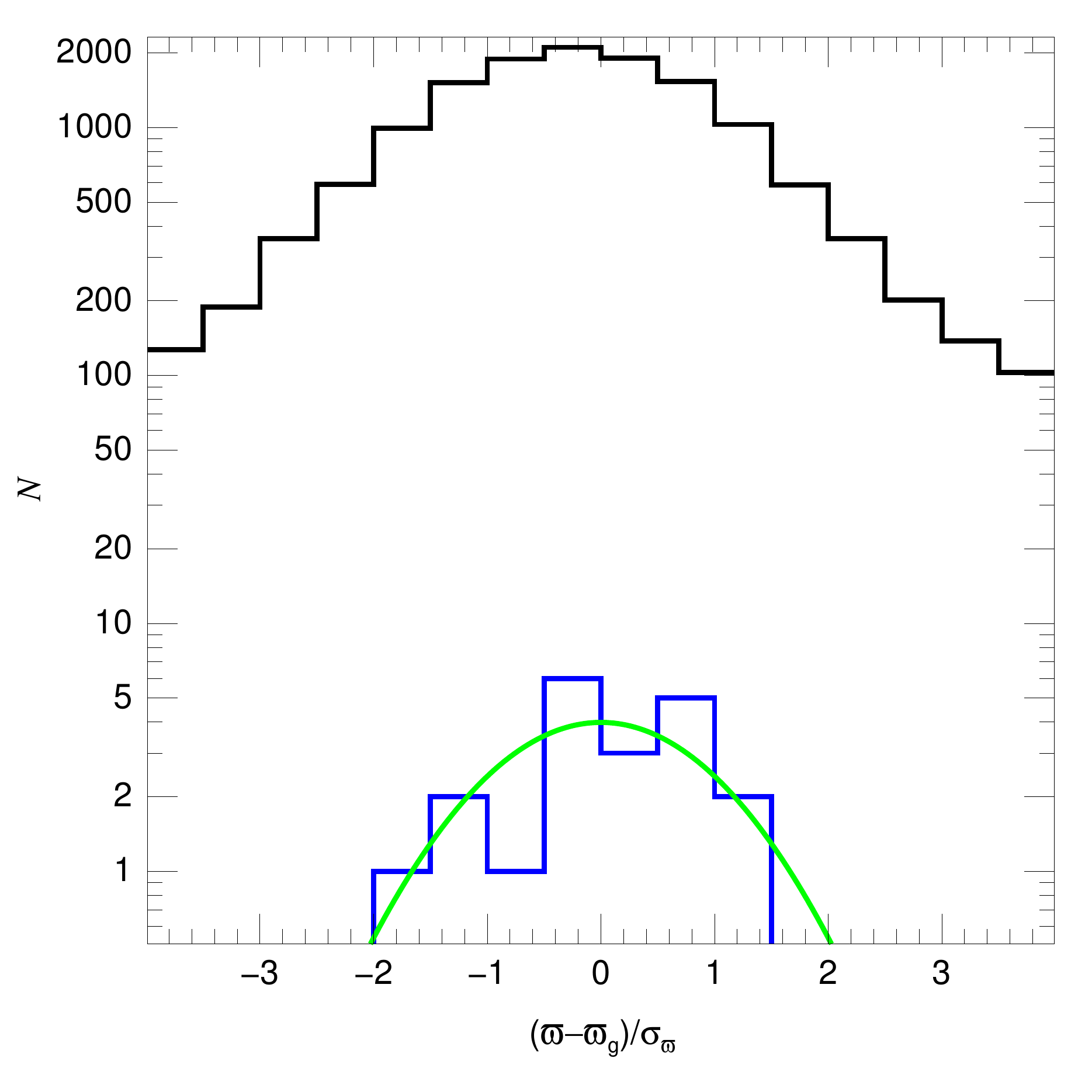}}
\caption{Same as Fig.~\ref{Trumpler_14_Gaia} for 
         Trumpler~16~W (\VO{003}) % REFEREE \VO{003}
         The partial dashed green circle in the top three panels shows the position of the neighbor
         Trumpler~14 (\VO{002}). % REFEREE \VO{002}
         }
\label{Trumpler_16_W_Gaia}
\end{figure*}   

\begin{figure*}
\centerline{\includegraphics*[width=0.34\linewidth, bb=0 0 538 522]{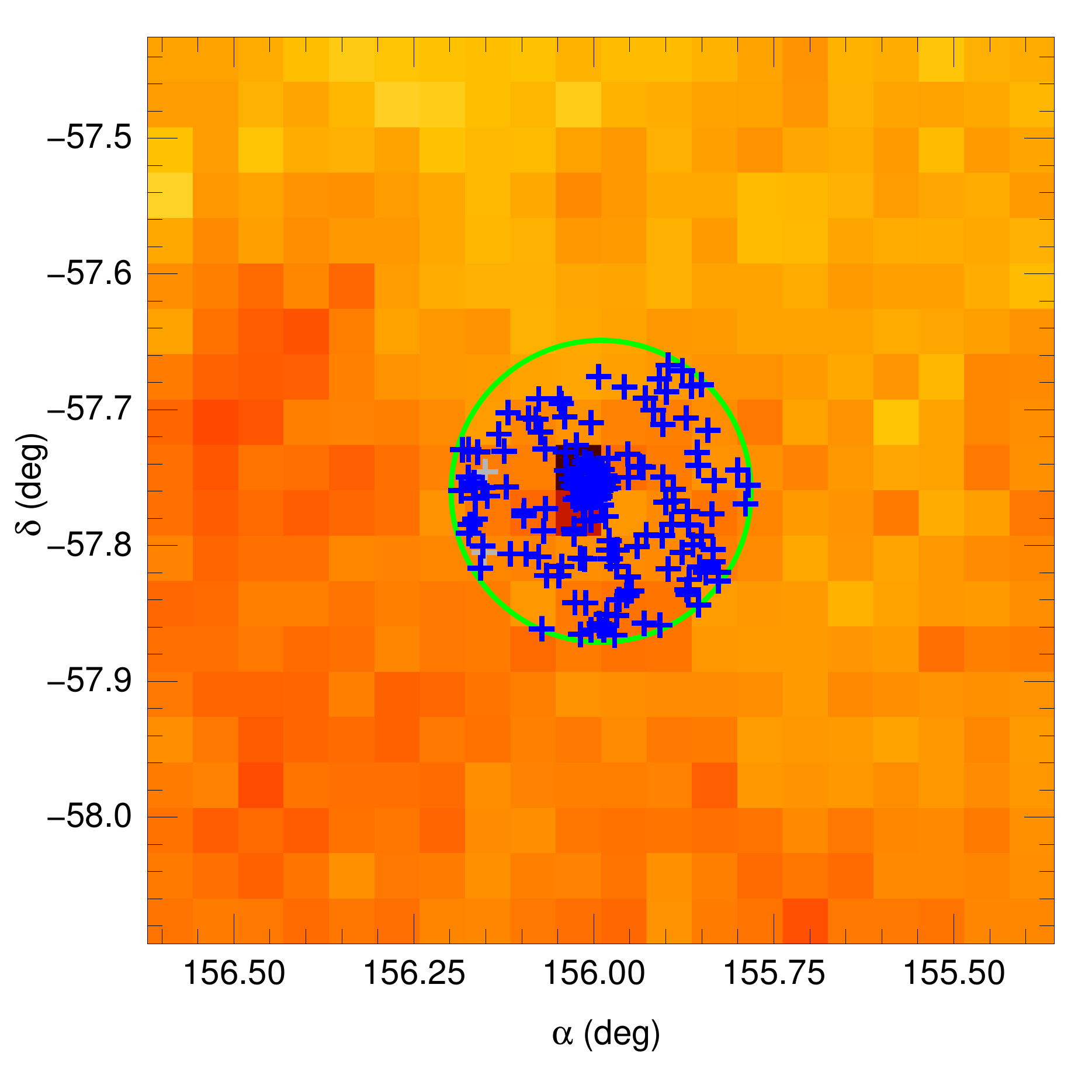} \
            \includegraphics*[width=0.34\linewidth, bb=0 0 538 522]{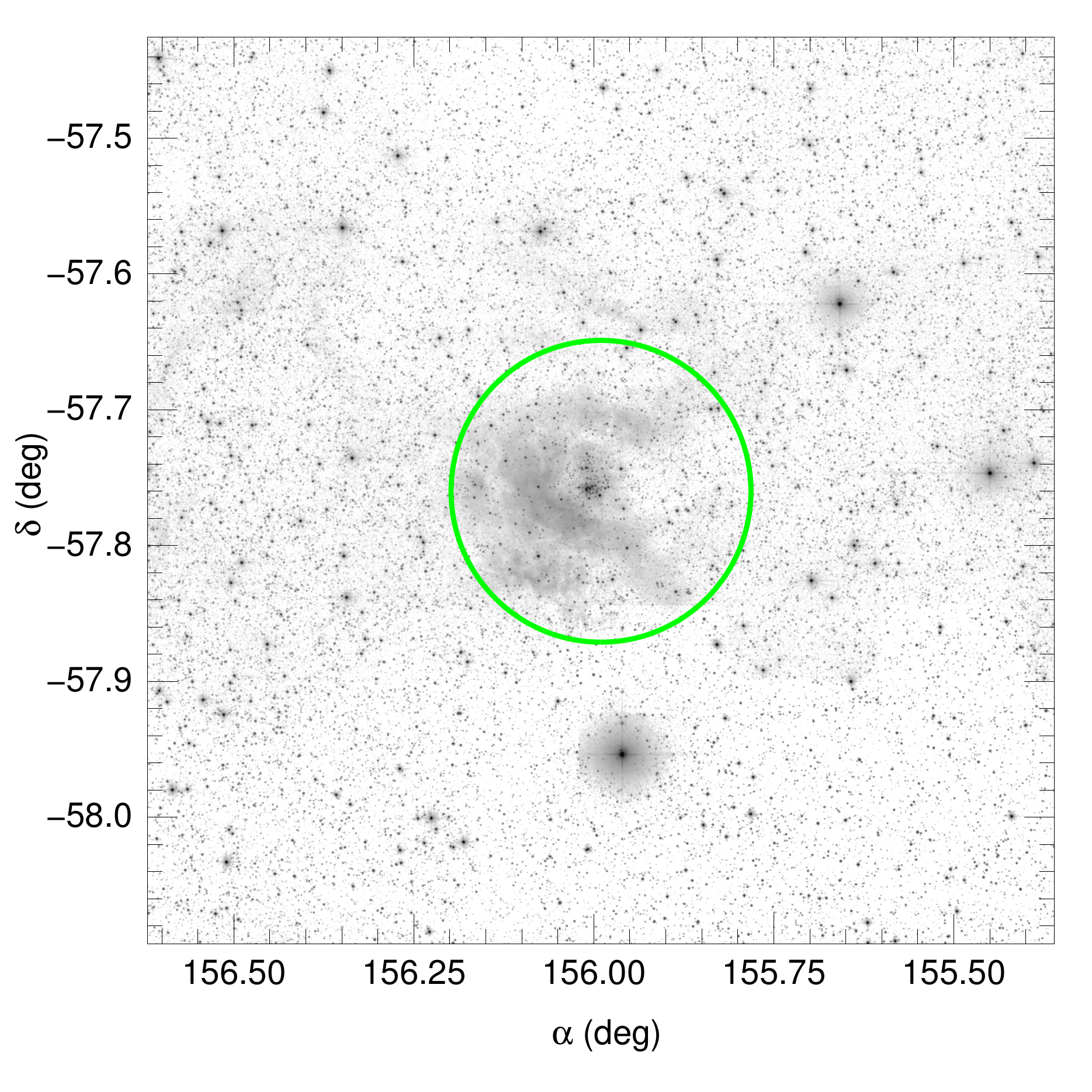} \
            \includegraphics*[width=0.34\linewidth, bb=0 0 538 522]{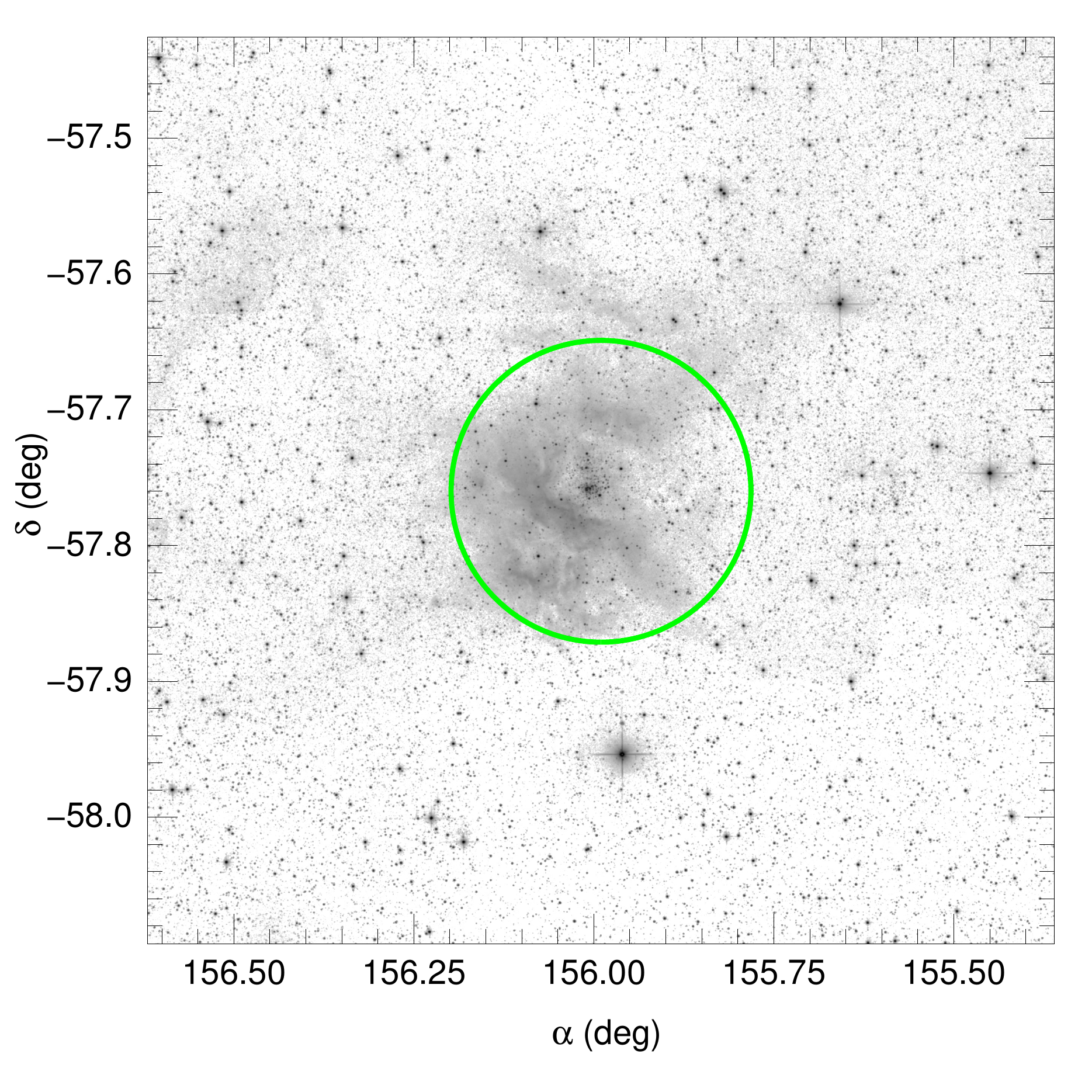}}
\centerline{\includegraphics*[width=0.34\linewidth, bb=0 0 538 522]{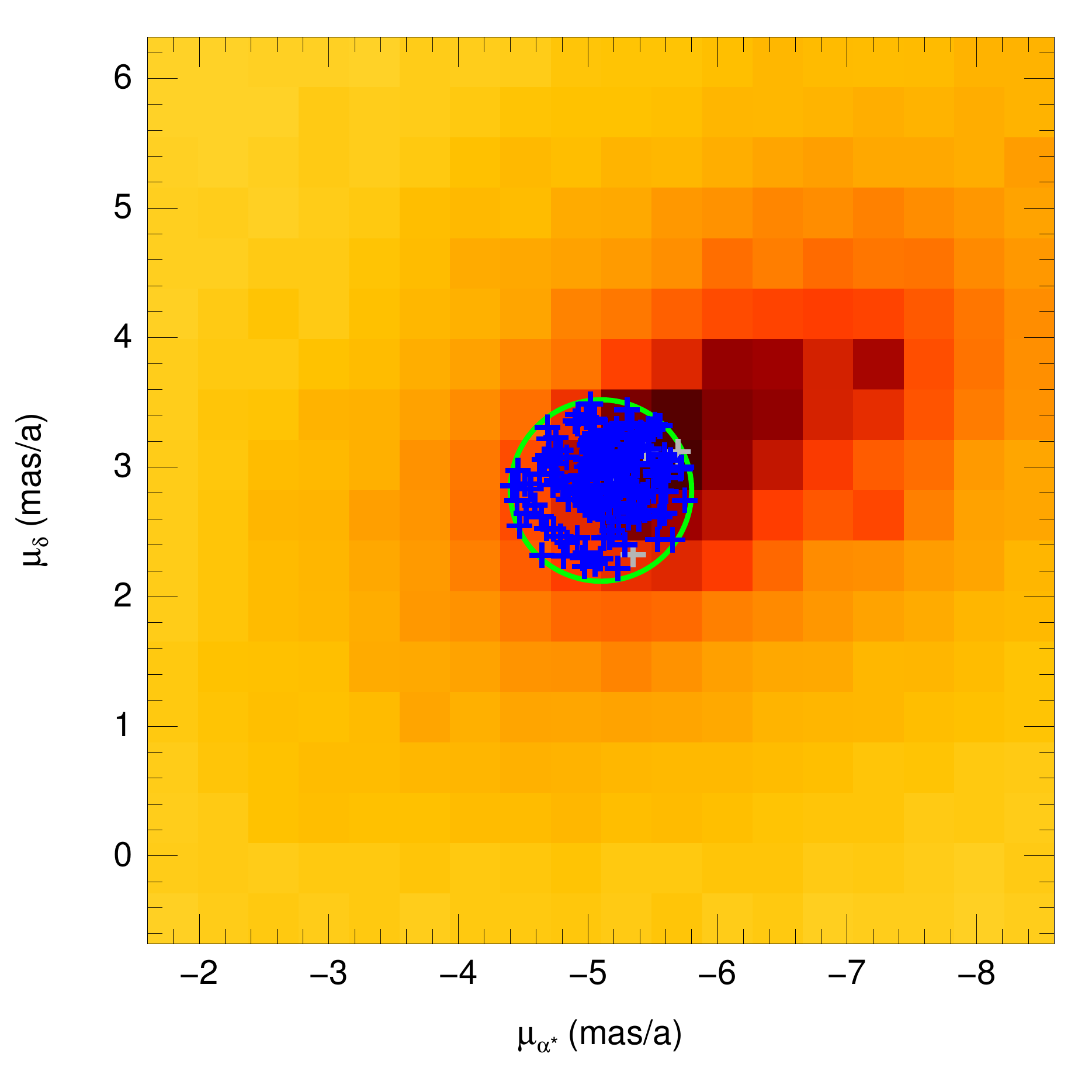} \
            \includegraphics*[width=0.34\linewidth, bb=0 0 538 522]{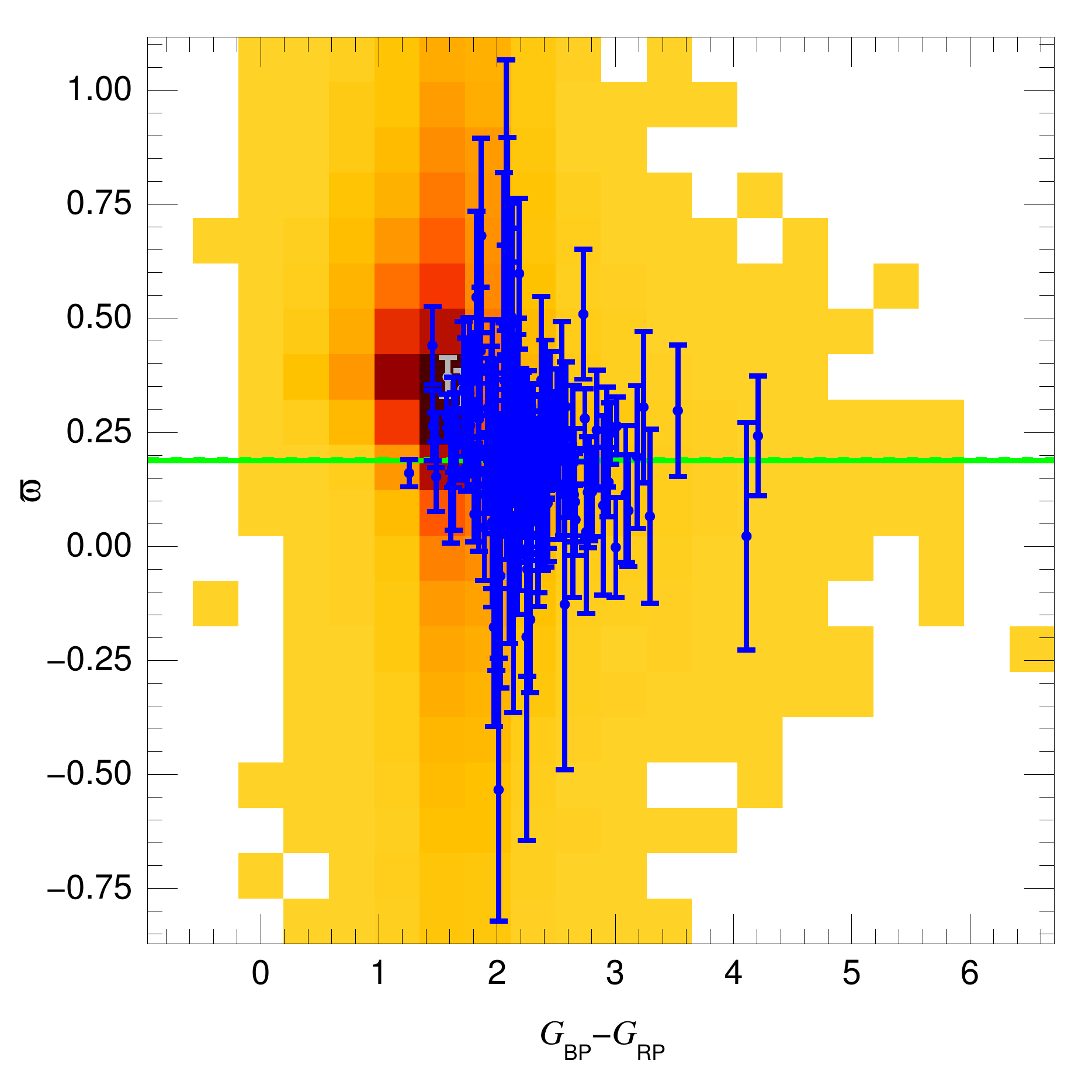} \
            \includegraphics*[width=0.34\linewidth, bb=0 0 538 522]{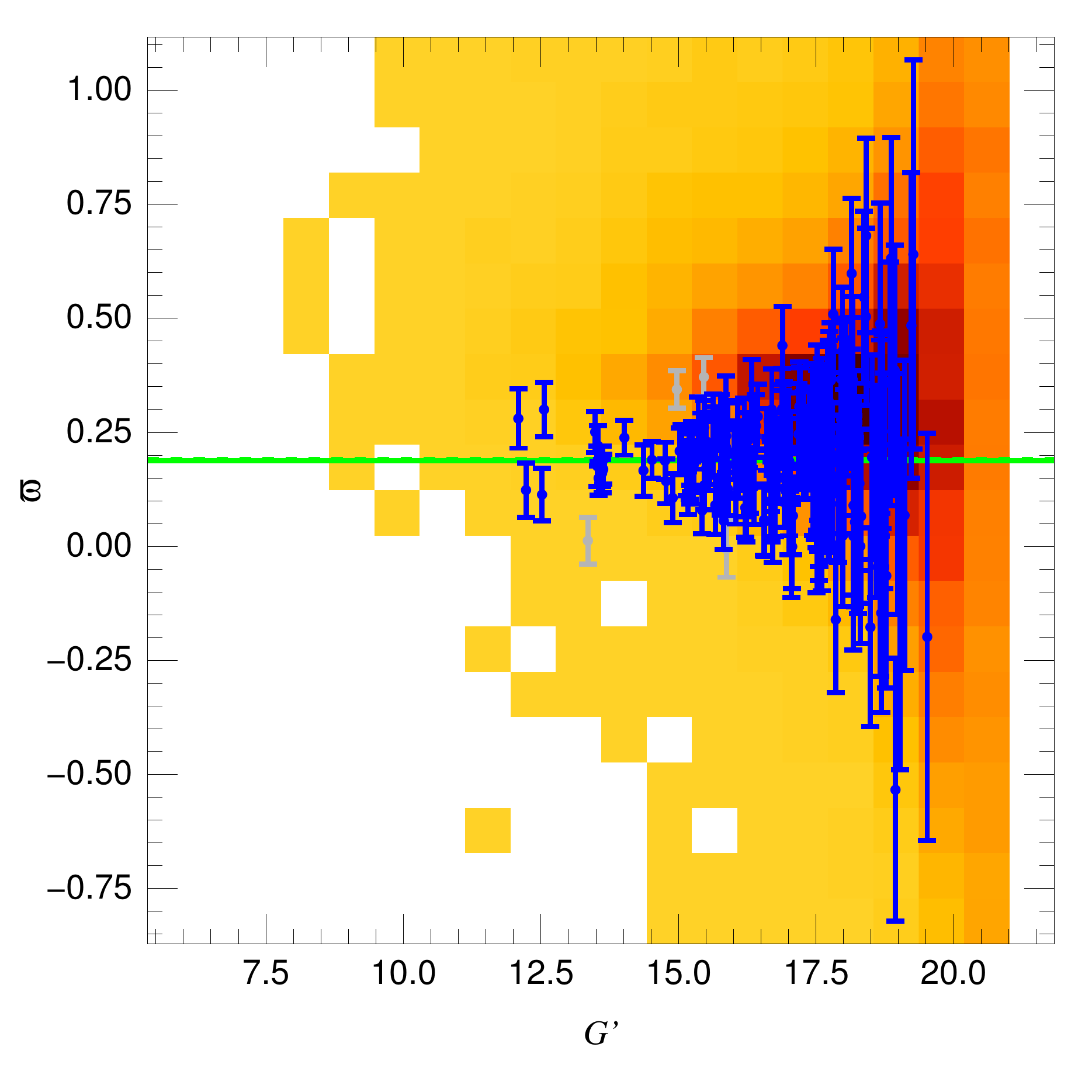}}
\centerline{\includegraphics*[width=0.34\linewidth, bb=0 0 538 522]{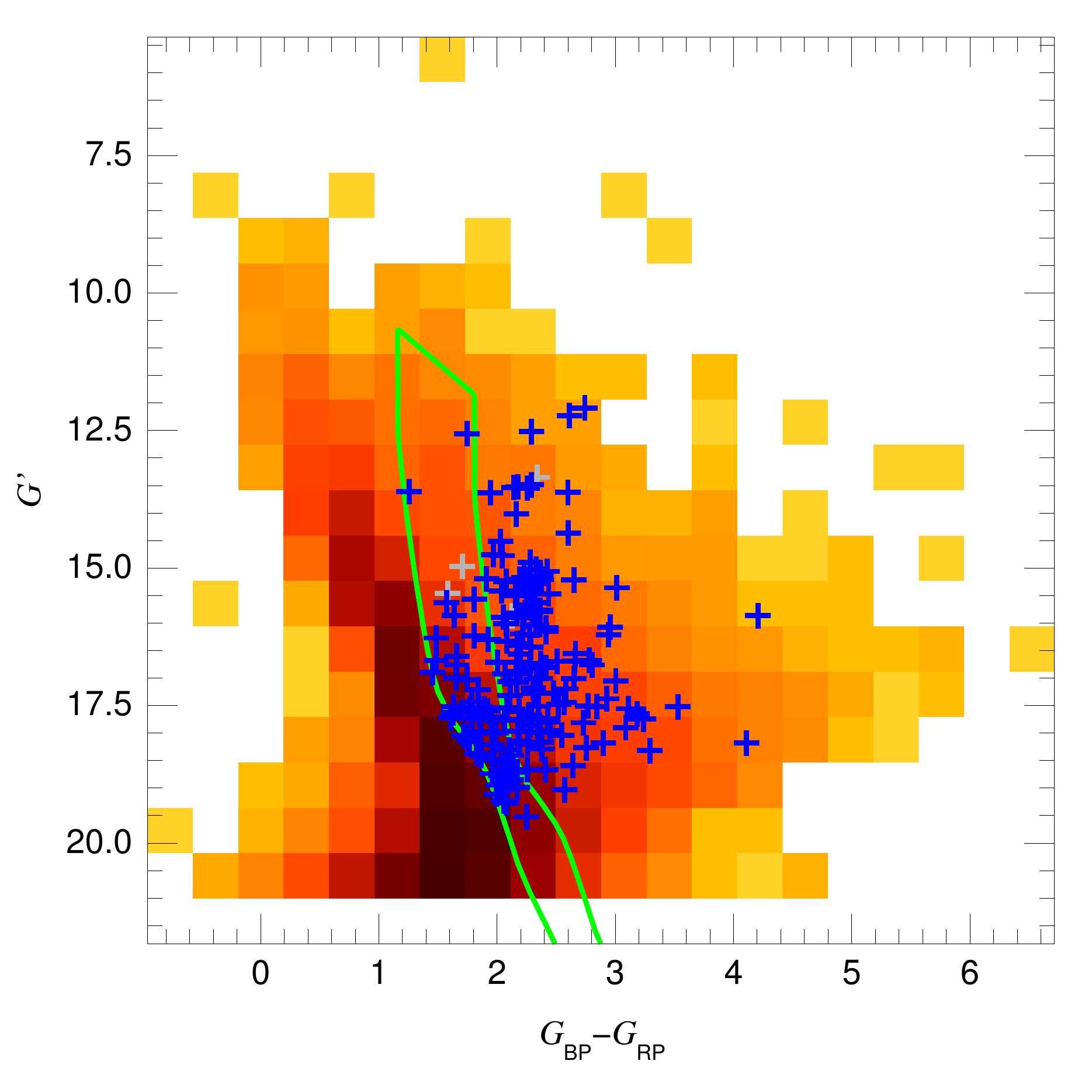} \
            \includegraphics*[width=0.34\linewidth, bb=0 0 538 522]{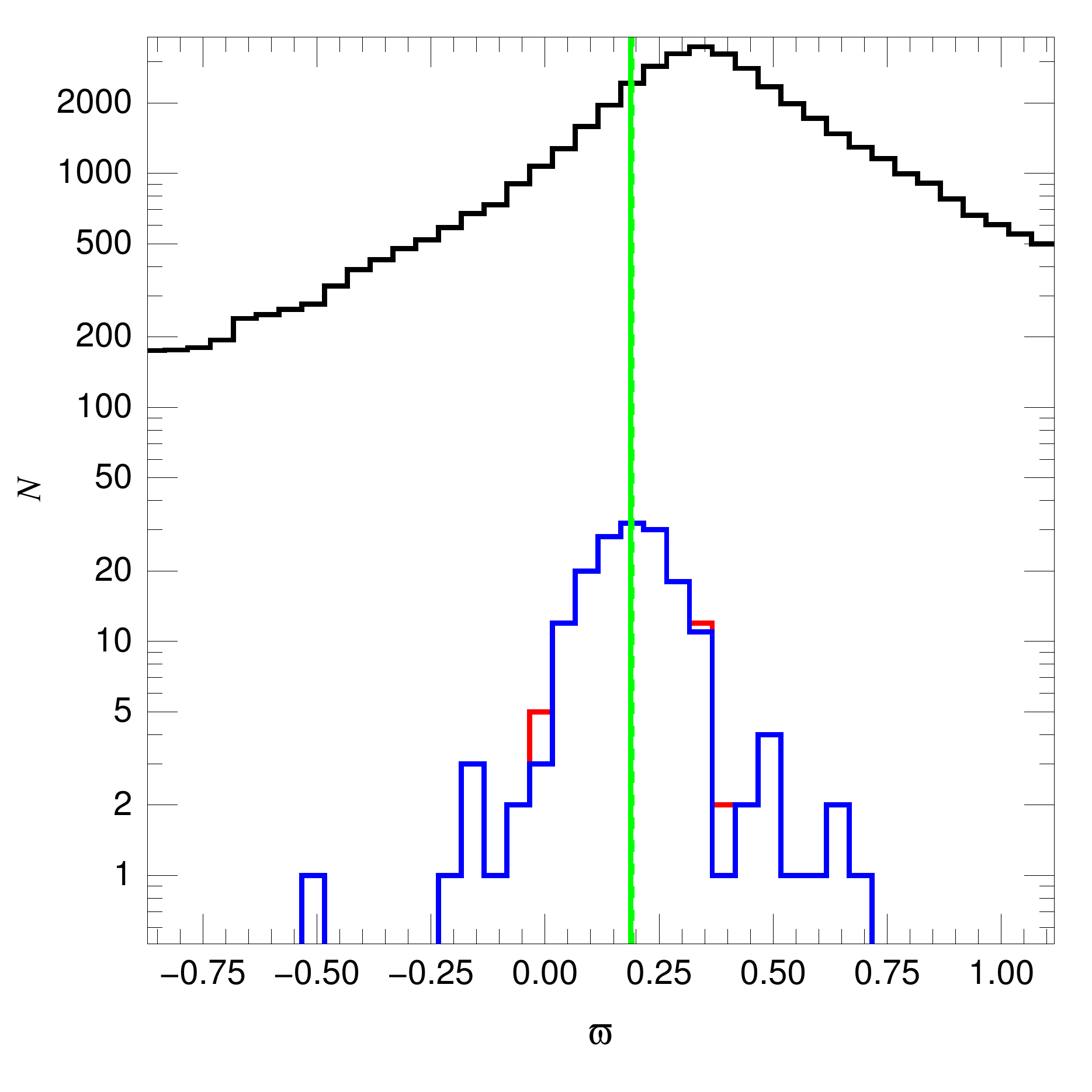} \
            \includegraphics*[width=0.34\linewidth, bb=0 0 538 522]{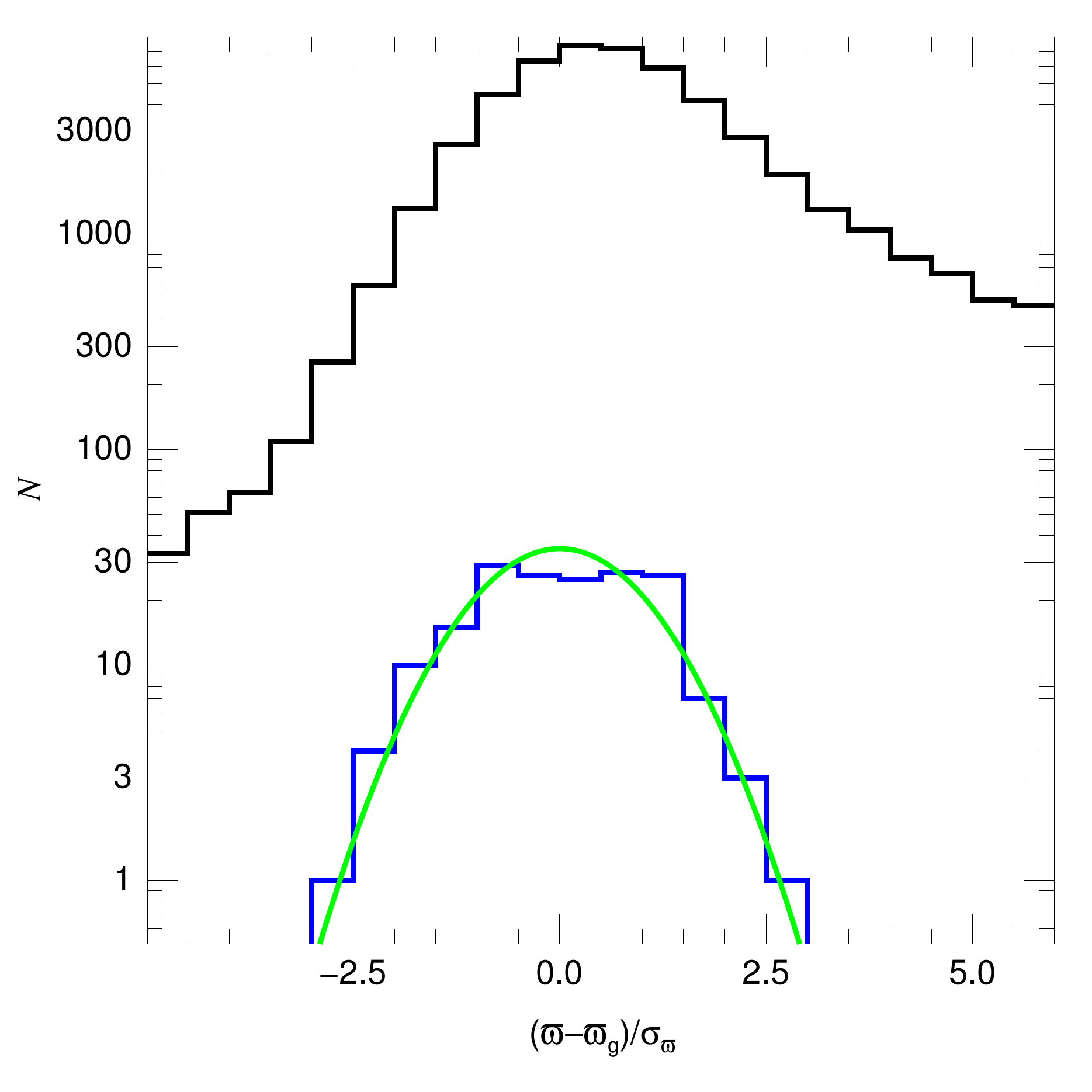}}
\caption{Same as Fig.~\ref{NGC_3603_Gaia} for 
         Westerlund~2 (\VO{004}) % REFEREE \VO{004}
         but with the 2MASS~$J$ image in the top center panel and the 2MASS~$K$ image in the top right one.}
\label{Westerlund_2_Gaia}
\end{figure*}   

\begin{figure*}
\centerline{\includegraphics*[width=0.34\linewidth, bb=0 0 538 522]{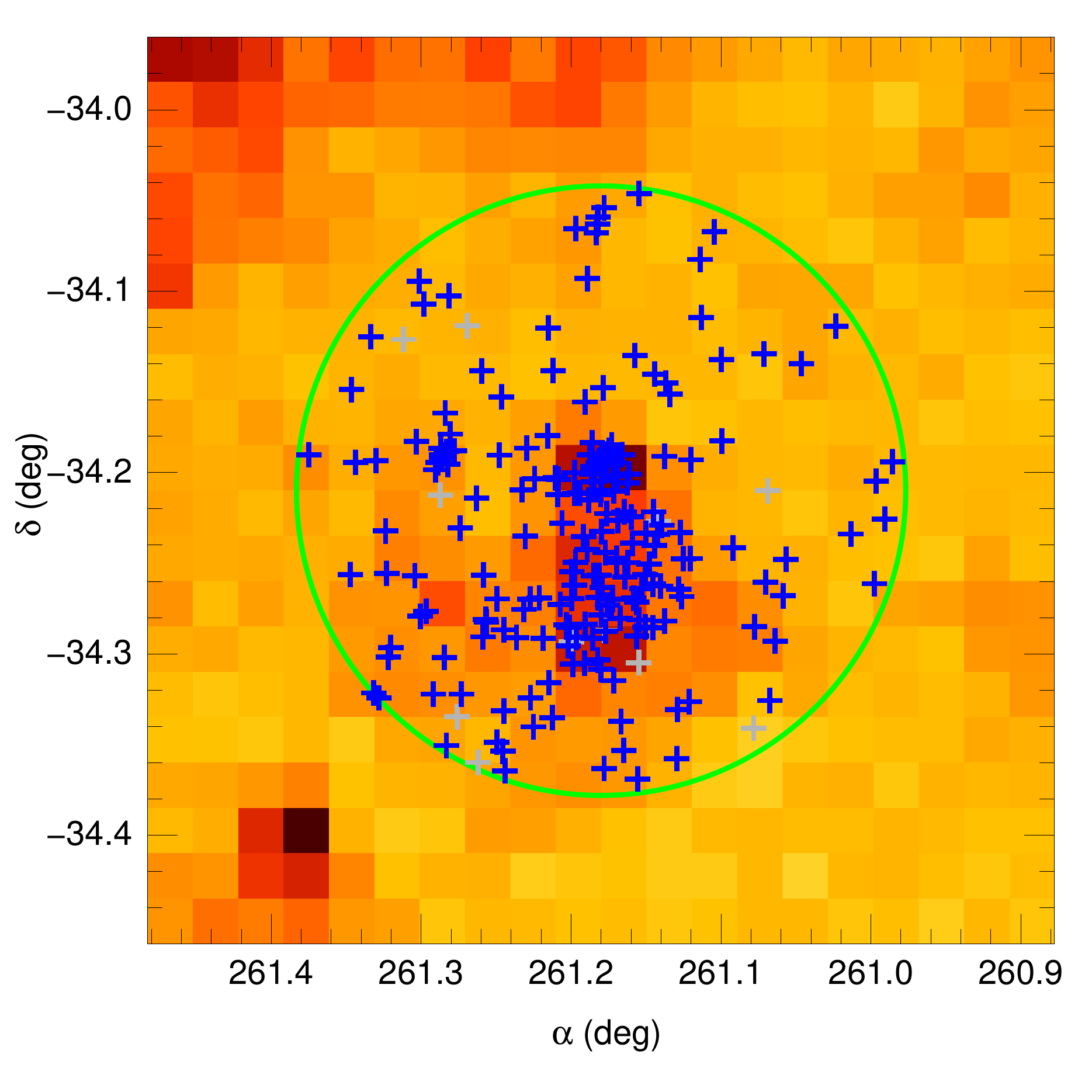} \
            \includegraphics*[width=0.34\linewidth, bb=0 0 538 522]{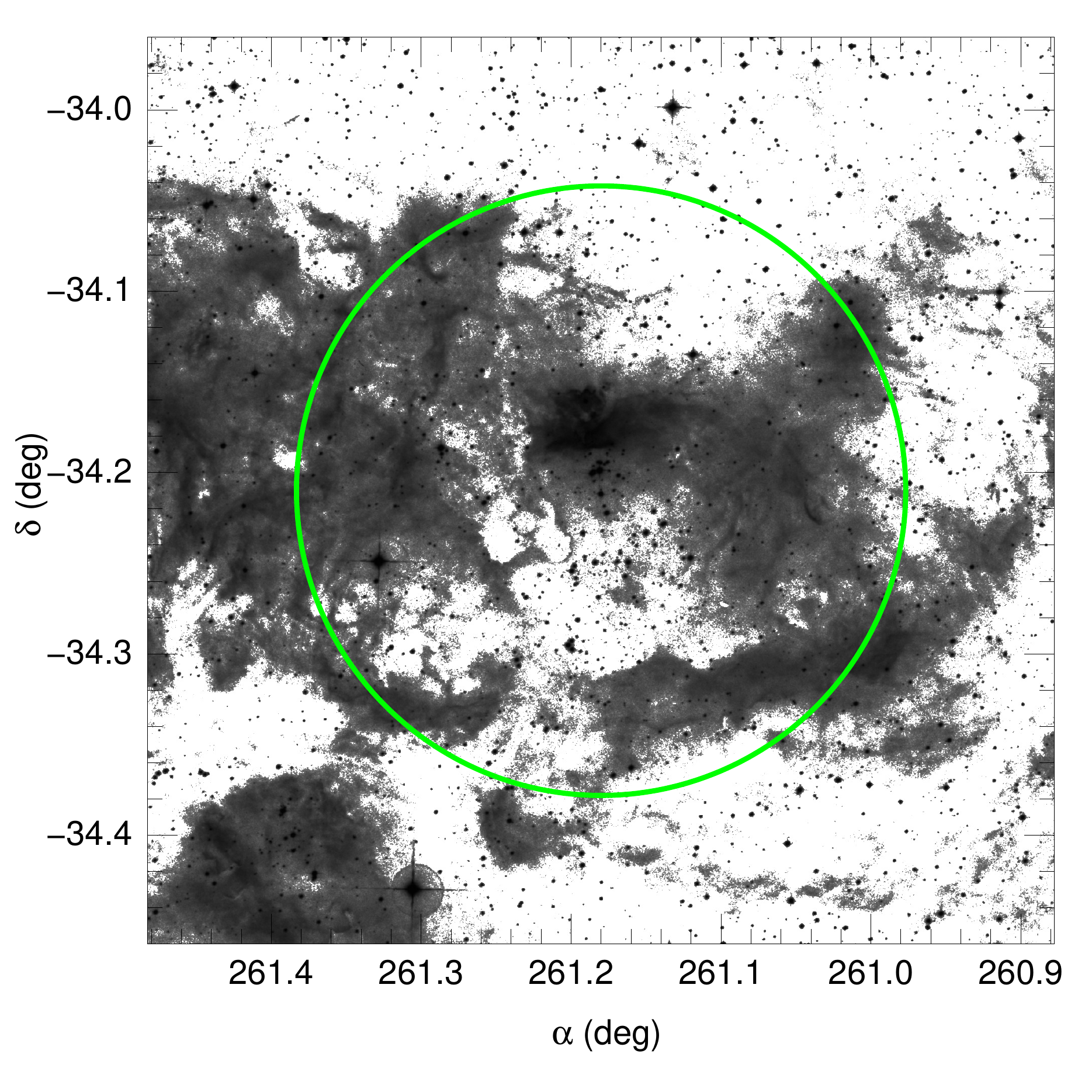} \
            \includegraphics*[width=0.34\linewidth, bb=0 0 538 522]{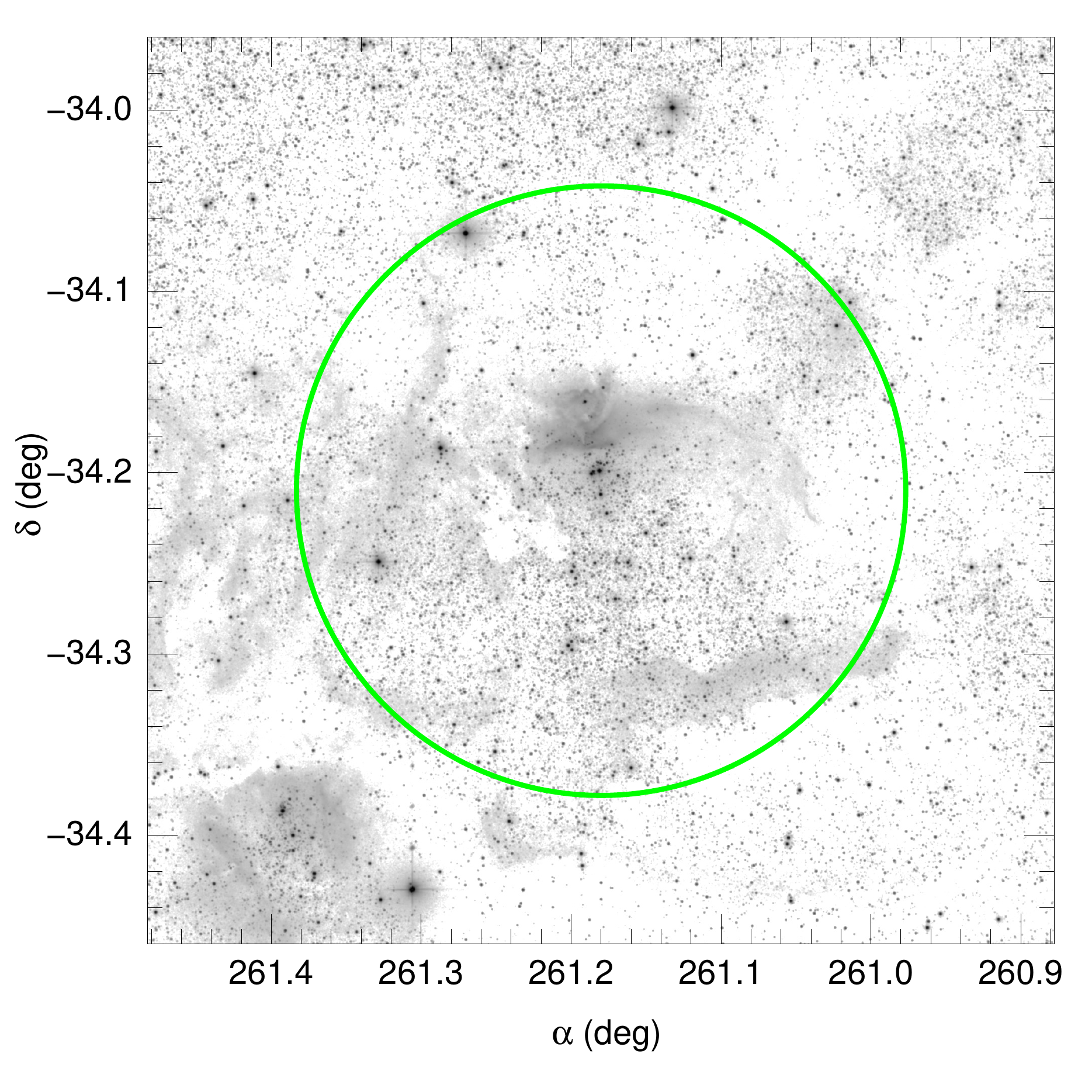}}
\centerline{\includegraphics*[width=0.34\linewidth, bb=0 0 538 522]{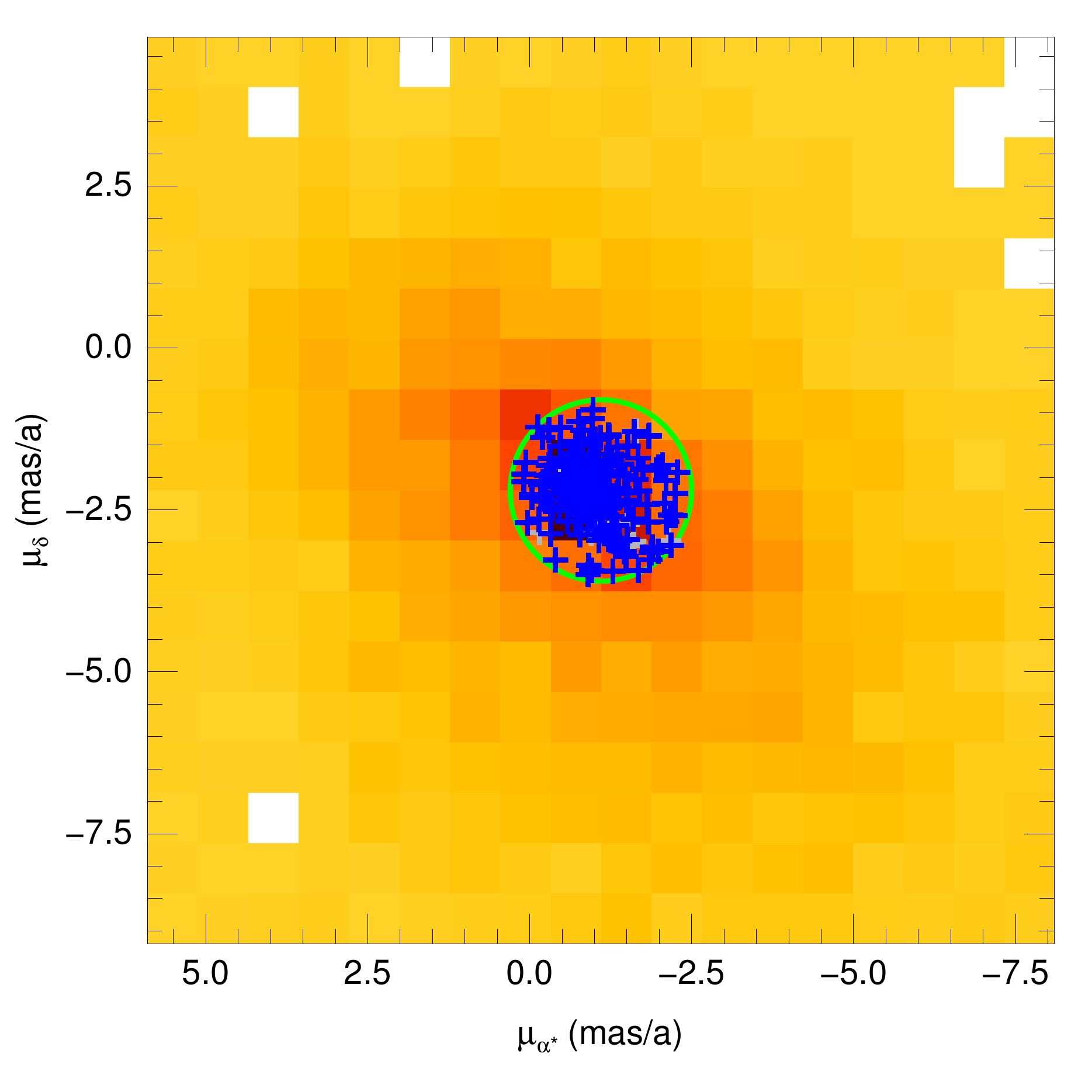} \
            \includegraphics*[width=0.34\linewidth, bb=0 0 538 522]{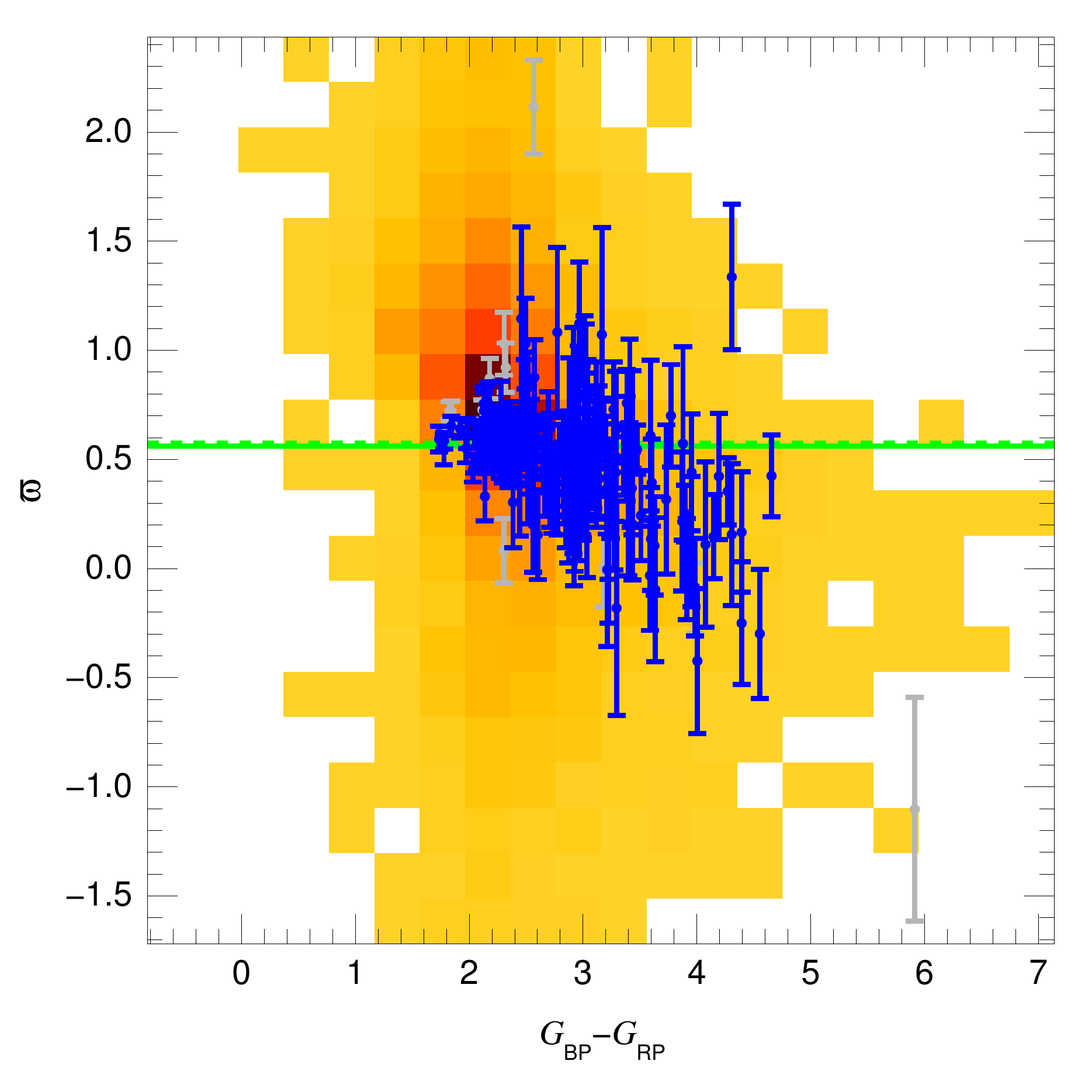} \
            \includegraphics*[width=0.34\linewidth, bb=0 0 538 522]{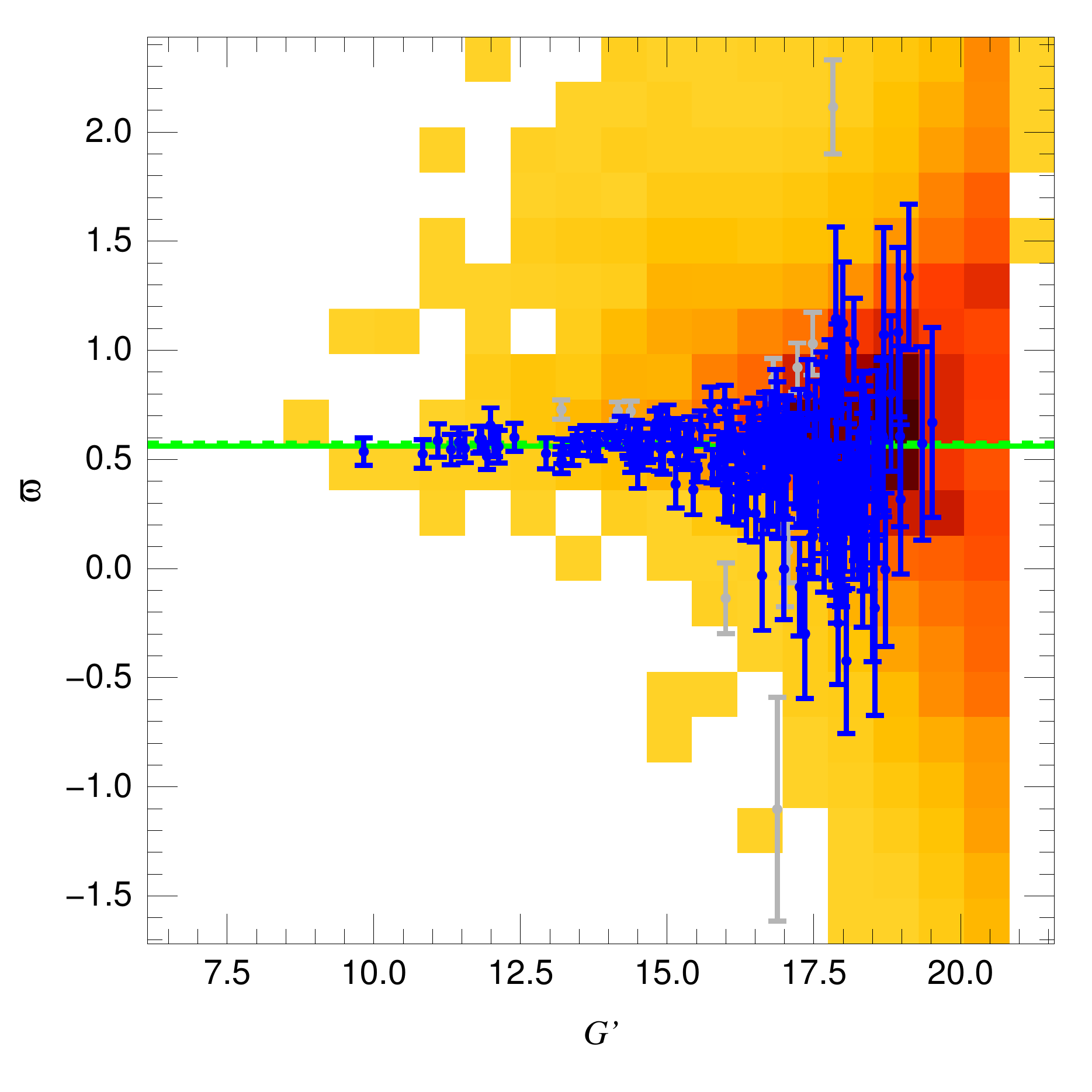}}
\centerline{\includegraphics*[width=0.34\linewidth, bb=0 0 538 522]{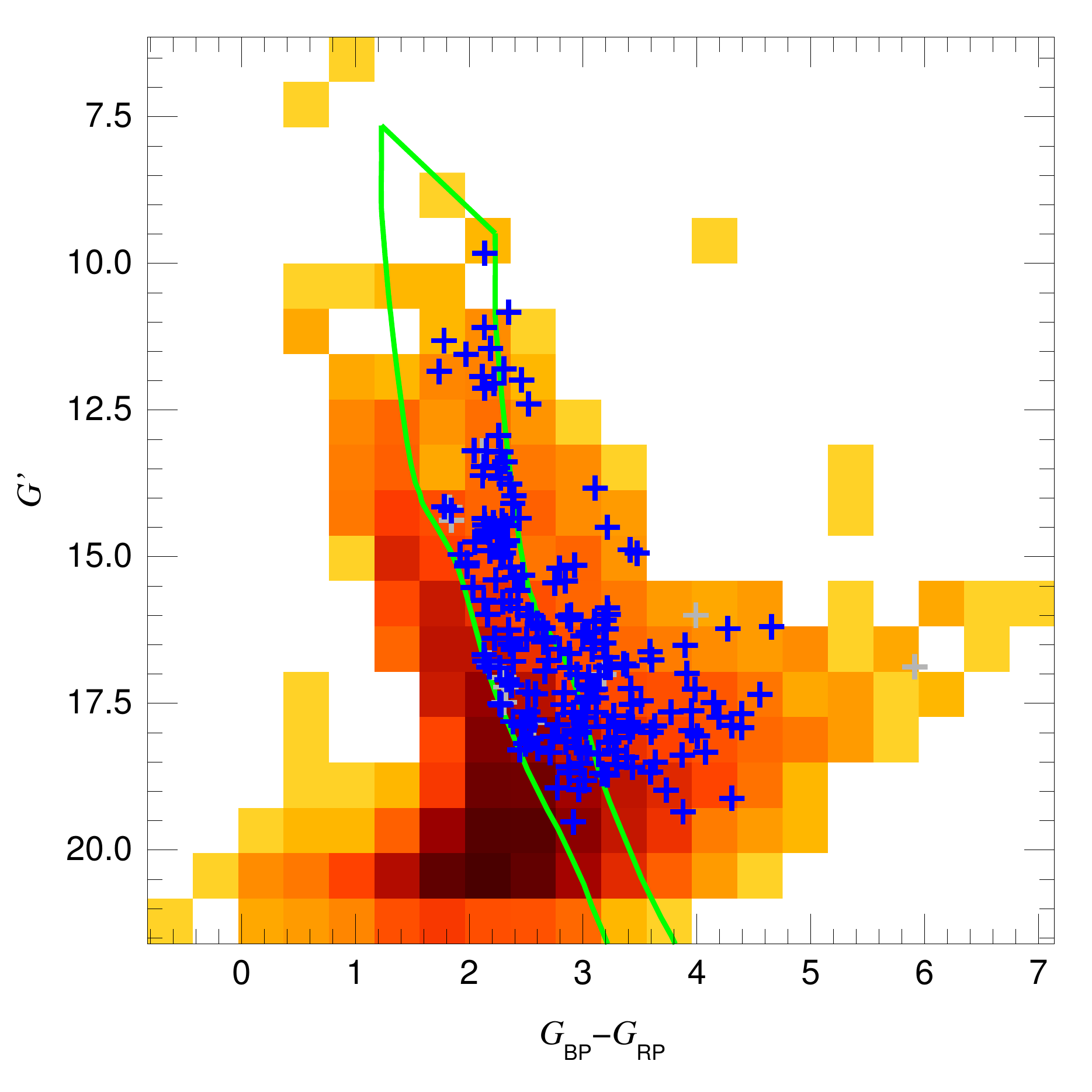} \
            \includegraphics*[width=0.34\linewidth, bb=0 0 538 522]{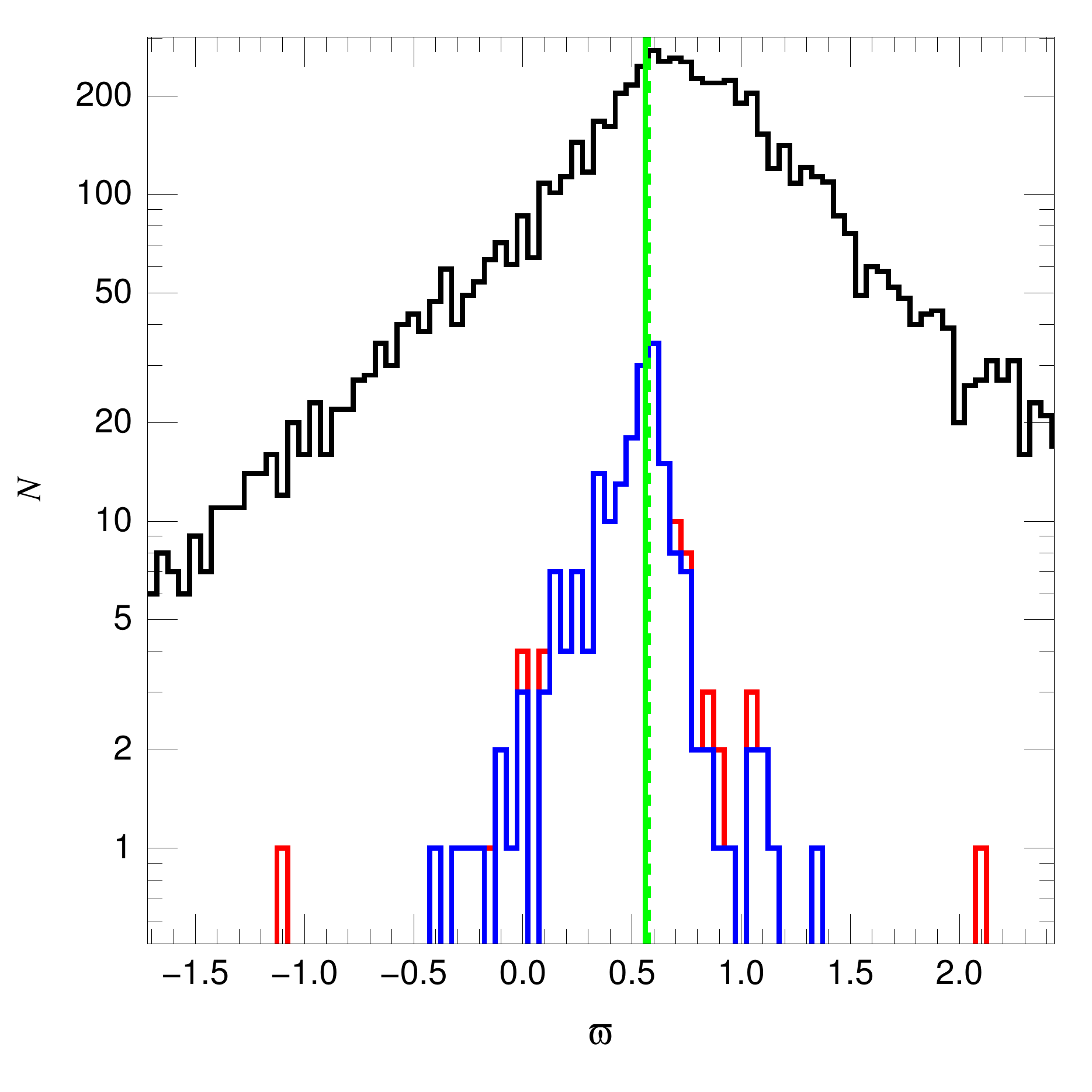} \
            \includegraphics*[width=0.34\linewidth, bb=0 0 538 522]{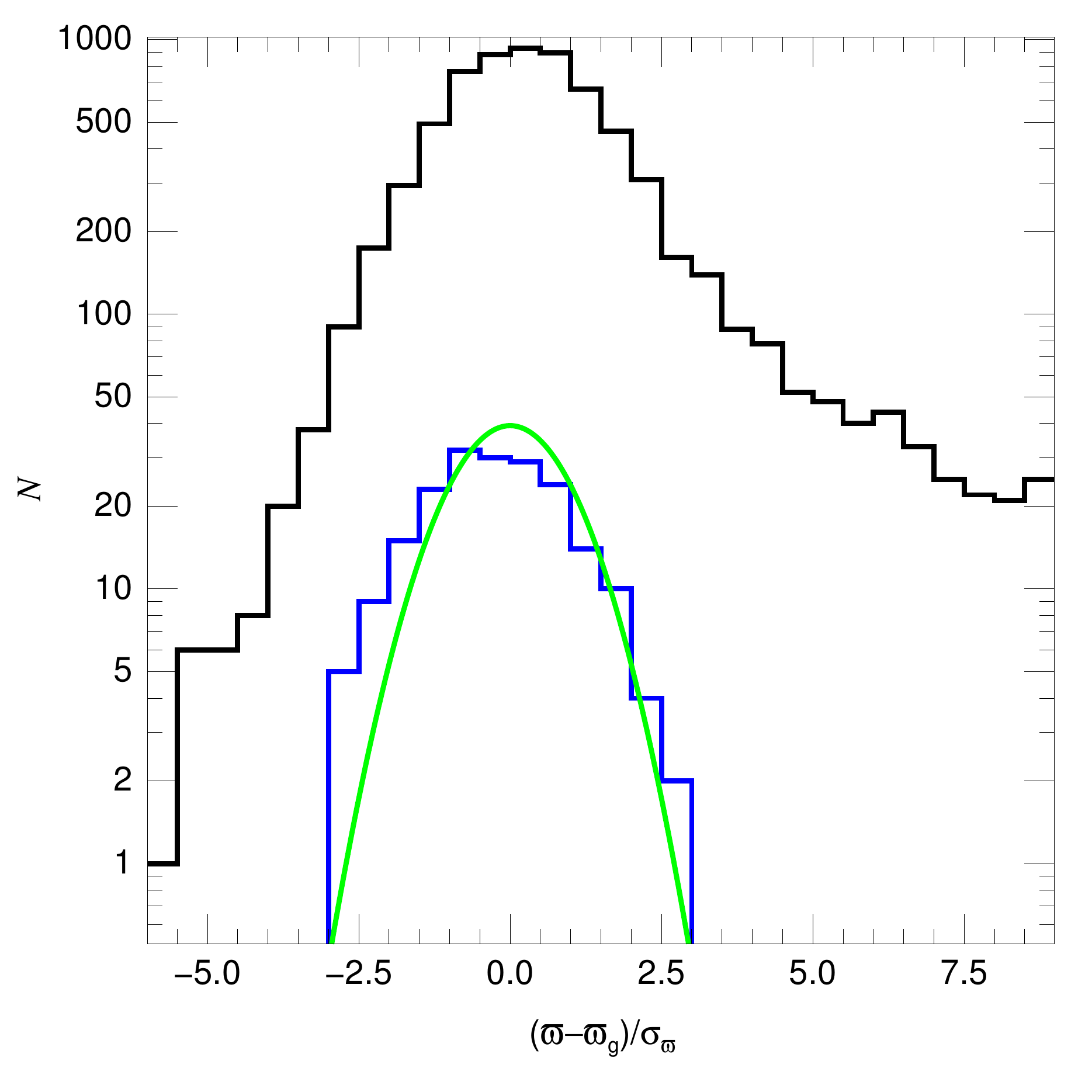}}
\caption{Same as Fig.~\ref{NGC_3603_Gaia} for 
         Pismis~24 (\VO{005}). % REFEREE \VO{005}
         }
\label{Pismis_24_Gaia}
\end{figure*}   

\begin{figure*}
\centerline{\includegraphics*[width=0.34\linewidth, bb=0 0 538 522]{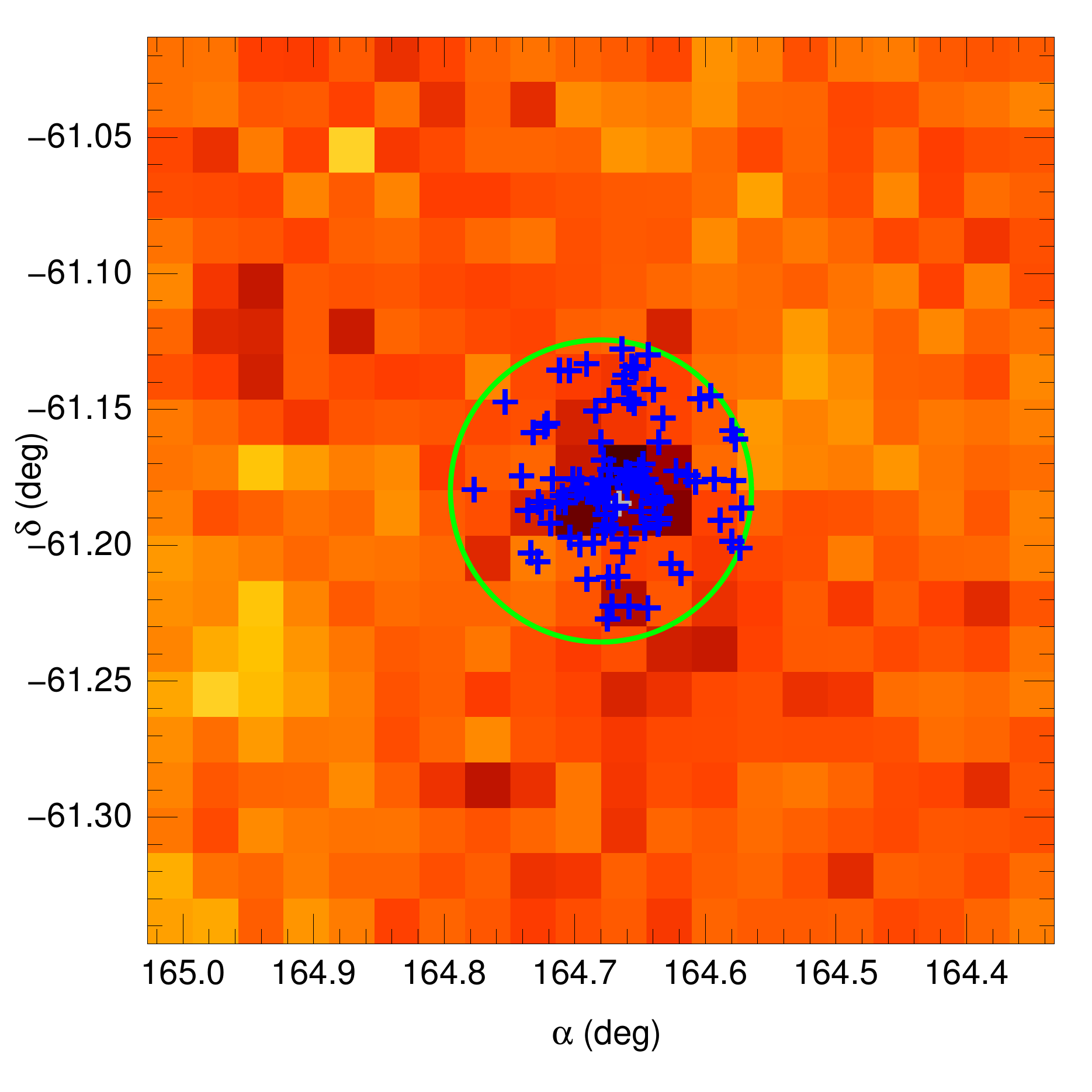} \
            \includegraphics*[width=0.34\linewidth, bb=0 0 538 522]{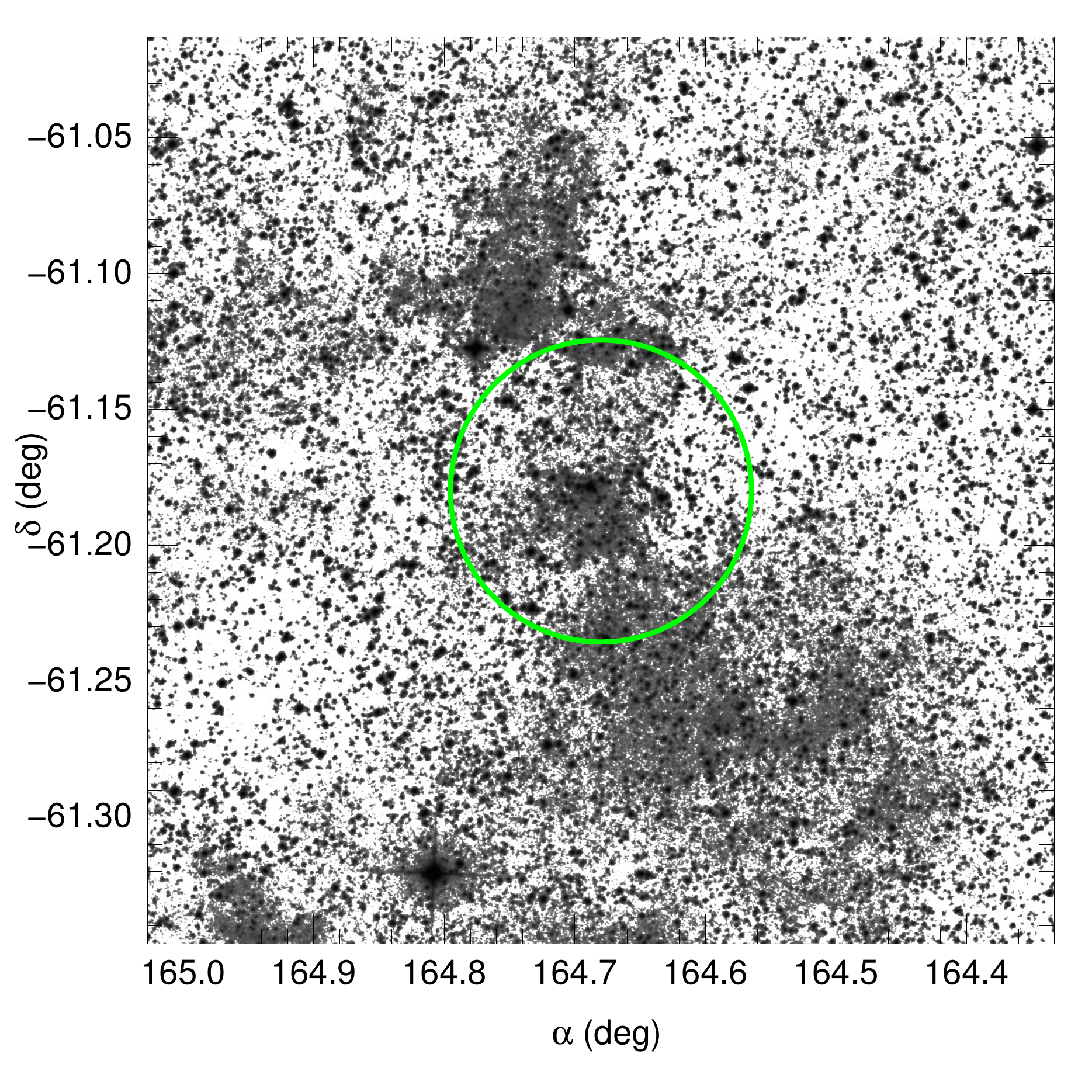} \
            \includegraphics*[width=0.34\linewidth, bb=0 0 538 522]{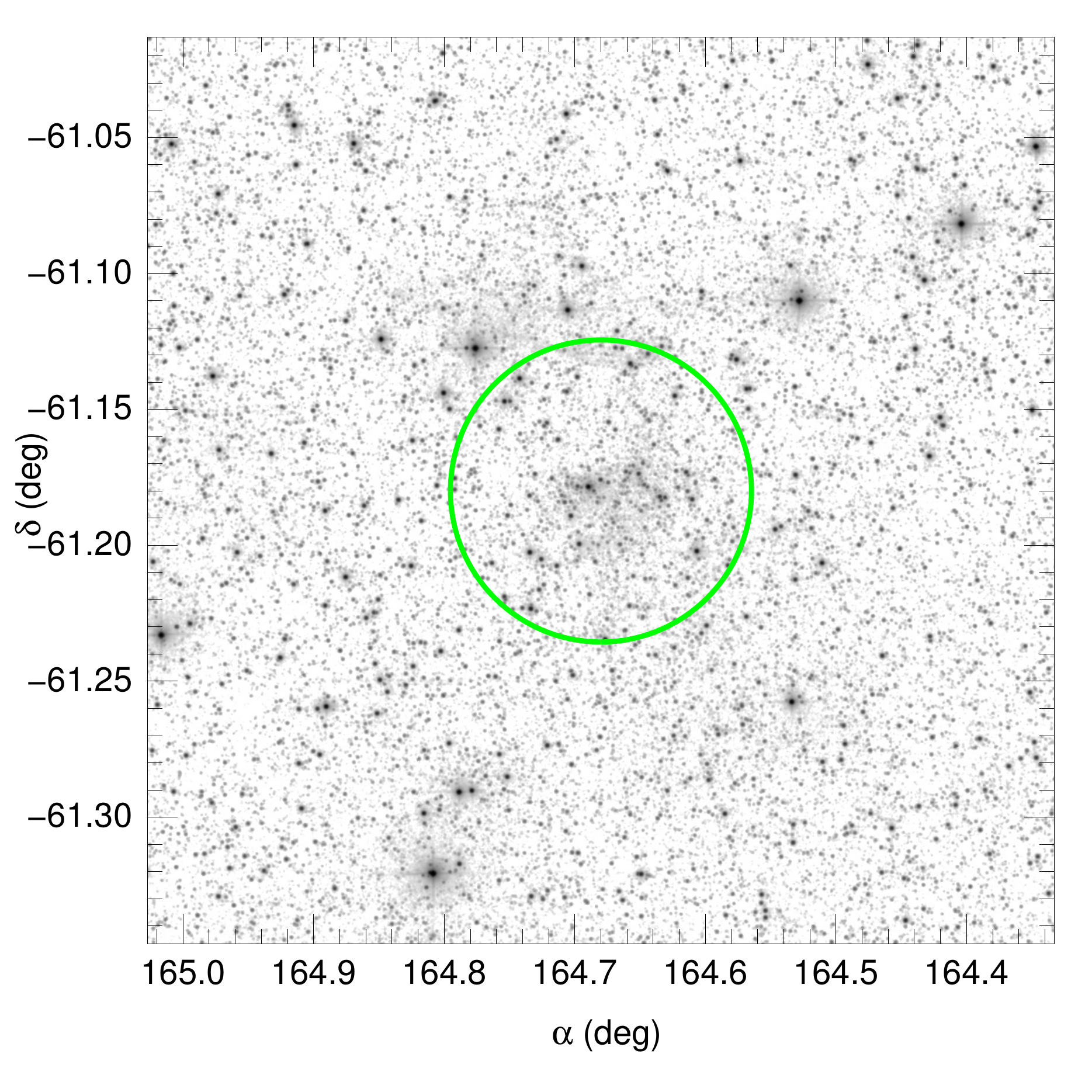}}
\centerline{\includegraphics*[width=0.34\linewidth, bb=0 0 538 522]{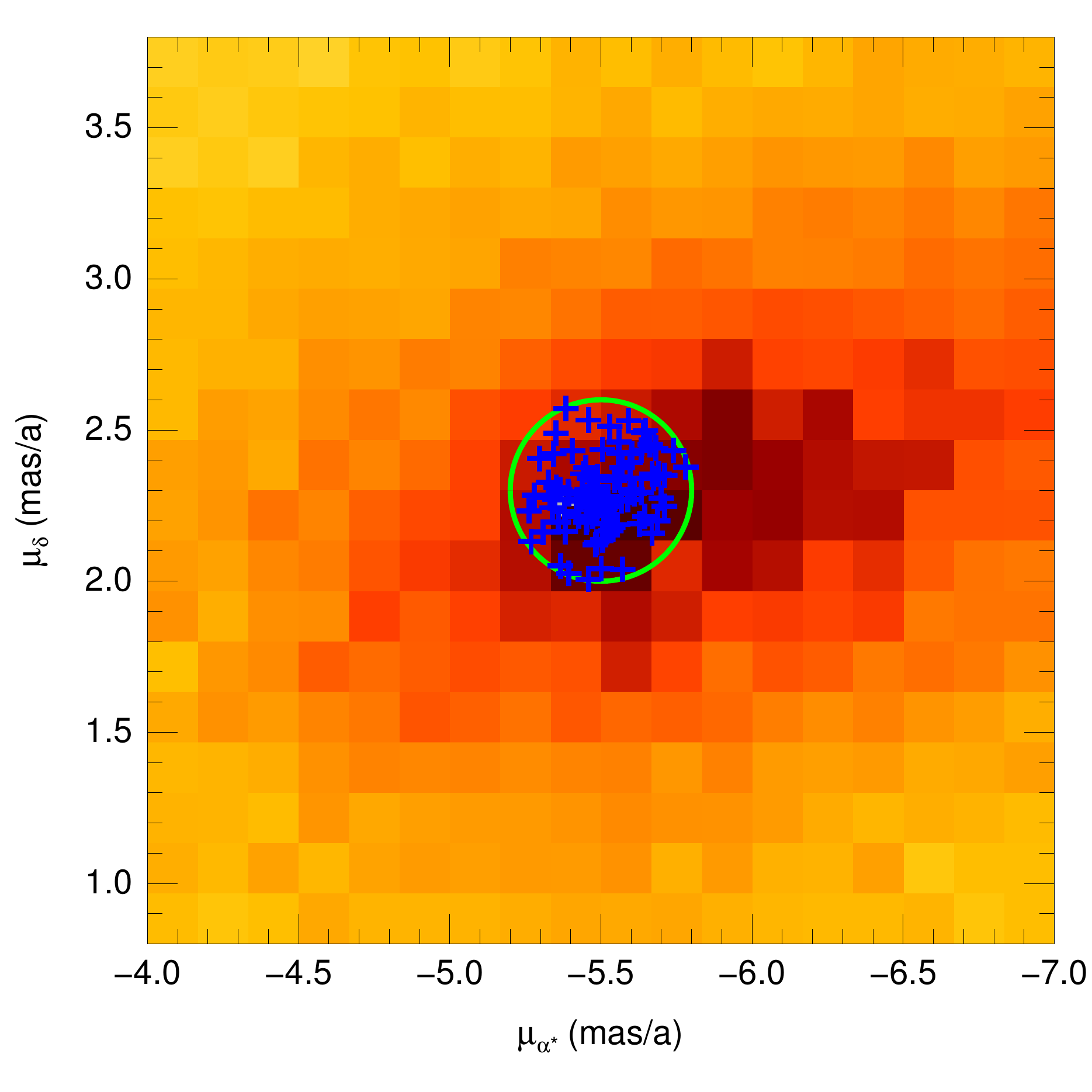} \
            \includegraphics*[width=0.34\linewidth, bb=0 0 538 522]{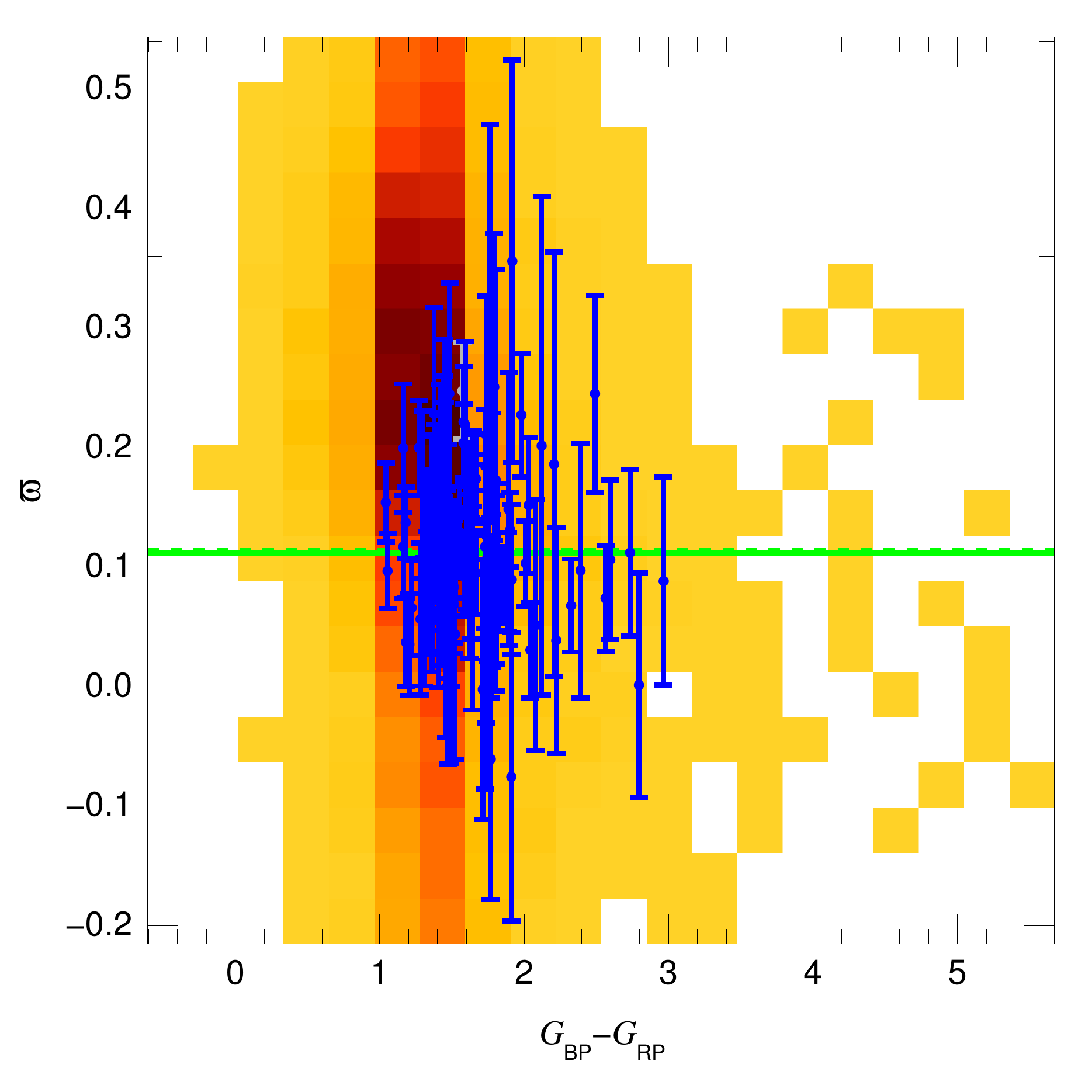} \
            \includegraphics*[width=0.34\linewidth, bb=0 0 538 522]{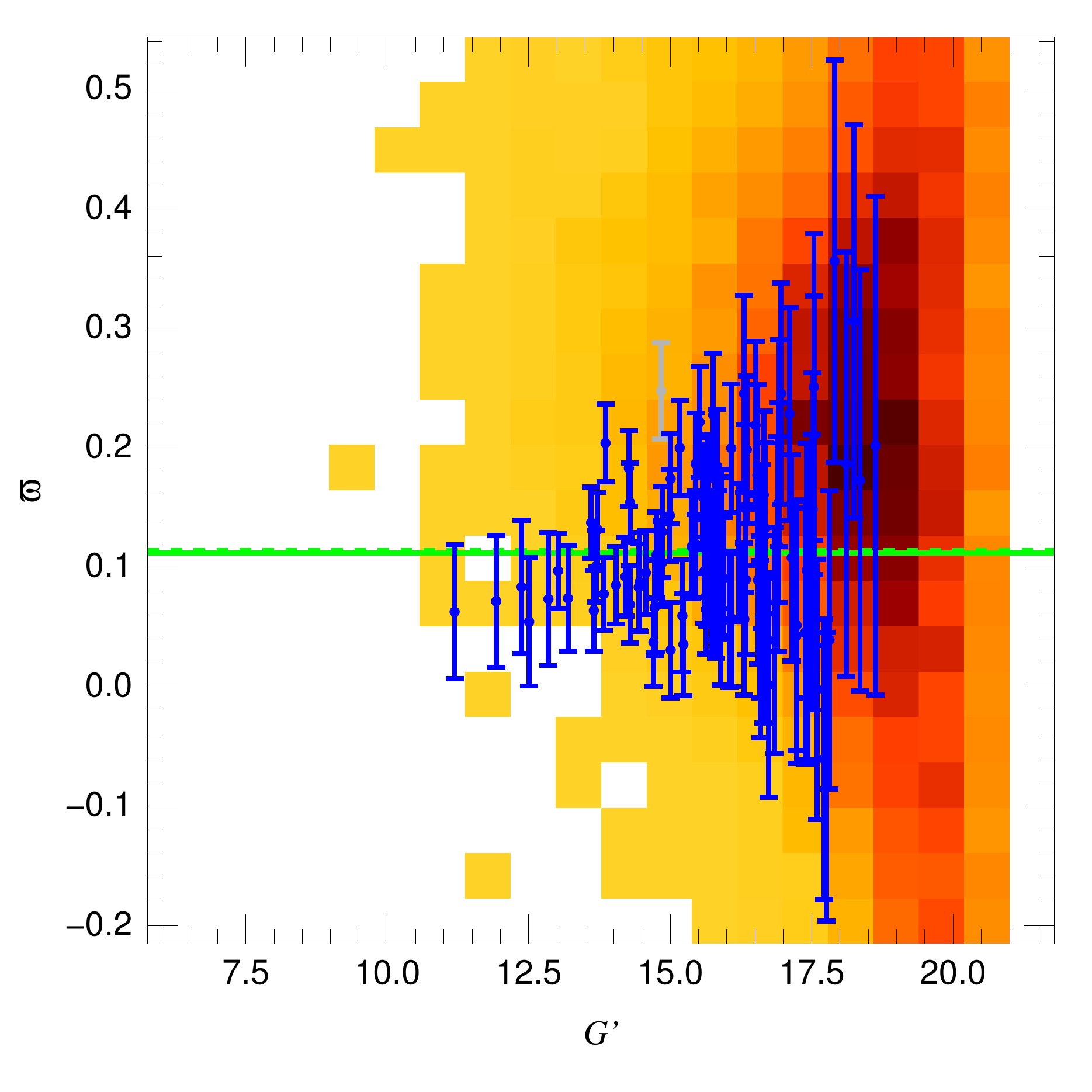}}
\centerline{\includegraphics*[width=0.34\linewidth, bb=0 0 538 522]{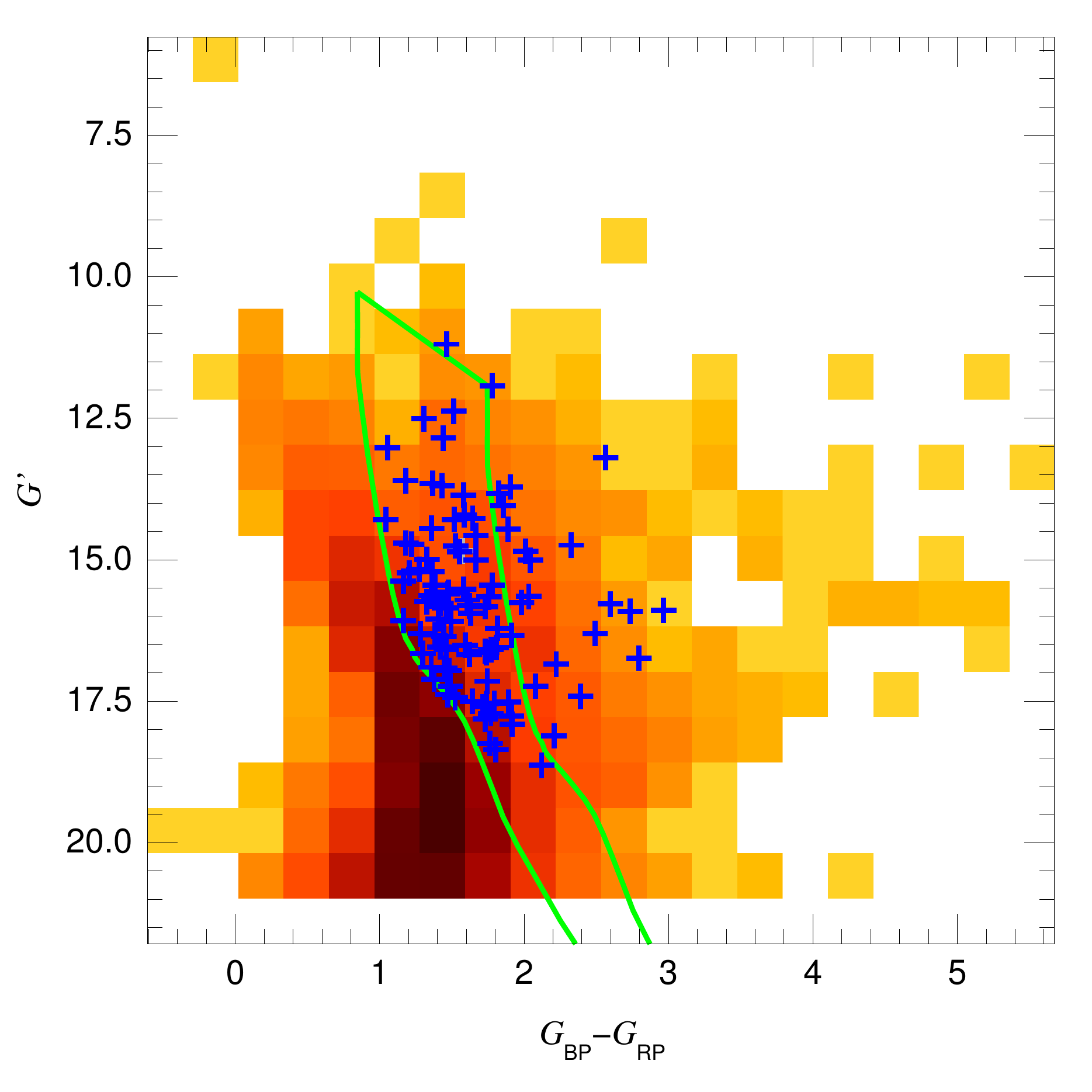} \
            \includegraphics*[width=0.34\linewidth, bb=0 0 538 522]{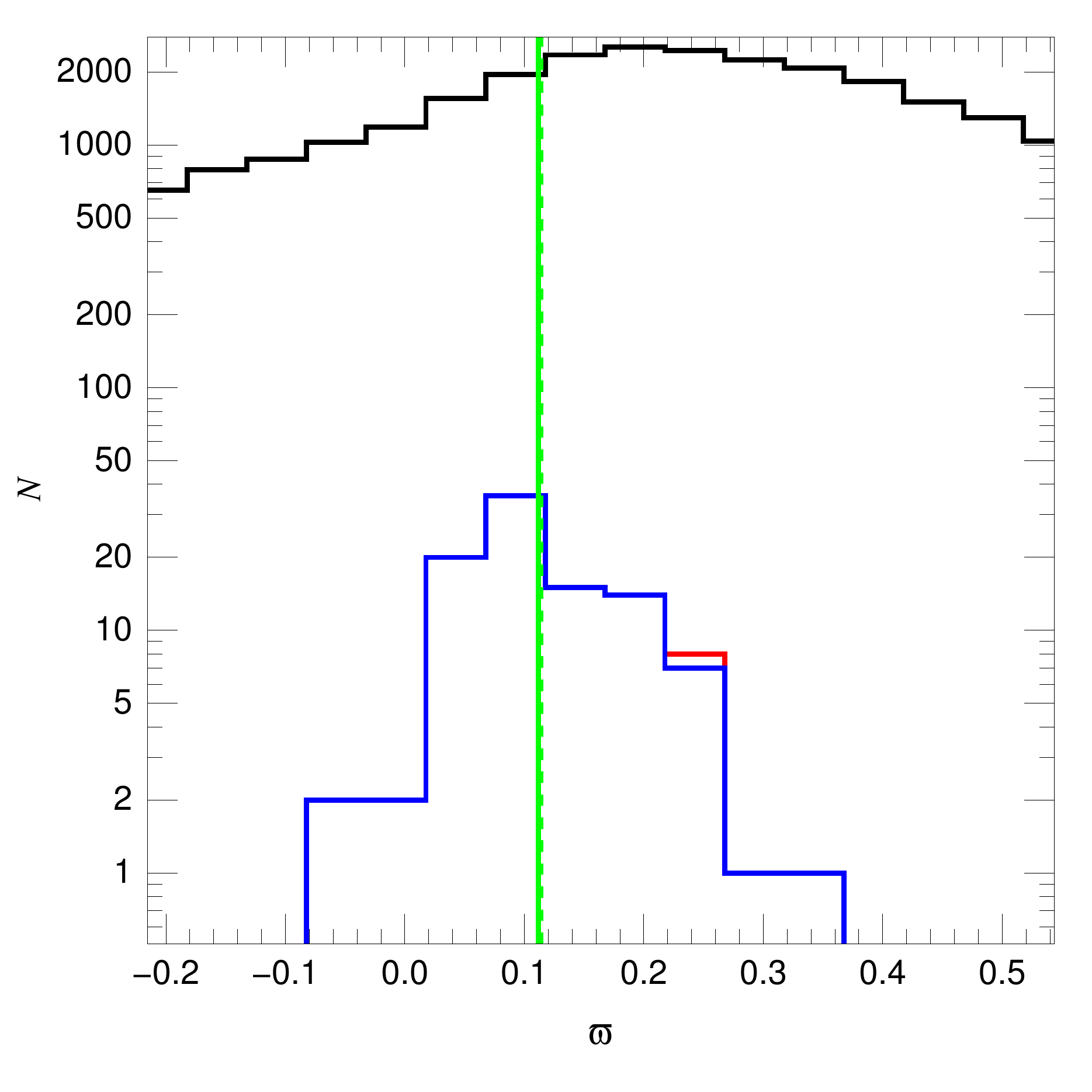} \
            \includegraphics*[width=0.34\linewidth, bb=0 0 538 522]{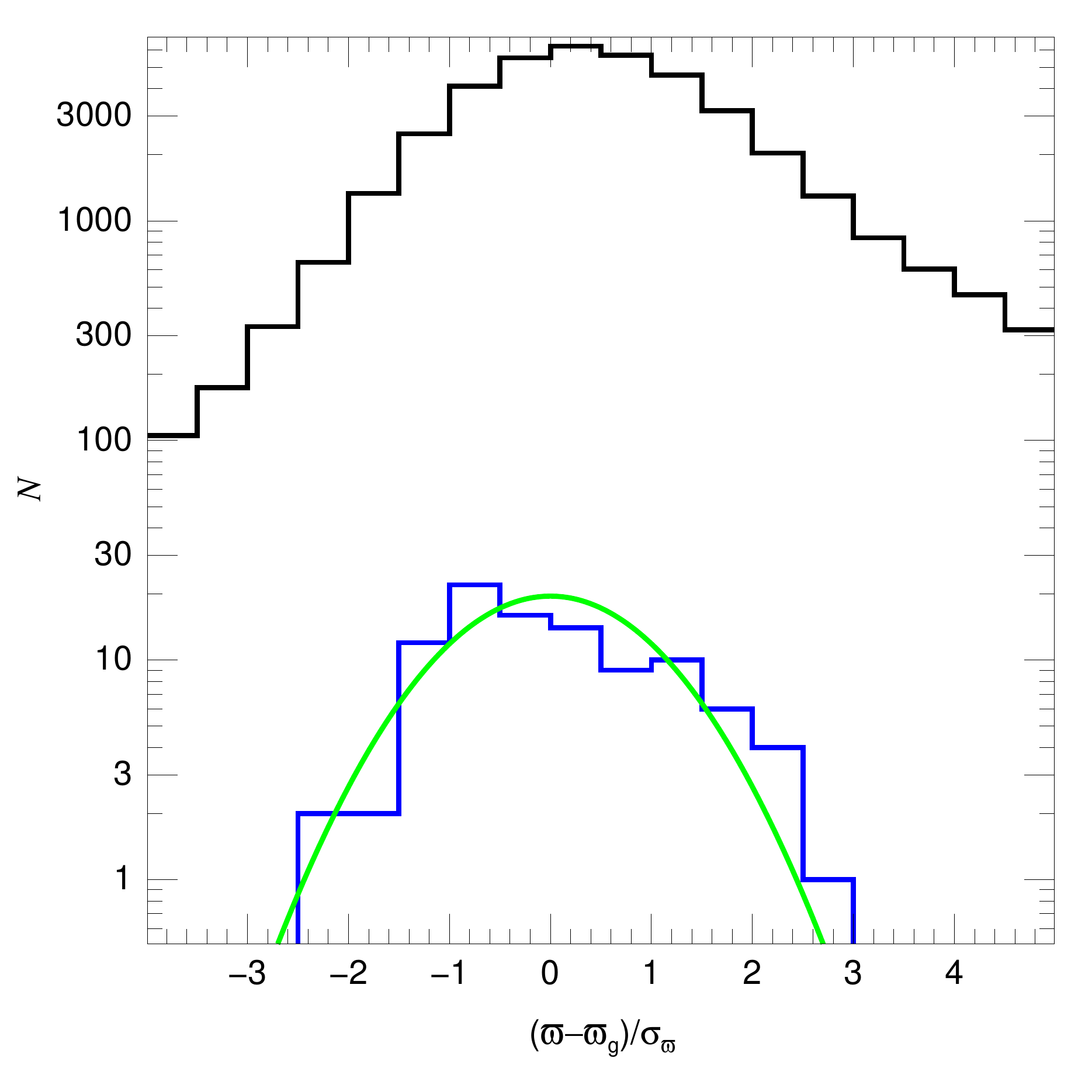}}
\caption{Same as Fig.~\ref{NGC_3603_Gaia} for \VO{006}.}
\label{Gum_35_Gaia}
\end{figure*}   

\begin{figure*}
\centerline{\includegraphics*[width=0.34\linewidth, bb=0 0 538 522]{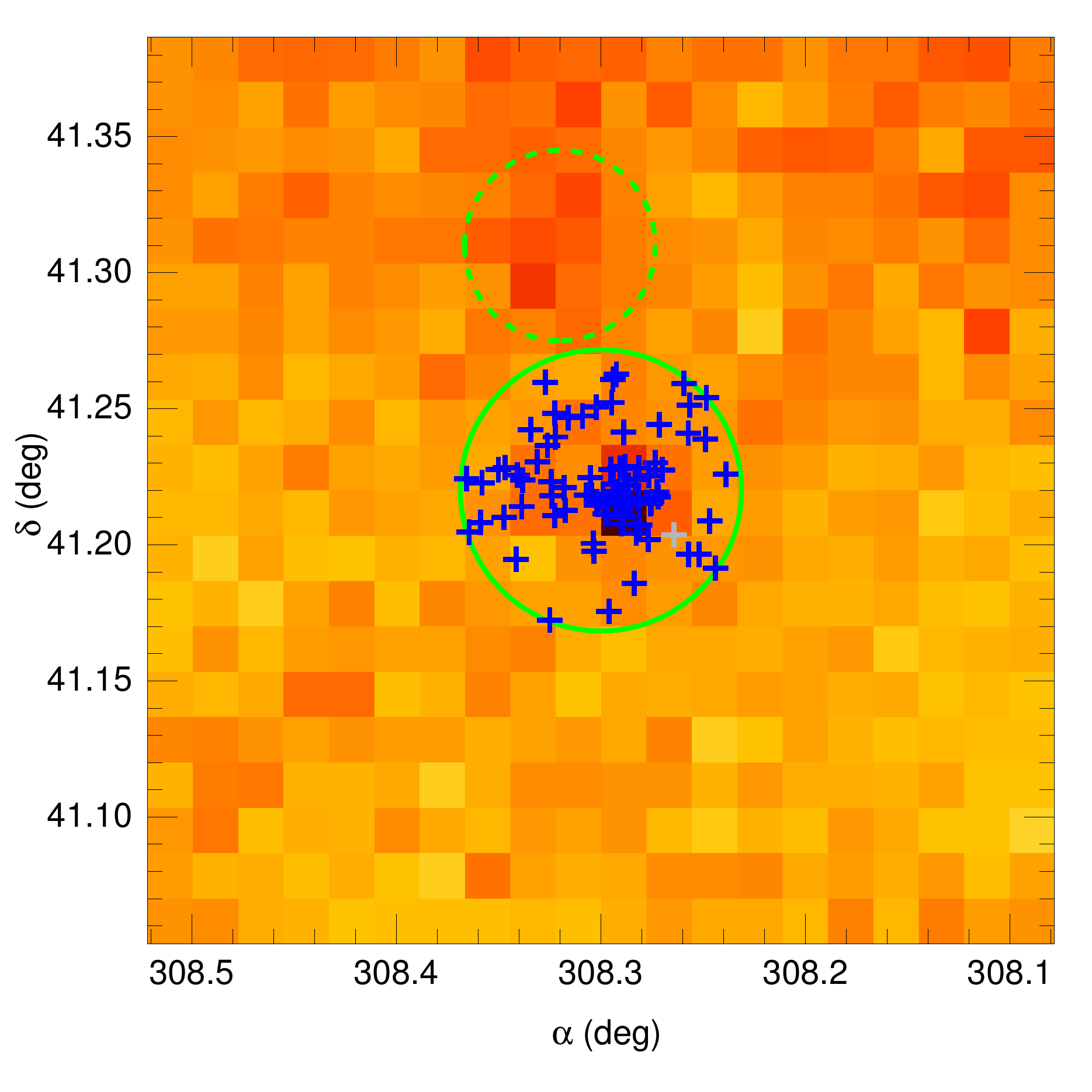} \
            \includegraphics*[width=0.34\linewidth, bb=0 0 538 522]{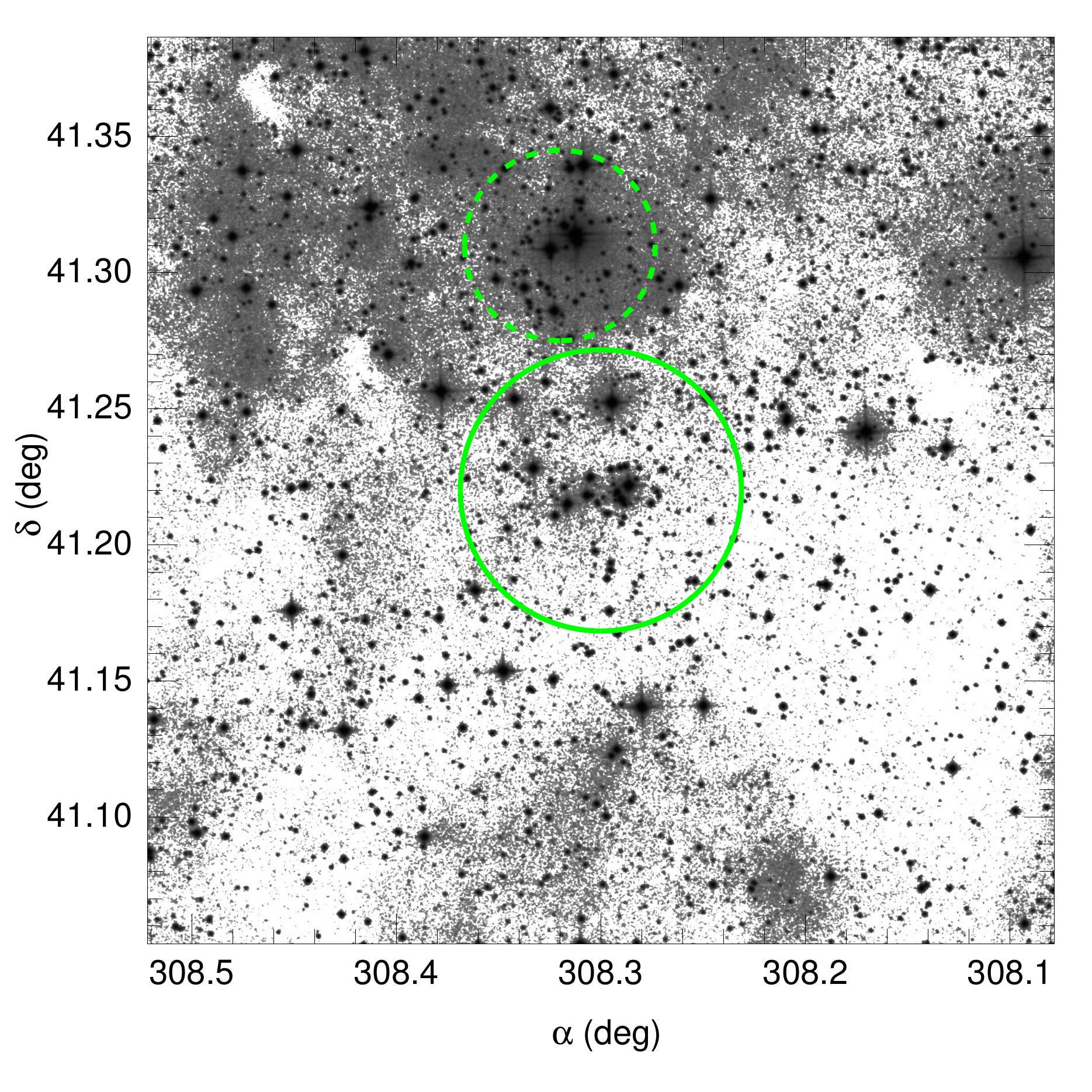} \
            \includegraphics*[width=0.34\linewidth, bb=0 0 538 522]{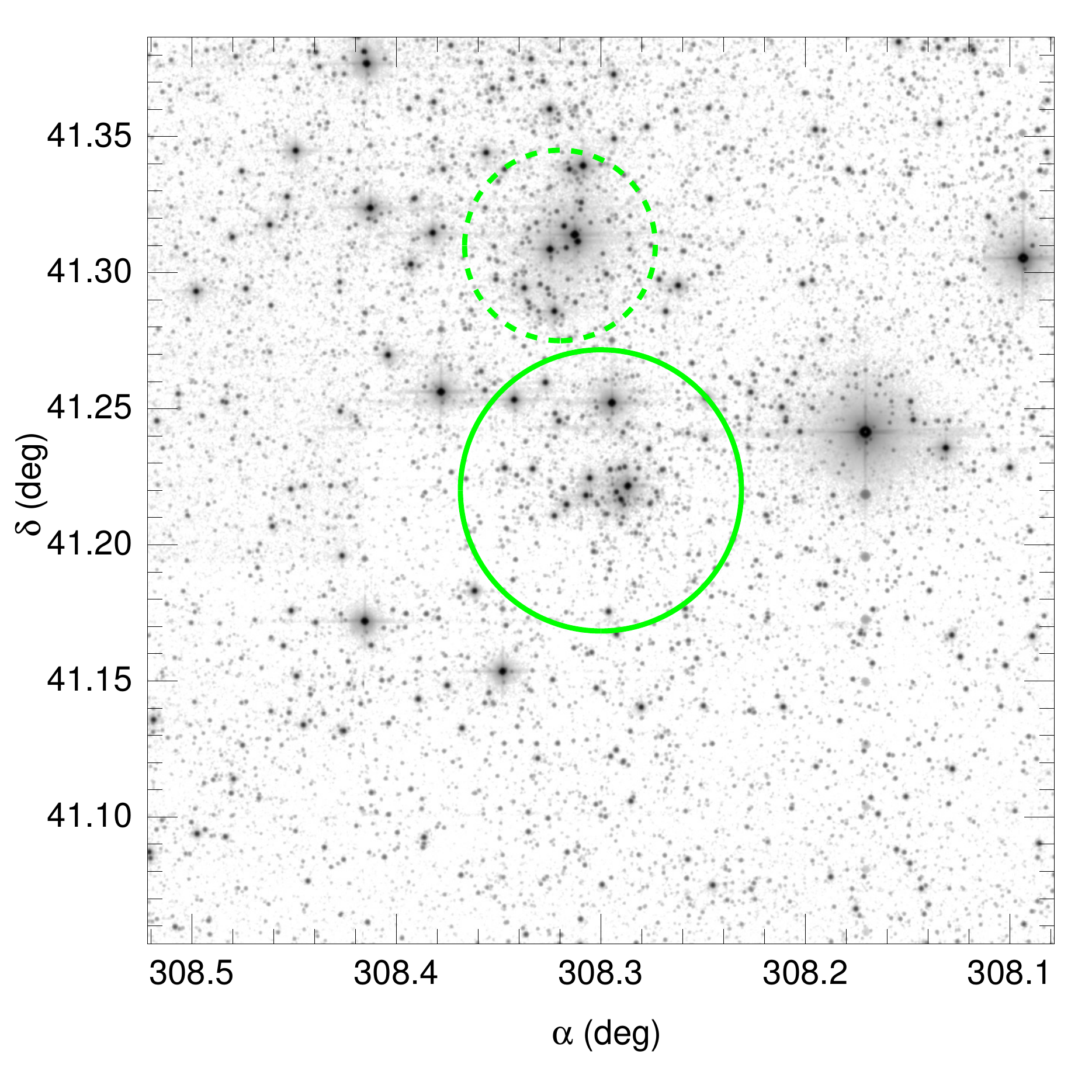}}
\centerline{\includegraphics*[width=0.34\linewidth, bb=0 0 538 522]{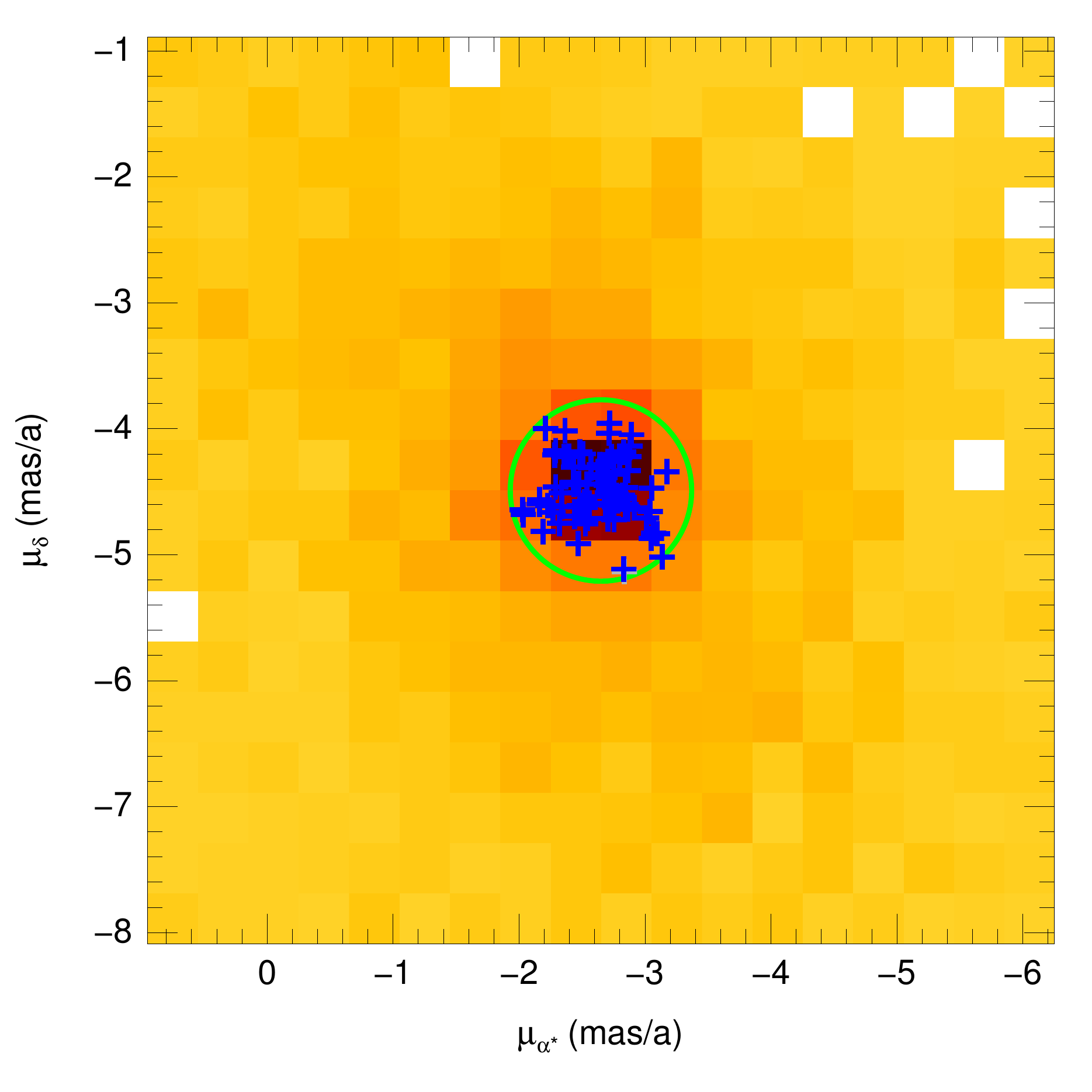} \
            \includegraphics*[width=0.34\linewidth, bb=0 0 538 522]{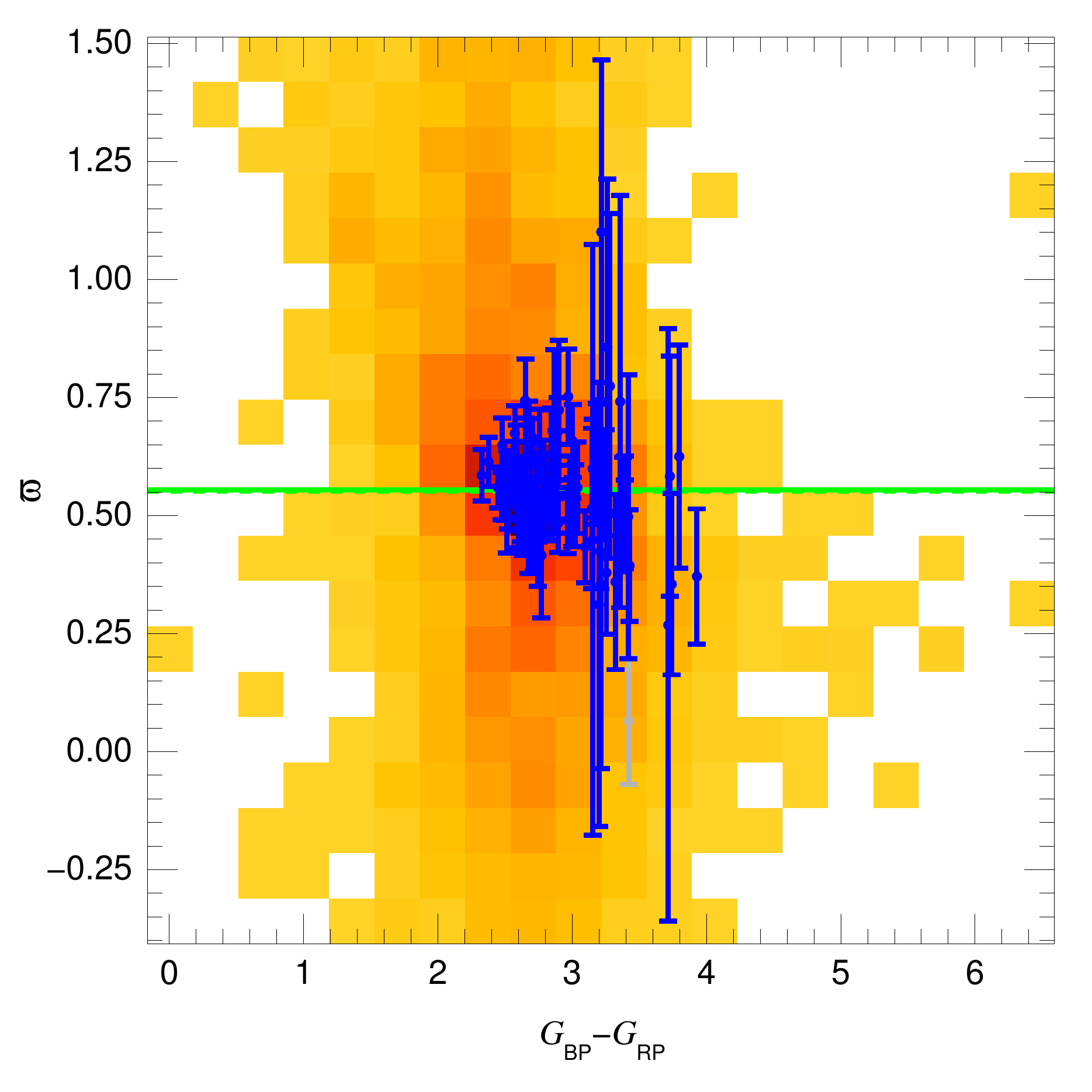} \
            \includegraphics*[width=0.34\linewidth, bb=0 0 538 522]{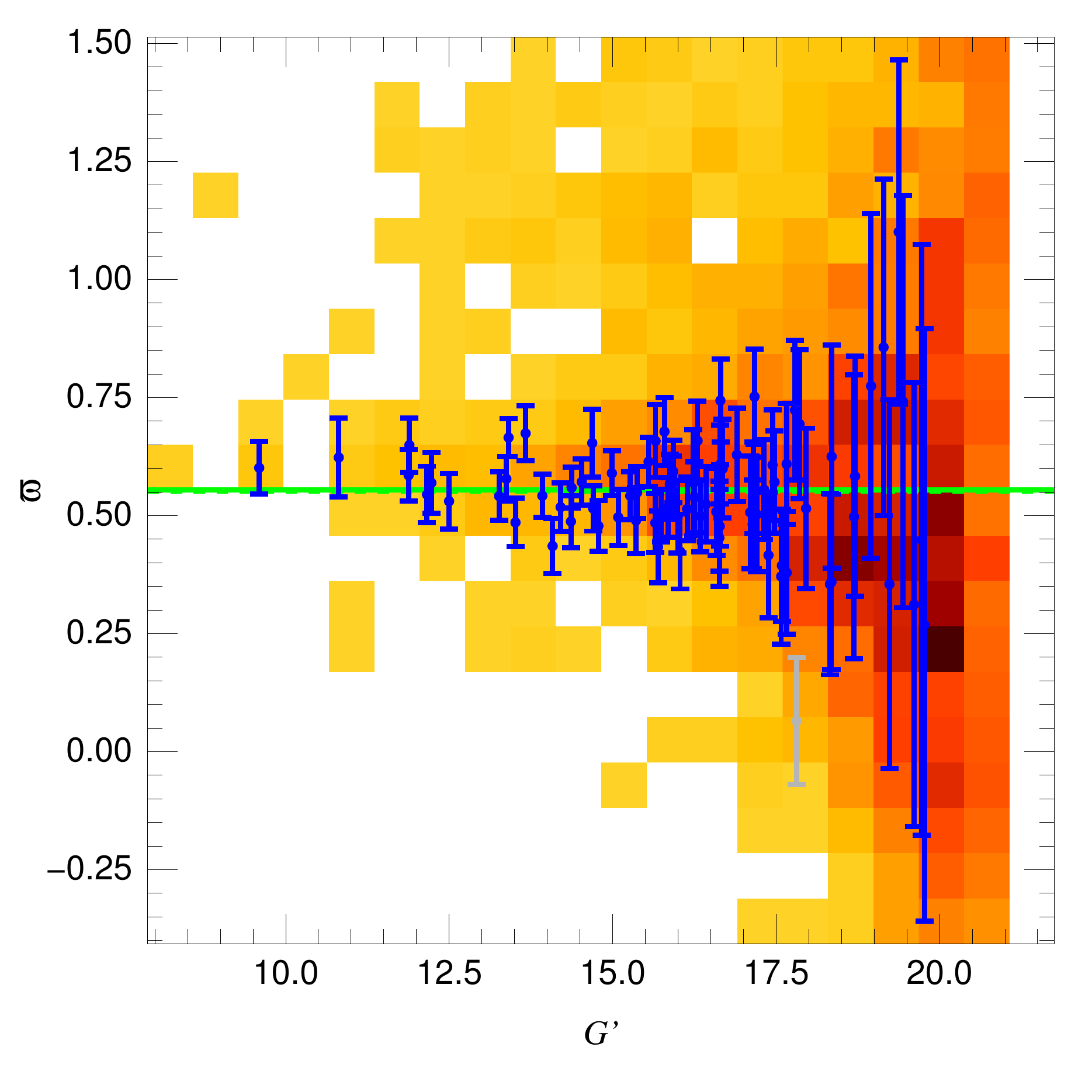}}
\centerline{\includegraphics*[width=0.34\linewidth, bb=0 0 538 522]{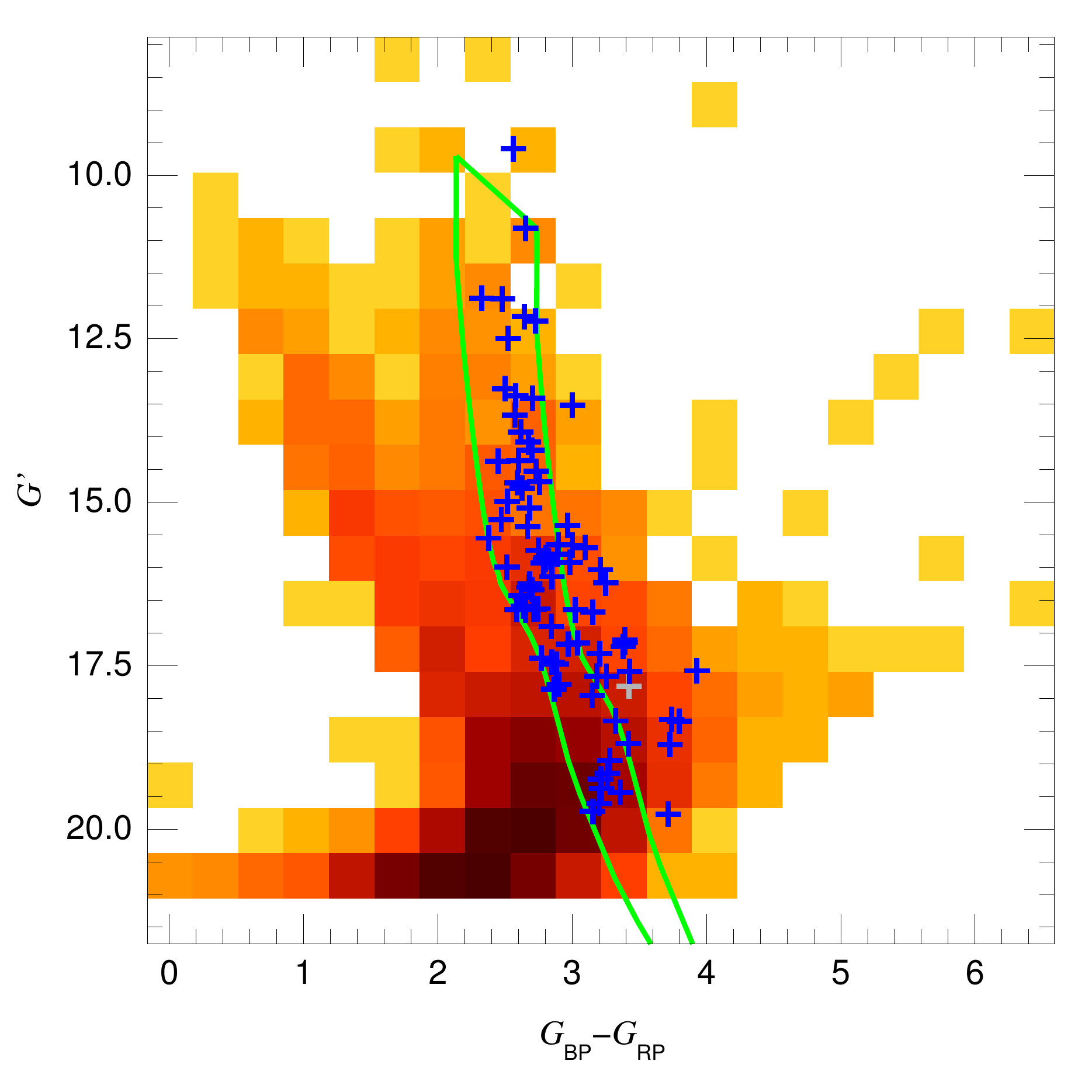} \
            \includegraphics*[width=0.34\linewidth, bb=0 0 538 522]{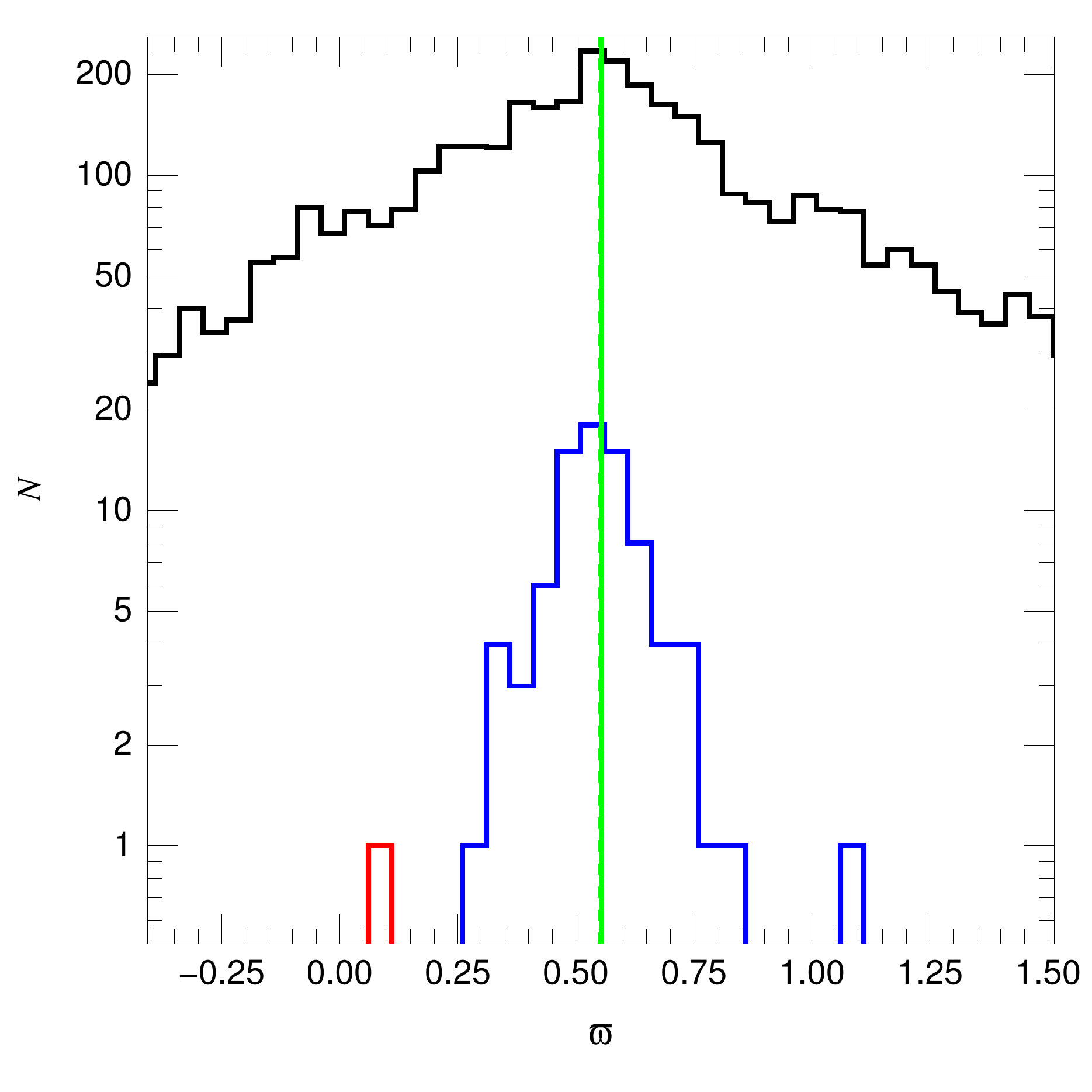} \
            \includegraphics*[width=0.34\linewidth, bb=0 0 538 522]{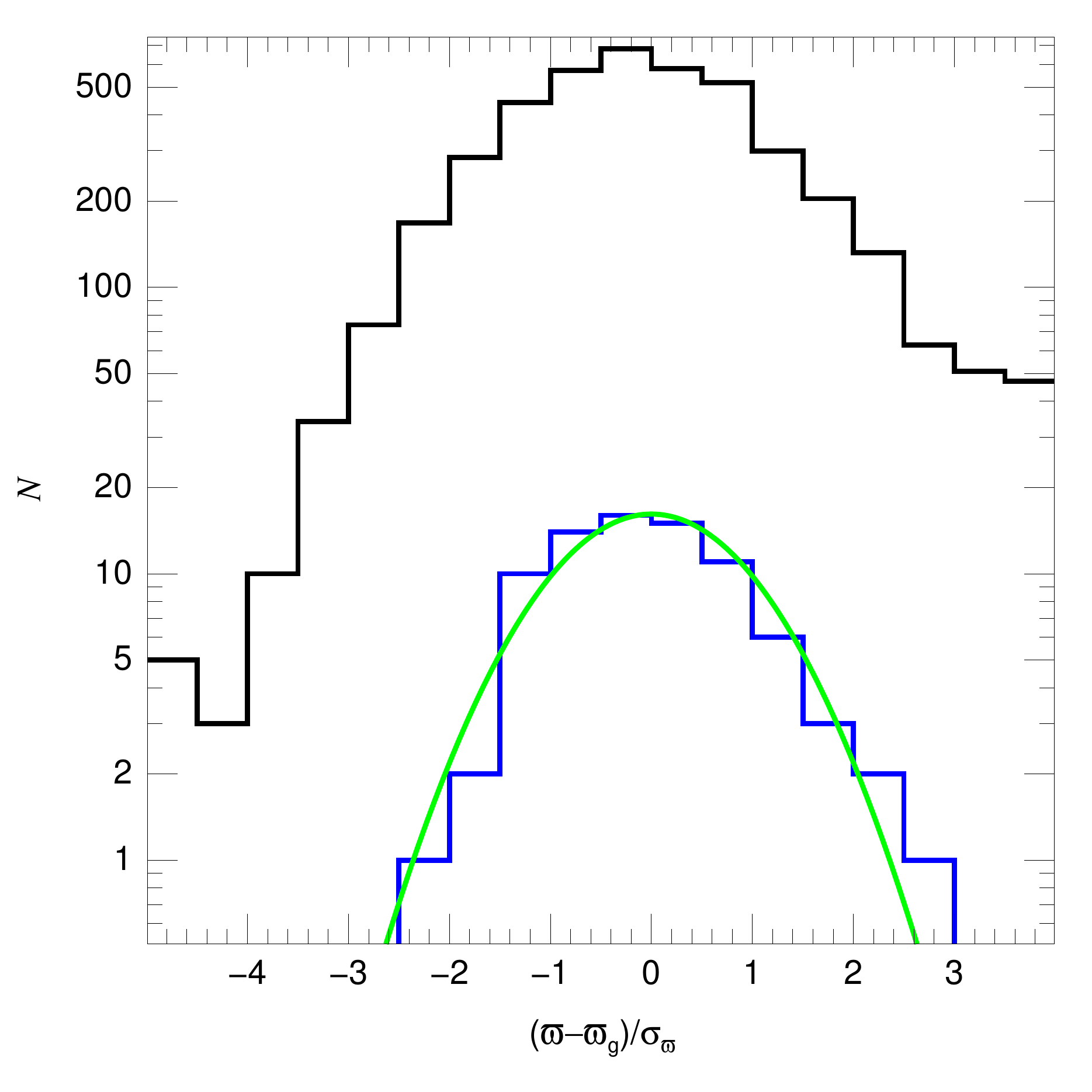}}
\caption{Same as Fig.~\ref{NGC_3603_Gaia} for 
         Bica~1 (\VO{007}). % REFEREE \VO{007}
         The dashed green circle in the top three panels shows the position of the neighbor 
         Bica~2 (\VO{008}). % REFEREE \VO{008}
         }
\label{Cyg_OB2-22_Gaia}
\end{figure*}   

\begin{figure*}
\centerline{\includegraphics*[width=0.34\linewidth, bb=0 0 538 522]{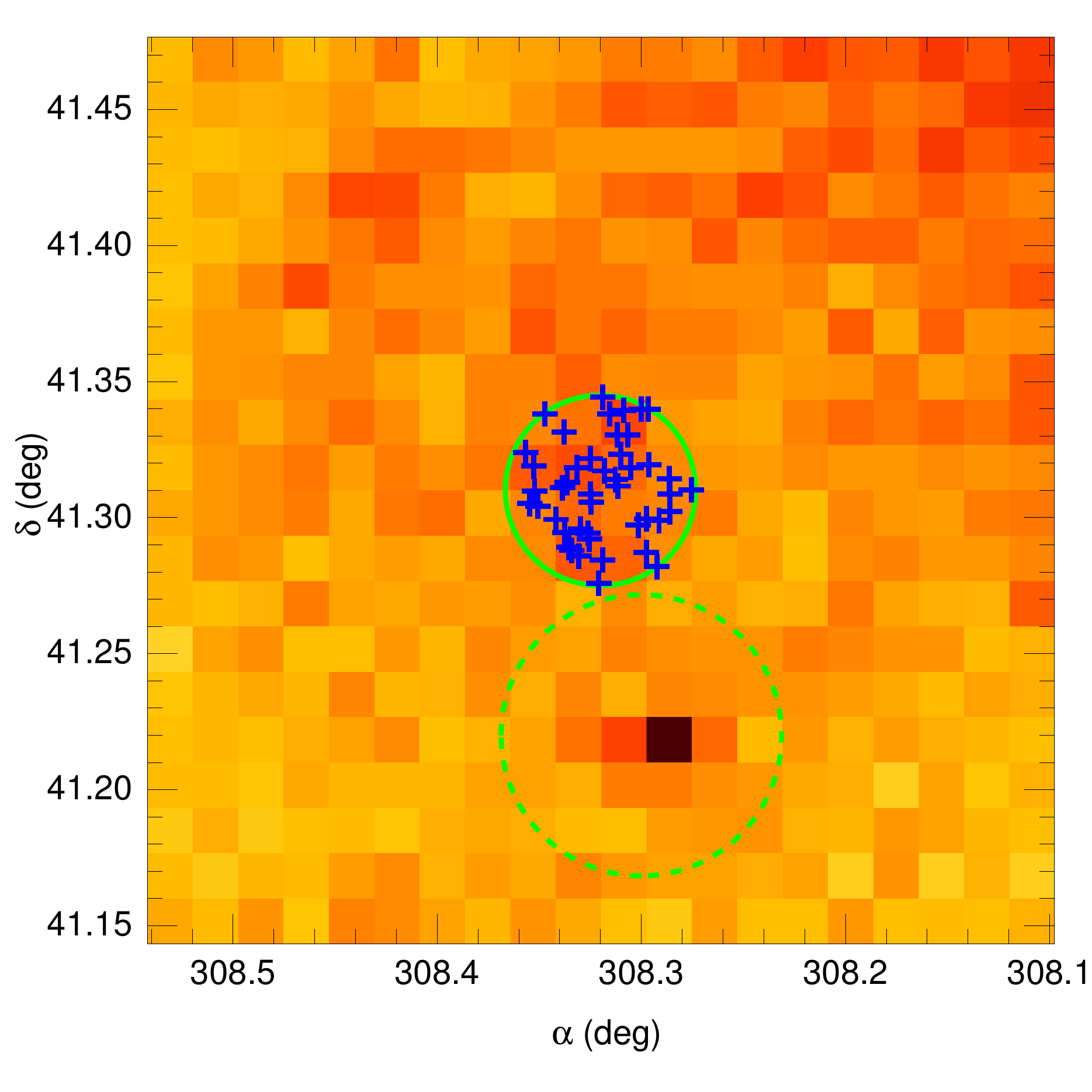} \
            \includegraphics*[width=0.34\linewidth, bb=0 0 538 522]{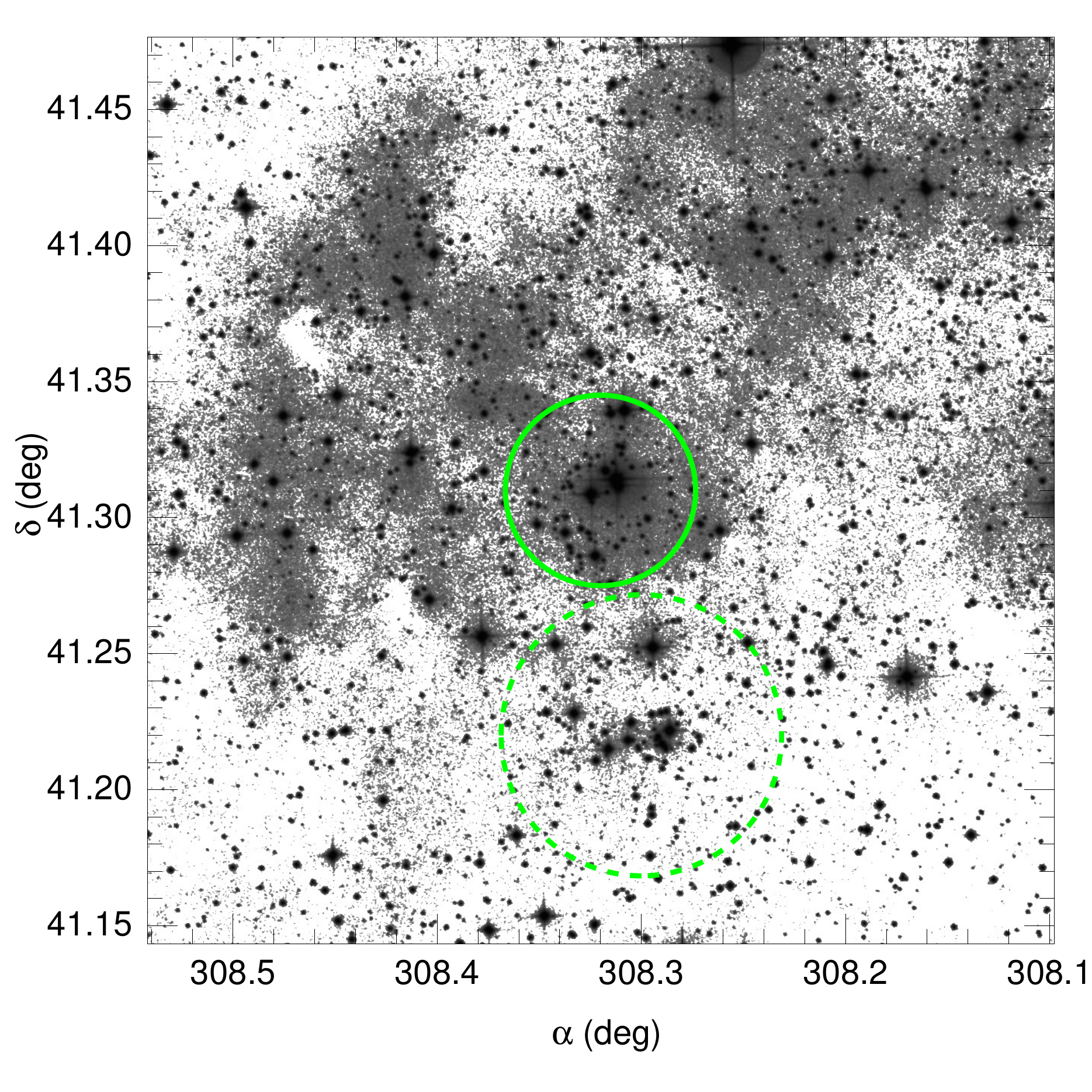} \
            \includegraphics*[width=0.34\linewidth, bb=0 0 538 522]{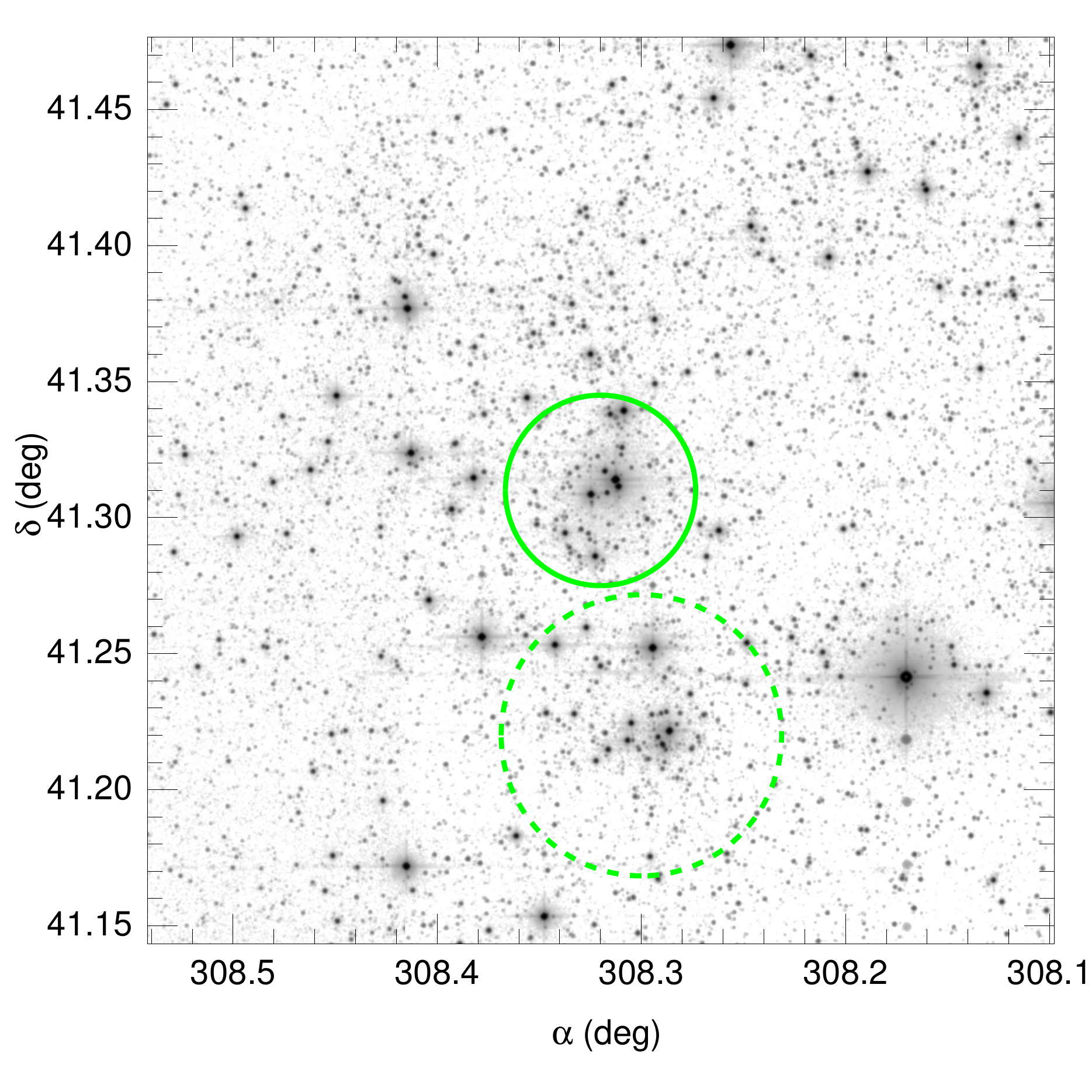}}
\centerline{\includegraphics*[width=0.34\linewidth, bb=0 0 538 522]{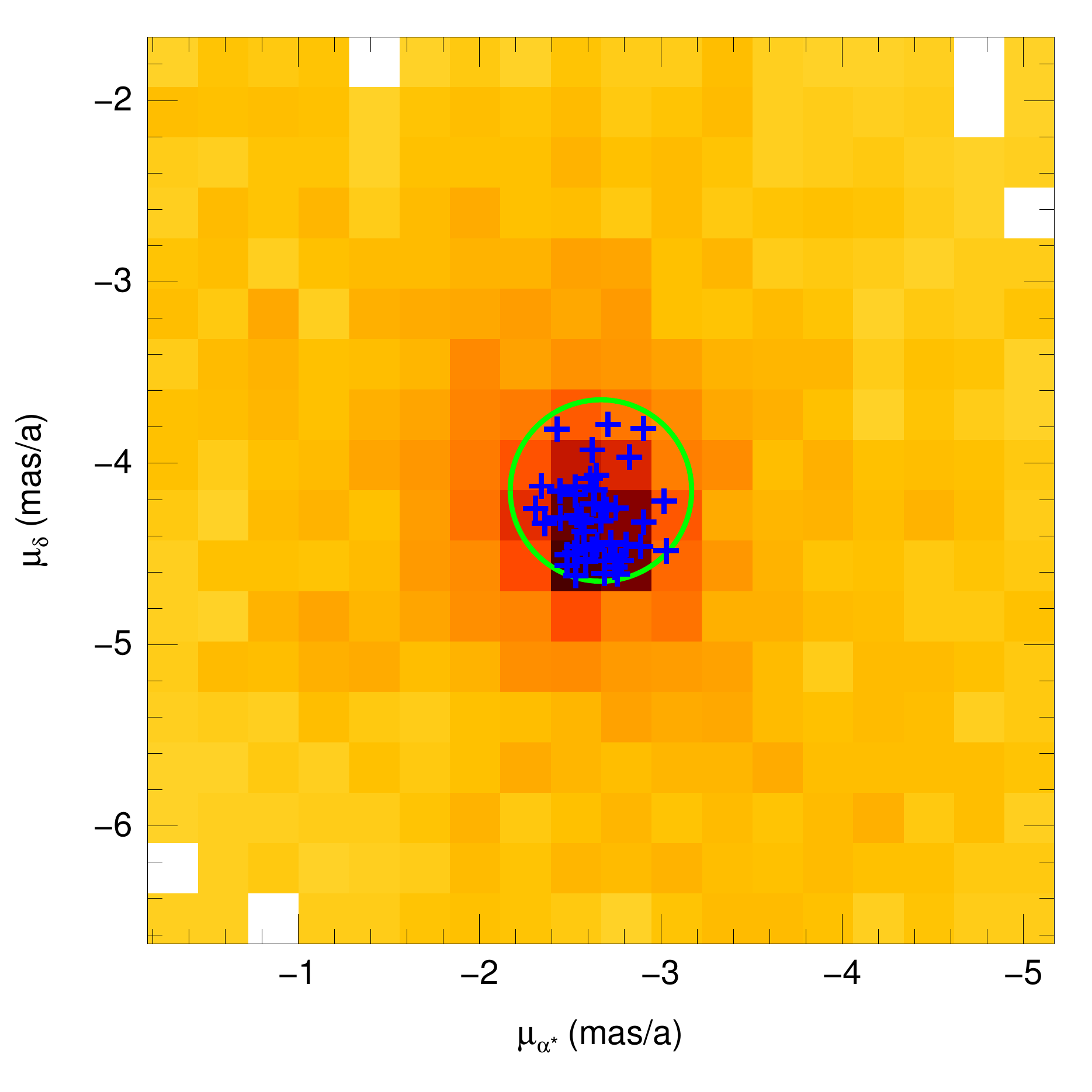} \
            \includegraphics*[width=0.34\linewidth, bb=0 0 538 522]{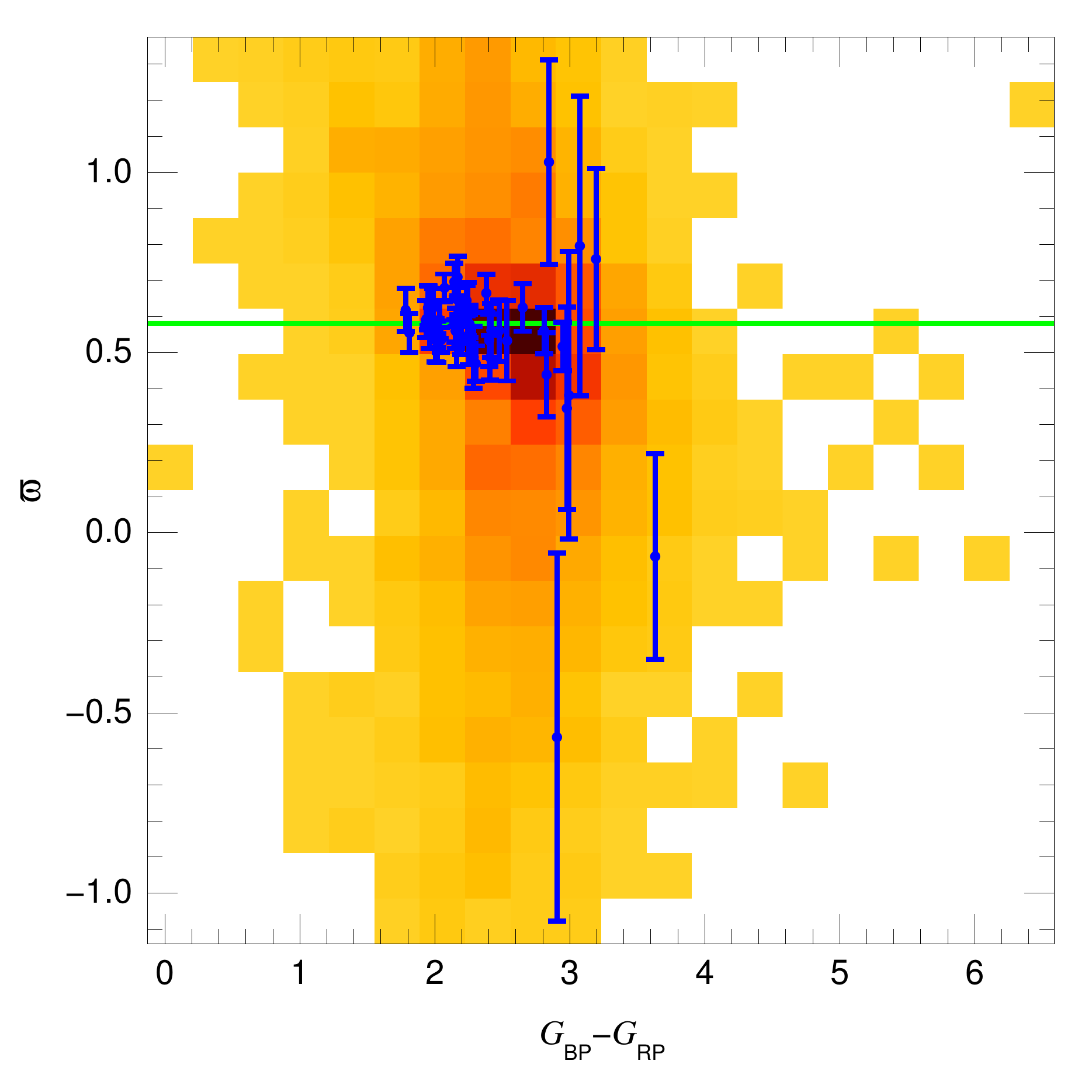} \
            \includegraphics*[width=0.34\linewidth, bb=0 0 538 522]{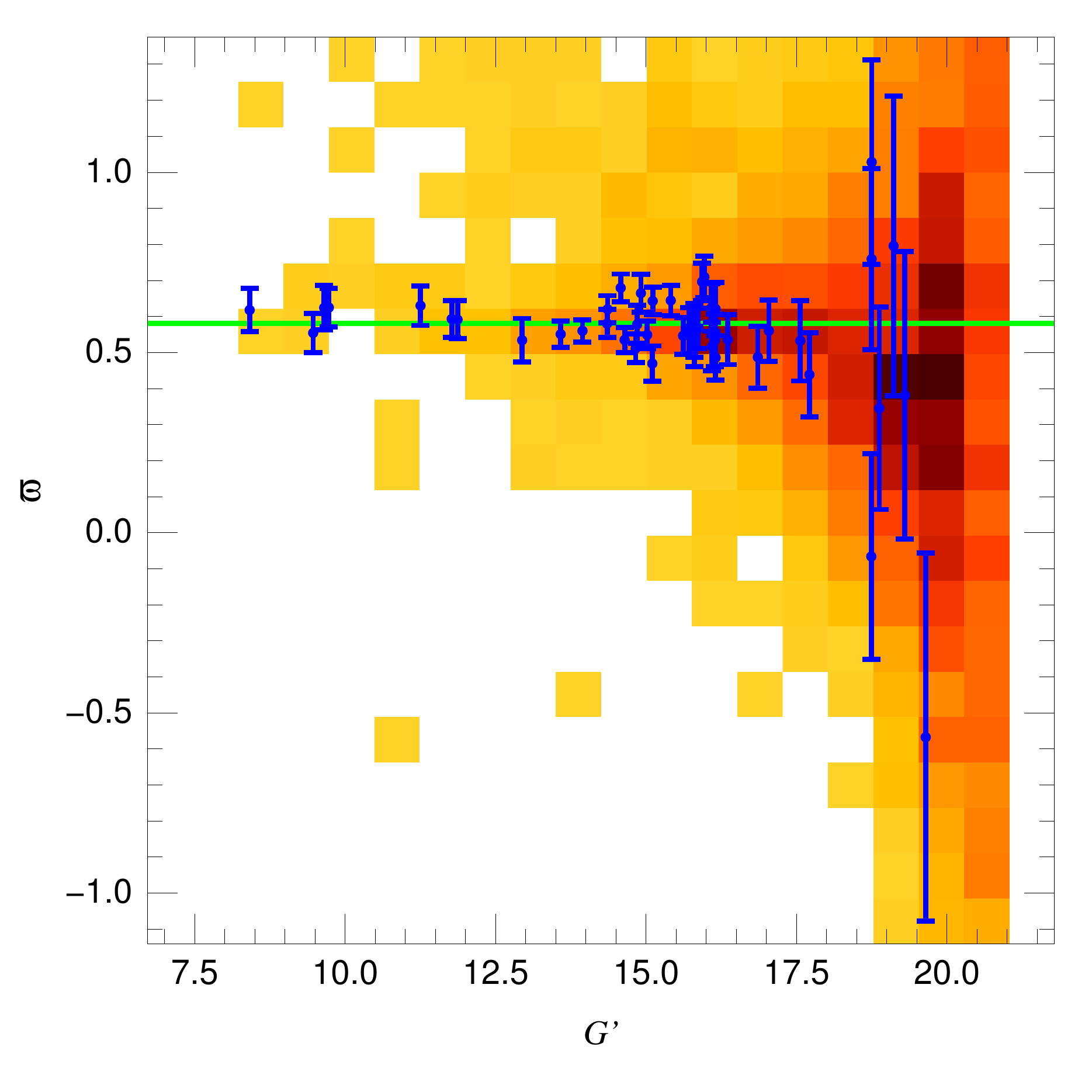}}
\centerline{\includegraphics*[width=0.34\linewidth, bb=0 0 538 522]{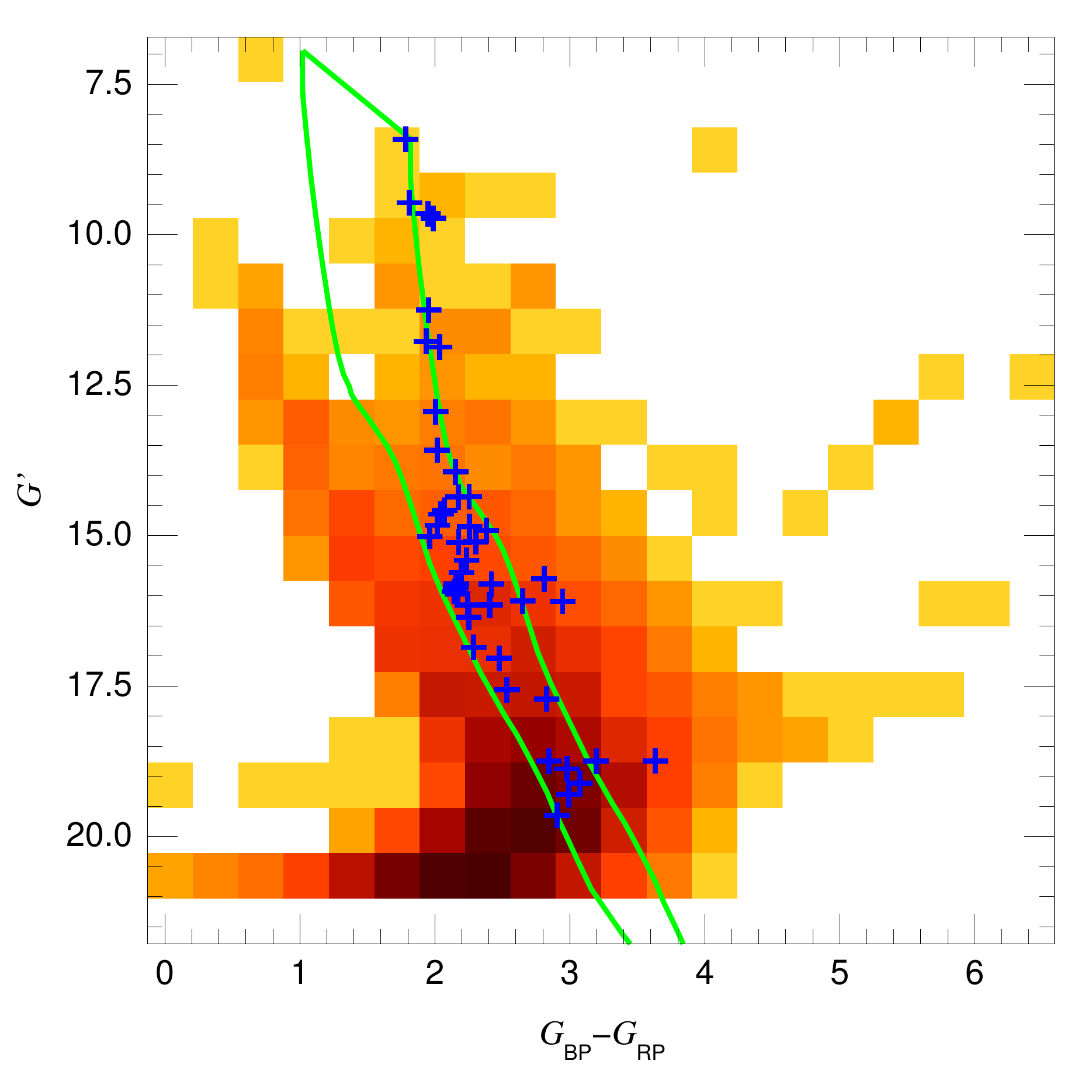} \
            \includegraphics*[width=0.34\linewidth, bb=0 0 538 522]{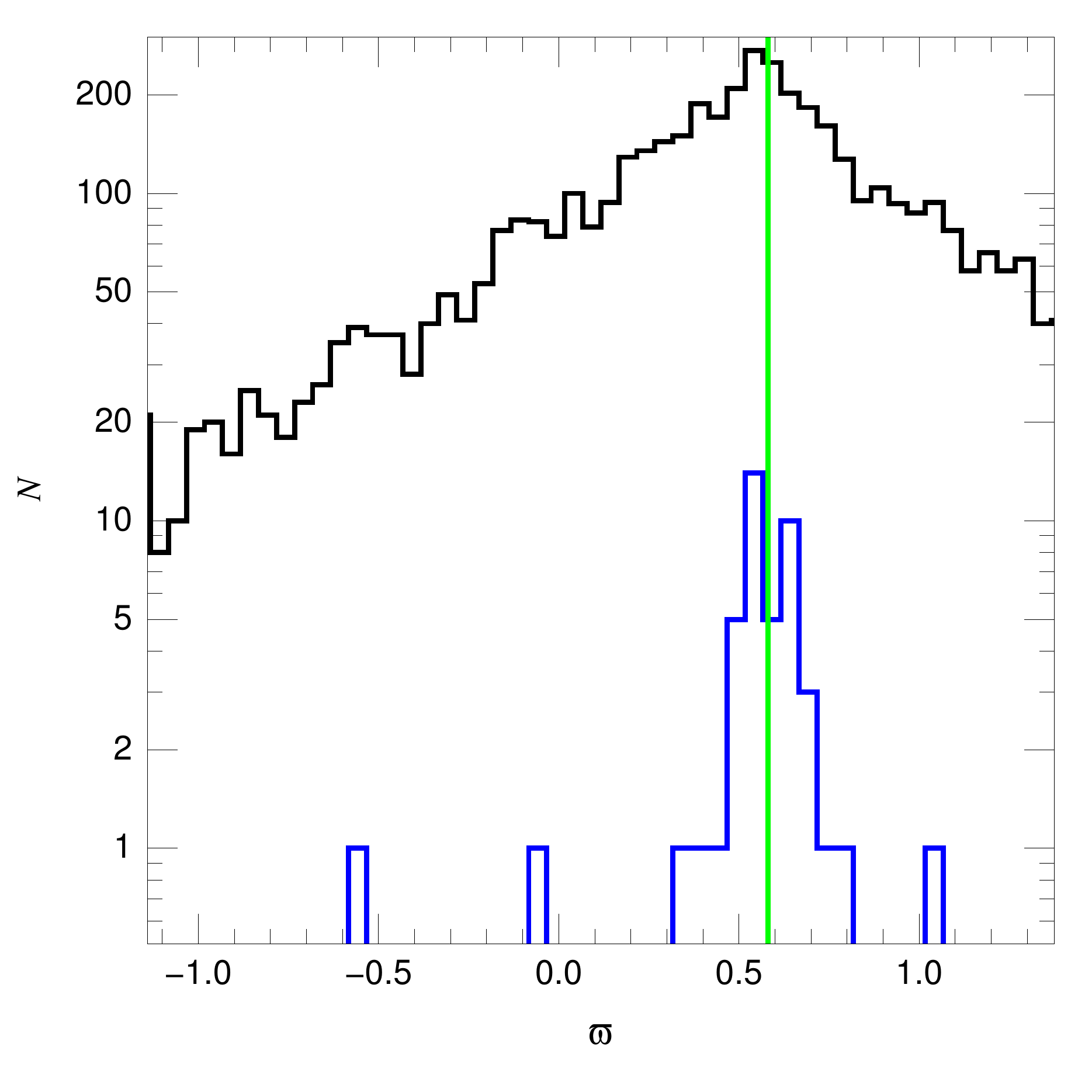} \
            \includegraphics*[width=0.34\linewidth, bb=0 0 538 522]{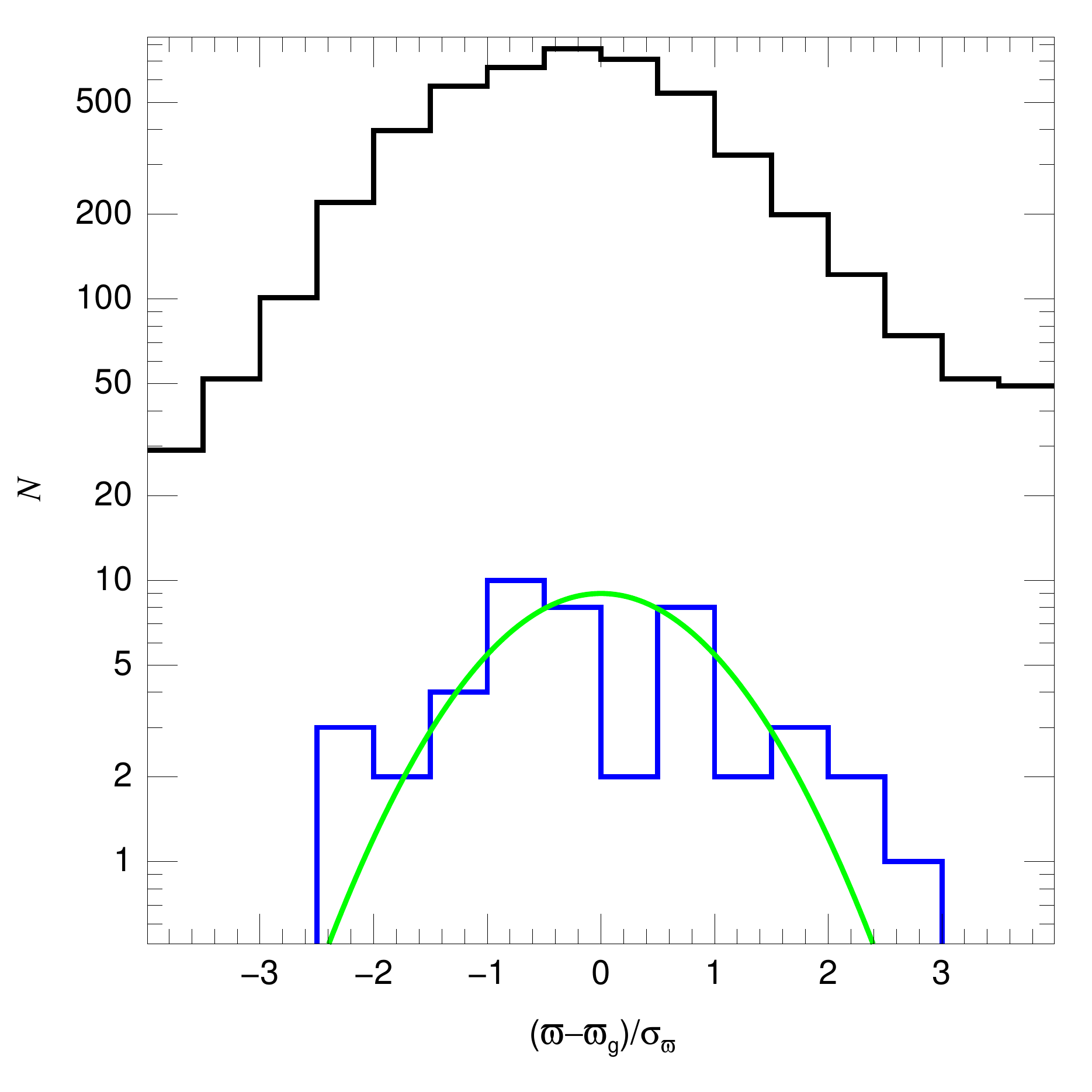}}
\caption{Same as Fig.~\ref{Cyg_OB2-22_Gaia} for 
         Bica~2 (\VO{008}). % REFEREE \VO{008}
         The dashed green circle in the top three panels shows the position of the neighbor
         Bica~1 (\VO{007}). % REFEREE \VO{007}
         }
\label{Cyg_OB2-8_Gaia}
\end{figure*}   

\begin{figure*}
\centerline{\includegraphics*[width=0.34\linewidth, bb=0 0 538 522]{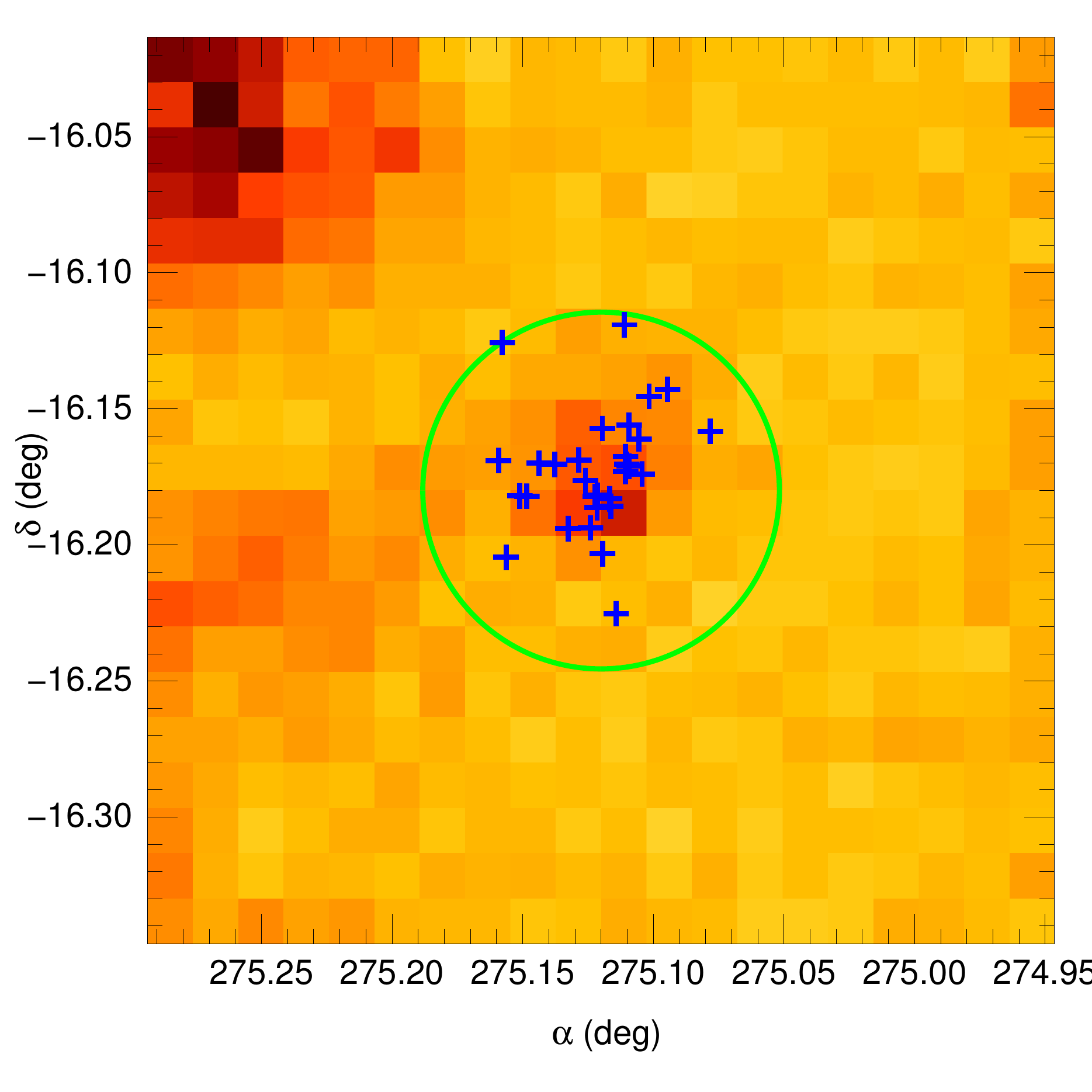} \
            \includegraphics*[width=0.34\linewidth, bb=0 0 538 522]{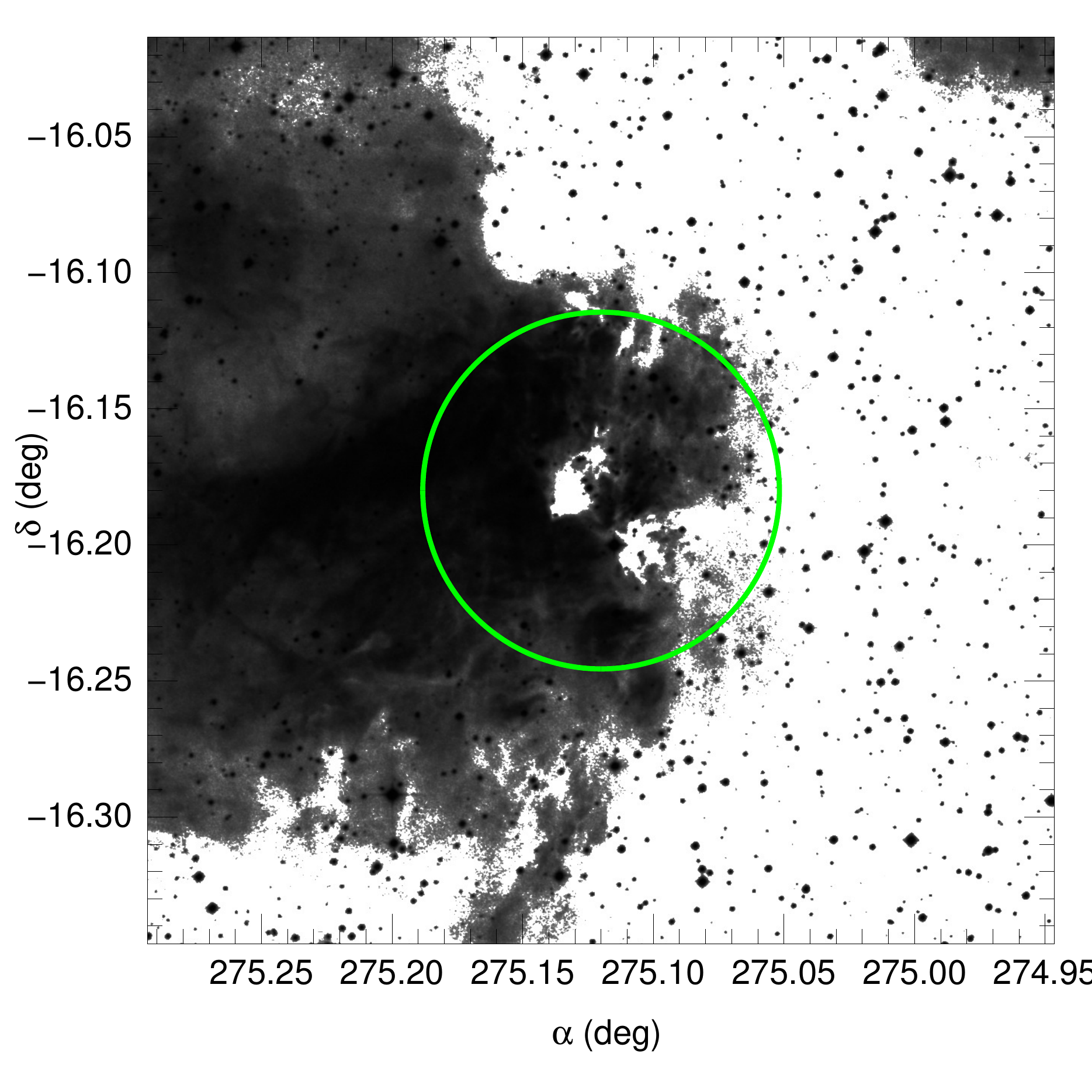} \
            \includegraphics*[width=0.34\linewidth, bb=0 0 538 522]{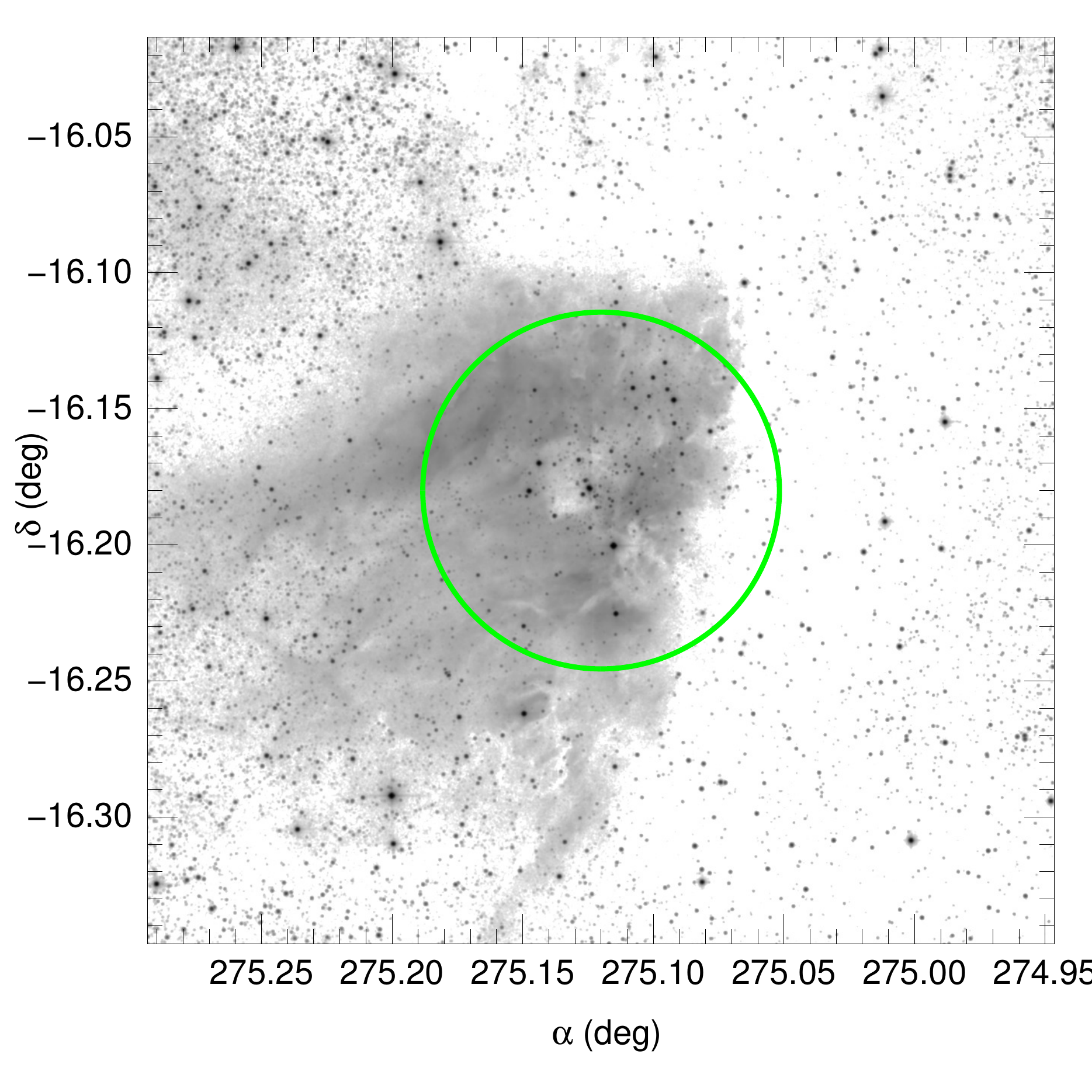}}
\centerline{\includegraphics*[width=0.34\linewidth, bb=0 0 538 522]{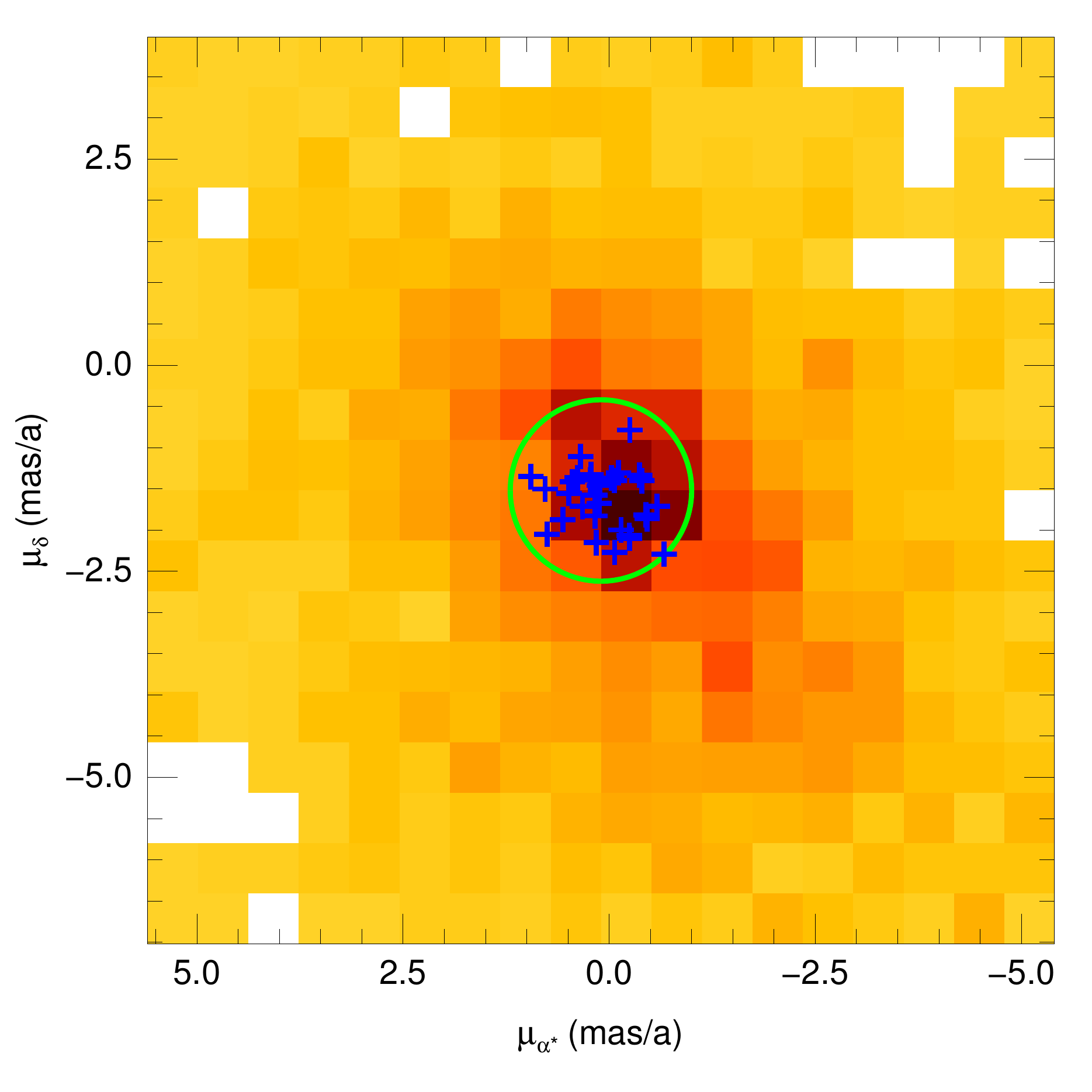} \
            \includegraphics*[width=0.34\linewidth, bb=0 0 538 522]{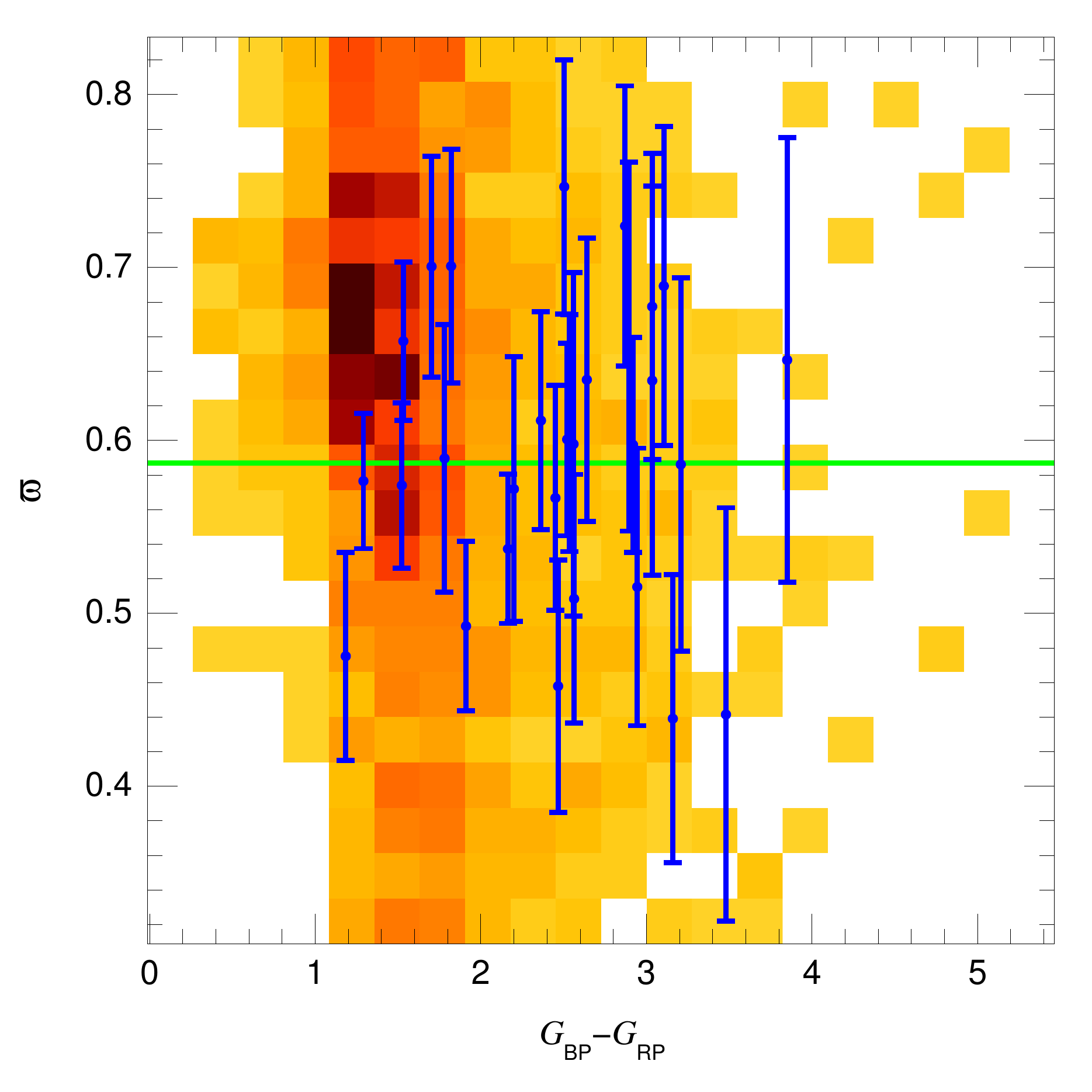} \
            \includegraphics*[width=0.34\linewidth, bb=0 0 538 522]{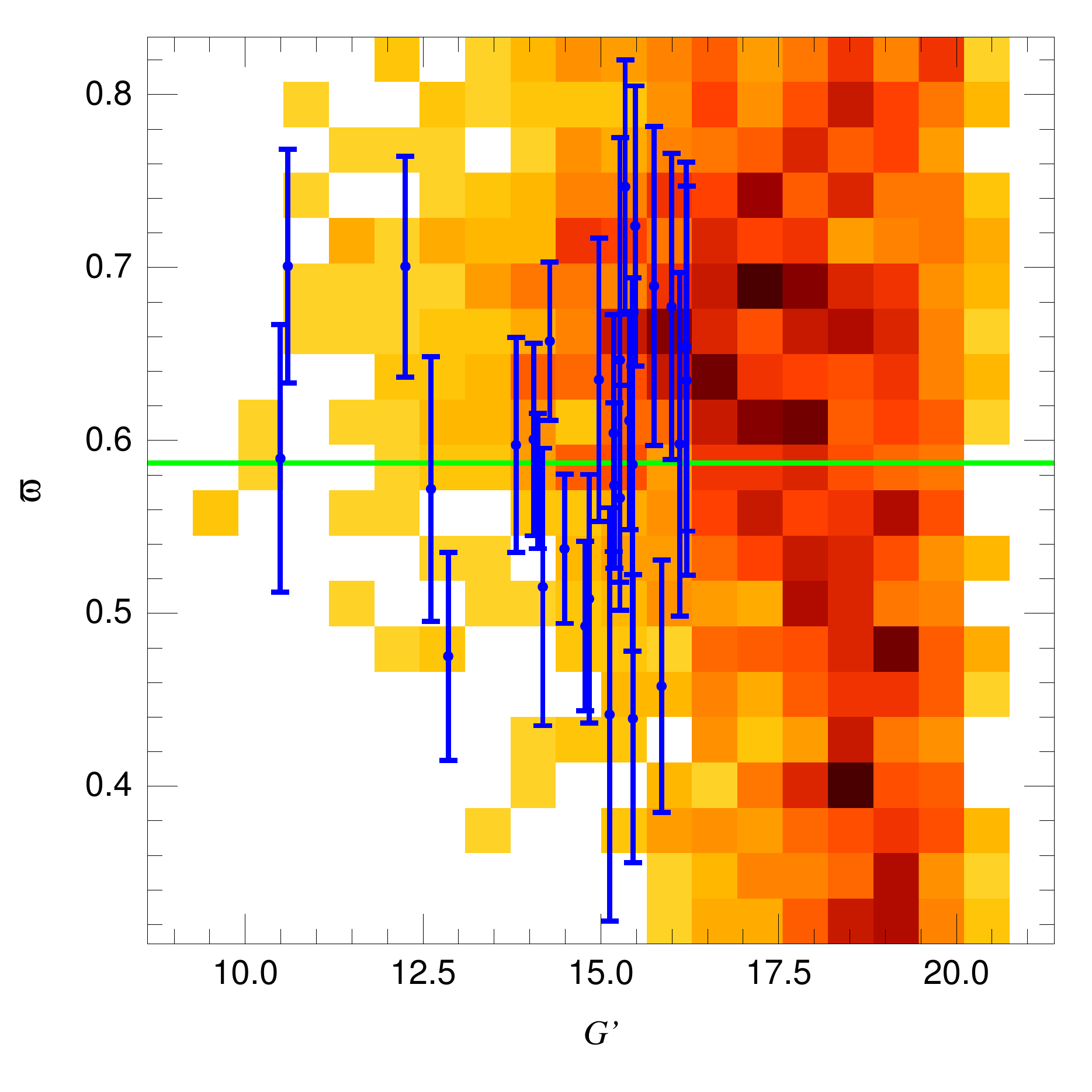}}
\centerline{\includegraphics*[width=0.34\linewidth, bb=0 0 538 522]{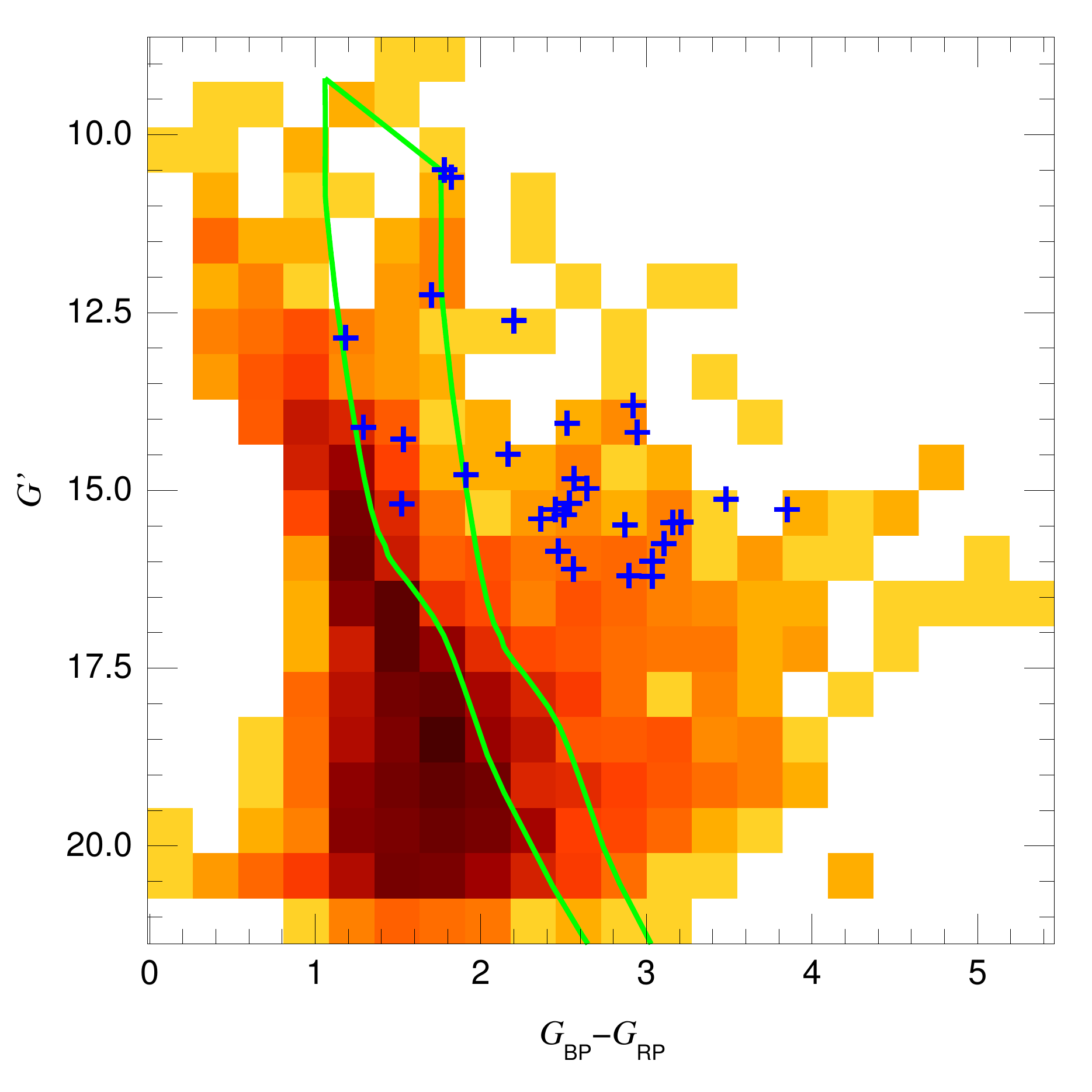} \
            \includegraphics*[width=0.34\linewidth, bb=0 0 538 522]{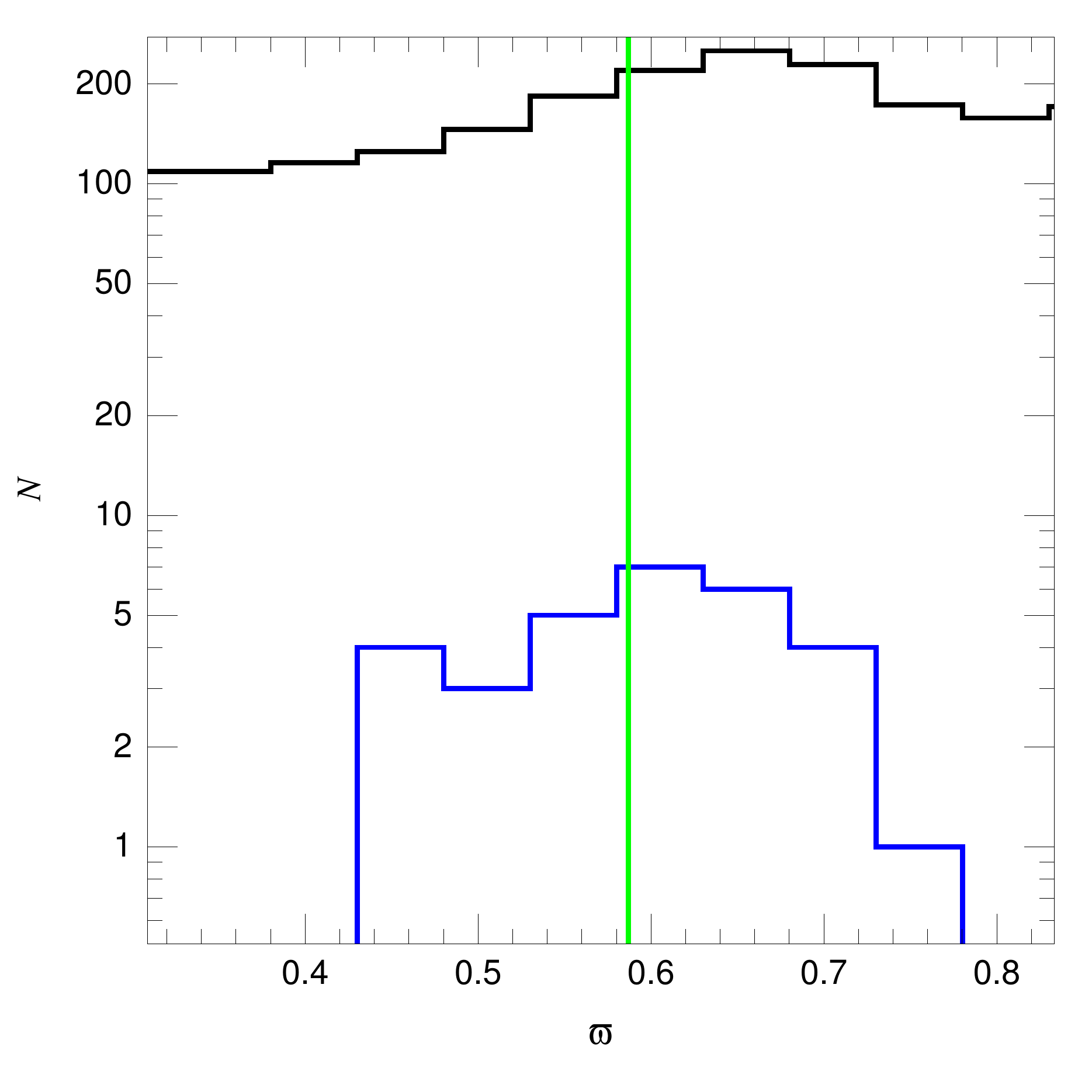} \
            \includegraphics*[width=0.34\linewidth, bb=0 0 538 522]{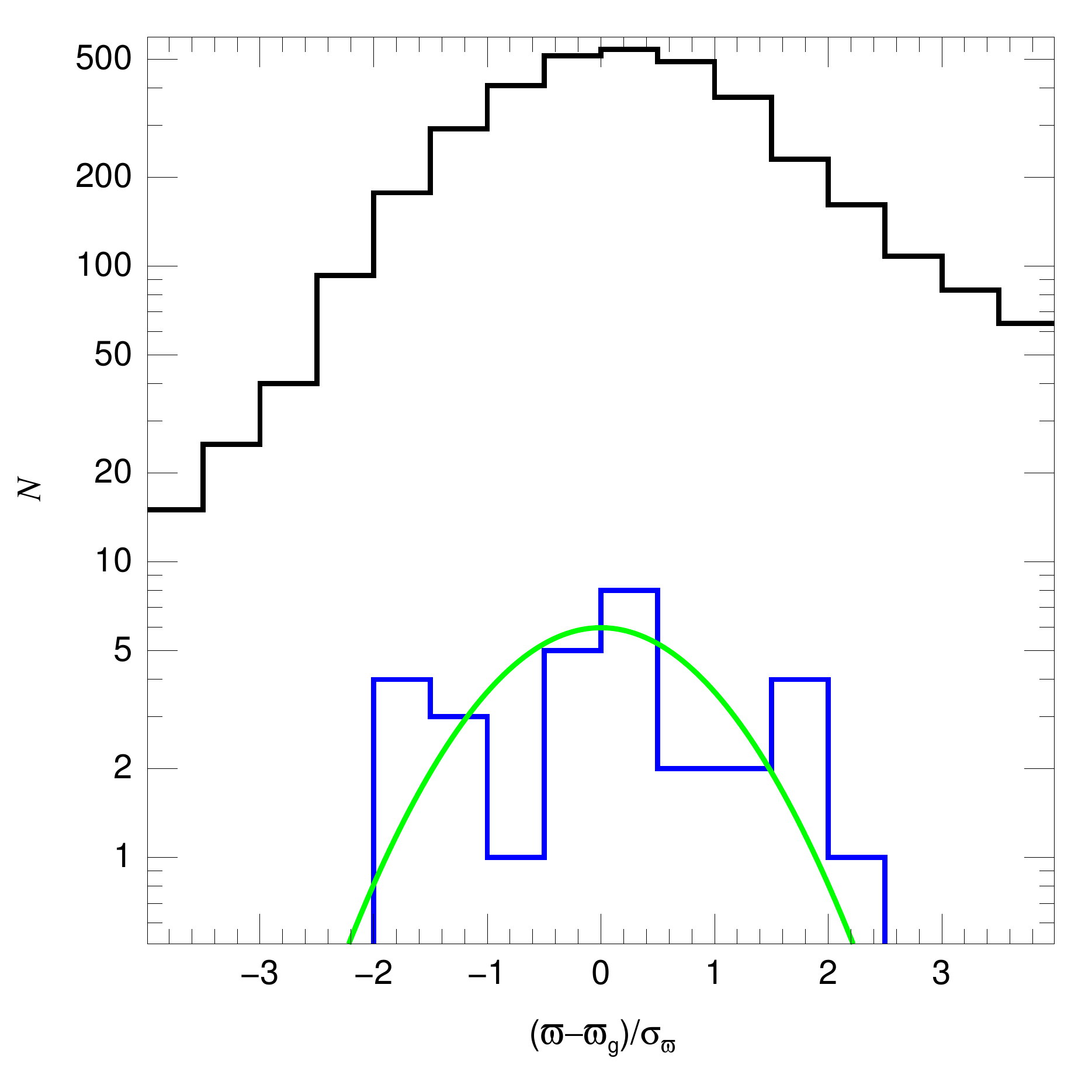}}
\caption{Same as Fig.~\ref{NGC_3603_Gaia} for 
         M17 (\VO{009}). % REFEREE \VO{009}
         }
\label{M17_Gaia}
\end{figure*}   

\begin{figure*}
\centerline{\includegraphics*[width=0.34\linewidth, bb=0 0 538 522]{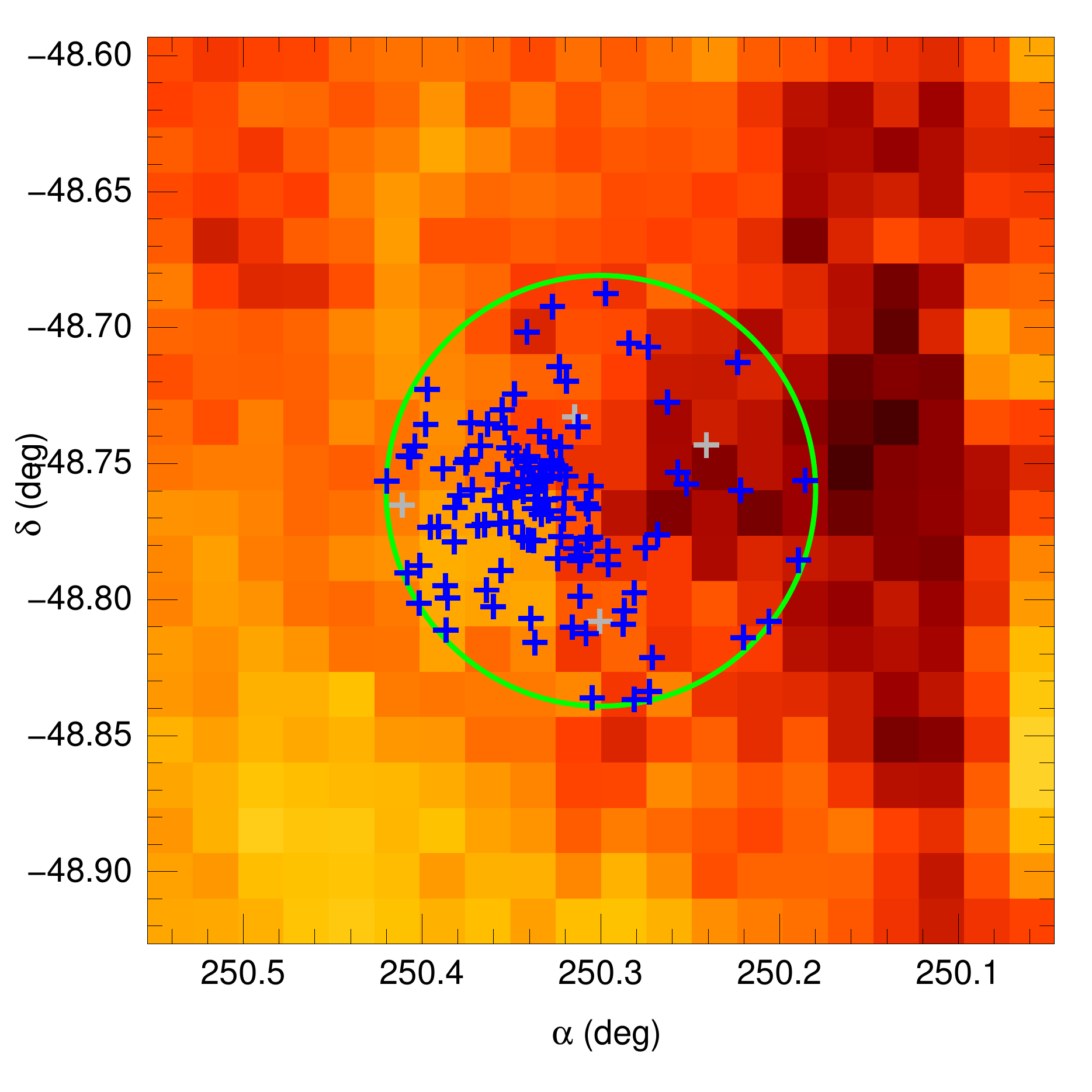} \
            \includegraphics*[width=0.34\linewidth, bb=0 0 538 522]{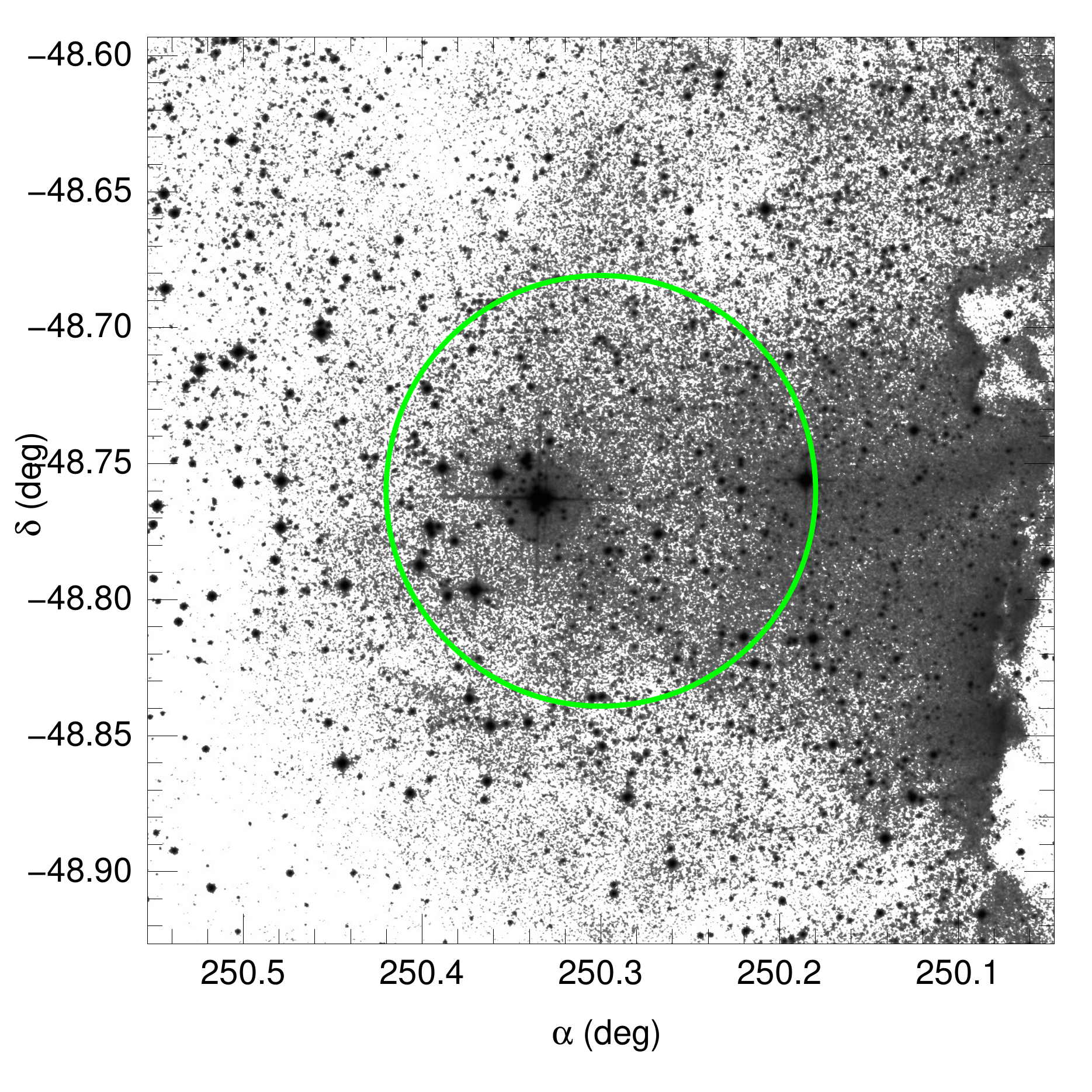} \
            \includegraphics*[width=0.34\linewidth, bb=0 0 538 522]{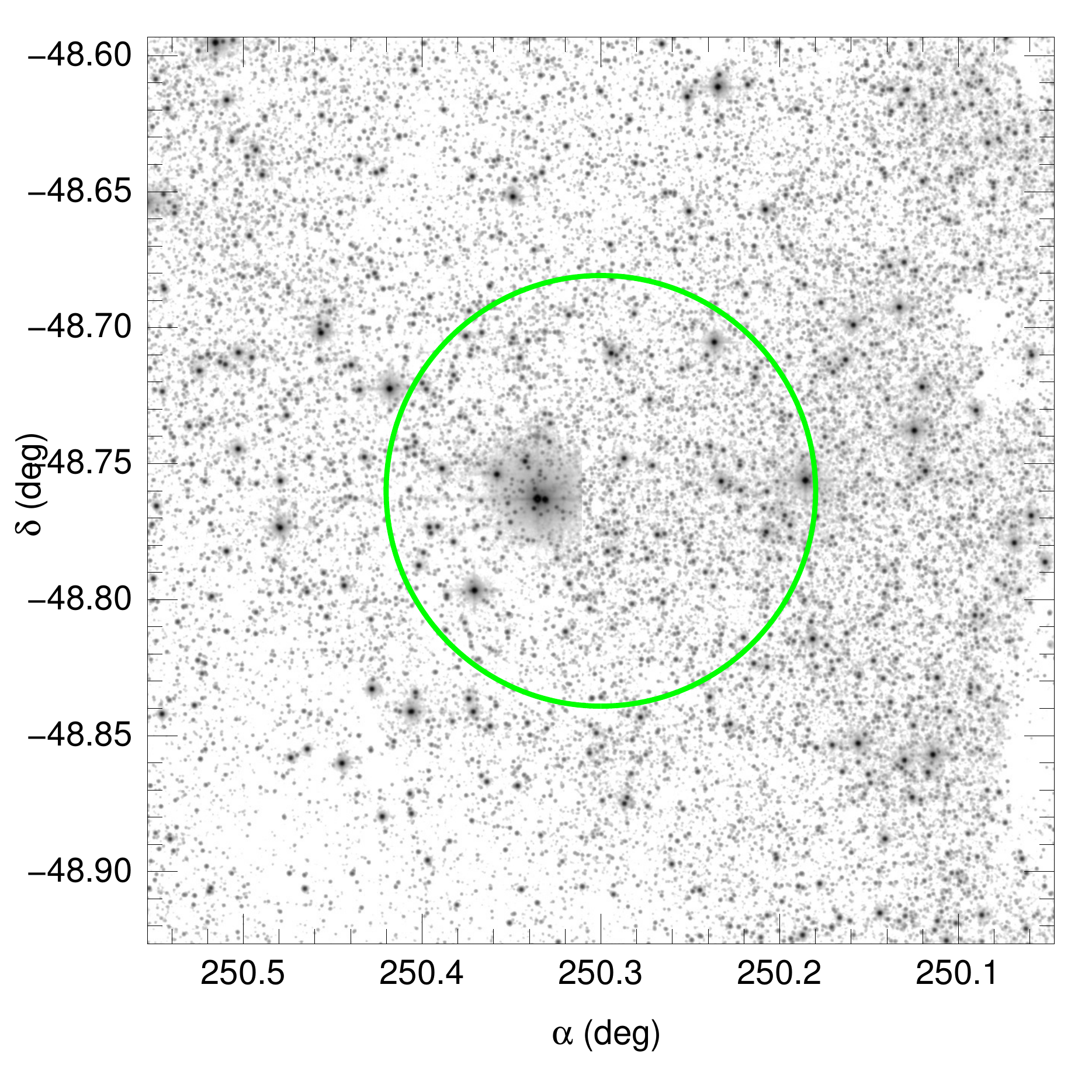}}
\centerline{\includegraphics*[width=0.34\linewidth, bb=0 0 538 522]{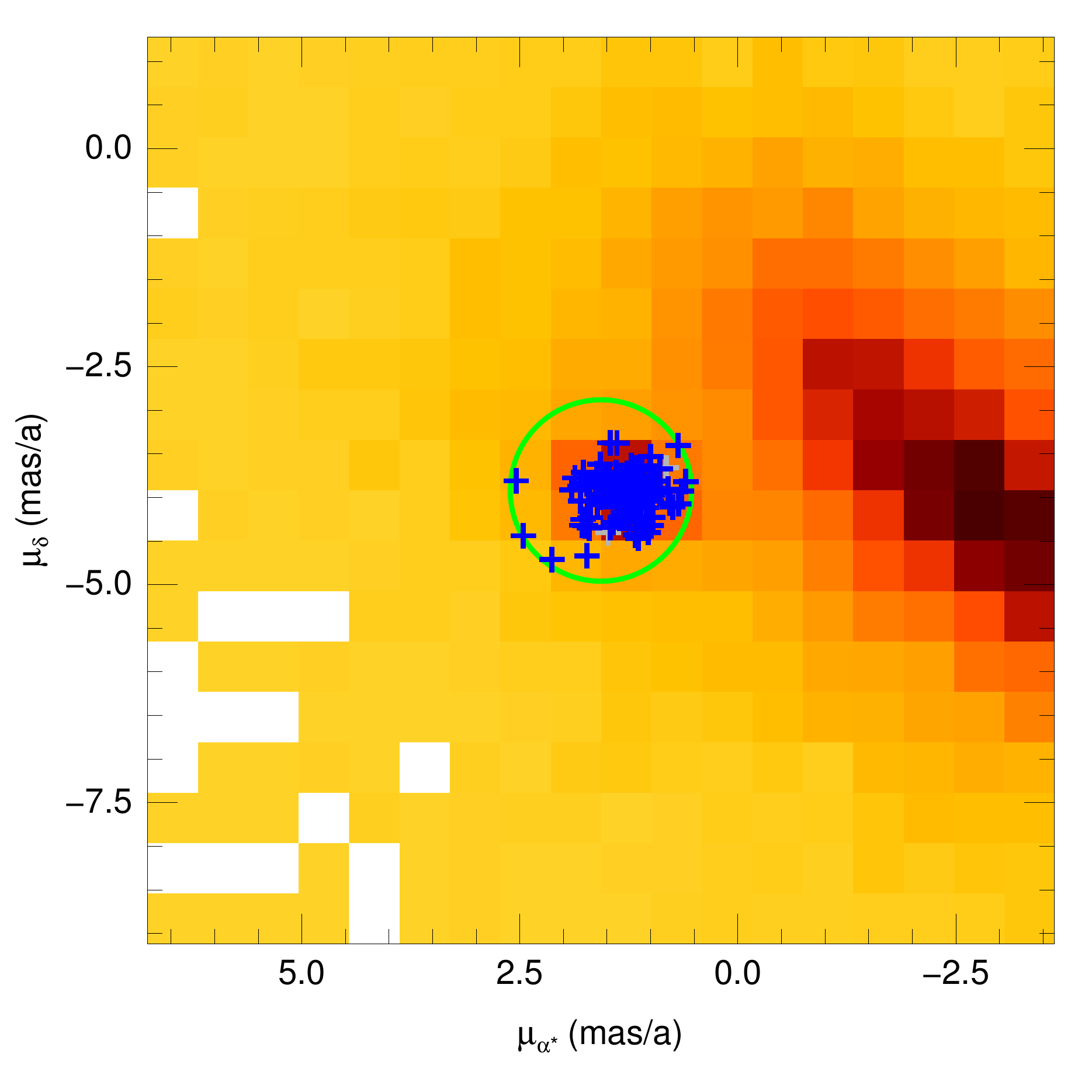} \
            \includegraphics*[width=0.34\linewidth, bb=0 0 538 522]{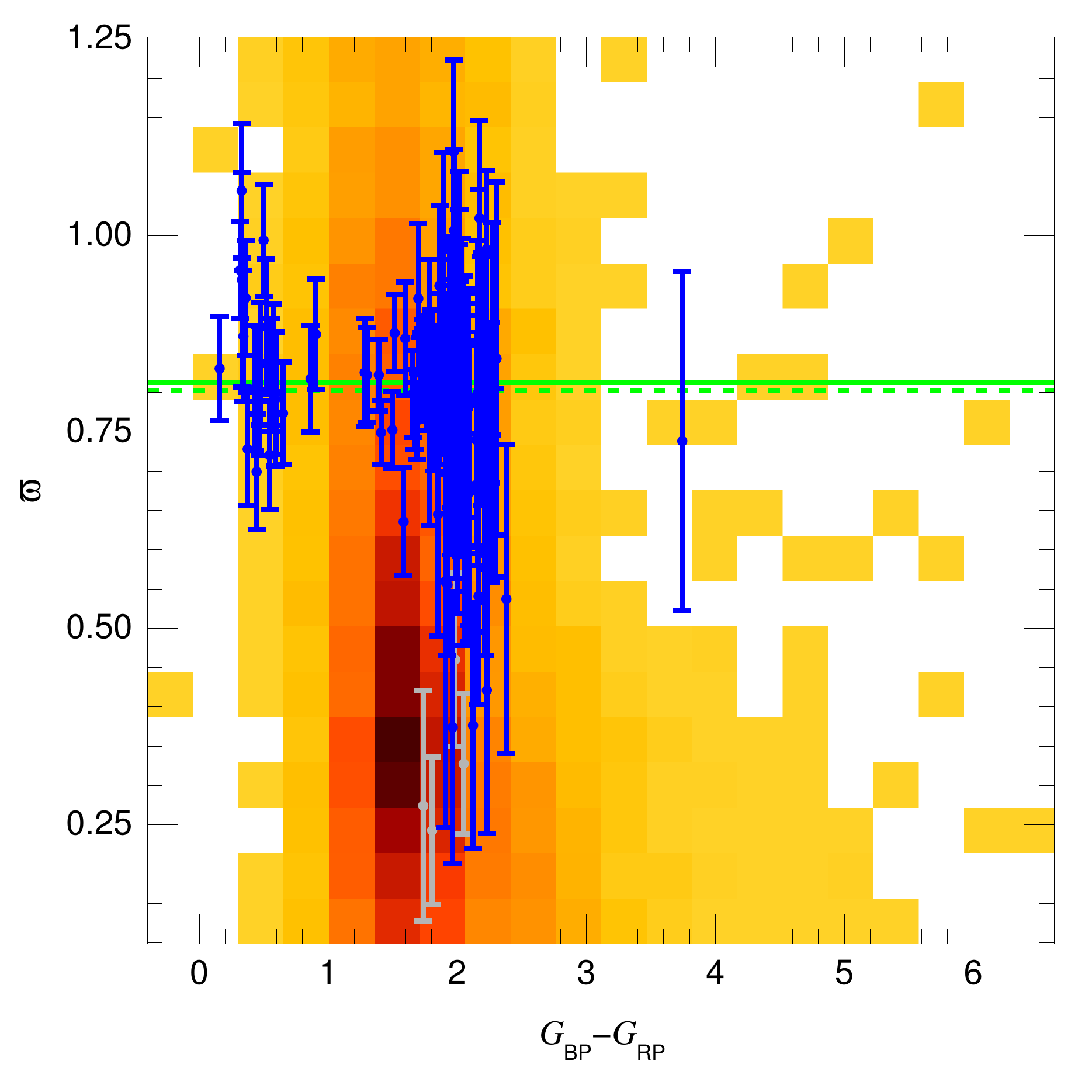} \
            \includegraphics*[width=0.34\linewidth, bb=0 0 538 522]{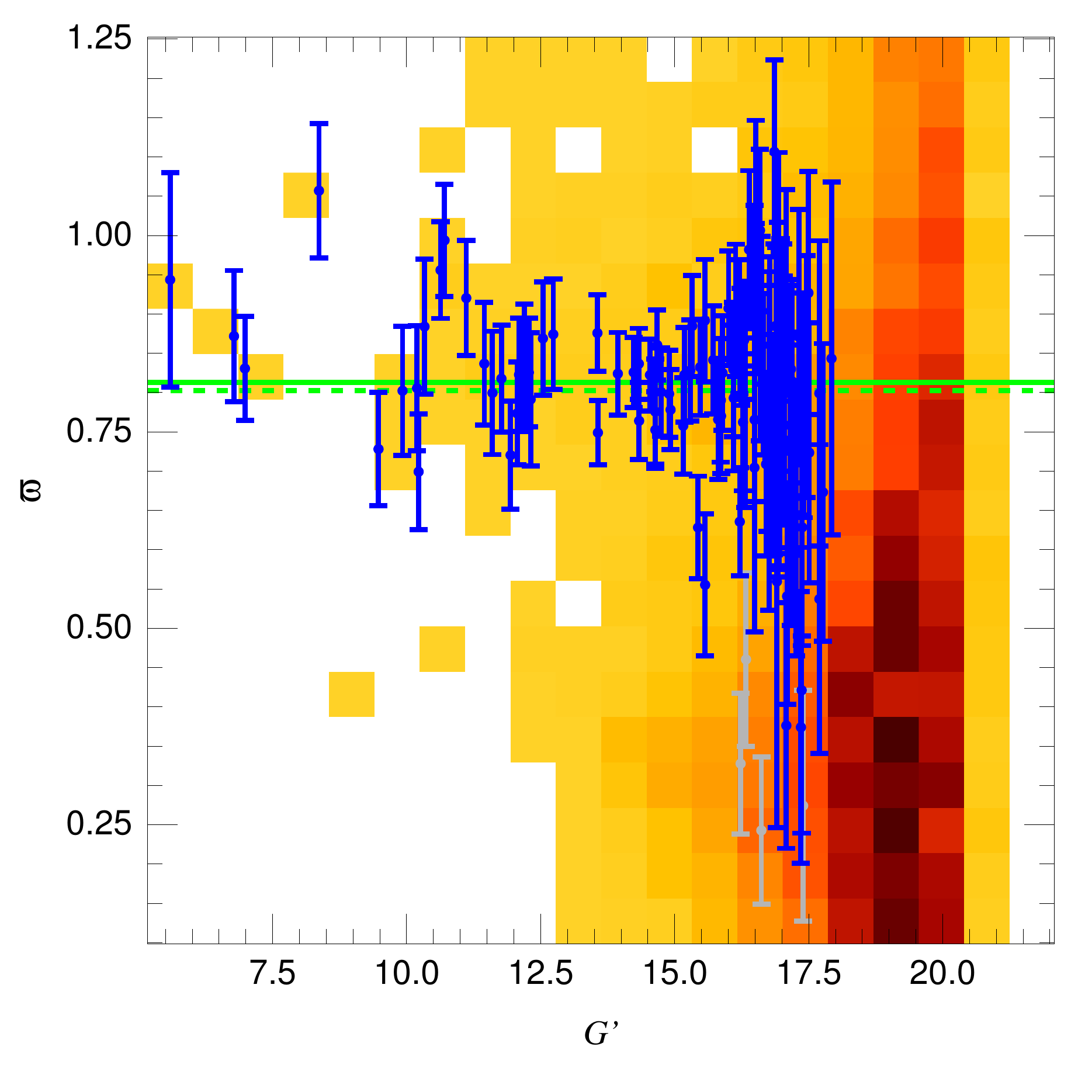}}
\centerline{\includegraphics*[width=0.34\linewidth, bb=0 0 538 522]{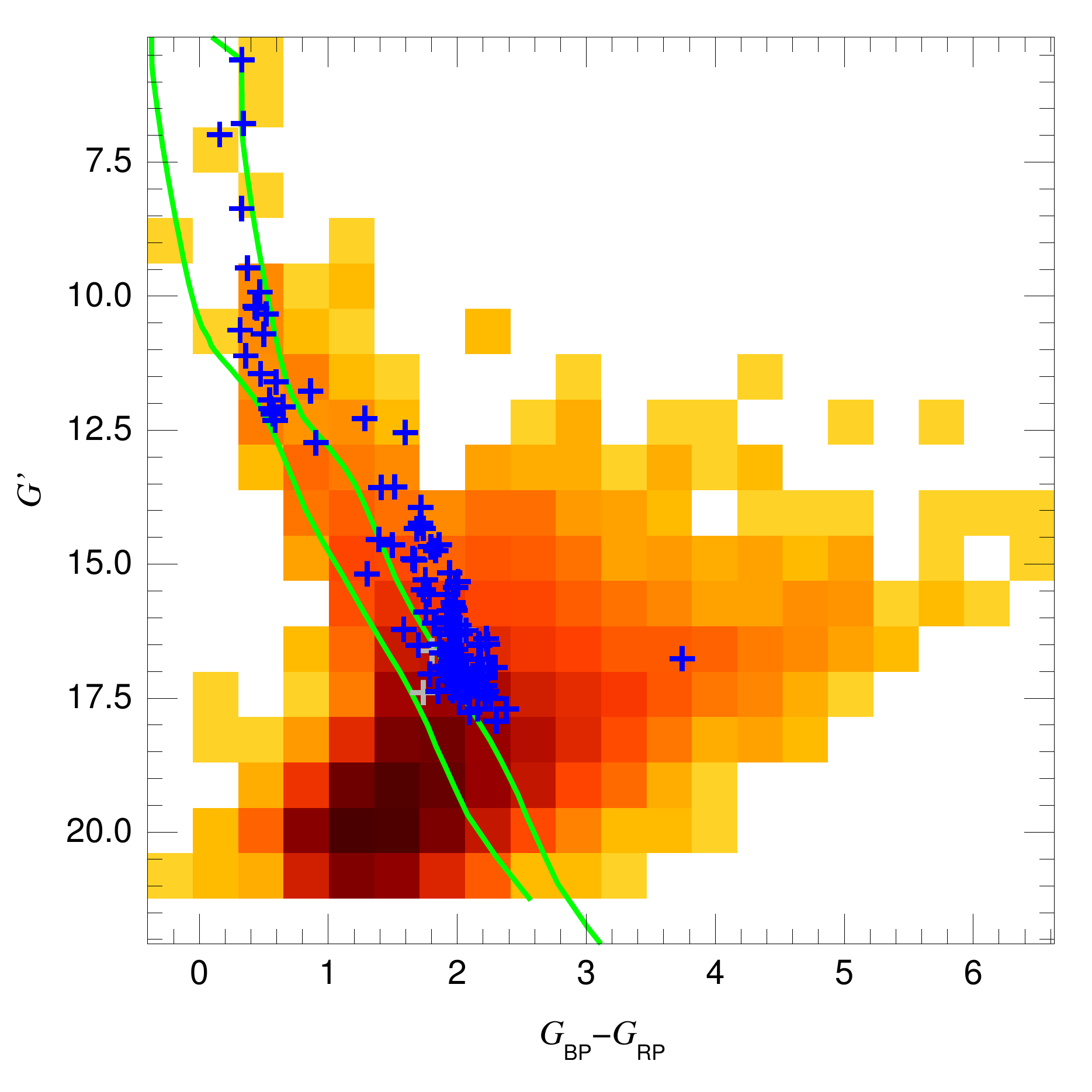} \
            \includegraphics*[width=0.34\linewidth, bb=0 0 538 522]{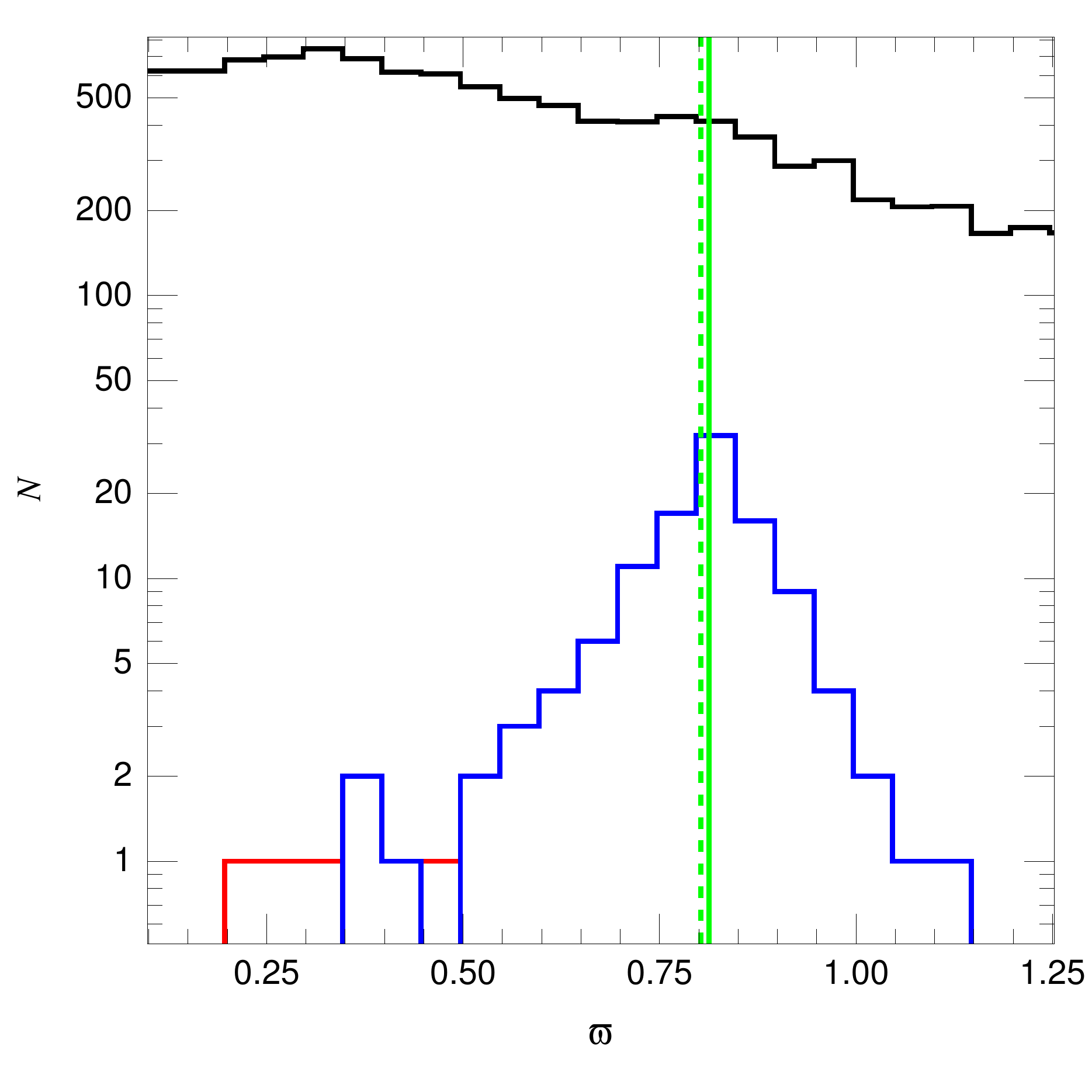} \
            \includegraphics*[width=0.34\linewidth, bb=0 0 538 522]{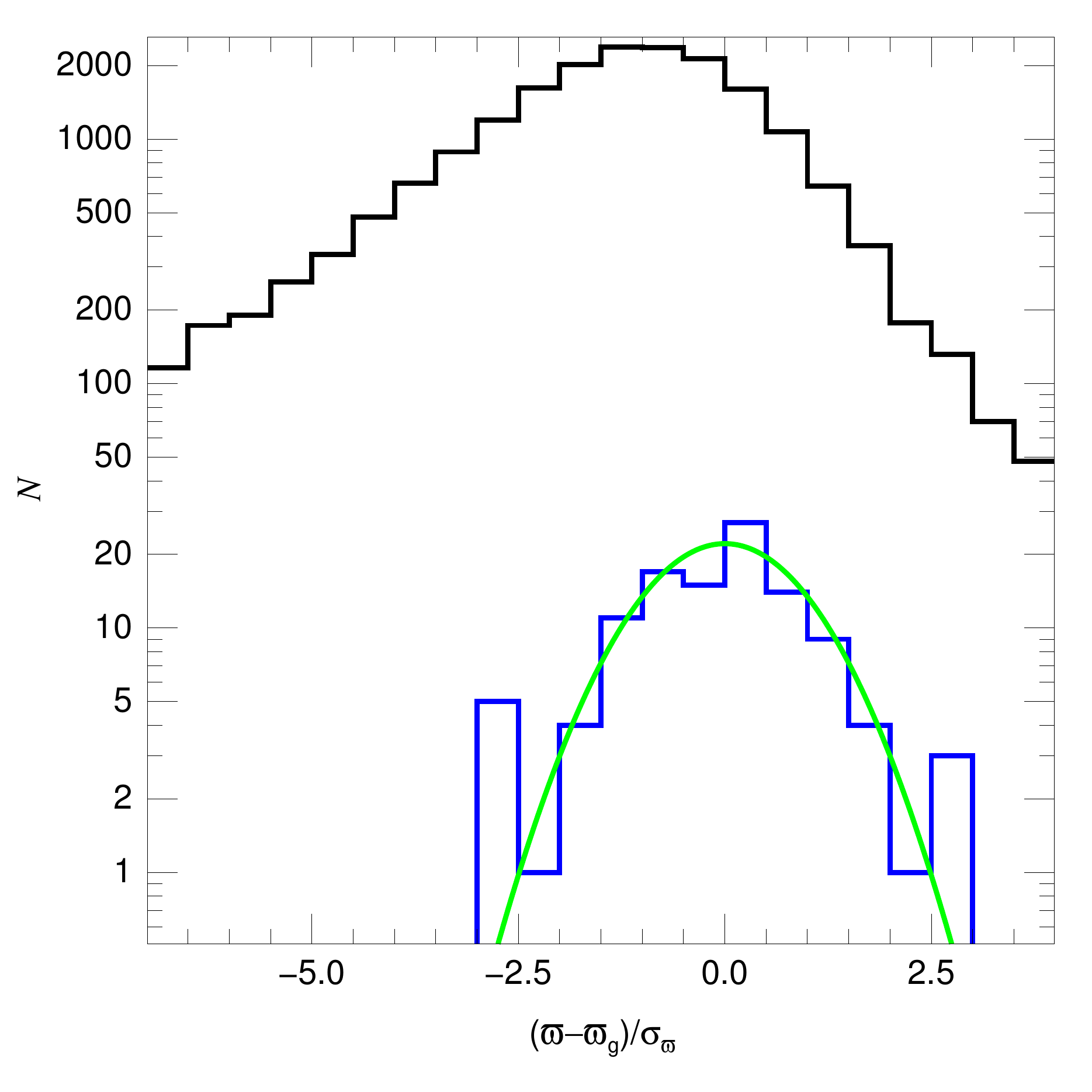}}
\caption{Same as Fig.~\ref{NGC_3603_Gaia} for 
         NGC~6193 (\VO{010}). % REFEREE \VO{010}
         }
\label{NGC_6193_Gaia}
\end{figure*}   

\begin{figure*}
\centerline{\includegraphics*[width=0.34\linewidth, bb=0 0 538 522]{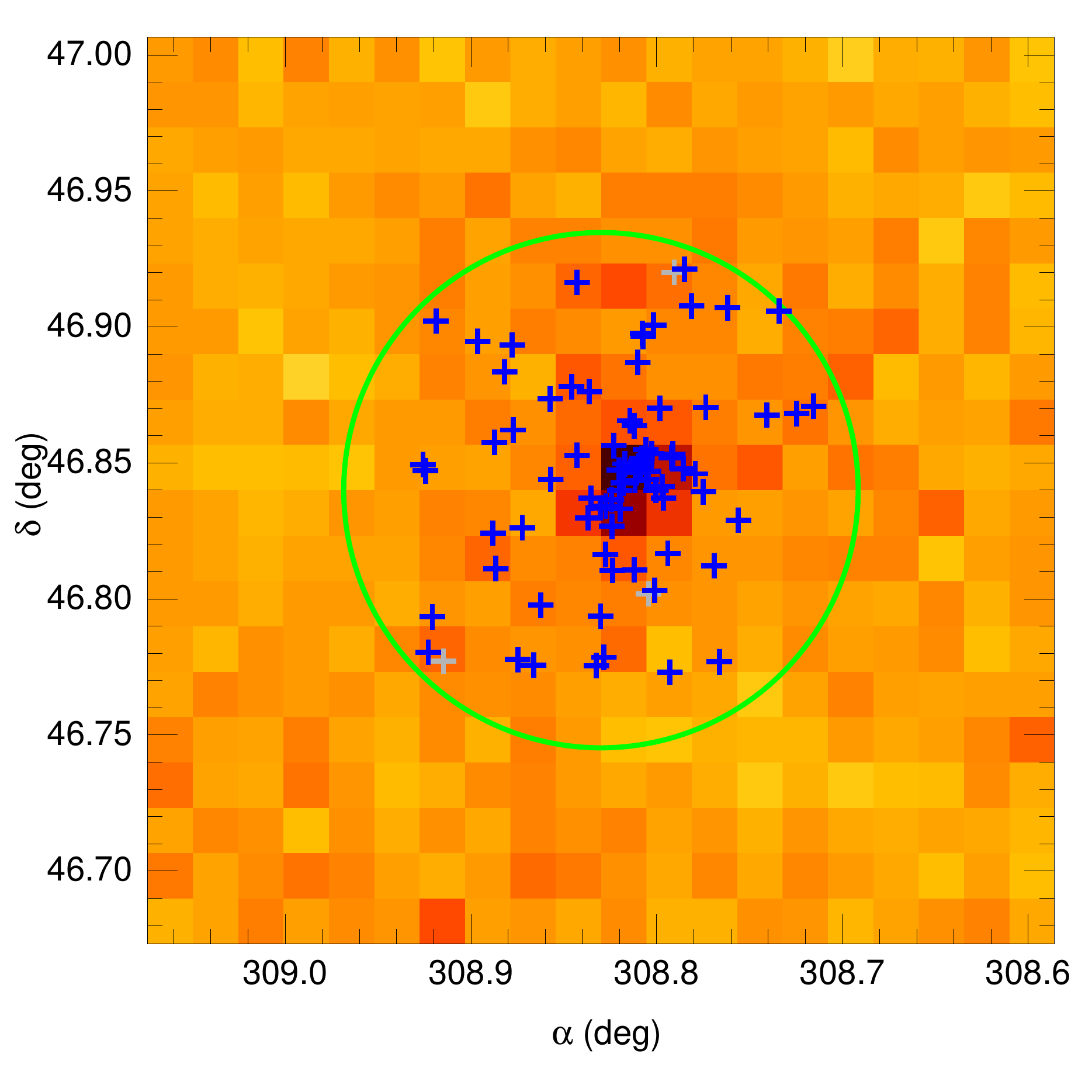} \
            \includegraphics*[width=0.34\linewidth, bb=0 0 538 522]{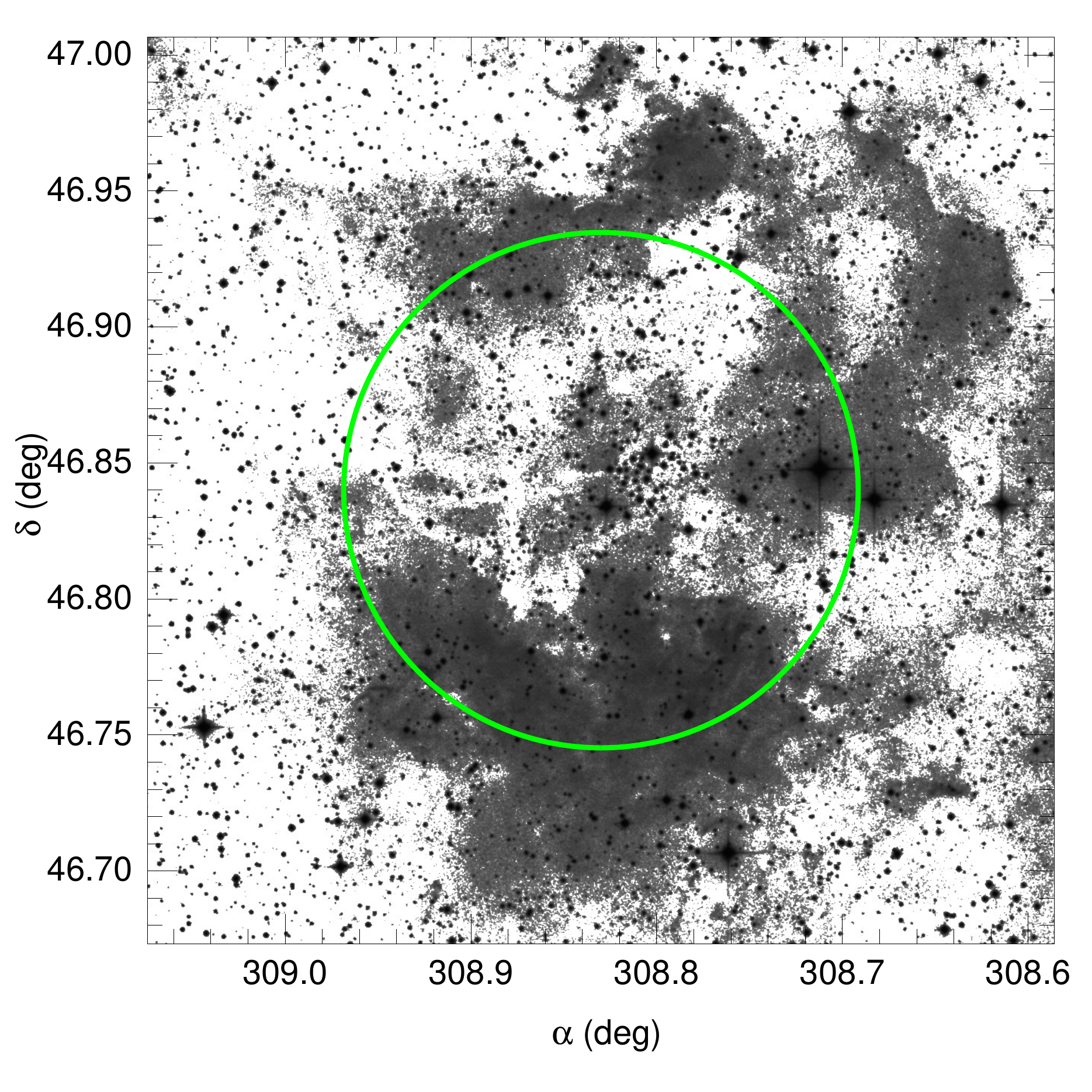} \
            \includegraphics*[width=0.34\linewidth, bb=0 0 538 522]{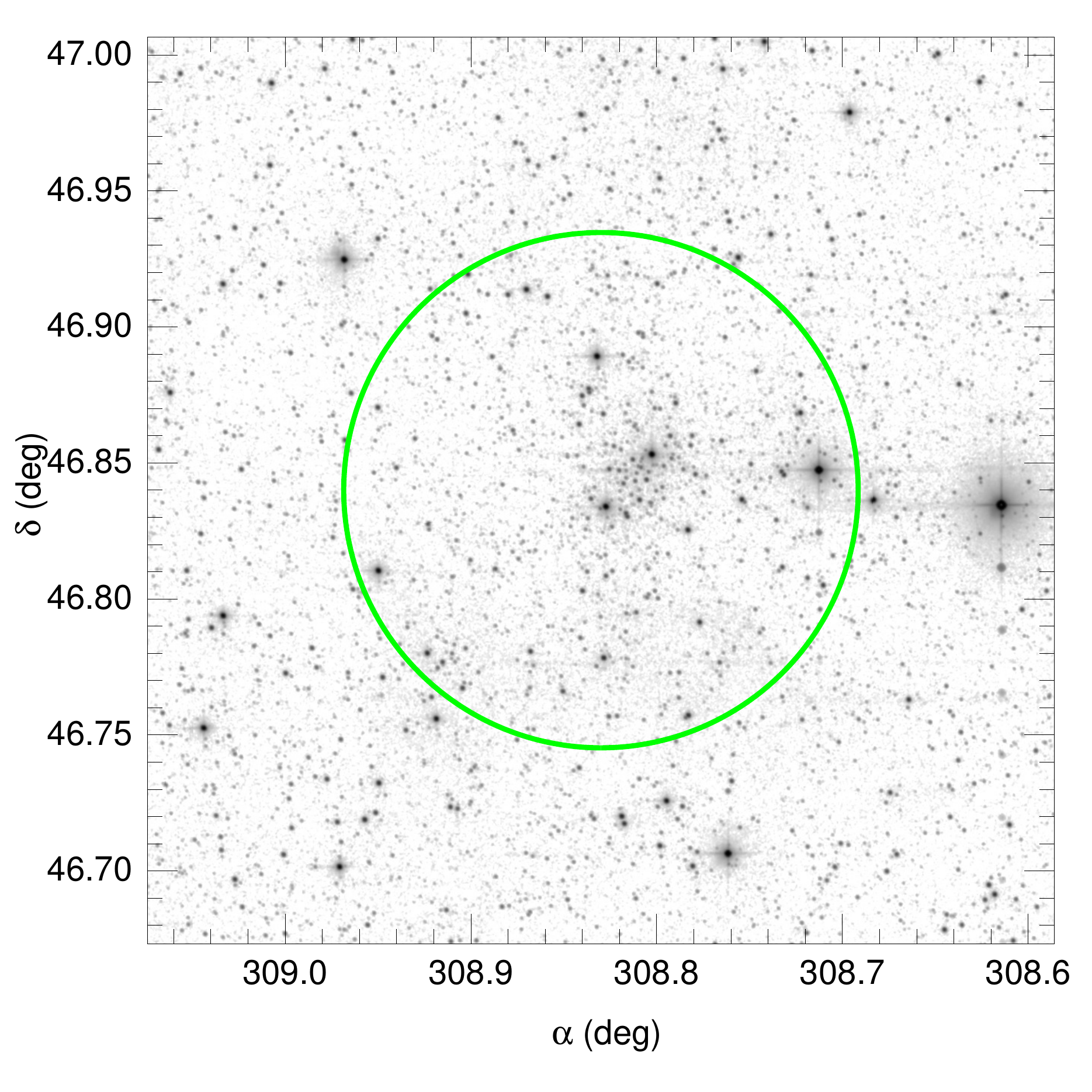}}
\centerline{\includegraphics*[width=0.34\linewidth, bb=0 0 538 522]{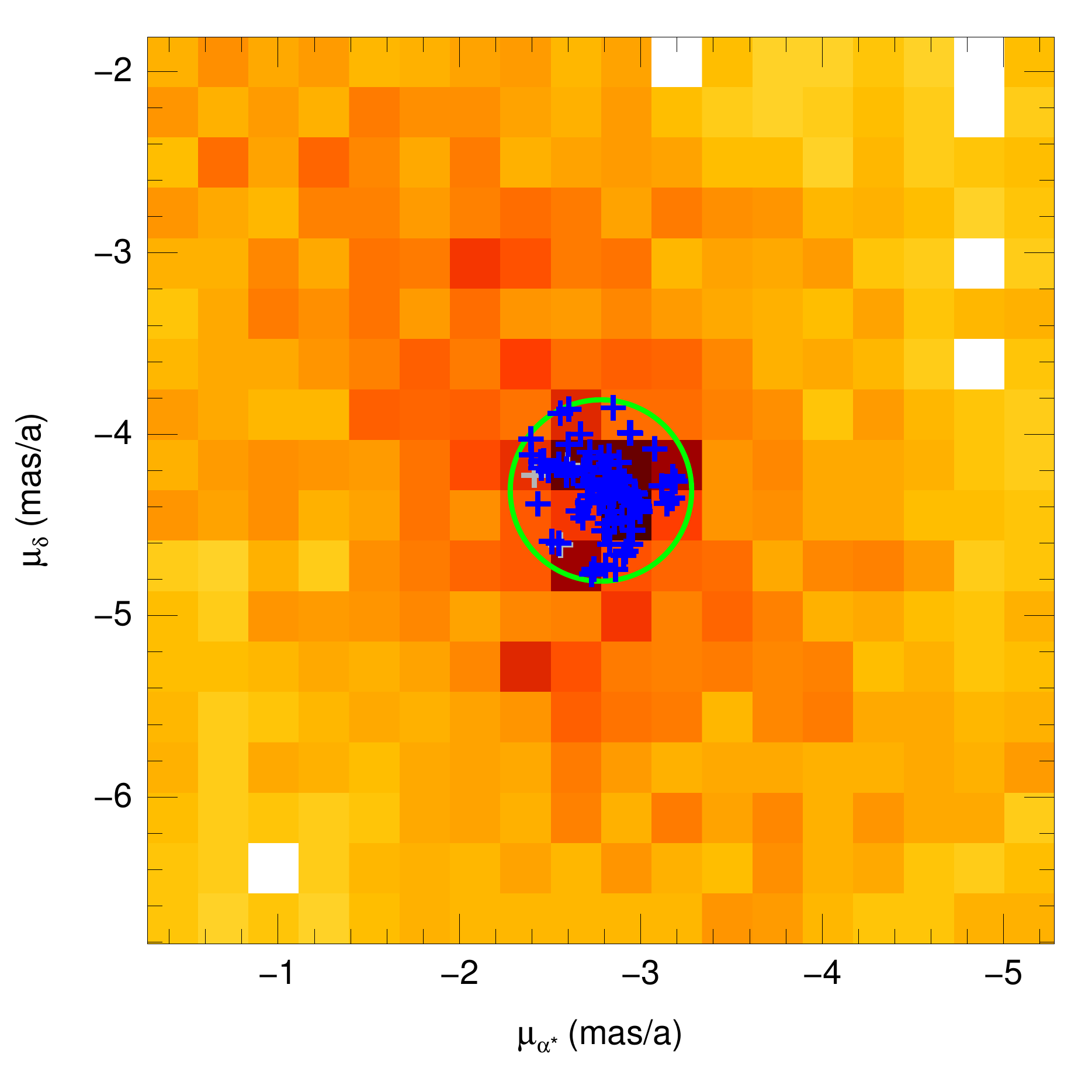} \
            \includegraphics*[width=0.34\linewidth, bb=0 0 538 522]{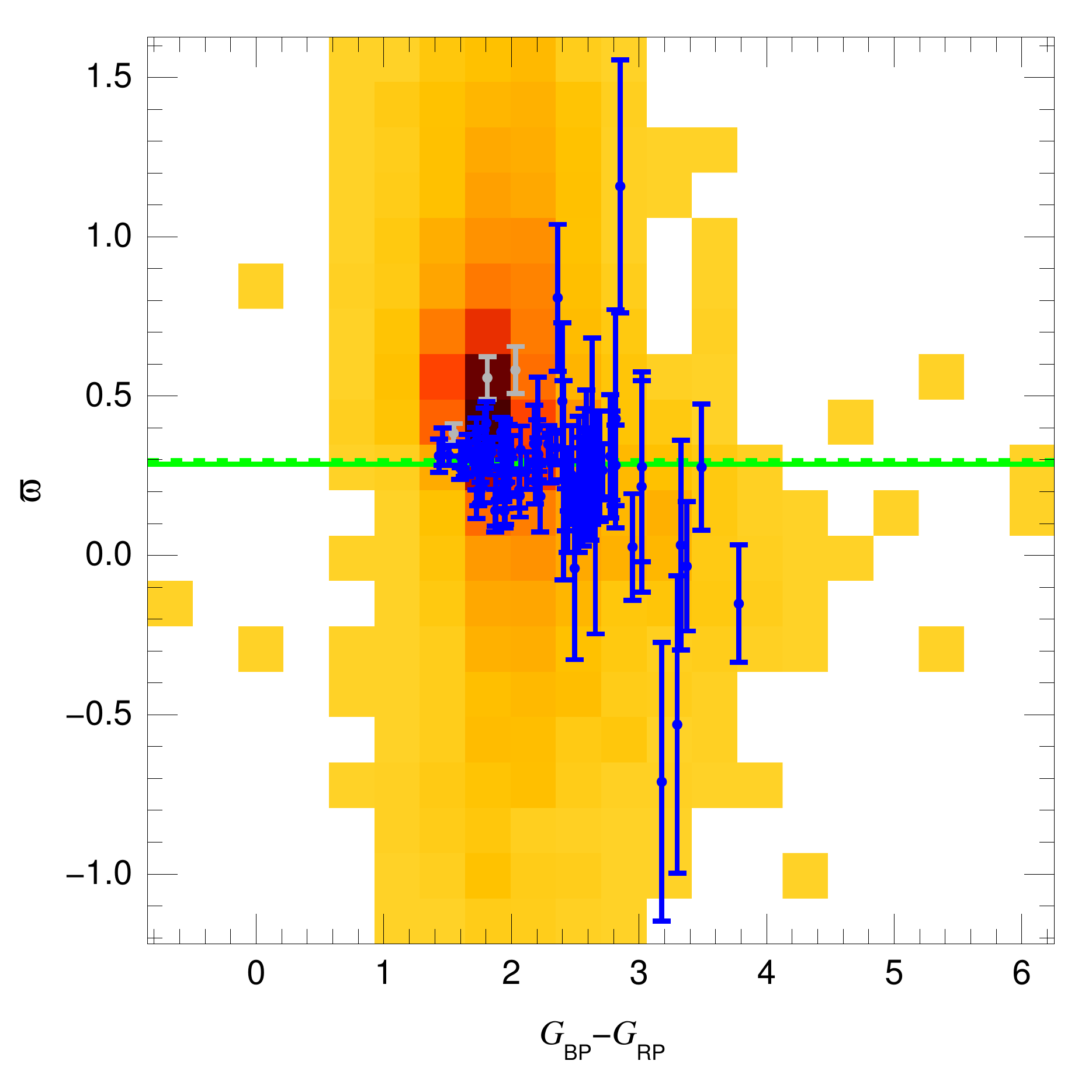} \
            \includegraphics*[width=0.34\linewidth, bb=0 0 538 522]{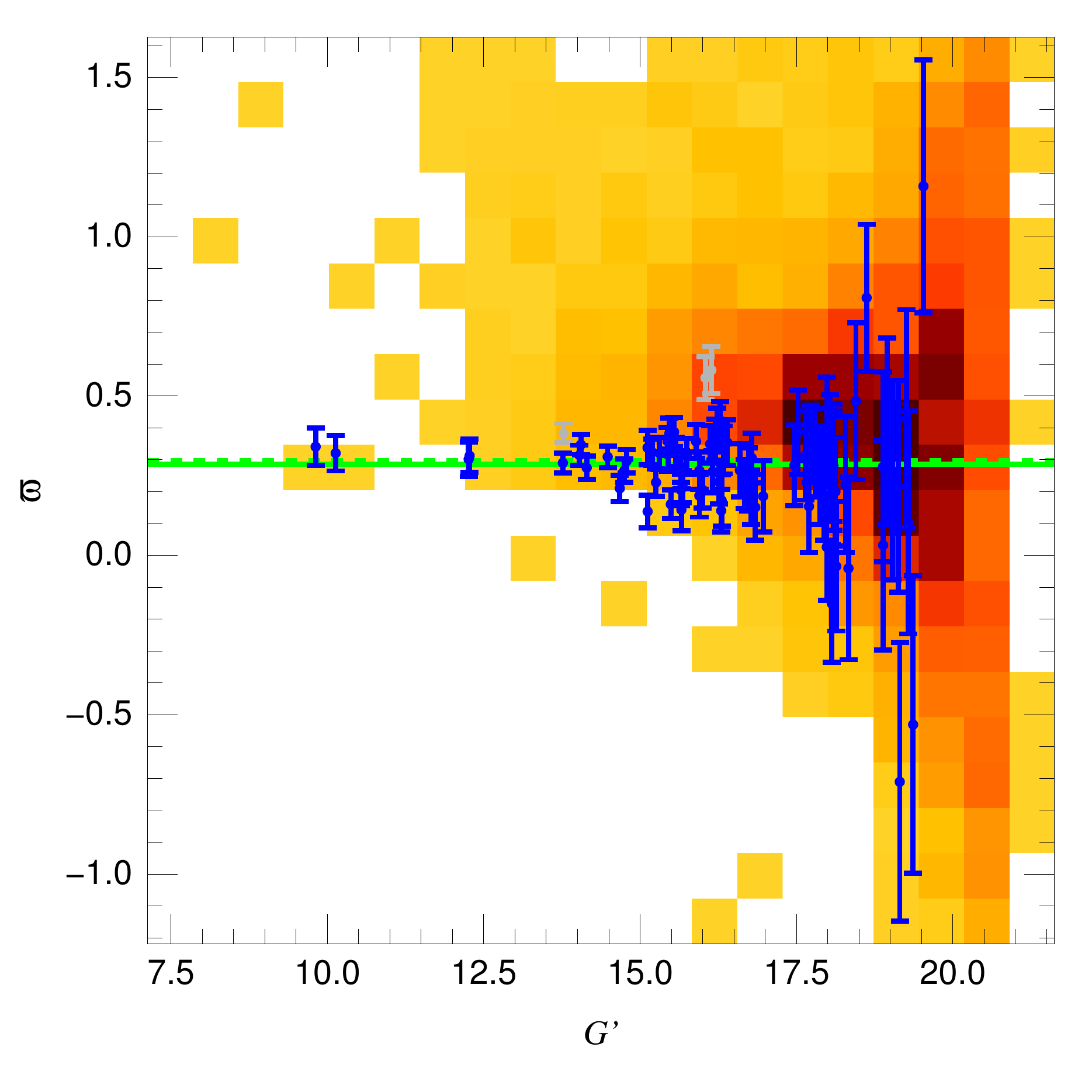}}
\centerline{\includegraphics*[width=0.34\linewidth, bb=0 0 538 522]{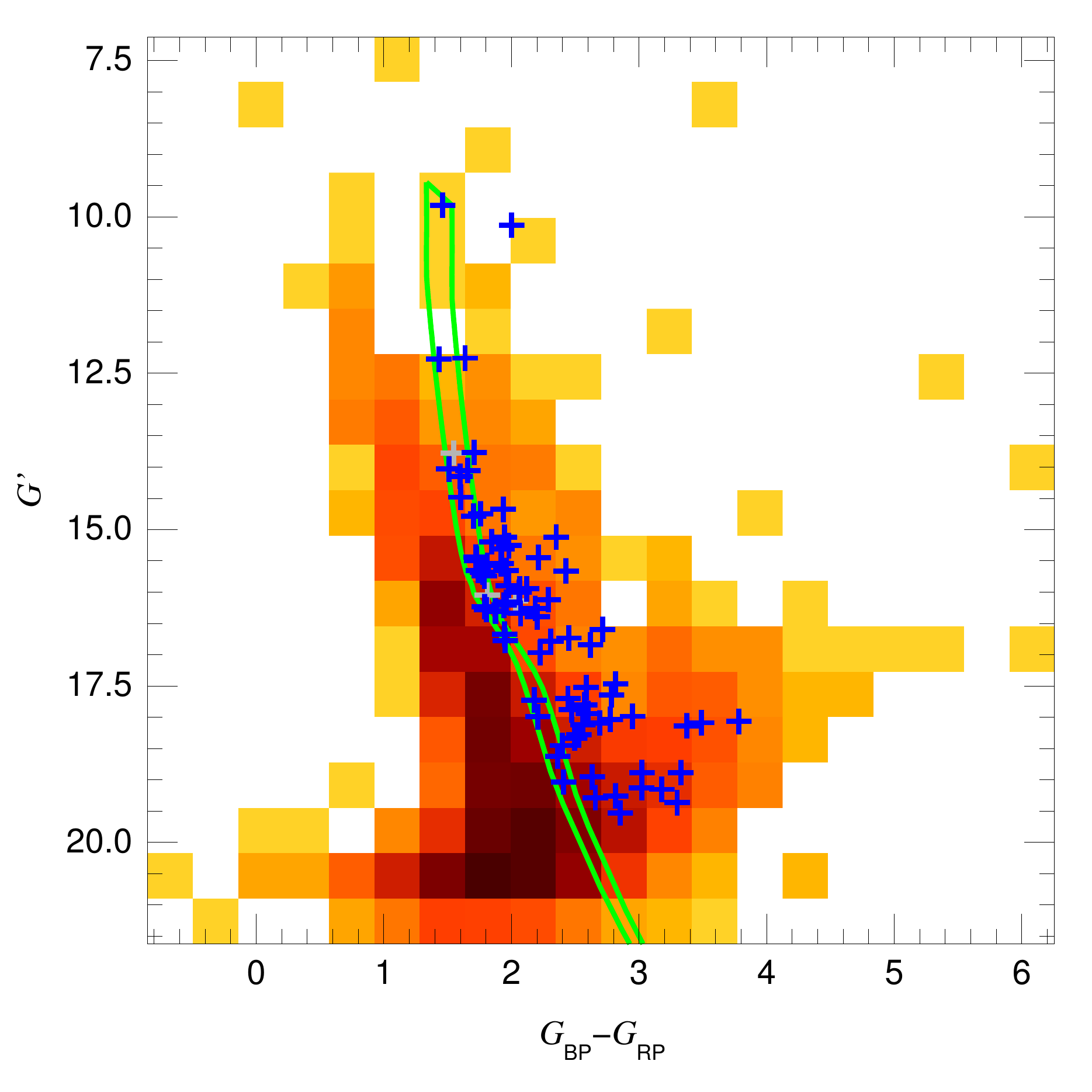} \
            \includegraphics*[width=0.34\linewidth, bb=0 0 538 522]{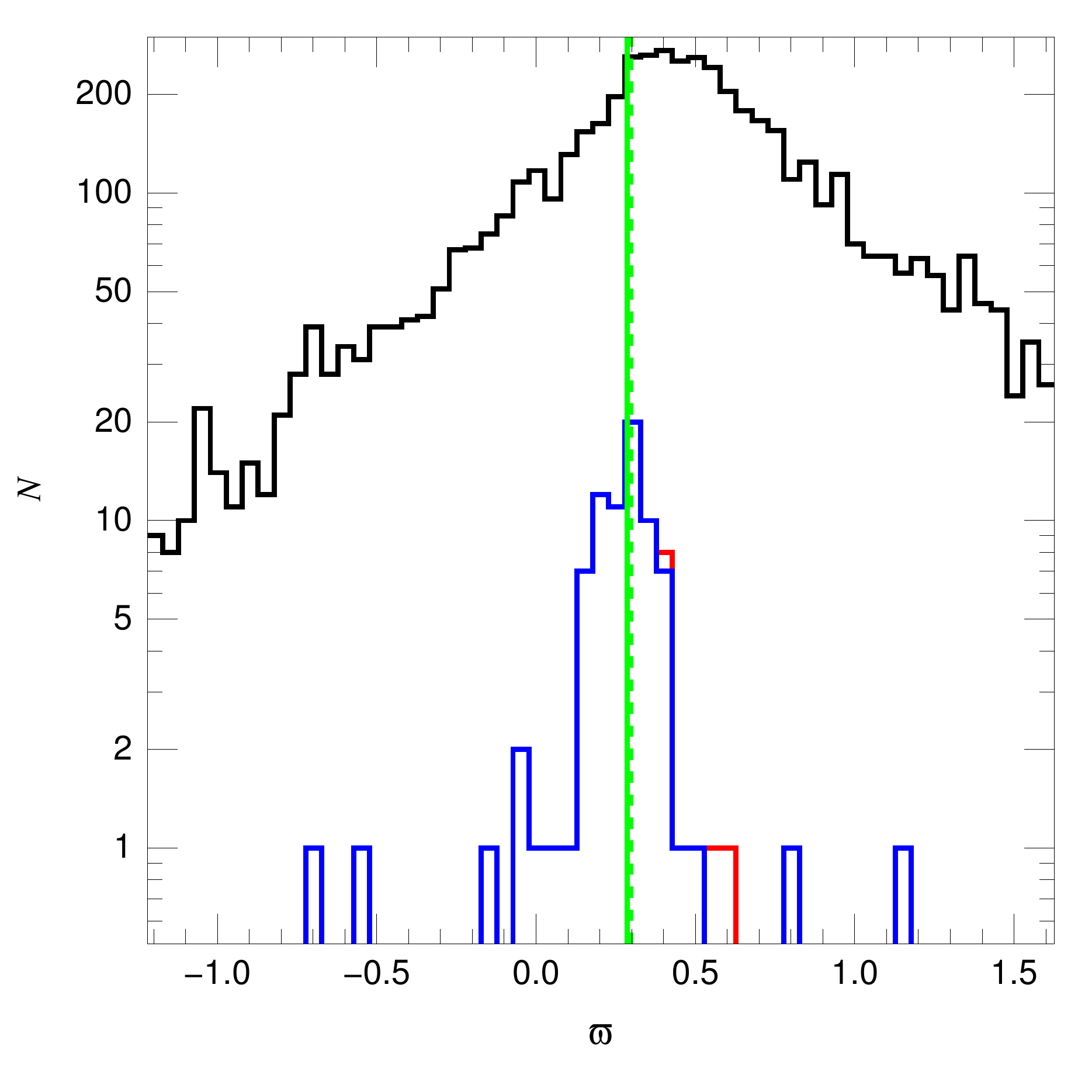} \
            \includegraphics*[width=0.34\linewidth, bb=0 0 538 522]{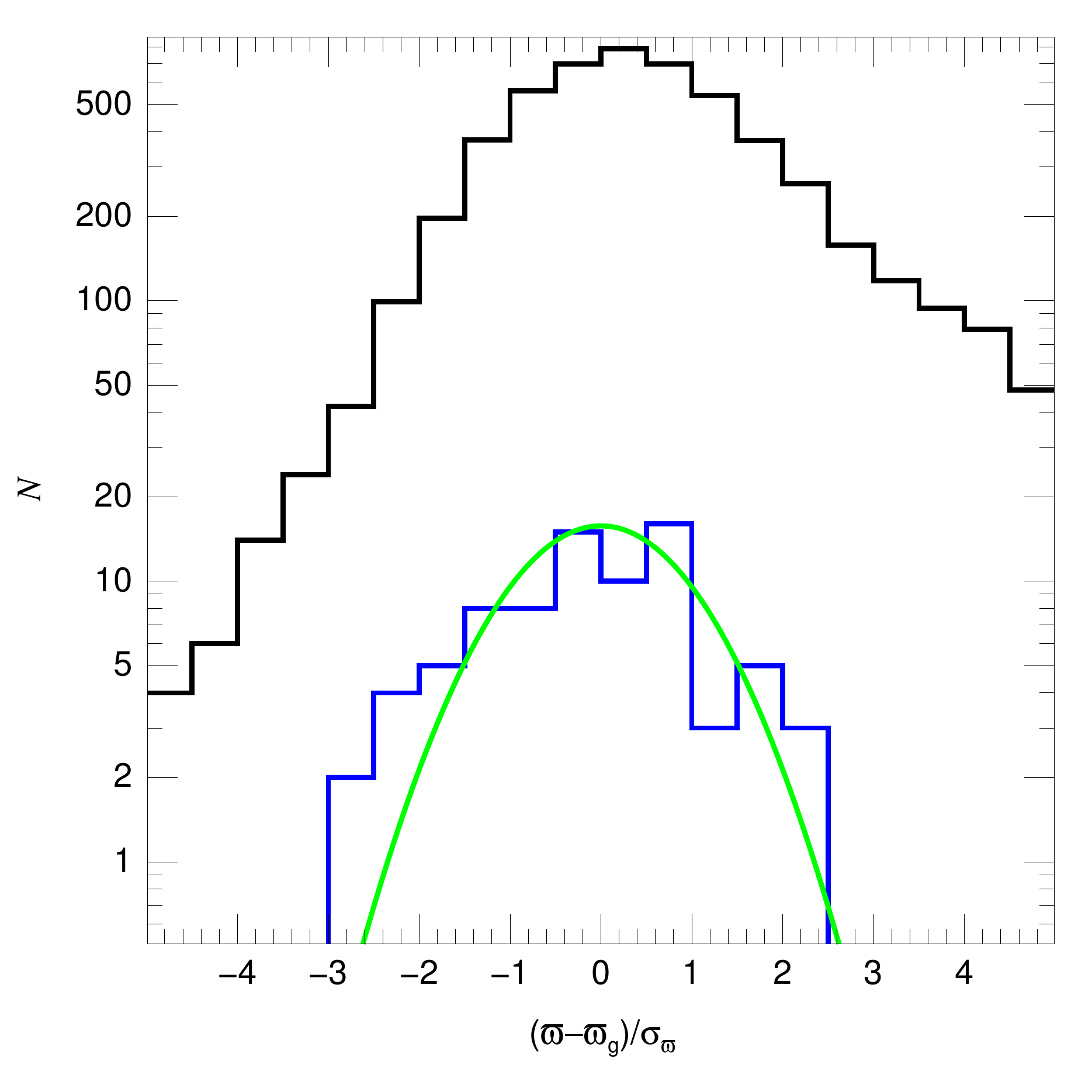}}
\caption{Same as Fig.~\ref{NGC_3603_Gaia} for 
         Berkeley~90 (\VO{011}). % REFEREE \VO{011}
         }
\label{Berkeley_90_Gaia}
\end{figure*}   

\begin{figure*}
\centerline{\includegraphics*[width=0.34\linewidth, bb=0 0 538 522]{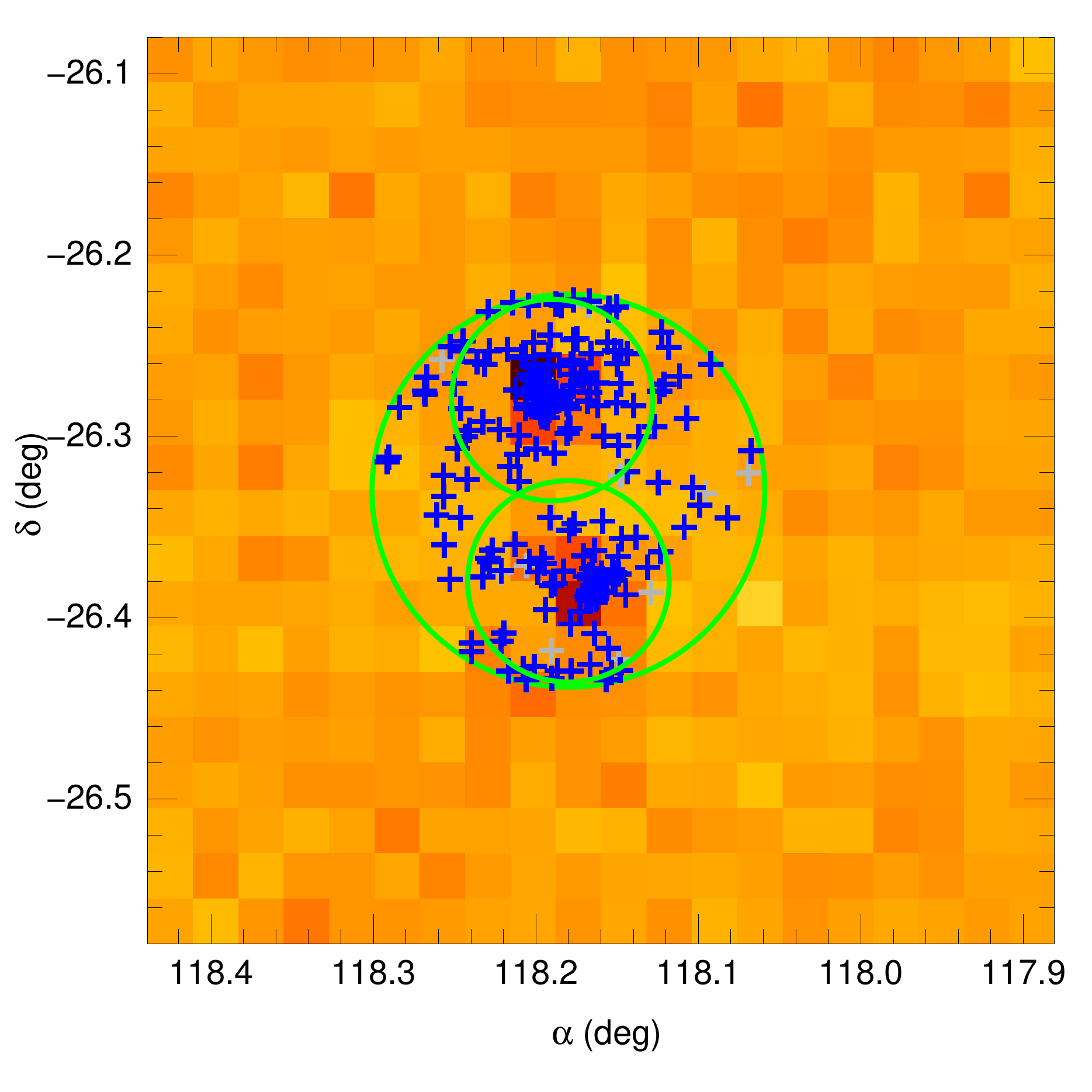} \
            \includegraphics*[width=0.34\linewidth, bb=0 0 538 522]{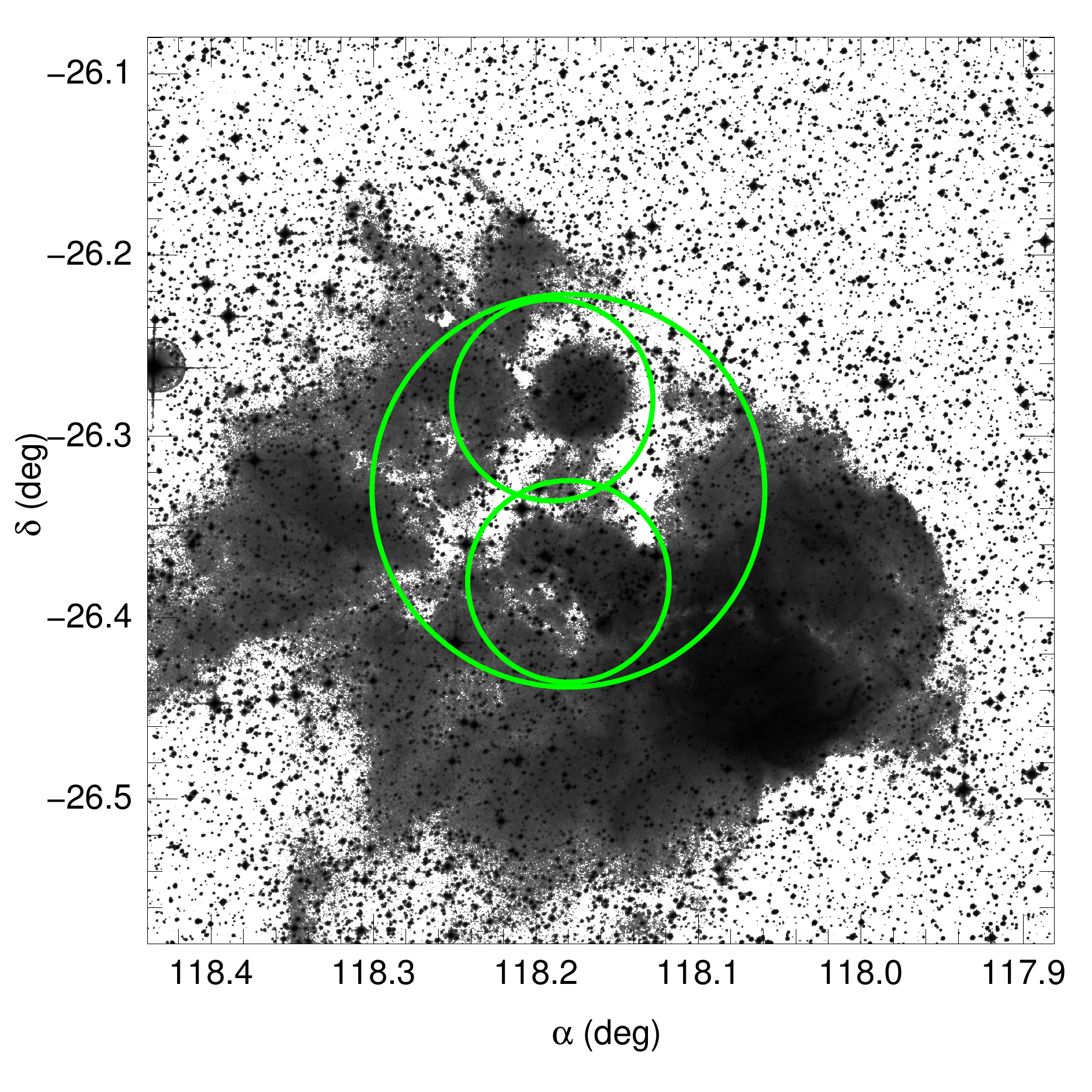} \
            \includegraphics*[width=0.34\linewidth, bb=0 0 538 522]{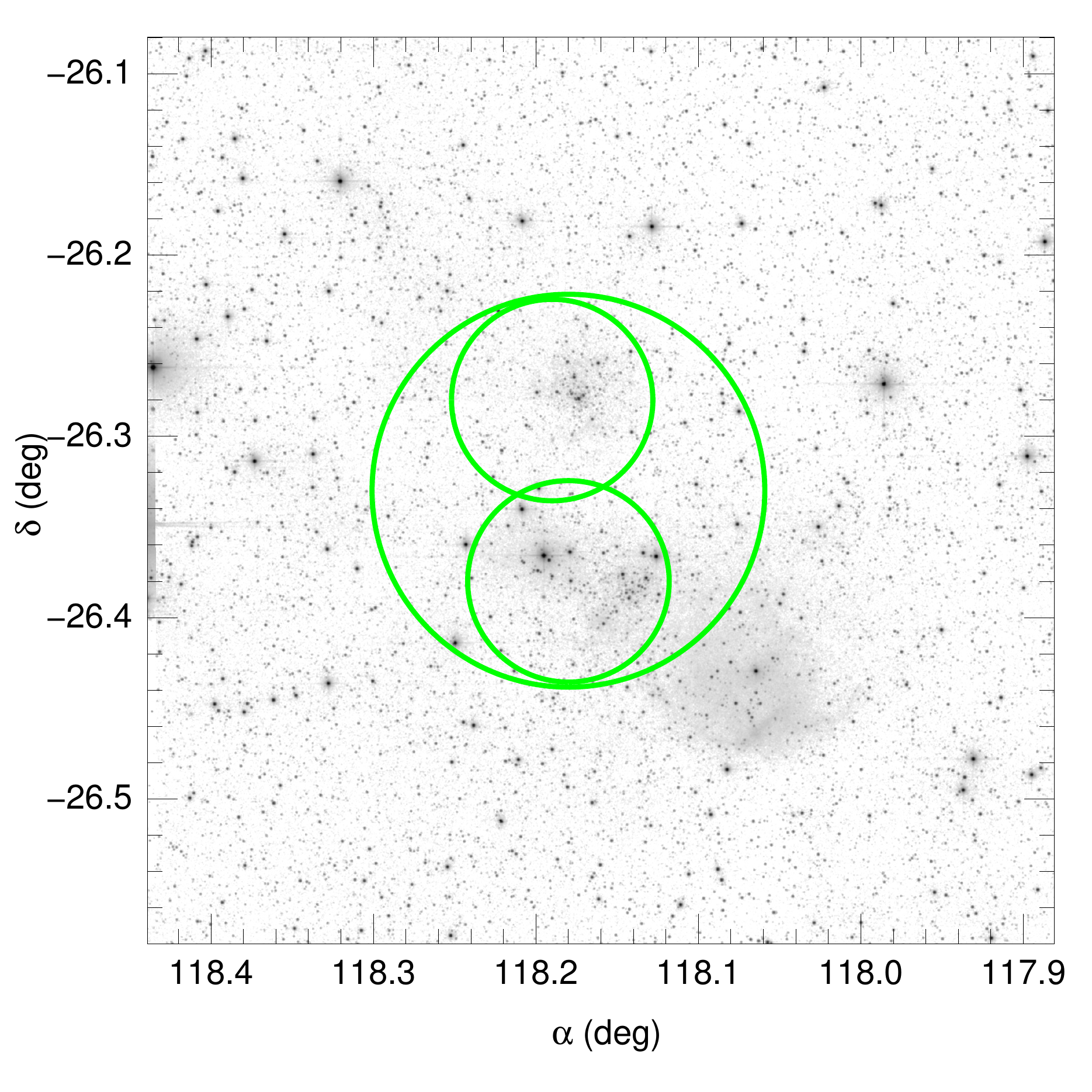}}
\centerline{\includegraphics*[width=0.34\linewidth, bb=0 0 538 522]{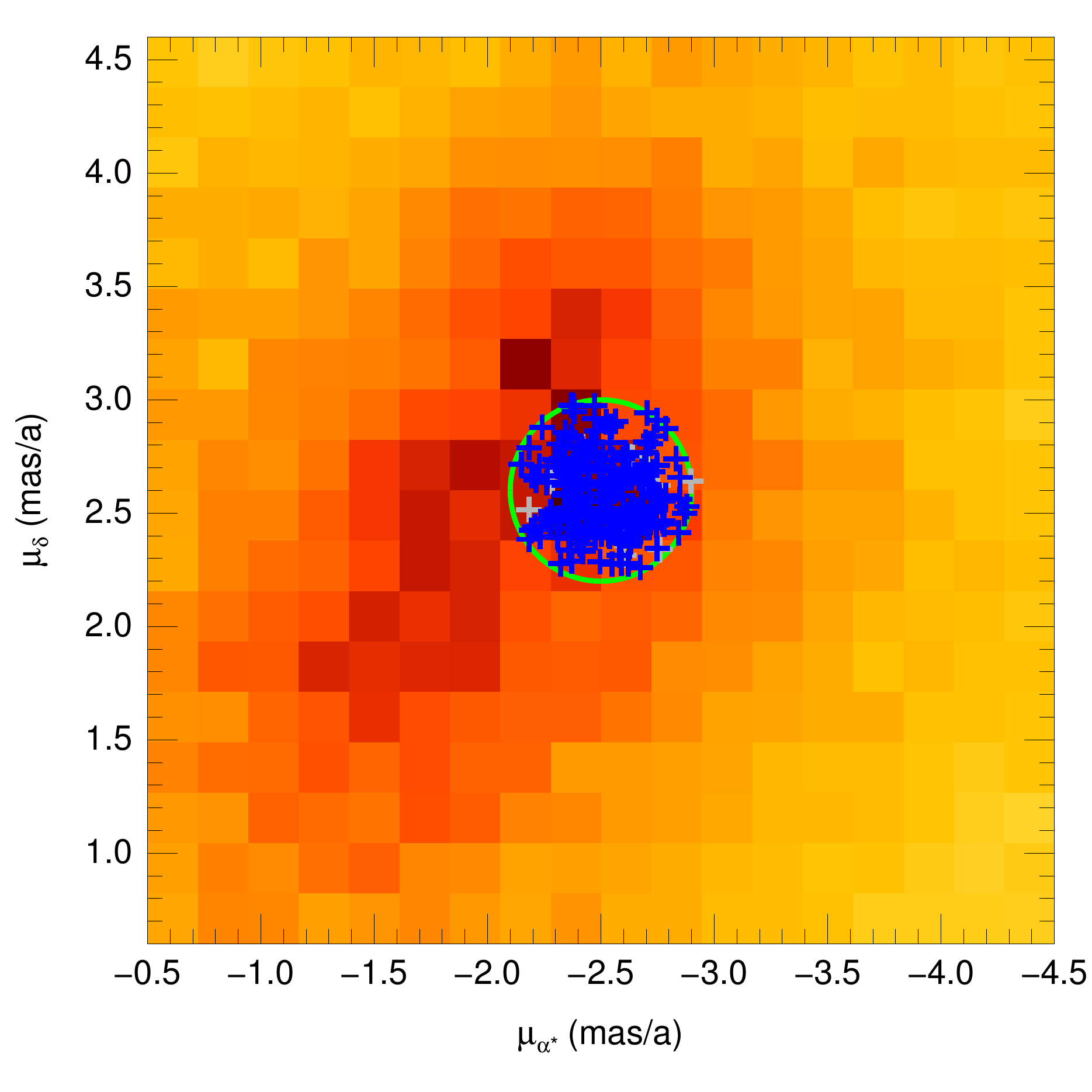} \
            \includegraphics*[width=0.34\linewidth, bb=0 0 538 522]{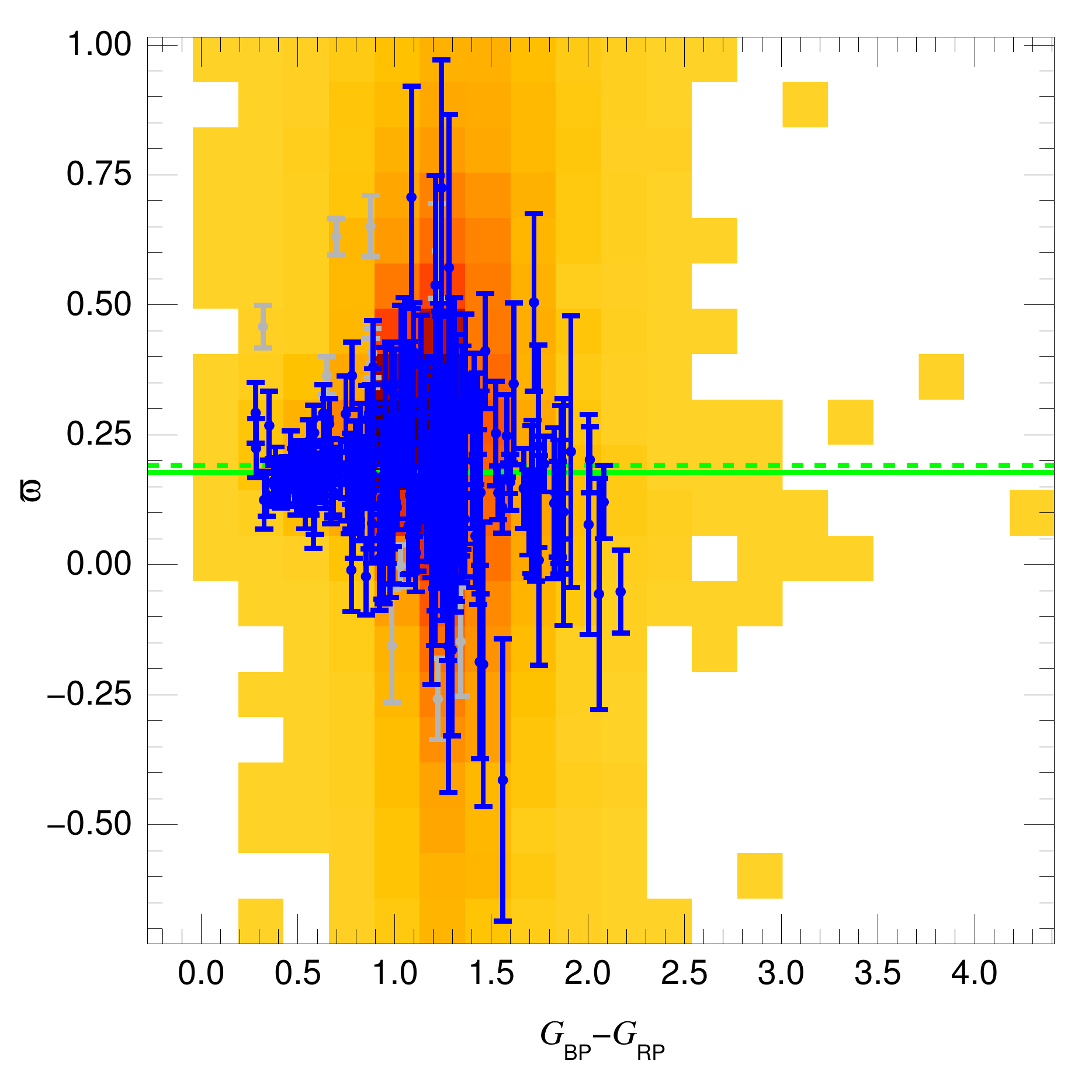} \
            \includegraphics*[width=0.34\linewidth, bb=0 0 538 522]{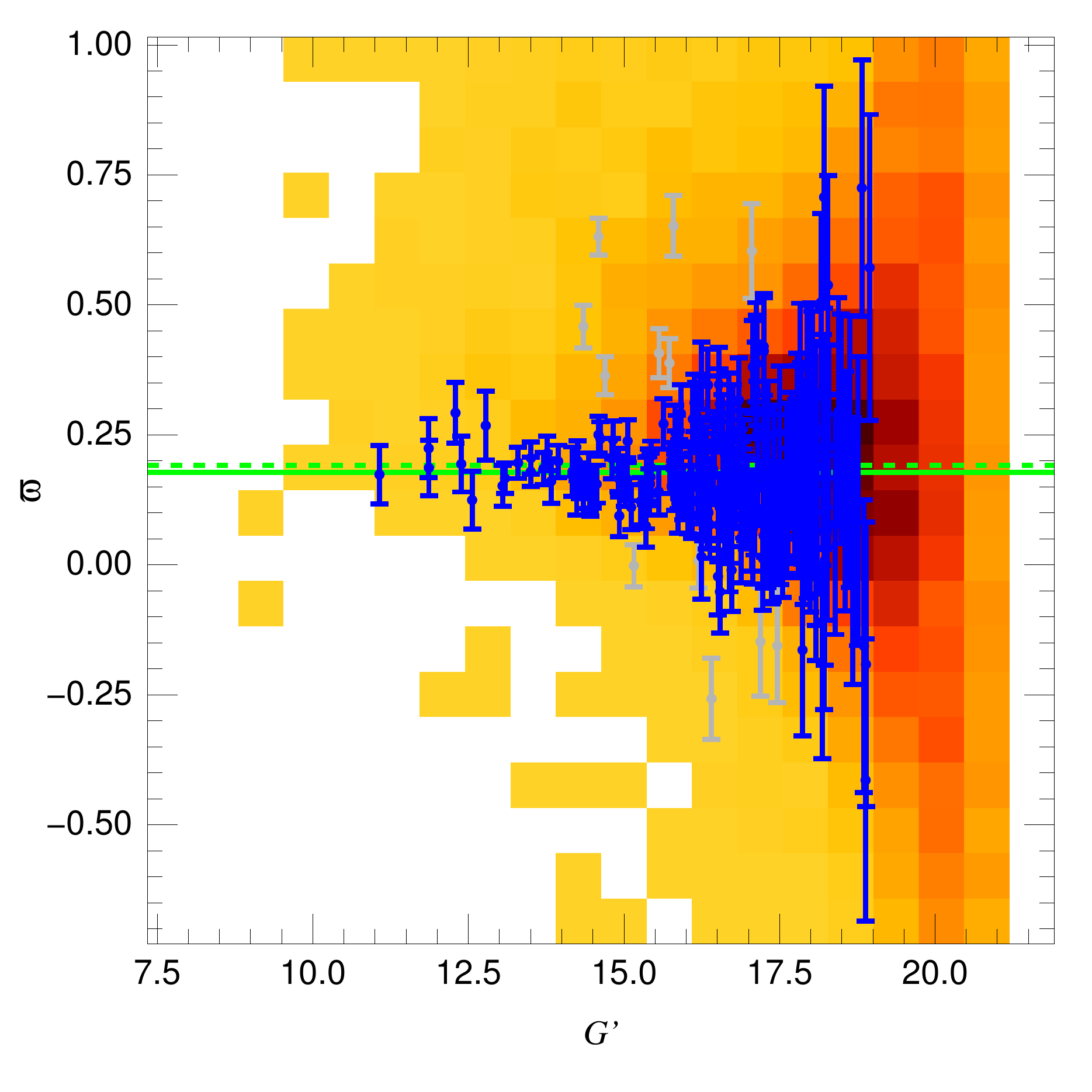}}
\centerline{\includegraphics*[width=0.34\linewidth, bb=0 0 538 522]{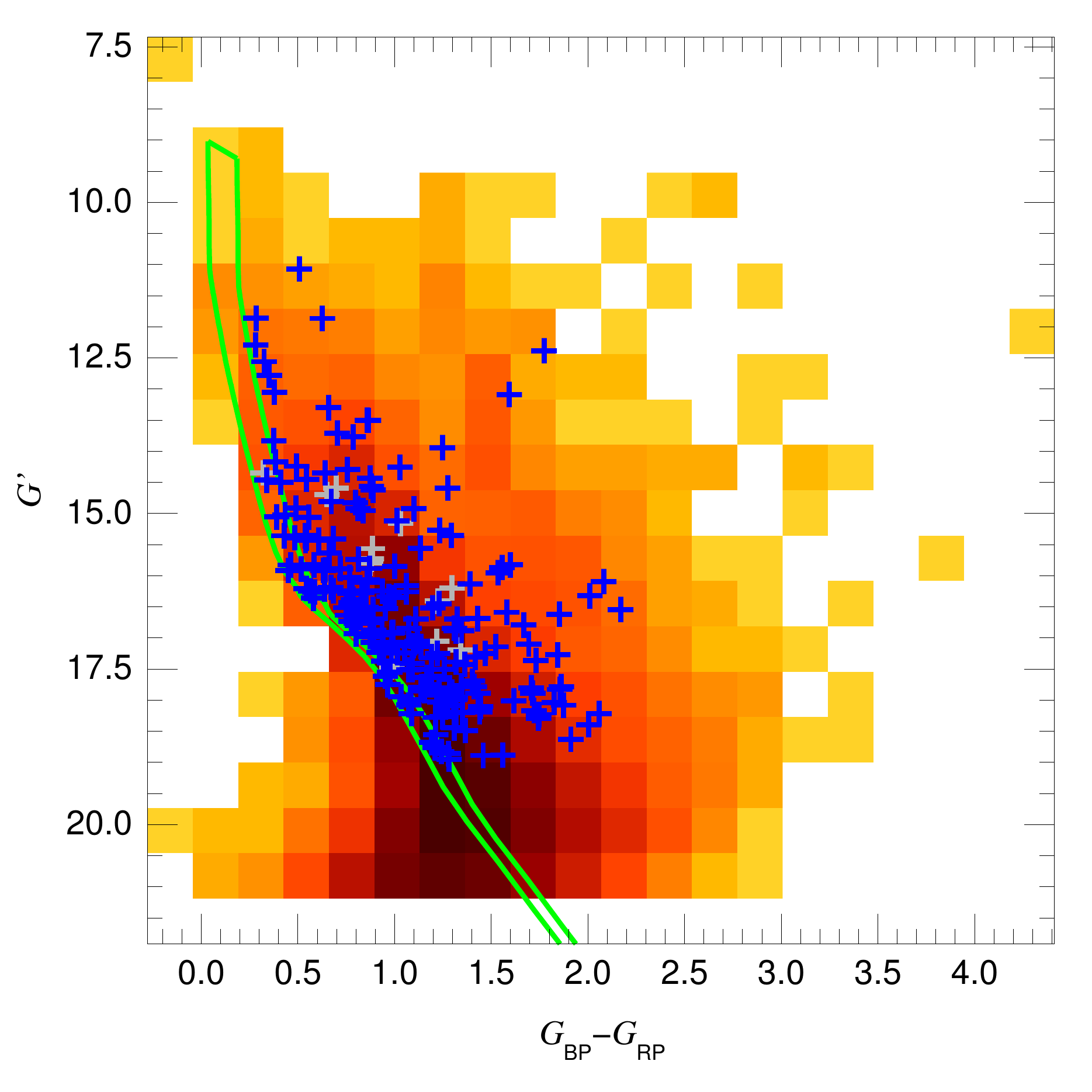} \
            \includegraphics*[width=0.34\linewidth, bb=0 0 538 522]{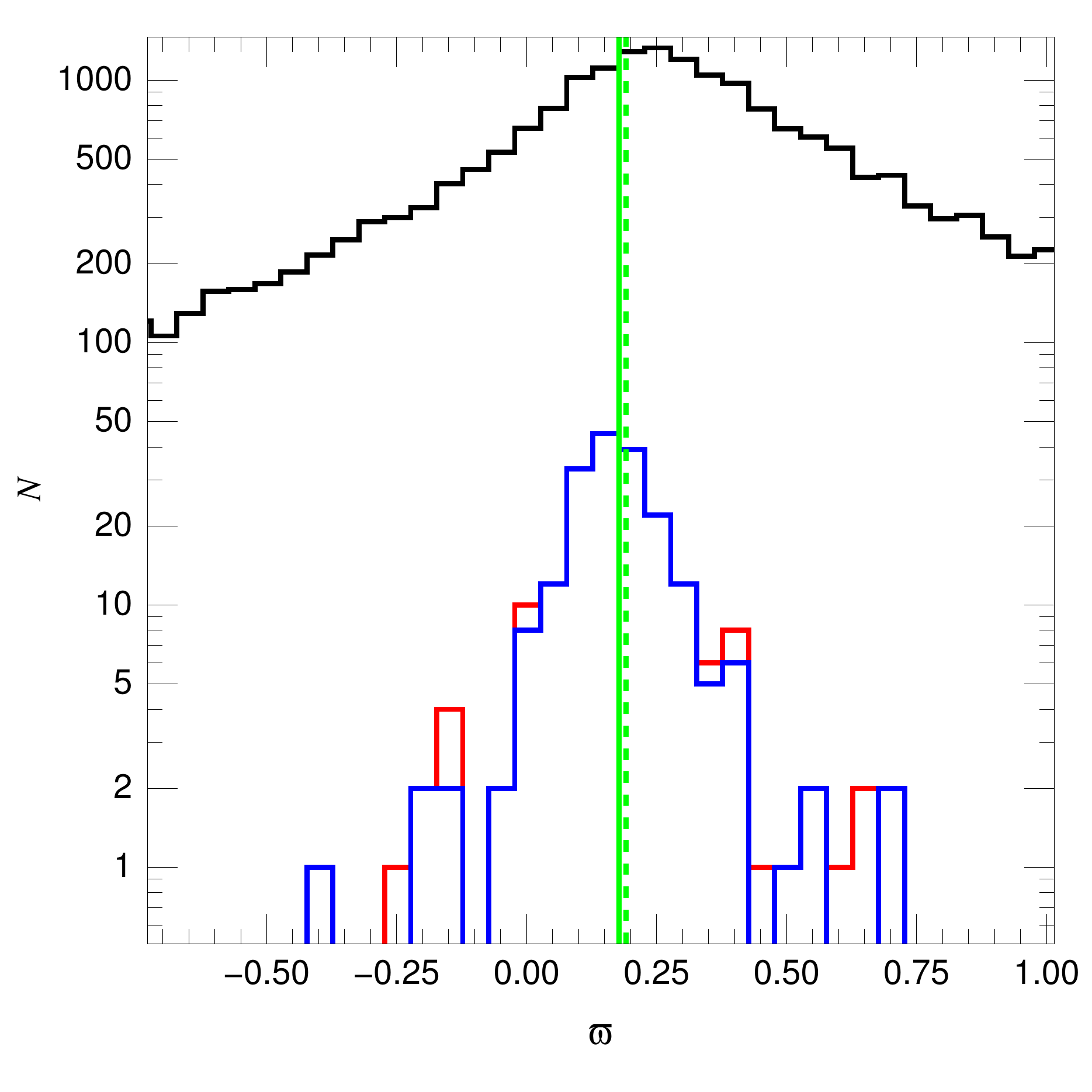} \
            \includegraphics*[width=0.34\linewidth, bb=0 0 538 522]{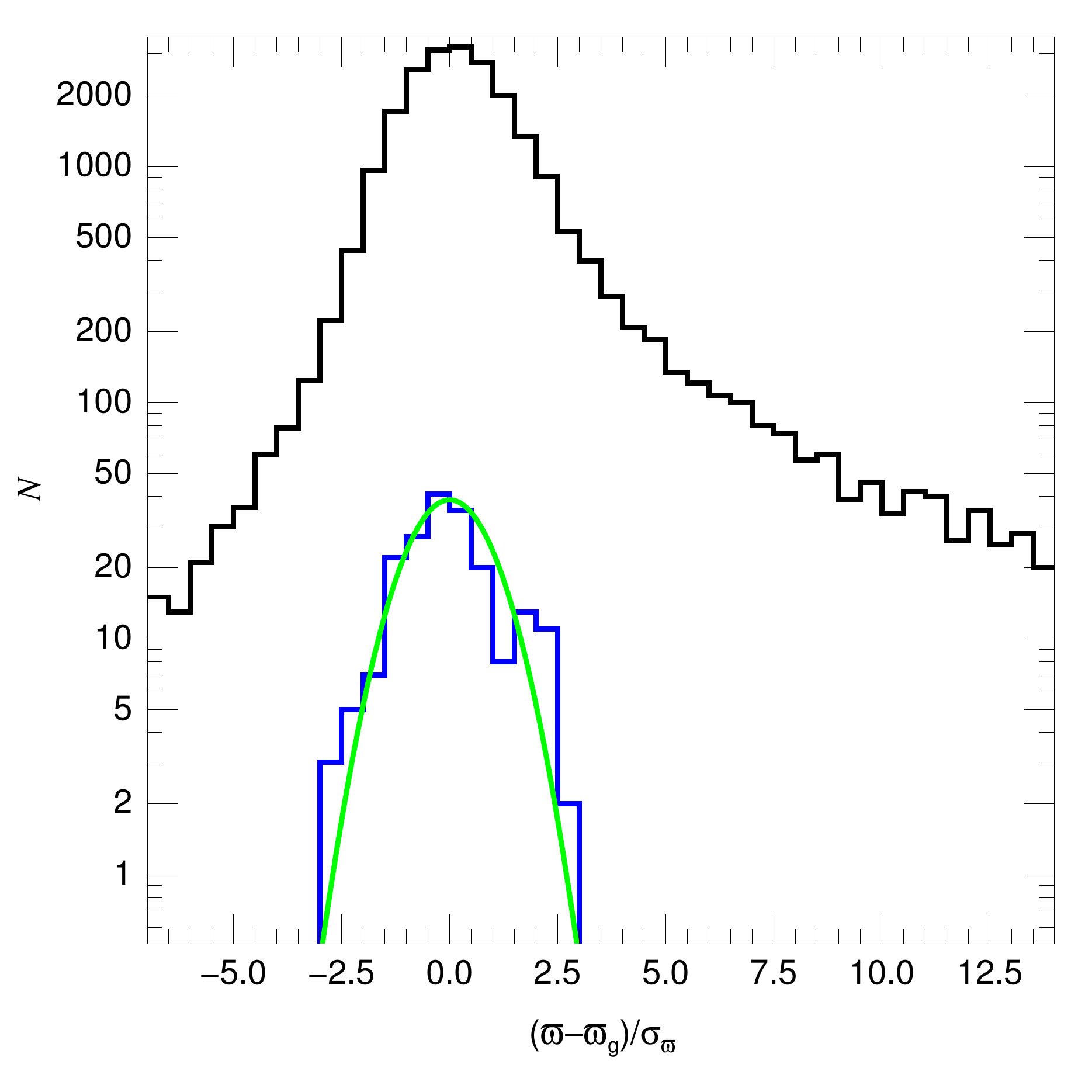}}
\caption{Same as Fig.~\ref{NGC_3603_Gaia} for \VO{012}. The two small green circles in the top panels correspond to the 
         Haffner~18 (\VO{O12}~S) and Haffner~19 (\VO{012}~N) % REFEREE \VO{O12}~N and \VO{012}~S 
         subgroups.}
\label{NGC_2467_Gaia}
\end{figure*}   

\begin{figure*}
\centerline{\includegraphics*[width=0.34\linewidth, bb=0 0 538 522]{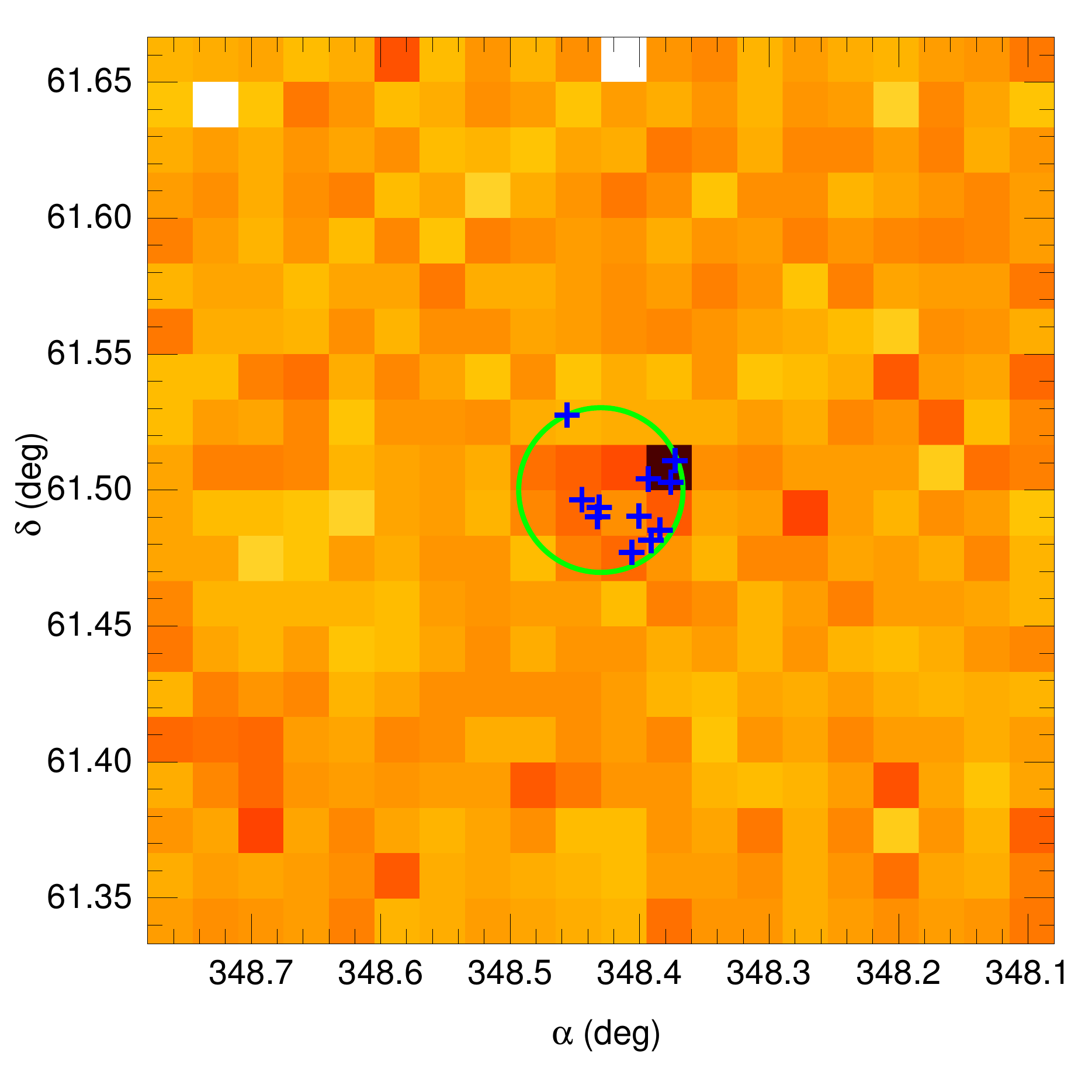} \
            \includegraphics*[width=0.34\linewidth, bb=0 0 538 522]{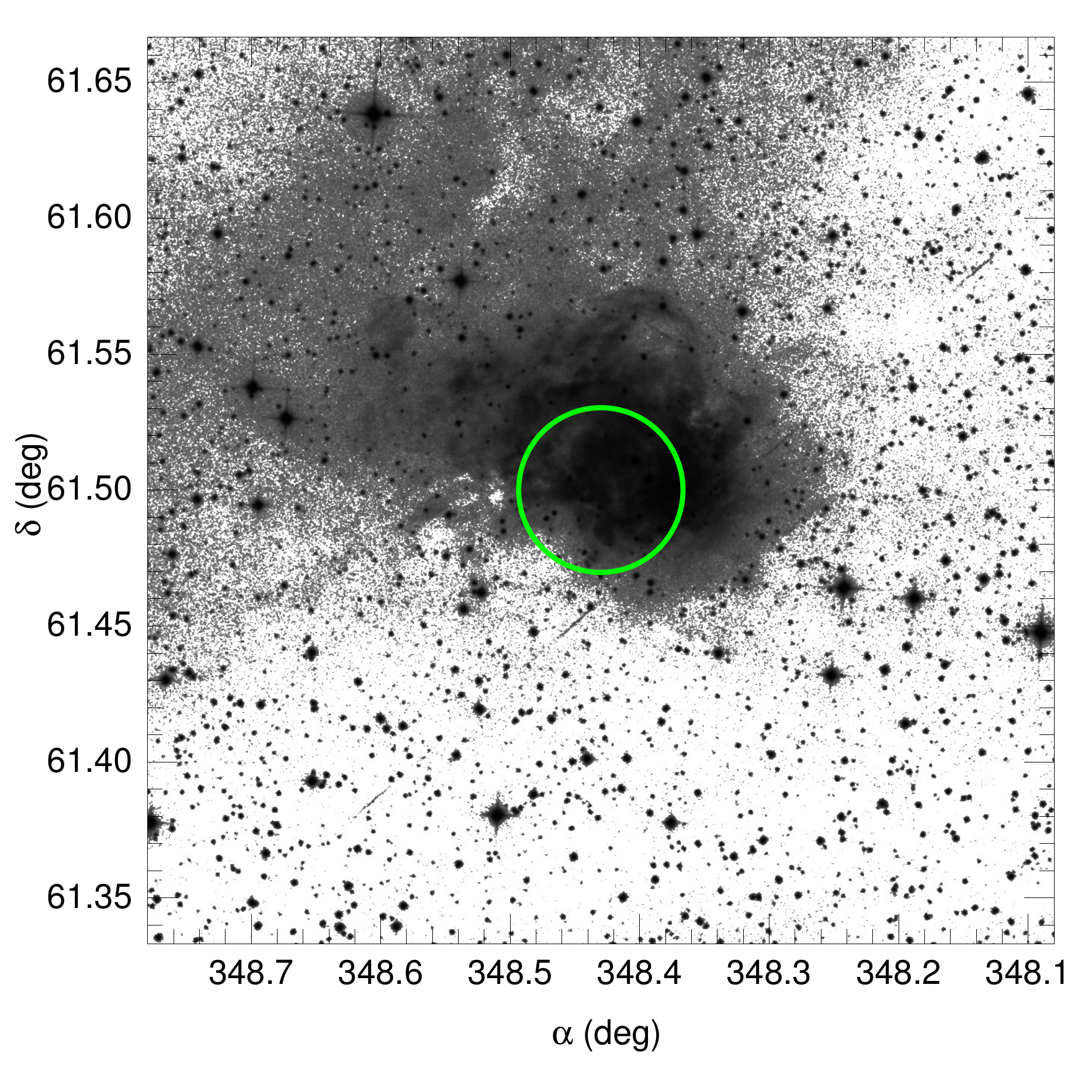} \
            \includegraphics*[width=0.34\linewidth, bb=0 0 538 522]{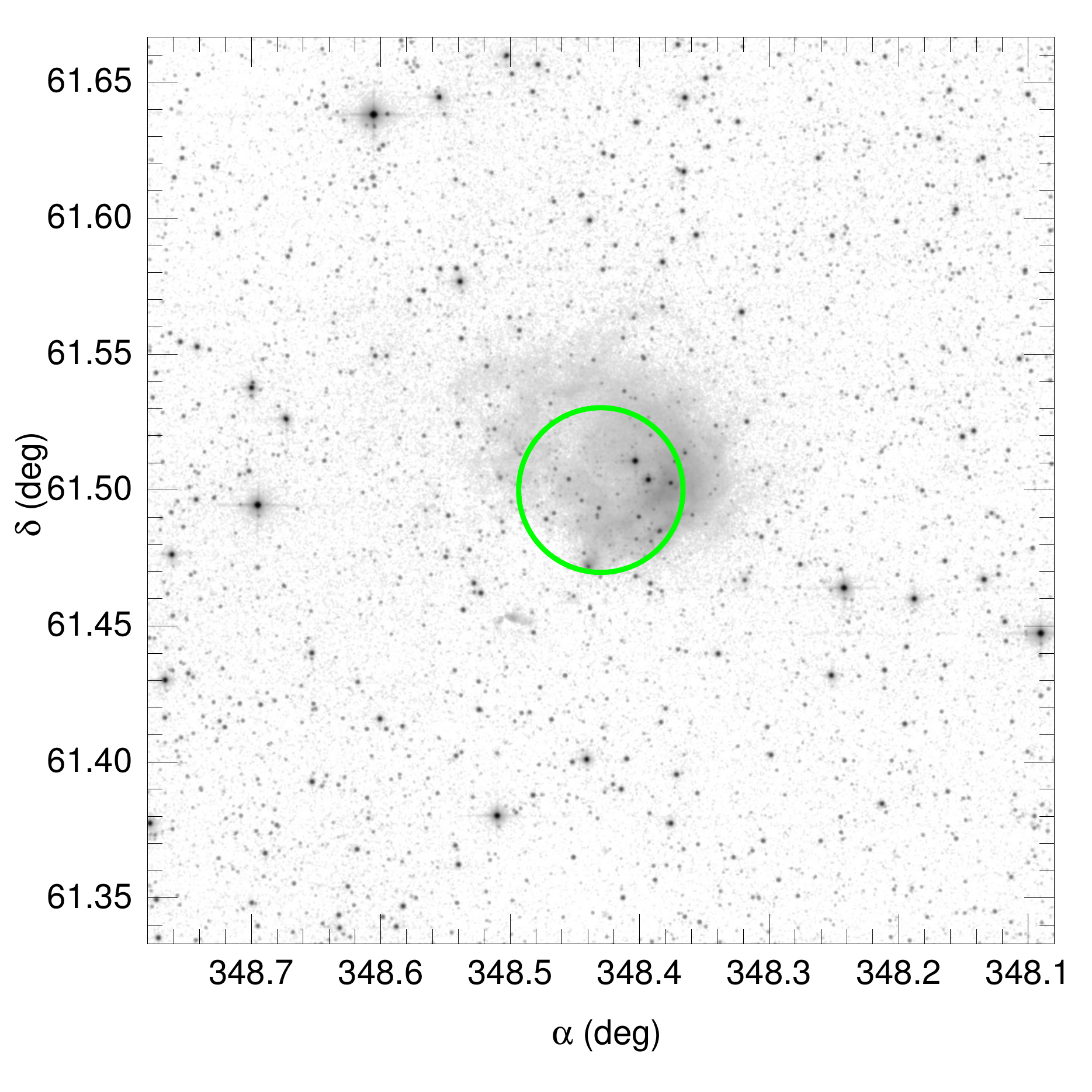}}
\centerline{\includegraphics*[width=0.34\linewidth, bb=0 0 538 522]{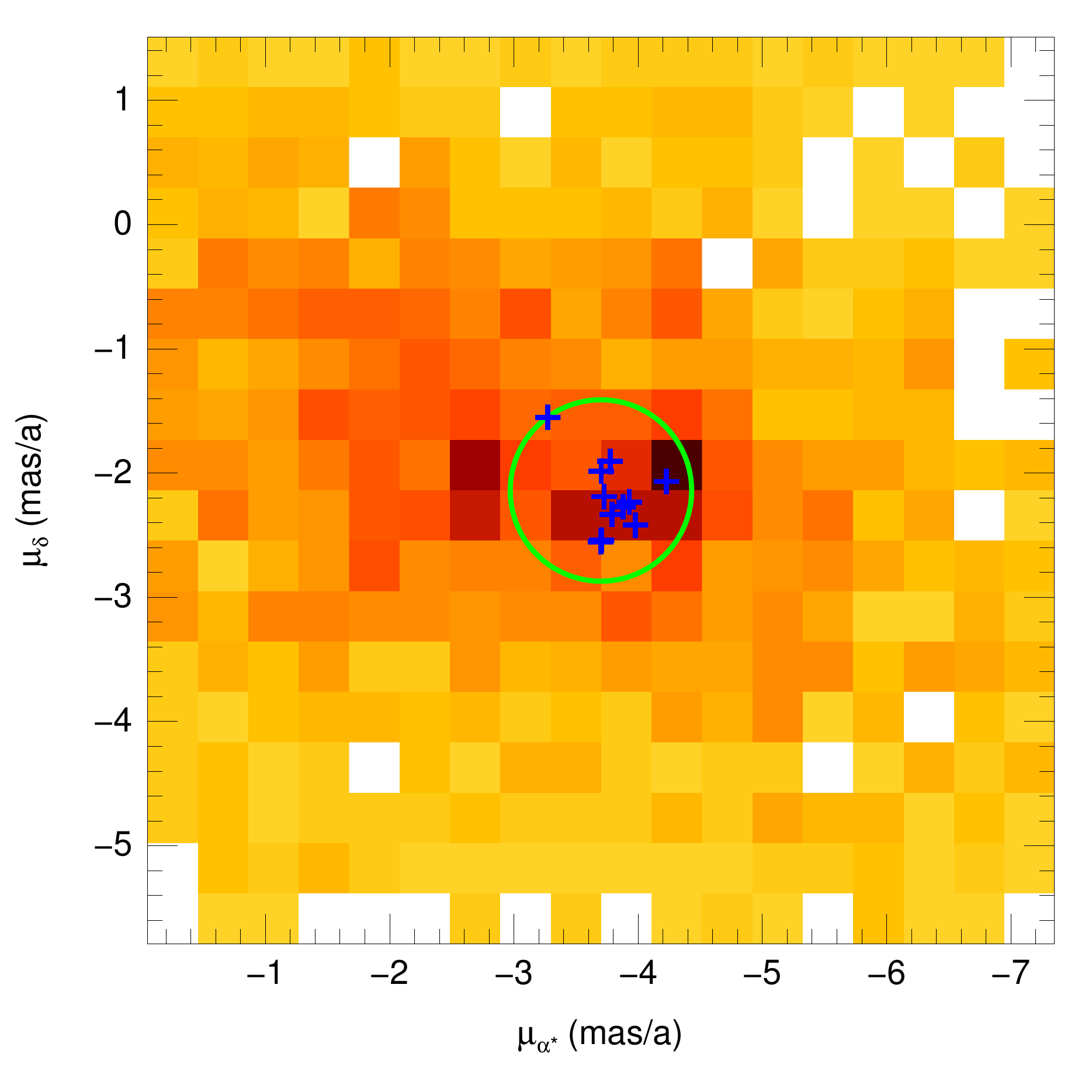} \
            \includegraphics*[width=0.34\linewidth, bb=0 0 538 522]{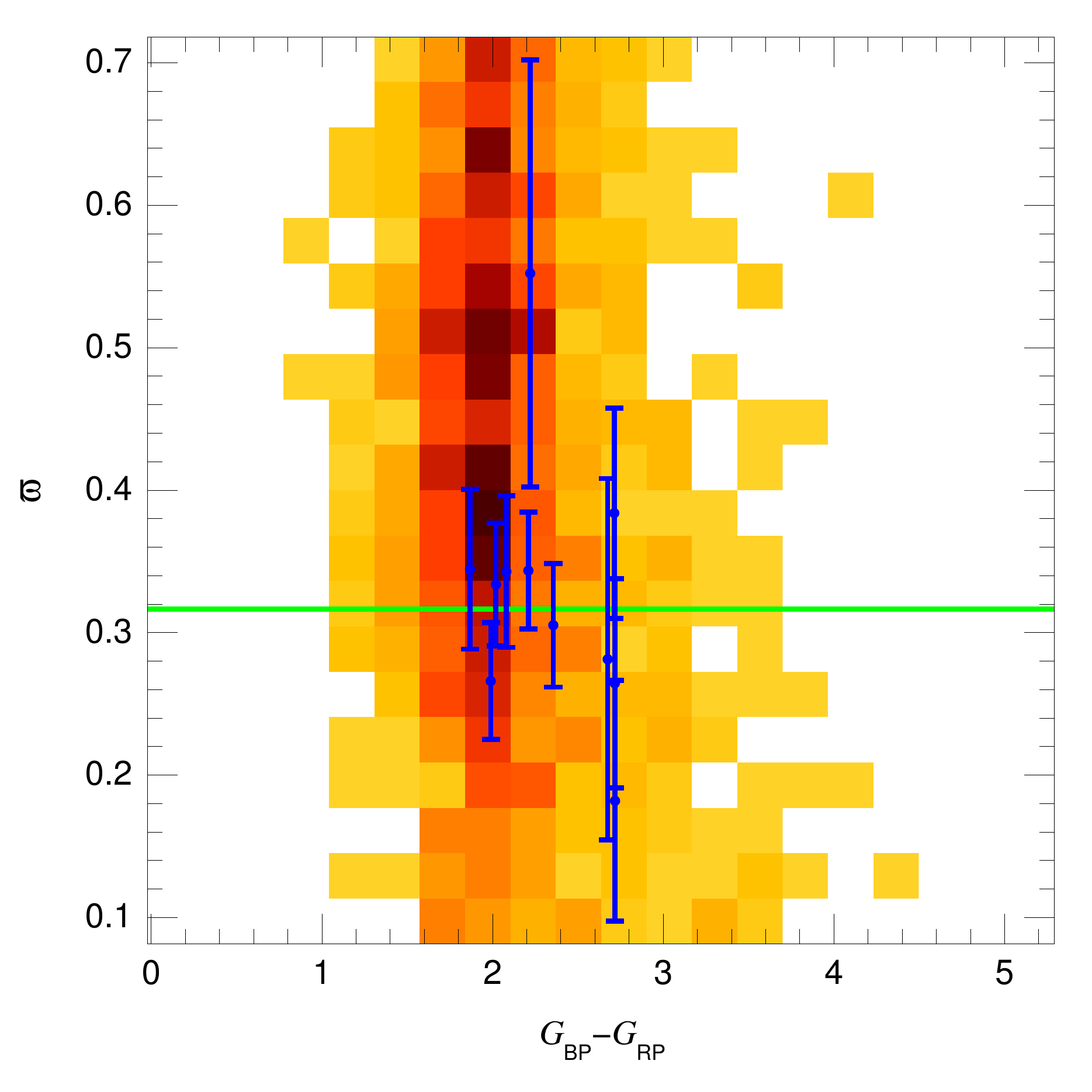} \
            \includegraphics*[width=0.34\linewidth, bb=0 0 538 522]{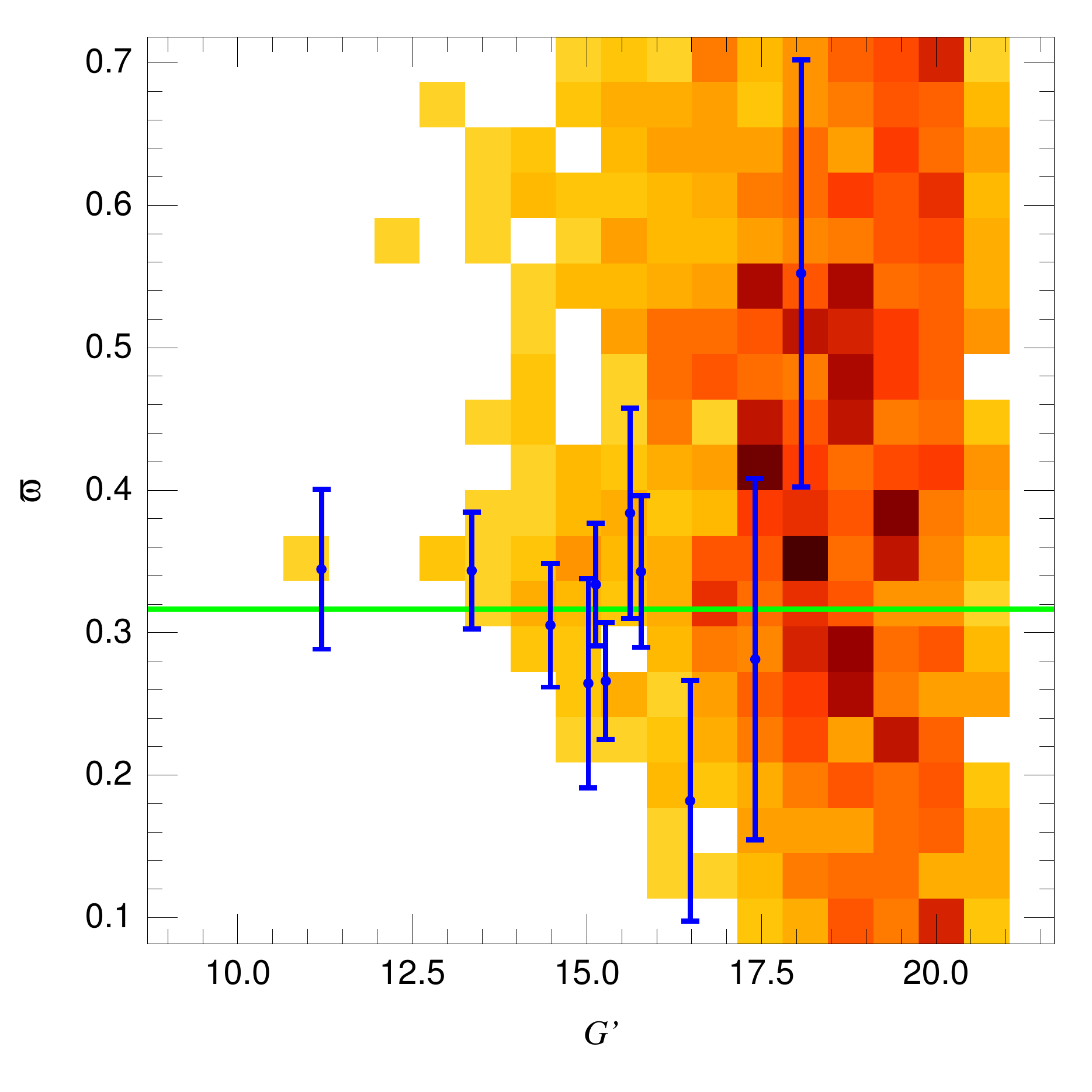}}
\centerline{\includegraphics*[width=0.34\linewidth, bb=0 0 538 522]{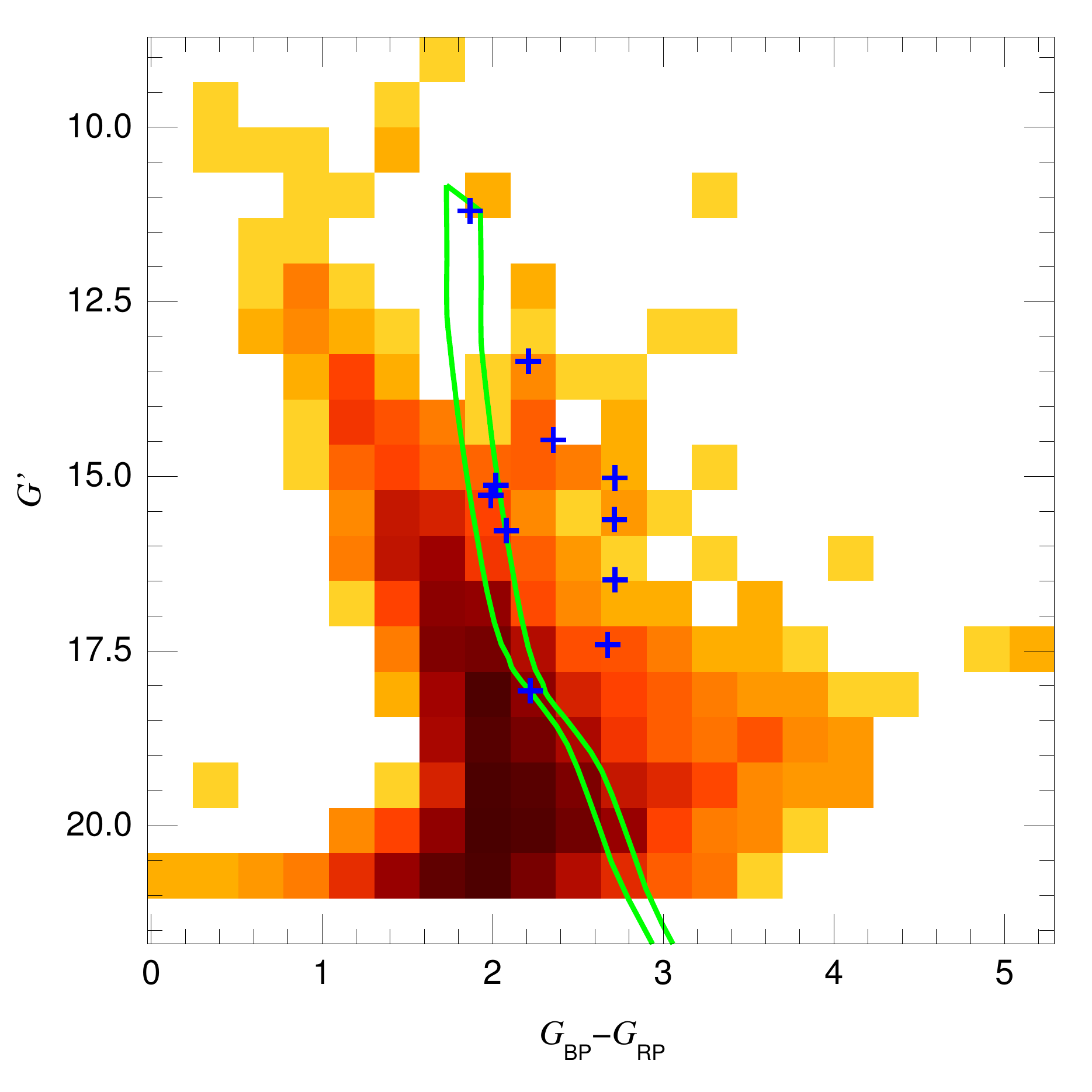} \
            \includegraphics*[width=0.34\linewidth, bb=0 0 538 522]{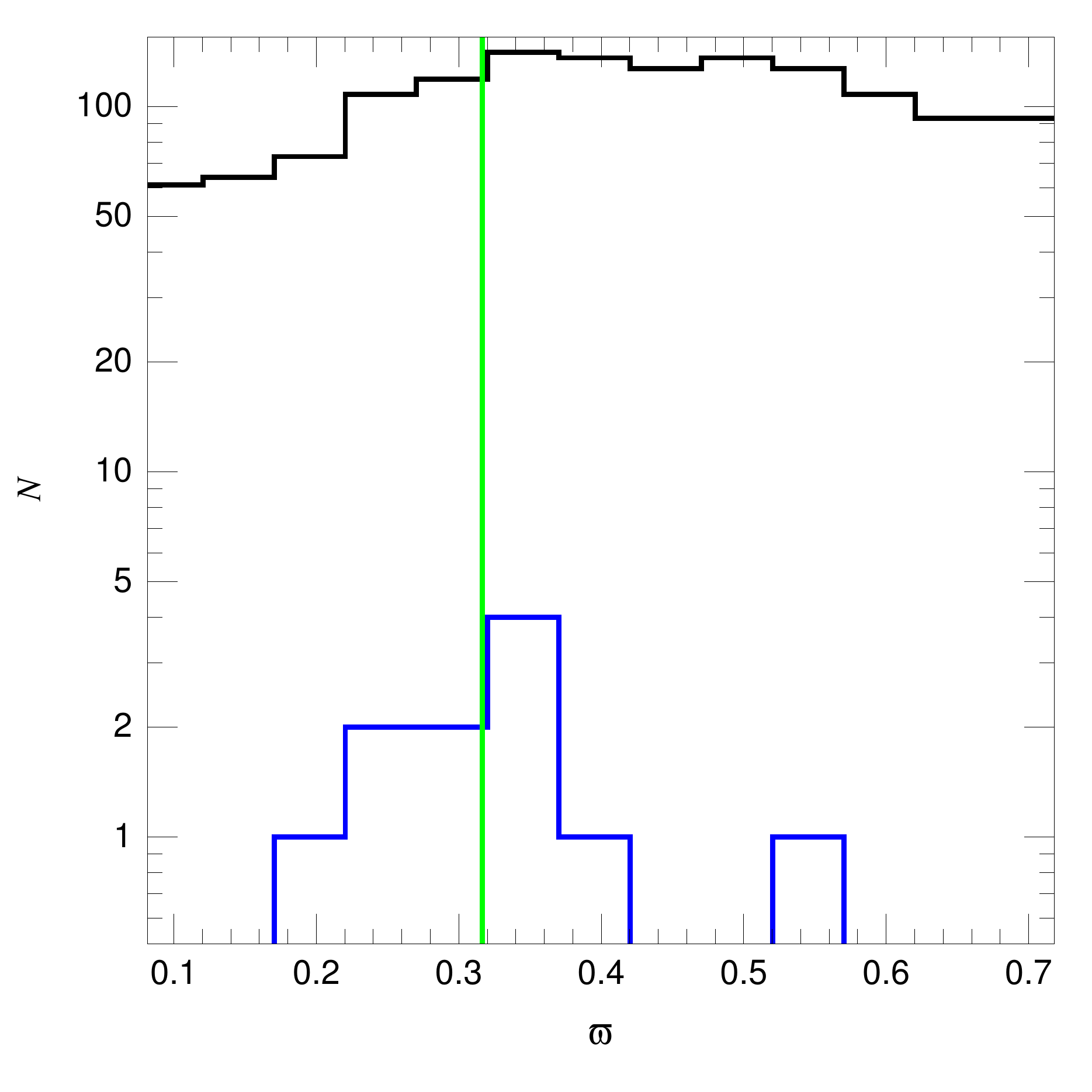} \
            \includegraphics*[width=0.34\linewidth, bb=0 0 538 522]{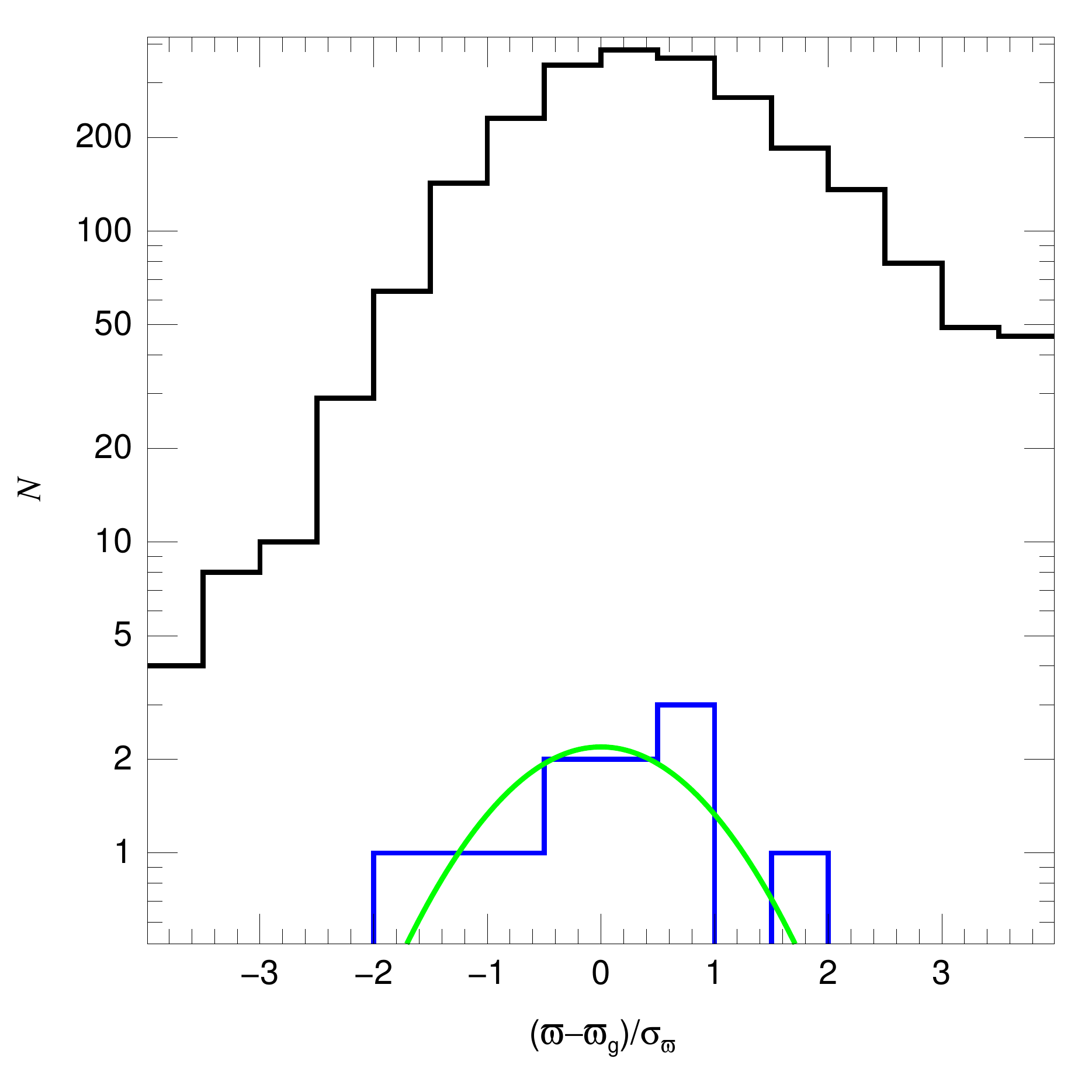}}
 \caption{Same as Fig.~\ref{NGC_3603_Gaia} for \VO{013}.}
\label{Sh_2-158_Gaia}
\end{figure*}   

\begin{figure*}
\centerline{\includegraphics*[width=0.34\linewidth, bb=0 0 538 522]{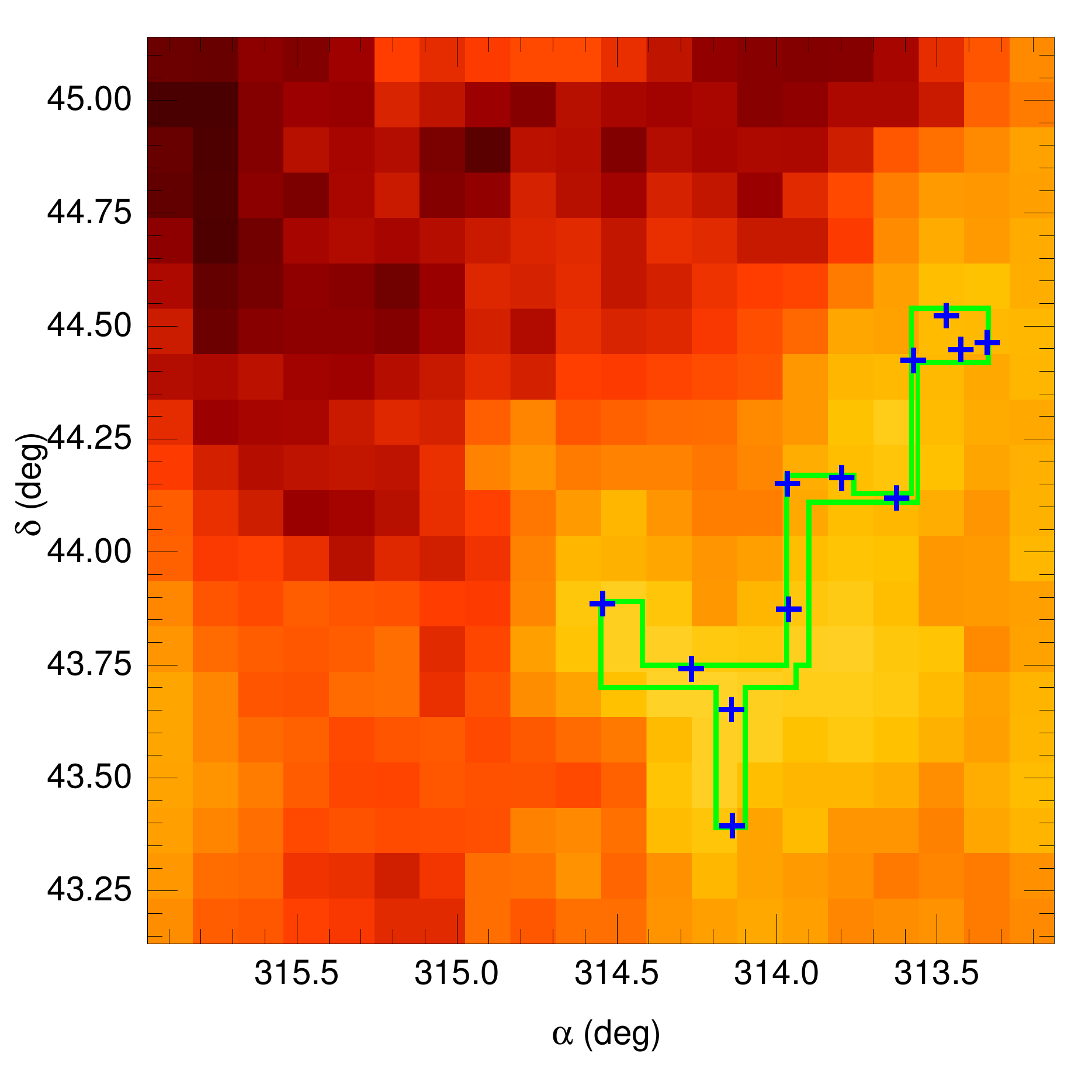} \
            \includegraphics*[width=0.34\linewidth, bb=0 0 538 522]{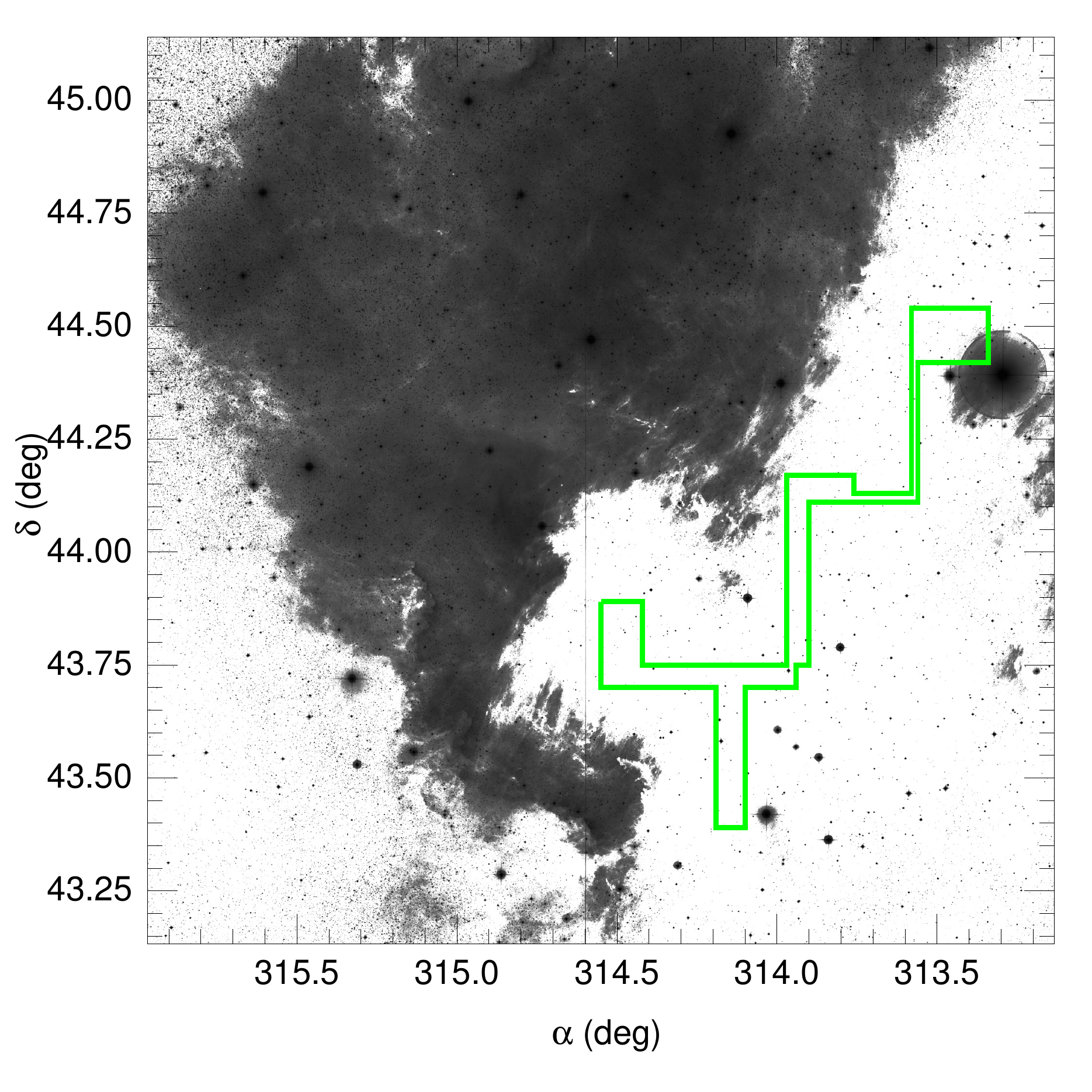} \
            \includegraphics*[width=0.34\linewidth, bb=0 0 538 522]{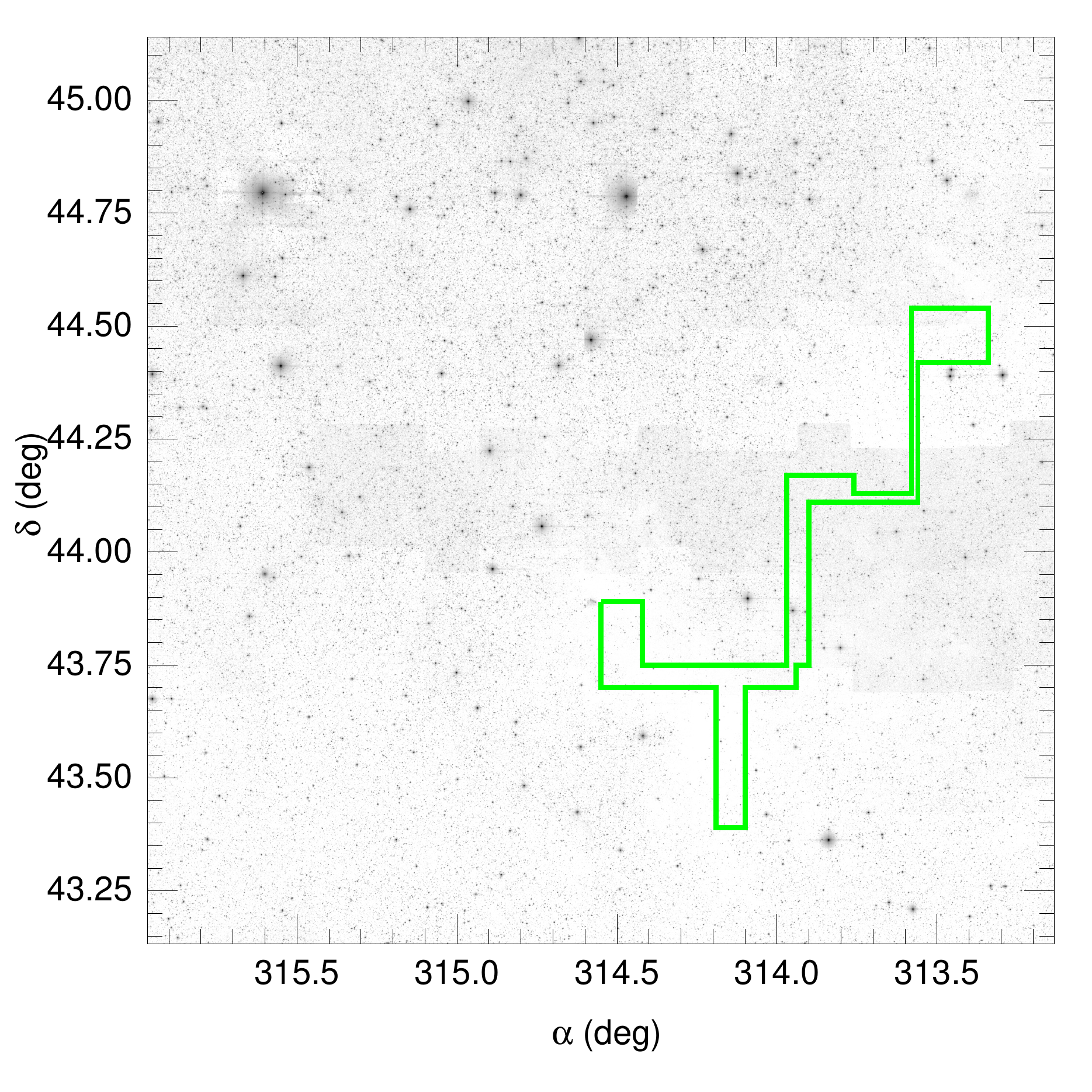}}
\centerline{\includegraphics*[width=0.34\linewidth, bb=0 0 538 522]{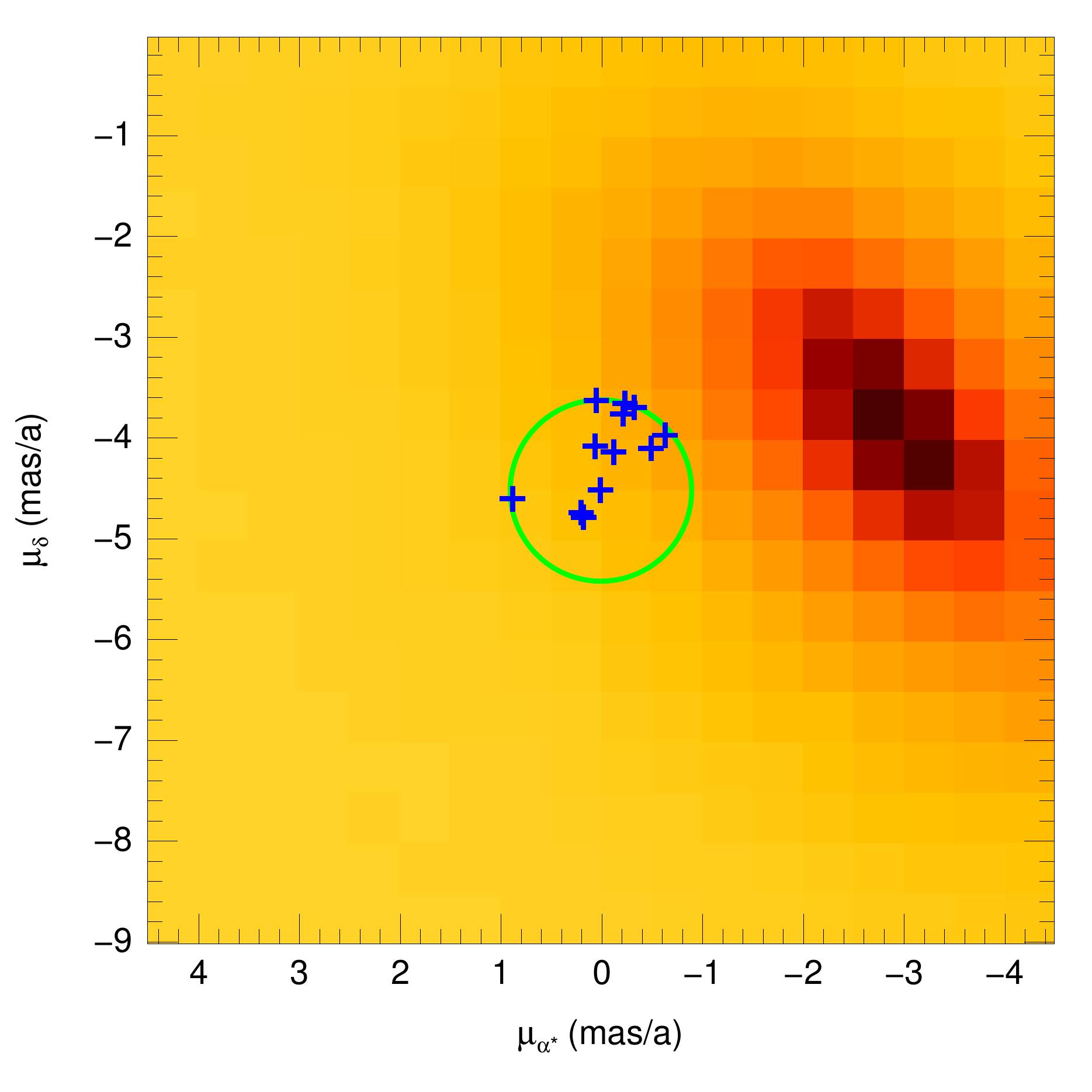} \
            \includegraphics*[width=0.34\linewidth, bb=0 0 538 522]{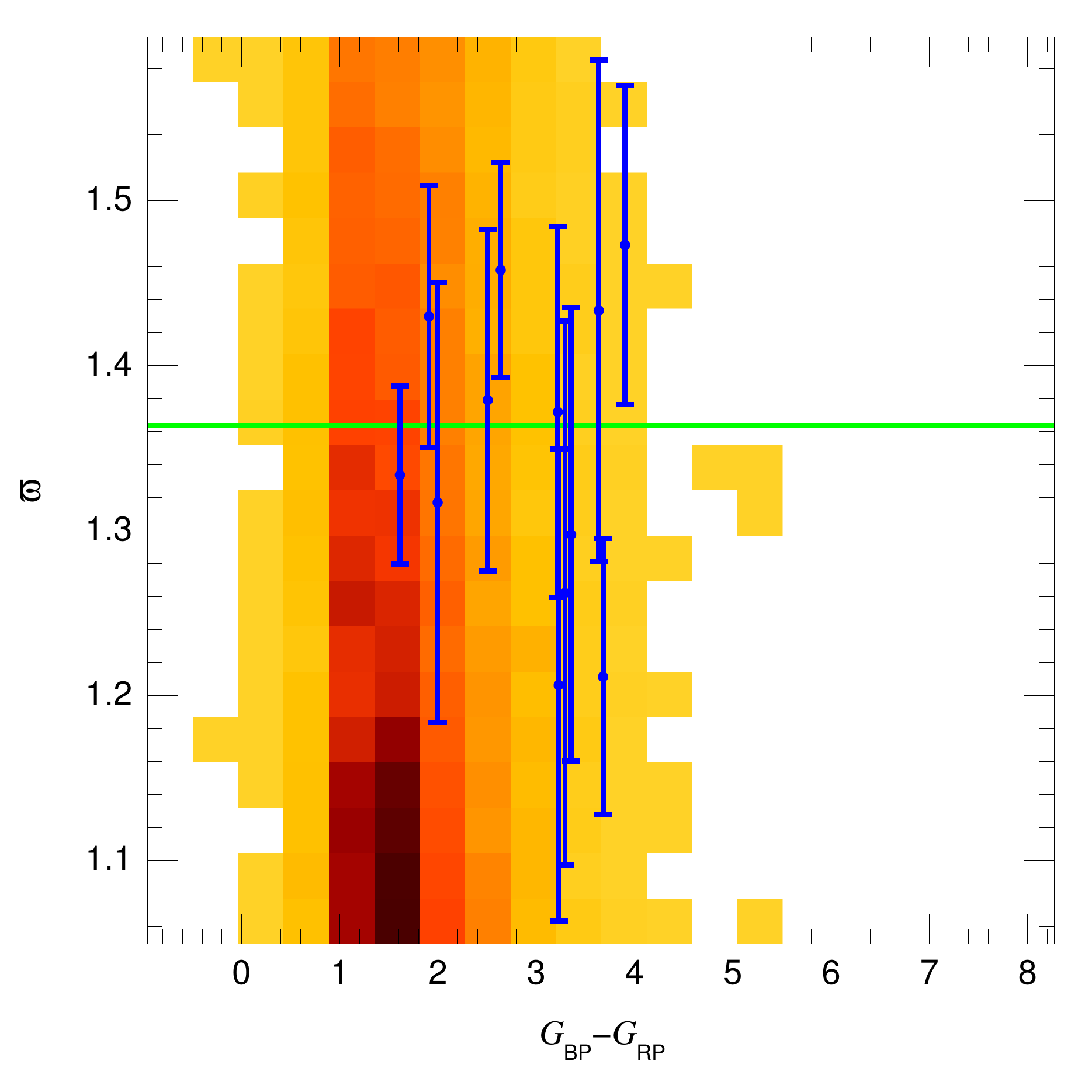} \
            \includegraphics*[width=0.34\linewidth, bb=0 0 538 522]{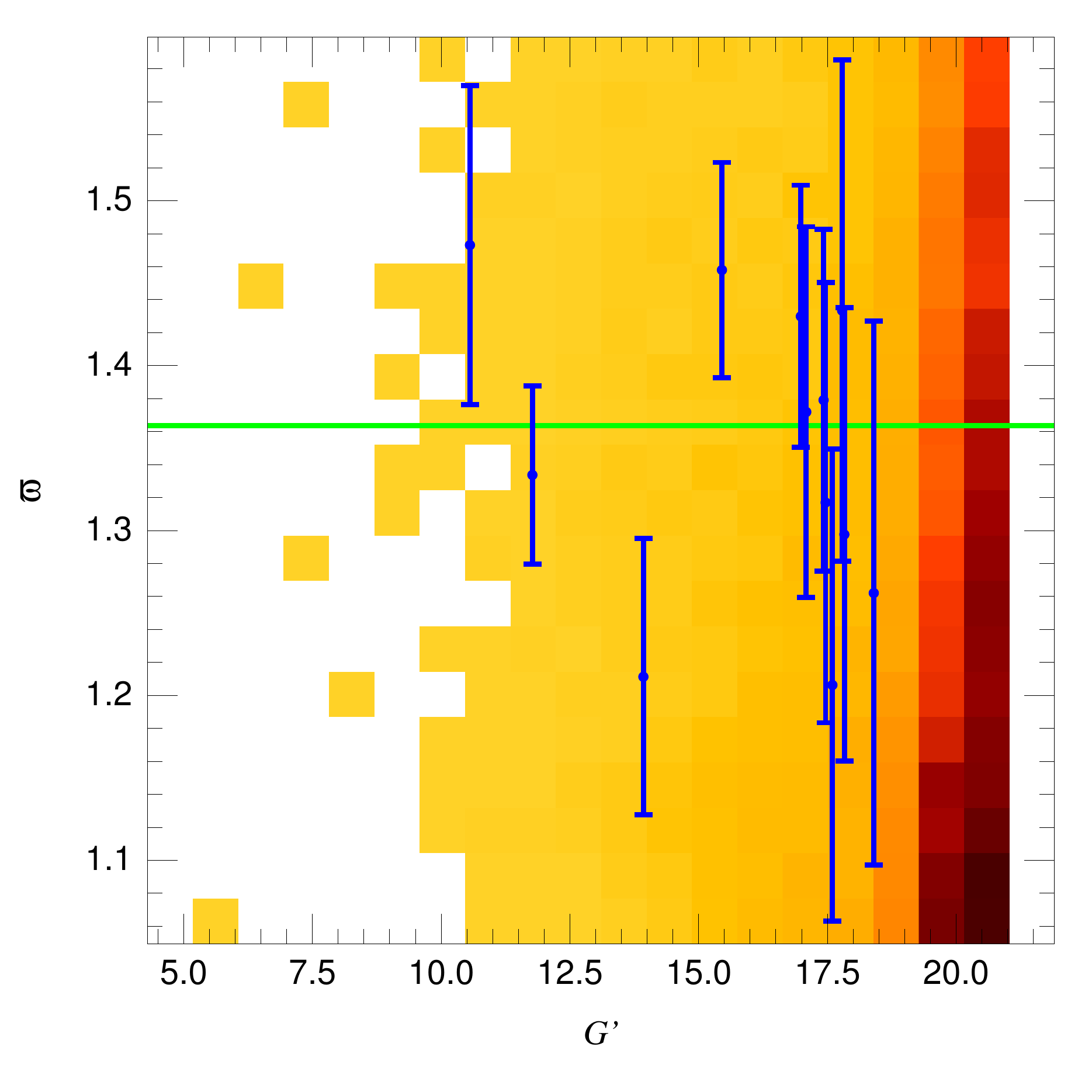}}
\centerline{\includegraphics*[width=0.34\linewidth, bb=0 0 538 522]{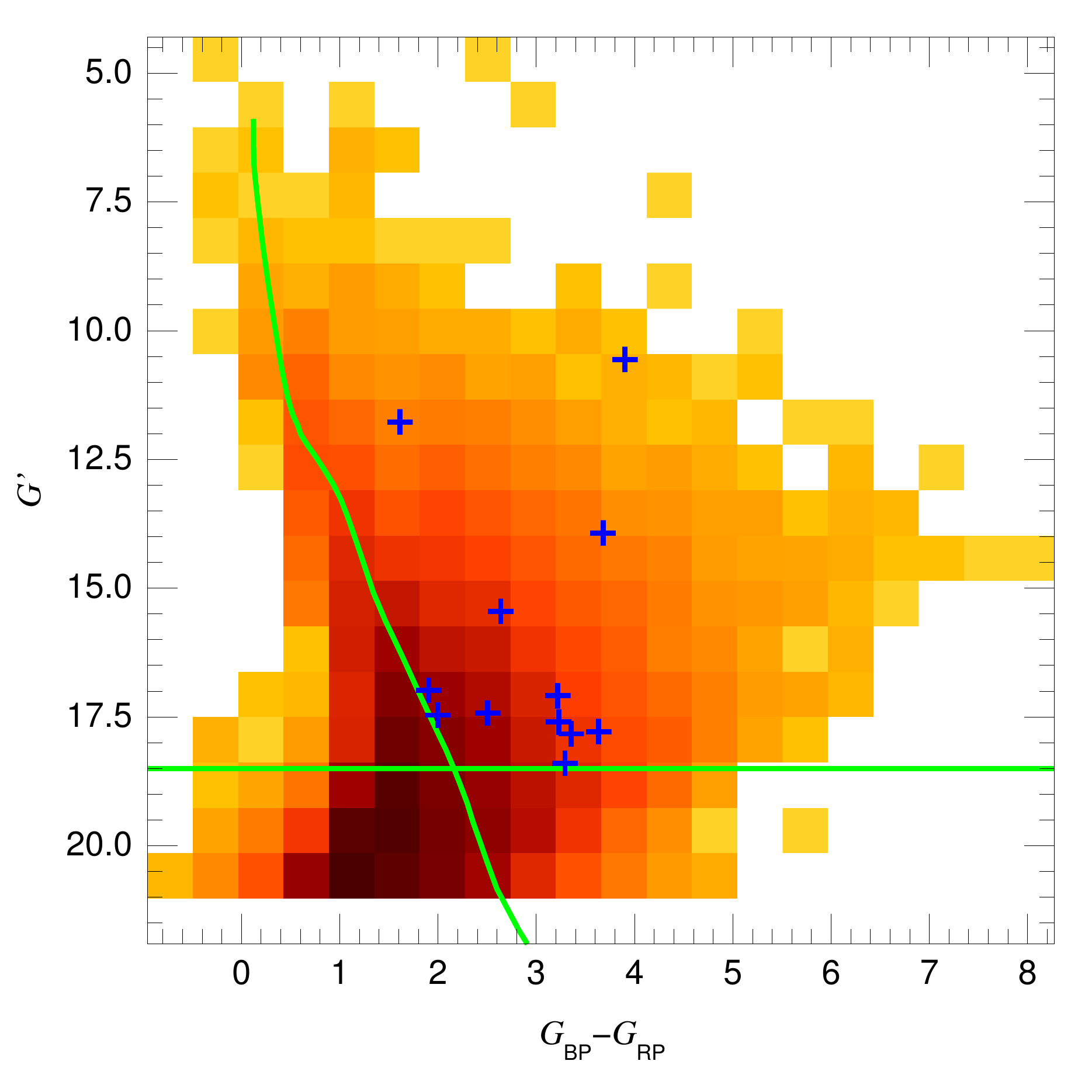} \
            \includegraphics*[width=0.34\linewidth, bb=0 0 538 522]{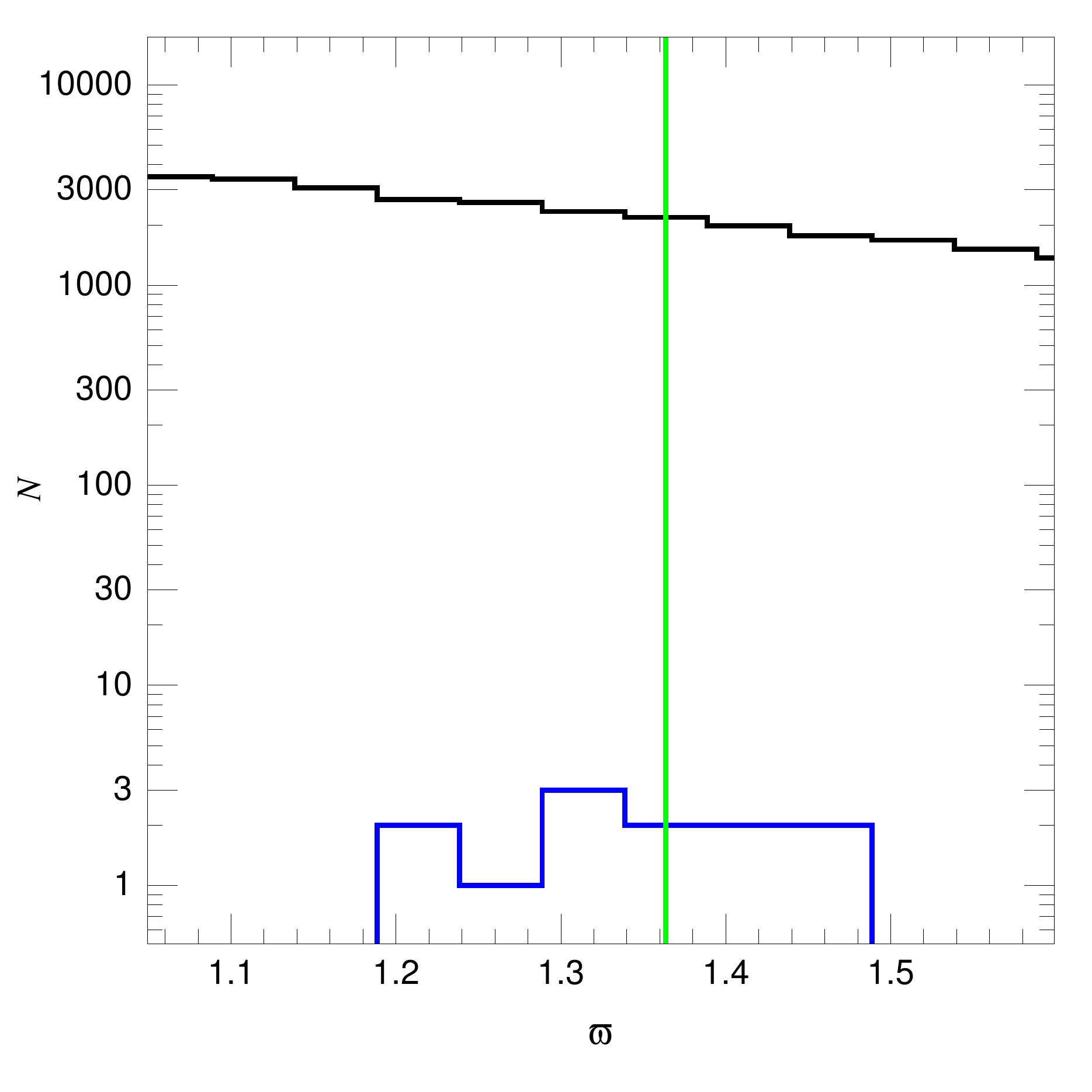} \
            \includegraphics*[width=0.34\linewidth, bb=0 0 538 522]{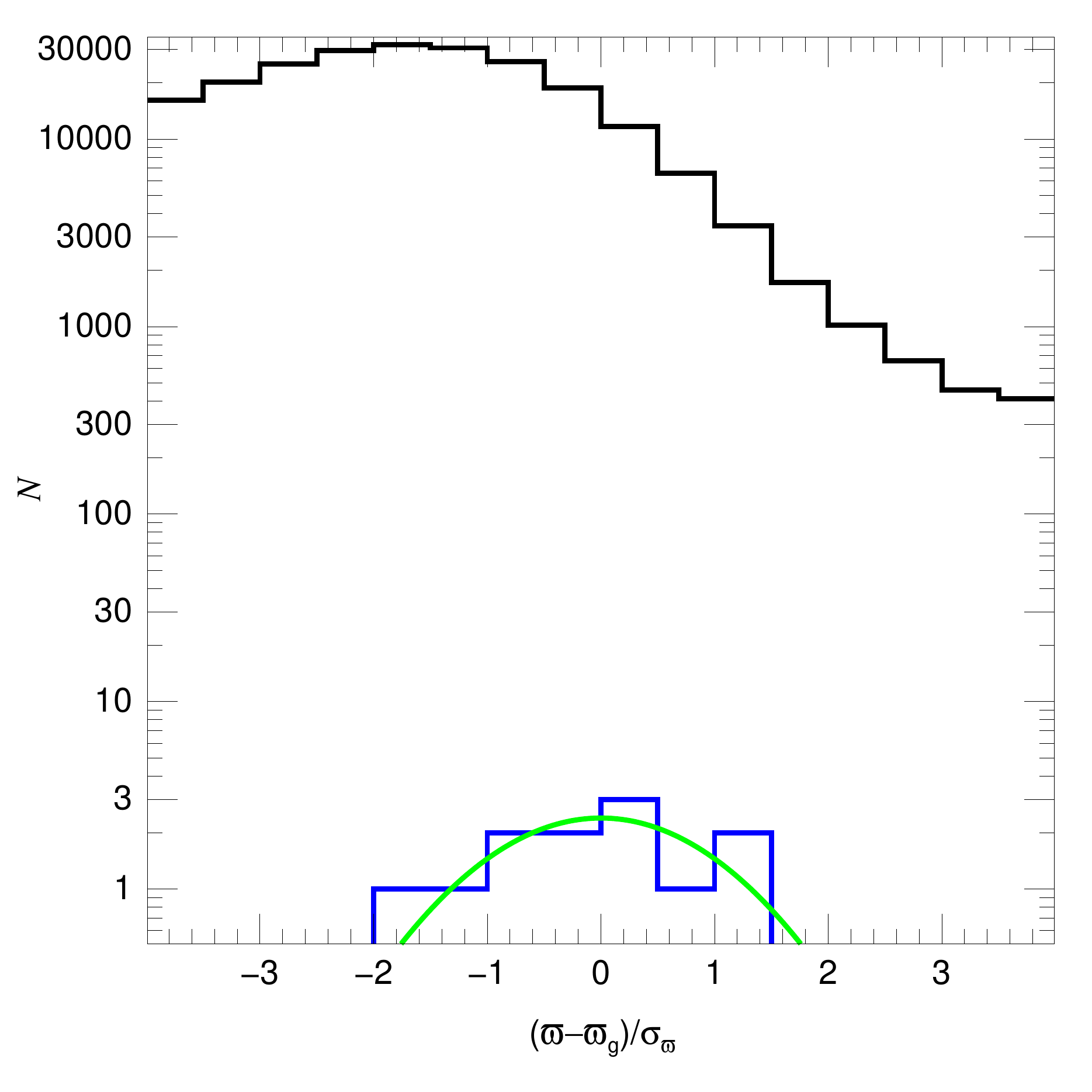}}
\caption{Same as Fig.~\ref{NGC_3603_Gaia} for \VO{014}. The shown polygon is the ad-hoc selection criterion used to select the stars in the
          foreground and vicinity of the Atlantic Ocean + Gulf of Mexico molecular cloud.}
\label{North_America_Gaia}
\end{figure*}   

\begin{table*}
\caption{Possible runaway stars.}
%\centerline{\scriptsize
%\centerline{\small
\centerline{
\begin{tabular}{cllr@{$\pm$}lrr}
\hline
group      & \mci{Gaia DR2 ID}         & \mci{other ID}             & \mcii{$\varpi$}  & \mci{$G$} & \GBP$-$\GRP        \\
ID         &                           &                            & \mcii{(mas)}     &           &                    \\
\hline
O-001      & \num{5337424376393813632} & 2MASS J11142369$-$6106042  &    0.1690&0.0601 &   12.4749 &             2.9503 \\
           & \num{5337422447912267008} & 2MASS J11135811$-$6109421  &    0.1542&0.0641 &   12.8501 &             2.9031 \\
           & \num{5337234538836917248} & 2MASS J11134605$-$6112487  &    0.1899&0.0319 &   13.5301 &             2.2358 \\
           & \num{5337046316178212864} & 2MASS J11134629$-$6116011  &    0.0878&0.0275 &   13.7263 &             1.7215 \\
           & \num{5337043739197624064} & 2MASS J11135770$-$6124267  &    0.0122&0.0407 &   13.9993 &             2.5442 \\
           & \num{5337418088561807872} & 2MASS J11150238$-$6115077  &    0.2157&0.0840 &   16.4556 &\vspace{1mm} 2.0834 \\
O-002      & \num{5350377447993712896} & HDE \num{303313}           &    0.3995&0.0544 &   10.2248 &             0.2503 \\
           & \num{5350400022343131008} & ALS \num{16078}            &    0.3709&0.0526 &   12.4273 &             0.4198 \\
           & \num{5350399811863167744} & 2MASS J10432367$-$5925594  &    0.4049&0.0515 &   12.4363 &\vspace{1mm} 0.4084 \\
O-003      & \num{5350358545816438272} & Trumpler 16-201            &    0.3592&0.0712 &   12.6082 &\vspace{1mm} 2.7149 \\
O-004      & \num{5351703390282380800} & THA 35-II-42               &    0.1819&0.0529 &   11.6903 &             2.2842 \\
           & \num{5255667681036173568} & SS 215                     &    0.1177&0.0542 &   11.8627 &\vspace{1mm} 1.9798 \\
O-005      & \num{5976057078081522048} & Pismis 24-18               &    0.7387&0.2707 &   13.0083 &             2.3063 \\
           & \num{5976155007639115264} & 2MASS J17252943$-$3424044  &    0.6617&0.0645 &   13.5357 &             3.1500 \\
           & \num{5976064057399569408} & 2MASS J17250098$-$3358378  &    0.6234&0.0558 &   14.5181 &             2.7384 \\
           & \num{5976046808810701056} & 2MASS J17252444$-$3358157  &    0.5355&0.0415 &   14.5611 &             1.9811 \\
           & \num{5976154762809358848} & 2MASS J17235197$-$3412298  &    0.6627&0.0441 &   14.5998 &             2.3270 \\
           & \num{5976033339793078272} & 2MASS J17252376$-$3359548  &    0.7444&0.1110 &   15.0023 &             3.2382 \\
           & \num{5976153908127437568} & 2MASS J17251438$-$3358077  &    0.6100&0.0513 &   15.2246 &             2.0040 \\
           & \num{5976054093074786048} & 2MASS J17254496$-$3406488  &    0.5698&0.0878 &   15.2442 &\vspace{1mm} 2.8499 \\
O-006      & \num{5337977980510724480} & 2MASS J10584671$-$6105512  &    0.0729&0.0556 &   12.1401 &             2.1680 \\
           & \num{5337971688334583680} & 2MASS J10592362$-$6114575  &    0.1982&0.0530 &   12.6812 &\vspace{1mm} 2.0841 \\
O-007      & \num{2067783799613328128} & Cyg OB2-24                 &    0.5413&0.0616 &   10.9932 &\vspace{1mm} 2.1264 \\
O-008      & \num{2067926564325711872} & [MT91] 453                 &    0.6022&0.0309 &   13.7738 &\vspace{1mm} 1.7636 \\
O-009      & \num{4098005975631742720} & BD $-$16 4826              &    0.5681&0.0636 &    9.4999 &             1.2675 \\
           & \num{4097817409389450624} & ALS 4943                   &    0.6086&0.0643 &   10.4758 &             0.9175 \\
           & \num{4097816275518137216} & 2MASS J18210951$-$1608186  &    0.7550&0.0665 &   11.2572 &             1.5734 \\
           & \num{4097803596774515584} & Tyc 6265-01474-1           &    0.6954&0.0679 &   11.5804 &             0.6549 \\
           & \num{4097808024877708160} & NGC 6618 B-373             &    0.4797&0.0917 &   11.8704 &             3.0881 \\
           & \num{4098007895470142080} & 2MASS J18200299$-$1602068  &    0.5604&0.0616 &   11.9439 &             2.5256 \\
           & \num{4097817924785524864} & 2MASS J18211029$-$1603505  &    0.6448&0.0384 &   13.8902 &\vspace{1mm} 1.7953 \\
O-010      & \num{5940956341923632896} & 2MASS J16403254$-$4846296  &    0.7897&0.0348 &   13.2440 &\vspace{1mm} 1.1712 \\
O-011      & \num{2071525987444459904} & 2MASS J20354794$+$4655566  &    0.3251&0.0190 &   13.2255 &             1.2004 \\
           & \num{2071522242233085824} & 2MASS J20351422$+$4650118  &    0.2333&0.0319 &   13.5376 &             1.8256 \\
           & \num{2071530041893546240} & 2MASS J20351898$+$4659543  &    0.2994&0.0196 &   13.8215 &             0.9733 \\
           & \num{2071516882113611776} & 2MASS J20351160$+$4642344  &    0.2999&0.0283 &   14.0482 &             2.4964 \\
           & \num{2071330136937555328} & 2MASS J20354818$+$4643174  &    0.3770&0.0265 &   14.3933 &\vspace{1mm} 2.2597 \\
O-012      & \num{5602025904044961536} & HD \num{64315} AB$\dagger$ & $-$0.0539&0.0892 &    9.1365 &             0.4058 \\
           & \num{5602033390154015744} & HD \num{64568}             &    0.1367&0.0599 &    9.3061 &             0.1610 \\
           & \num{5602048542798630784} & CPD $-$25 5194             &    0.2484&0.0994 &   11.0859 &             0.1442 \\
           & \num{5601982473333181184} & 2MASS J07532352$-$2626112  &    0.2533&0.0605 &   11.3573 &             2.3334 \\
           & \num{5602234016670302848} & Tyc 6557-03393-1           &    0.2250&0.0544 &   12.4050 &             0.1545 \\
           & \num{5602002741267436288} & 2MASS J07513825$-$2628576  &    0.3530&0.0617 &   12.7415 &             1.4881 \\
           & \num{5602033291388495616} & 2MASS J07533255$-$2614513  &    0.1431&0.0588 &   12.8183 &\vspace{1mm} 2.0944 \\
O-013      & \num{2014962779980826240} & 2MASS J23141320$+$6138397  &    0.3382&0.0499 &   13.1733 &             3.1796 \\
           & \num{2014867191189025920} & 2MASS J23122869$+$6127057  &    0.4154&0.0419 &   14.5214 &             2.1011 \\
           & \num{2014866808933891584} & [MO2001] 77                &    0.3424&0.0343 &   14.7460 &             1.5634 \\
           & \num{2014961852267620608} & 2MASS J23133905$+$6133222  &    0.3191&0.0379 &   15.1941 &             1.5426 \\
           & \num{2014915569700615296} & 2MASS J23122004$+$6132246  &    0.3595&0.0417 &   15.4022 &\vspace{1mm} 1.1392 \\
O-014      & \num{2162214599862429056} & V354 Cyg                   &    1.2783&0.1033 &    7.7724 &             4.1208 \\
           & \num{2162063898039698816} & Tyc~3179-00416-1           &    1.3422&0.0581 &    9.4345 &             1.9894 \\
           & \num{2162048058200239616} & Tyc~3179-00023-1           &    1.4521&0.0541 &   10.0277 &             0.3824 \\
           & \num{2162872004724035456} & [SKV93] 2-72               &    1.3890&0.0521 &   11.7749 &\vspace{1mm} 1.5588 \\
O-015      & \num{2068356950109724288} & 2MASS J20191747$+$4057512  &    0.9501&0.0528 &   12.2638 &             2.0731 \\
           & \num{2068352659441818368} & 2MASS J20192974$+$4051474  &    0.9808&0.0520 &   12.4495 &             0.8636 \\
           & \num{2068362791265246464} & 2MASS J20192209$+$4058234  &    0.9418&0.0540 &   12.5043 &             2.1756 \\
           & \num{2068369633152677760} & 2MASS J20184783$+$4101193  &    0.8784&0.0293 &   13.5034 &             1.2614 \\
           & \num{2068357194927339136} & 2MASS J20191645$+$4059011  &    0.9434&0.0288 &   13.6283 &\vspace{1mm} 1.5131 \\
O-016      & \num{3326734332924414976} & 2MASS J06423798$+$0958062  &    1.2625&0.0693 &   12.9885 &             0.7344 \\
           & \num{3326951215889632128} & 2MASS J06395491$+$1010070  &    1.3160&0.0334 &   13.1739 &             0.9582 \\
\hline
$\dagger$: & \multicolumn{6}{l}{Target with bad RUWE.} \\
\end{tabular}
}
\label{runaways}                 
\end{table*}

\end{appendix}

\end{document}